\pdfoutput=1
\documentclass [11pt]{article}

\usepackage{jheppub}

\usepackage[utf8]{inputenc}

\usepackage{amsfonts}

\usepackage{amsmath,amssymb}

\usepackage{mathtools}

\usepackage{graphicx,subfigure}
\usepackage[space]{grffile}

\usepackage{simplewick}

\usepackage{array}

\usepackage{float}

\usepackage{xcolor}

\usepackage{mathrsfs}

\usepackage{titlesec}

\usepackage{slashed}
\def\nn{\nonumber}
\usepackage{tikz}
\usetikzlibrary{calc,snakes, patterns,angles,quotes,arrows.meta,shapes.misc,decorations.pathmorphing,decorations.markings}
\tikzset{
  branch point/.style={cross out,draw=black,fill=none,minimum size=(#1-\pgflinewidth),inner sep=0pt,outer sep=0pt}, 
  branch point/.default=5
}

\newcommand{\cO}{\mathcal{O}}

\newcommand{\<}{\langle}
\renewcommand{\>}{\rangle}

\newcommand{\mat}[1]{\mathtt{#1}}

\newcommand{\parity}{{\mathcal{P}}}
\renewcommand{\time}{{\mathcal{T}}}

\newcommand{\spin}{{\ell}}
\newcommand{\hel}{{\lambda}}
\newcommand{\myRho}{{\mathfrak{r}}}

\newcommand{\myP}{\boldsymbol{p}}

\newcommand{\aux}{h}

\titleformat{\paragraph}
{\normalfont\normalsize\bfseries}{\theparagraph}{1em}{}
\titlespacing*{\paragraph}
{0pt}{3.25ex plus 1ex minus .2ex}{1.5ex plus .2ex}

\title{\boldmath Spinning S-matrix Bootstrap in 4d}

\author[a]{Aditya Hebbar,}
\author[a,b]{Denis Karateev,}
\author[a]{and Jo\~ao Penedones}

\affiliation[a]{Fields and Strings Laboratory, Institute of Physics\\ École Polytechnique Fédéral de Lausanne (EPFL)
\\ Route de la Sorge, CH-1015 Lausanne, Switzerland}

\affiliation[b]{
Philippe Meyer Institute, Physics Department\\
\'Ecole Normale Sup\'erieure (ENS), Universit\'e PSL\\
 24 rue Lhomond, F-75231 Paris, France}

\abstract{
We review unitarity and crossing constraints on scattering amplitudes for particles with spin in four dimensional quantum field theories. As an application we study two to two scattering of neutral spin 1/2 fermions in detail. Assuming Mandelstam analyticity of its scattering amplitude, we use the numerical S-matrix bootstrap method to estimate various non-perturbative  bounds on quartic and cubic (Yukawa) couplings. 
}

\dedicated{Dedicated to the memory of Andr\'e Martin}

\begin{document}

\maketitle
\section{Introduction}
\label{sec:intro}

The space of Quantum Field Theories (QFT) is vast and uncharted.
The numerical S-matrix Bootstrap is a nonperturbative approach to explore this space  \cite{Paulos:2016fap, Paulos:2016but, Paulos:2017fhb, Doroud:2018szp, He:2018uxa, Cordova:2018uop, Guerrieri:2018uew, Homrich:2019cbt, EliasMiro:2019kyf, Cordova:2019lot, Bercini:2019vme, Karateev:2019ymz, Karateev:2020axc, Correia:2020xtr, Bose:2020shm}. 
This recent approach is a practical implementation of the decades old idea of S-matrix Theory (see for example the book \cite{Eden:1966dnq}).
So far this method has only been applied to scattering amplitudes of scalar particles.\footnote{In fact, most recent papers studied two dimensional QFT where there is no spin.}
In this work, we develop the formalism to study 2 to 2 scattering amplitudes of particles with spin in four dimensional QFTs. 

The main idea of the numerical S-matrix Bootstrap \cite{Paulos:2016but, Paulos:2017fhb} is to write a generic analytic and Lorentz invariant ansatz for the scattering amplitude and then impose the constraints from crossing symmetry and unitarity.
In the case of identical scalar particles of mass $m$, the interacting part of the scattering amplitude denoted by $T(s,t,u)$, where $s$, $t$ and $u$ are the Mandelstam variables, can be written as
\begin{equation}
T(s,t,u) = \sum_{a} 
\alpha_{i}\;F_i(s,t,u)\,,
\label{Tansatz}
\end{equation}
where $\alpha_i$ are undetermined real coefficients and $F_i(s,t,u)$ are functions with appropriate analyticity properties.\footnote{In practice, we assume maximal analyticity of the scattering amplitude, \emph{i.e.} the only non-analyticities are the ones predicted by Landau diagrams. }
Crossing symmetry  
\begin{equation}
\label{eq:crossing_intro}
T(s,t,u) = 
T(t,s,u) =
T(u,t,s)\,,
\end{equation} 
imposes linear constraints on the  coefficients  $\alpha_i$ in the ansatz \eqref{Tansatz}.
Finally, the unitarity requirement can be cast as the following semidefinite positive condition\footnote{This is equivalent to requiring non-negativity of the determinant of the matrix which leads to the unitarity constraint in a more recognizable form $|\mathcal{S}_\ell(s)|^2\leq 1$.}
\begin{equation}
\label{eq:semidefiniteness_intro}
\begin{pmatrix}
1 & \mathcal{S}^*_\ell(s)\\
\mathcal{S}_\ell(s) & 1
\end{pmatrix}\succeq 0,
\end{equation}
where $\ell=0,2,4,\ldots$ and  $s\geq 4m^2$. The partial amplitudes $\mathcal{S}_\ell(s)$  are defined as the projection of the scattering amplitude onto definite spin via the Legendre Polynomials $P_\ell(\cos\theta)$ as
\begin{equation}
\label{eq:partial_amplitudes_intro}
\mathcal{S}_\ell(s)  \equiv 1+ i
\frac{\sqrt{s-4m^2}}{32\pi \sqrt{s}} \times \int_{0}^{\pi} d\theta \sin\theta P_\ell(\cos\theta)
T(s,t(s,\cos\theta),u(s,\cos\theta)).
\end{equation}
The integration variable $\theta$ is  the scattering angle and its precise relation to the Mandelstam variables is given in \eqref{eq:COM_parameters_2}.
By keeping a finite number of terms in the ansatz \eqref{Tansatz} one can numerically explore the space of consistent scattering amplitudes, namely the ones obeying \eqref{eq:crossing_intro} and \eqref{eq:semidefiniteness_intro}. In particular, extremizing any physical parameter linearly related to the scattering amplitude translates into a numerical (primal semidefinite) optimization problem. 

In the presence of spin, the expressions \eqref{eq:crossing_intro} - \eqref{eq:partial_amplitudes_intro} immediately become more complicated, see \cite{Martin:102663} for a review. The goal of section \ref{sec:s_matrix_approach_QFT} is to setup the formalism which allows to study crossing and unitarity for amplitudes with generic masses and spins. This section is mostly a review of known results in a concise form.

Compared to the scalar case a generic scattering amplitude describing the process $12\rightarrow 34$ has the form
\begin{equation}
\label{eq:amplitude_generic}
T_{12\rightarrow 34}{}_{\lambda_1,\lambda_2}^{\lambda_3,\lambda_4}(p_1,p_2,p_3,p_4),
\end{equation}
where $p_i$ are four-momenta and $\lambda_i$ are helicities of scattering particles respectively. Given the spins $j_i$ of four particles, helicities take values from the range $\lambda_i= -j_i,\ldots +j_i$ with step one. There are thus
\begin{equation}
N_4\equiv(2j_1+1)(2j_2+1)(2j_3+1)(2j_4+1)
\end{equation}
components describing the scattering process $12\rightarrow 34$.\footnote{When the scattering process is parity and/or time-reversal invariant, when it contains identical and/or massless particles, the counting of independent amplitudes becomes much more complicated. We discuss it in section \ref{sec:counting}.} We would then like to define $N_4$ scattering amplitudes which depend only on the Mandelstam variables.
This can be done in two different ways: using the center of mass frame or  using tensor structures. We denote these two options respectively by
\begin{equation}
\label{eq:options}
T_{12\rightarrow 34}{}_{\lambda_1, \lambda_2}^{\lambda_3,\lambda_4}(s,t,u)
\qquad\text{and}\qquad
T_{12\rightarrow 34}^I(s,t,u),
\end{equation}
where $I=1,\ldots N_4$. The two sets of amplitudes in \eqref{eq:options} are related by a linear transformation which depends on the Mandelstam variables only. This relation is completely fixed once the basis of tensor structures is chosen.
Both options have their advantages and disadvantages. In the center of mass frame one can derive crossing equations once and for all spins and masses as summarized in section \ref{sec:scattering_amplitudes} (see appendix \ref{app:crossing_equations} for details).\footnote{Generically, crossing equations relate physical amplitudes to the analytic continuation of other amplitudes beyond their physical domain.
There is no general proof (especially for particles with spin) that this continuation exists and how big is the domain where crossing equations hold.  For a review of results for scalar amplitudes, see \cite{Sommer:1970mr}. In this work we do not address this issue and simply assume that all the amplitudes under consideration are maximally analytic.
}
However, the analyticity properties of the center of mass amplitudes are subtle and one is 
forced to deal with the issue of kinematic singularities discussed in section \ref{sec:kinematic_non_analiticities}. 
Analyticity is more straightforward when using tensor structures, however one needs to study crossing case by case due to non-trivial linear relations between covariant tensor structures.
 As a result we cannot give it a completely general treatment in section \ref{sec:s_matrix_approach_QFT} and instead employ it only in the particular example of identical neutral spin $\frac{1}{2}$ particles in section \ref{sec:crossing_alternative}. 
In addition, the construction of general spin tensor structures is discussed to some extent in appendix \ref{sec:general_spin_tensor_structures}.

The key element for imposing unitarity constraints are the partial amplitudes \eqref{eq:partial_amplitudes_intro}. In the case of generic spin, \eqref{eq:partial_amplitudes_intro} remains  valid for center of mass amplitudes, if the Legendre polynomial is replaced by the small Wigner d-matrix given in \eqref{eq:Wigner_d} in full generality. This is explained in detail in section \ref{sec:partial_amplitudes}.
Finally, the unitarity constraints for generic spin are given in section \ref{sec:unitarity}.

In section \ref{sec:application}, we specialize to the case of scattering of neutral  spin $\frac{1}{2}$  massive fermions, also known as Majorana fermions \cite{Majorana:1937vz}. This section should be seen as the simplest application of the formalism given in section \ref{sec:s_matrix_approach_QFT}. In section \ref{sec:results} we write an ansatz for the scattering of identical Majorana particles in a parity invariant QFT (assuming this is the lightest particle).
We then setup the optimization problem using crossing and unitarity constraints derived in section \ref{sec:application}. Finally, we present our numerical results for the allowed values of the non-perturbative quartic and cubic (Yukawa) couplings defined from the physical scattering amplitude in section \ref{sec:non-perturbative_couplings}. The universal bounds for the quartic coupling, given in \eqref{eq:bound_quartic}, and for the cubic couplings, shown in figures \ref{fig:Yukawa} and \ref{fig:pseudoYukawa}, are our main numerical results.

Our work opens the ground for many other interesting bootstrap studies in 4 dimensions. For example, photon-photon, pion-nucleon or spinning glueball scattering.
We conclude in section \ref{sec:conclusions} with a discussion of  these examples  and other open questions for the future. We also include several appendices that fill in the details of the presentation in the main text.

\section{Review: spinning $S$-matrix approach}
\label{sec:s_matrix_approach_QFT}

In this paper we study quantum systems invariant under the restricted Poincar\'e symmetry group. In addition, we consider special situations where parity and/or time-reversal symmetries are also present. 
Scattering amplitudes are important physical observables in these systems.
They are constrained by many non-trivial consistency conditions.
The program of deriving and exploiting such theoretical constraints is referred to as the S-matrix bootstrap.

The S-matrix bootstrap program was extensively developed half a century ago. In this section we review its kinematic aspects (the ones fixed by  symmetries).\footnote{For other sources covering this topic see for example  \cite{Weinberg:1995mt}.} Our goal is to provide an updated and   easy to use practical summary of the basic ingredients.

\subsection{States}
\label{sec:states}
We work in Lorentzian metric with the mostly plus signature
\begin{equation}
\label{eq:signature}
\eta_{\mu\nu}=\{-+++\}.
\end{equation}
The unitary irreducible representations of the restricted Poincar\'e group were classified by Wigner.\footnote{For a textbook discussion see for example \cite{Weinberg:1995mt,Tung:1985na}.} Here we will work with a particular unitary representation which is positive energy time-like. The basis for such a representation is formed by the states
\begin{equation}
\label{poincare_irreps}
|c, \vec p;  \spin, \hel ; \gamma \>\,.
\end{equation}
where the 3-momentum $\vec p \in \mathbb{R}^3$ and the helicity $\lambda=-\ell,-\ell+1,\ldots,+\ell$.
The other labels are fixed within the irreducible representation and serve to specify a given state.
 The label $c>0$ is a continuous real parameter related to the energy $p^0>0$ as
\begin{equation}
\label{eq:relation_c_energy}
c^2= - p_\mu p^\mu = (p^0)^2 - (\vec p\,)^2.
\end{equation}
It is defined as the eigenvalue of the first Casimir of the Poincar\'e group $c^2 = -P^2$, where $P^\mu$ are the generators of translations. Instead of $c$ one can equivalently label the basis of states \eqref{poincare_irreps} using the energy $p^0$.   We will often use both labels throughout the text interchangeably.
The spin label $\ell=0,\,\tfrac{1}{2},\,1,\ldots$ is a non-negative integer or half integer related to the eigenvalue of the second Casimir of the Poincar\'e group $W^2 = 4 c^2\, \ell(\ell+1)$, where $W^\mu$ is the Pauli–Lubanski pseudovector.\footnote{In our convention $W_\mu\equiv \epsilon_{\mu\nu\rho\sigma}M^{\nu\rho} P^\sigma$, where $M^{\nu\rho}$ are the generators of the Lorentz group.
}
The helicity label $\lambda$ is the projection of the spin vector along $\vec p$. 
Finally the label $\gamma$ stands for any other additional (discrete or continuous) labels which might be required to fully specify the state.
We will see one particular case of their importance in section \ref{sec:two_particle_states}.

We will often need to use spherical coordinates in which case the three momentum is parametrized as
\begin{equation}
\label{eq:spherical_coordinates_momentum}
\vec p =
(\myP \sin\theta\cos\phi,\; \myP\sin\theta\sin\phi,\; \myP\cos\theta),\quad
\myP\equiv |\vec p|,\quad
0 \leq \phi < 2\pi,\quad 0 \leq \theta \leq \pi,
\end{equation} 
The helicity basis states \eqref{poincare_irreps} with non-zero three momentum $\vec p$ are defined as follows
\begin{equation}
\label{poincare_irreps_definition}
|c, \vec p;  \spin, \hel ; \gamma \> =
e^{-i\phi J_3}e^{-i\theta J_2}e^{+i\phi J_3} e^{-i\eta K_3}
|c, \vec 0;  \spin, \hel ; \gamma \>,
\end{equation}
where $J_a$ and $K_a$ with $a=1,2,3$ are the generators of spatial rotations and boosts respectively. The rapidity parameter $\eta$ is related to the energy and momentum of the states as
\begin{equation}
\label{eq:rapidity}
\cosh\eta = \frac{p^0}{c},\qquad
\sinh\eta = \frac{\myP}{c}.
\end{equation}
Lastly the helicity $\hel$ in the zero three momentum state $|c, \vec 0;  j, \hel; \gamma \>$ is a projection of the spin $\spin$ onto the $z$ axis. In other words these states are eigenstates of the $J_3$ generators. Given a generic 3d rotation characterized by the three Euler angles $(\alpha,\beta,\gamma)$ (where $\alpha$ and $\gamma$ parametrize rotations around the $z$-axis and $\beta$ parametrizes rotations around the $y$-axis) we have the following transformation property
\begin{equation}
\label{eq:transformation}
R(\alpha,\beta,\gamma)|c, \vec 0;  \spin, \hel ; \gamma \> =
\sum_{\lambda'}\mathscr D^{(\ell)}_{\lambda' \lambda}(\alpha,\beta,\gamma)
|c, \vec 0;  \spin, \hel' ; \gamma \>,\quad
R(\vec a)\equiv e^{-i \alpha J_3} e^{-i \beta J_2} e^{-i \gamma J_3}.
\end{equation}
The object $\mathscr D^{(\ell)}_{\lambda' \lambda}$ is known as the large Wigner D matrix and reads as
\begin{equation}
\label{D_matrix_large}
\mathscr D^{(\ell)}_{\lambda'\lambda} (\alpha,\beta,\gamma) =
e^{-i\alpha \lambda'}
d^{(\ell)}_{\lambda'\lambda} (\beta)
e^{-i\gamma \lambda}.
\end{equation}
Here the object $d^{(\spin)}_{\lambda'\lambda}$ is the small Wigner d matrix, its explicit expression reads\footnote{Note that Mathematica implements the small Wigner d matrices \eqref{eq:Wigner_d} with indices $\lambda$ and $\lambda'$ flipped. In other words $d^{(\spin)}_{\lambda'\lambda} (\beta)$ is generated by the command $\text{WignerD}[\{\spin,\lambda,\lambda'\},\beta]$.}
\begin{align}
d^{(\spin)}_{\lambda^\prime\lambda} (\beta) &=
\sqrt{(\spin+\lambda)!(\spin-\lambda)!(\spin+\lambda')!(\ell-\lambda')!}\nn\\
&\times
\sum_{\nu=0}^{2j}
(-1)^\nu
\frac{\left(\cos(\beta/2)\right)^{2j+\lambda-\lambda'-2\nu}
	\left(-\sin(\beta/2)\right)^{\lambda'-\lambda+2\nu}}
{\nu!(\ell-\lambda'-\nu)!(\spin+\lambda-\nu)!(\nu+\lambda'-\lambda)!}.
\label{eq:Wigner_d}
\end{align}
In case both helicities $\lambda$ and $\lambda^\prime$ are equal to zero, the small Wigner d matrix reduces to the usual Legendre polynomial
\begin{equation}
d^{(\ell)}_{0 0}(\beta) = P_\ell (\cos \beta).
\end{equation}
The Wigner D matrices satisfy orthogonality. They also obey various useful relations, see for example appendix A.2 in \cite{Martin:102663}.

Consider now a generic Poincar\'e transformation which consists of a translation fixed by four Lie parameters $a^\mu$ and the Lorentz transformation $\Lambda^\mu{}_\nu$ fixed by six Lie parameters $\rho_{[\mu\nu]}$. It transforms the 4-momentum $p^\mu$ into $p'^{\mu}=\Lambda^\mu{}_\nu p^\nu$. Under the Poincar\'e transformation $(a,\rho)$ the states \eqref{poincare_irreps} transform according to
\begin{equation}
\label{poincare_transformation}
U(a,\rho)|c,\vec p;\spin,\hel;\gamma\rangle = 
e^{ia\cdot p'}\times
\sum_{\lambda'}\mathscr D^{(\ell)}_{\lambda' \lambda}(\alpha,\beta,\gamma)|c,\vec{p}^{\,\prime};\spin,\hel'; \gamma\>,
\end{equation}
where the three Wigner angles $(\alpha,\beta,\gamma)$  can be expressed in terms of the six Lie parameters $\rho_{[\mu \nu]}$. We provide an example of a Wigner angles computation in appendix \ref{app:wigner_angles}. For the special case of pure 3d rotations, the Wigner D matrix reduces to a phase and therefore the helicity of the state remains unchanged\footnote{Notice that this is compatible with \eqref{eq:transformation} because the limit $\vec p \to 0$ of $|c, \vec p;  j, \hel; \gamma \>$  is only equal to $|c, \vec 0;  j, \hel; \gamma \>$ if $\vec p$ is parallel to the $z$-axis. 
}
\begin{equation}
\label{eq:rotphase}
R(\vec a) |c,\vec p;\spin,\hel;\gamma\rangle = e^{-i\lambda\, \xi(\vec a,\phi,\theta)}
|c,\vec p\,';\spin,\hel;\gamma\rangle,
\end{equation}
where the phase $\xi$ has a complicated dependence on its arguments.

\paragraph{Particles}

So far we discussed states and their properties in a generic Poincar\'e invariant quantum theory.
The related notion of particle does not exist in every QFT. The simplest context where particles  are well defined is in free QFTs. 
A free theory is described by a set of one particle states (describing freely propagating particles with a given mass and spin). Taking tensor products of these states we form a complete basis of states spanning the Hilbert space of the theory.
In interacting QFTs, particles can still be defined in the asymptotic far past and future if interactions decay sufficiently fast with distance. This is the case for theories with a mass gap and   massless theories with soft interactions like Goldstone bosons or photons (in the absence of charged particles in the asymptotic states).  
In this paper,  we will only consider those theories where particles do exist.
Then, one can define two types of asymptotic states: the $in$ states (far past) and the $out$ states (far future). 
Both the $in$ states and the $out$ states span the full Hilbert space of the theory.
A natural basis of $in/out$ states is given by the tensor product of one particle $in/out$ states.

\subsubsection{One particle states (1PS)}
\label{sec:one_particle_states}
 
One particle states (1PS) are special cases of \eqref{poincare_irreps} where $c$ takes only discrete values corresponding to the masses of stable particles in the theory.
For this reason, we will set $c=m$ from now on.
If global symmetries are present, the label $\gamma$ describes the charge or more generically the representation under the global symmetry group. 
For simplicity, in this section we  ignore global symmetries, thus the 1PS will not carry any extra labels $\gamma$.

\noindent We introduce a shorthand notation for one particle states
\begin{equation}
\label{eq:shorthand_1PS}
|\kappa\rangle \equiv |m,\vec p;j,\lambda \>.
\end{equation}
For indicating spin we use the label $j$ instead of $\spin$ here to ease the visual distinction between the 1PS and generic irreps. 
We normalize 1PS as follows
\begin{eqnarray}
\< m', \vec{p}^{\,\prime} ; j', \lambda' | m , \vec p; j, \lambda \> &=& (2 \pi)^3 \,2 p^0 \,\delta^3(\vec{p}^{\,\prime} -\vec p)\,\delta_{m' m} \delta_{j' j} \delta_{\lambda' \lambda} \nn\\
 &\equiv& \delta(\kappa' - \kappa),
\label{eq:1PS_normalization}
\end{eqnarray} 
where in accordance with \eqref{eq:relation_c_energy} the energy of the states is
\begin{equation}
p^0 = \sqrt{m^2 + p^2}.
\end{equation}
In the second line of \eqref{eq:1PS_normalization} we have introduced the shorthand notation $\delta(\kappa' - \kappa)$ for the set of Kronecker and Dirac delta functions.
The transformation rule for a one particle state under a Poincare transformation remains the same as in \eqref{poincare_transformation}.

The case of massless particles should be treated separately since $c\neq 0$ in \eqref{poincare_irreps}. Skipping details, the following statement holds: massless particles can be described by the states \eqref{eq:shorthand_1PS} with $m=0$ and the range of helicities restricted to only two values $\lambda=-j$ and $\lambda=+j$.  

Finally we state the transformation properties of 1PS under parity and time-reversal. For further details see appendix \ref{app:parity_and_timereversal}. One has
\begin{align}
\label{eq:parity_1PS_main}
\mathcal P | m , \vec p; j, \lambda \> &= \eta (-1)^{j - \lambda}  \exp(2 i \phi \lambda) | m , -\vec p; j, -\lambda\rangle,\\
\label{eq:time_reversal_1PS_main}
\mathcal T | m , \vec p; j, \lambda \> &= \varepsilon (-1)^{2j}  \exp(-2 i \phi \lambda) | m , -\vec p; j, \lambda\rangle,
\end{align}
where $\phi$ is the spherical angle defined in \eqref{eq:spherical_coordinates_momentum}, see also \eqref{eq:negative_spherical}, $\eta$ and $\epsilon$ are pure phases. 
Their values are yet other quantities characterizing the state. Let us briefly discuss the phase $\eta$ called intrinsic parity. One can argue as in section 3.3 of \cite{Weinberg:1995mt}, that one can always define $\parity$ in such a way that $\parity^2=+1$ or $\parity^2=-1$. As a result one has
\begin{equation}
\label{eq:intrinsic_parities}
\eta=\pm 1
\quad\text{or}\quad
\eta=\pm i.
\end{equation}
The imaginary values of intrinsic parities are only possible for fermions, however no such fermions have been discovered so far in the nature.

\subsubsection{Two particle states (2PS)}
\label{sec:two_particle_states}
We define the two particle states (2PS) by taking the ordered tensor product of two 1PS
\begin{equation}
\label{eq:2PS_tensor_product}
 |\kappa_1,\kappa_2 \> \equiv
 |m_1,\vec p_1;j_1,\lambda_1\>\otimes|m_2,\vec p_2;j_2,\lambda_2\>.
\end{equation}
The normalization of the 2PS defined above follows from that of the 1PS:
 \begin{align}
\label{eq:normalization_2PS}
\<\kappa_1,\kappa_2 |\kappa_3,\kappa_4 \> &= \delta(\kappa_1 - \kappa_3)\delta(\kappa_2 - \kappa_4).
\end{align}

Two particle states do not form   an irreducible representation of the restricted Poincar\'e group. However, they can be decomposed into a direct sum of states \eqref{poincare_irreps} transforming in irreducible representations. This is done by injecting the completeness relation into 2PS
\begin{equation}
\label{eq:2PS_decomposition_general}
|\kappa_1,\kappa_2\rangle = \int \frac{d^4p}{(2\pi)^4} \, \theta(p^0)\theta(-p^2)\sum_\gamma \sum_{\ell,\lambda} 
|c,\vec p;\spin,\hel;\gamma\rangle\,
\langle c,\vec p;\spin,\hel;\gamma|\kappa_1,\kappa_2\rangle,
\end{equation}
where we normalize the states \eqref{poincare_irreps} as follows\footnote{The Kronecker deltas follow from the fact that the states here are simultaneous eigenstates of the two Casimirs \eqref{eq:casimirs} and the helicity operator \eqref{eq:helicity}.}
\begin{equation}
\label{eq:irrep_normalization}
\langle c',\vec{p}^{\,\prime};\spin',\hel'; \gamma'
|c,\vec p;\spin,\hel;\gamma\rangle =
(2 \pi)^4 
\delta^{4}({ p'}^\mu - p^\mu)\,
\delta_{\ell' \ell}\,
\delta_{\lambda' \lambda}\,
\delta_{\gamma'\gamma},
\end{equation}
where  $p^\mu = (p^0,\,\vec p)$ is the 4-momentum and the symbolic expression $\delta_{\gamma'\gamma}$ will be properly specified when the additional labels $\gamma$ and $\gamma'$ are defined.
We use the normalization \eqref{eq:irrep_normalization} for all irreducible Poincar\'e representations, with the exception of 1PS which are the only states where the label $c$ takes particular discrete values.
Looking at the right-hand side of \eqref{eq:2PS_decomposition_general}, we see that the Clebsch-Gordon coefficients of this decomposition (due to translation invariance) obey
\begin{equation}
\label{eq:CG}
\langle c,\vec p;\spin,\hel;\gamma|\kappa_1,\kappa_2\rangle
\propto (2\pi)^4\delta^4(p^\mu-p_1^\mu-p_2^\mu).
\end{equation}
This delta function completely removes the integration over $p$ in \eqref{eq:2PS_decomposition_general}.
The label $\gamma$ is the multiplicity label. In the case of 2PS decomposition the multiplicity label $\gamma$ consists of only discrete parameters\footnote{This is no longer the case for the decompositions of three or more particle states. In these cases the multiplicity label $\gamma$ would also contain  continuous parameters associated with the relative momenta of the component particles.} and reads as
\begin{equation}
\label{eq:gamma_2PS}
\gamma = (m_1,j_1,\lambda_1;\, m_2,j_2,\lambda_2).
\end{equation}
Thus the label $\gamma$ keeps track of which particles and what helicities were used to make the two particle state. In what follows we will almost always drop the explicit mass and spin labels in order to simplify the formulas. However, when dealing with particles of different masses and spins, the mass and spin labels are important.
Finally we choose to normalize the states appearing in \eqref{eq:2PS_decomposition_general} according to \eqref{eq:irrep_normalization} with
\begin{equation}
\label{eq:kronecker_gamma}
\delta_{\gamma'\gamma} =
\delta_{m_1'm_1}\delta_{m_2'm_2}
\delta_{j_1'j_1}\delta_{j_2'j_2}
\times
\delta_{\lambda_1'\lambda_1}\delta_{\lambda_2'\lambda_2}.
\end{equation}

\paragraph{COM two particle states}
We can always use Lorentz invariance to go to the frame where the total momentum of the two particles is 0. This frame is called the centre of mass (COM) frame. Therefore, we do not need to know the most general decomposition \eqref{eq:2PS_decomposition_general}, and instead it is enough to focus on the special case of 2PS in the center of mass (COM) frame, namely the states \eqref{eq:2PS_tensor_product} obeying the constraint $\vec p_1 = - \vec p_2$. We give a special label to such two particles states
\begin{align}
\label{eq:2PS_com}
|(\myP,\theta,\phi); \lambda_1, \lambda_2 \rangle  \equiv
|m_1,+\vec p;j_1,\lambda_1\>\otimes|m_2,-\vec p;j_2,\lambda_2\>,
\end{align}
where the three-momenta have the following spherical coordinates\footnote{\label{foot:patches}
Notice that given the vector $+\vec p$ in spherical coordinates, the vector opposite to it is defined as $-\vec p =(\myP,\, \pi-\theta,\,\pi+\phi)$ for $\phi \in[0,\pi]$ and  $-\vec p = (\myP,\, \pi-\theta,\,-\pi+\phi)$ for $\phi \in(\pi,2\pi]$. In other words in order to describe the vector $-\vec p$ in spherical coordinates one needs two different descriptions, one for $\phi\in[0,\pi]$ and one for $\phi\in(\pi,2\pi]$. We do not indicate it in the main text, since all the consequent formulas remain uniform in the whole range $\phi\in[0,2\pi]$.
The reason for that is the choice of the helicity basis \eqref{poincare_irreps_definition} and the fact that $R(\phi,\theta,-\phi)$ is $2\pi$ periodic, see footnote \ref{foot:periodicity_R}.
Notice also that there is a special case when $\theta=\pi$. For this particular point, we choose  the spherical angles of the first state to be $+\vec p =(\myP,\,\theta=\pi,\,\phi=\pi)$, whereas the spherical angles of the second state are $-\vec p =(\myP,\,\theta=0,\, \phi=0)$.}
\begin{equation}
\label{eq:negative_spherical}
+\vec p =(\myP,\,   \theta,\,\phi),\quad
-\vec p =(\myP,\, \pi-\theta,\,\pi+\phi),\quad
\theta\in[0,\pi], \quad
\phi\in[0,2\pi].
\end{equation}
The states \eqref{eq:2PS_com} are normalized according to \eqref{eq:normalization_2PS}. By performing a change of variables we can rewrite the normalization in terms of spherical coordinates as
\begin{multline}
\label{eq:normalization_spherical}
\<(\myP,\theta,\phi); \lambda_1, \lambda_2
|(\myP', 0, 0); \lambda_1', \lambda_2' \rangle =
(2\pi)^4\delta^4(0) \times
\frac{16\pi^2 \sqrt{s}}{\sqrt{\myP \myP'}}
\frac{\delta(\theta)\delta(\phi)}{\sin\theta}\times\\
\delta_{m_1'm_1}\delta_{m_2'm_2}
\delta_{j_1'j_1}\delta_{j_2'j_2}\times
\delta_{\lambda_1\lambda_1'}\delta_{\lambda_2\lambda_2'},
\end{multline}
 where $s=(p_1^0+p_2^0)^2$ is the square of the COM energy.
 In appendix  \ref{app:clebsch_gordan_coefficients}, we compute the Clebsch-Gordon coefficients \eqref{eq:CG} for the COM states \eqref{eq:2PS_com}.
Here we present only the result which reads as
\begin{equation}
\label{eq:2PS_decomposition_com}
|(\myP,\theta,\phi); \lambda_1, \lambda_2 \rangle =  \sum_{\ell,\lambda} C_\ell(\myP) \,
e^{i\phi(\lambda_1+\lambda_2-\lambda)}
d^{(\ell)}_{\lambda \lambda_{12}}(\theta) |c, 0; \spin, \hel; \gamma \rangle,
\end{equation}
where we have
\begin{equation}
\lambda_{12}\equiv\lambda_1-\lambda_2,\quad
c =\sqrt{s}=\sqrt{m_1^2+\myP^2}+\sqrt{m_2^2+\myP^2},
\end{equation}
with the multiplicity labels $\gamma$ given in \eqref{eq:gamma_2PS} and $d^{(\ell)}_{\lambda \lambda_{12}}(\theta)$ given by \eqref{eq:Wigner_d}.
Using the orthogonality of the small Wigner $d$ matrix and the exponential function, we can invert equation \eqref{eq:2PS_decomposition_com} as follows
\begin{equation}
\label{eq:2PS_decomposition_com_inverse}
|c, 0; \spin, \hel; \gamma \rangle = \frac{2\ell +1}{4 \pi C_\ell(\myP)} \int_0^{2 \pi} d\phi \int_{0}^{\pi} d\theta \sin\theta \, e^{-i\phi(\lambda_1+\lambda_2-\lambda)} d^{(\ell)}_{\lambda \lambda_{12}}(\theta)|(\myP,\theta,\phi); \lambda_1, \lambda_2 \rangle.
\end{equation}
The coefficient $C_\ell$ is completely fixed by the consistency requirement that the left-hand side of \eqref{eq:2PS_decomposition_com} satisfies the normalization condition \eqref{eq:normalization_2PS} and the state in the right-hand side of \eqref{eq:2PS_decomposition_com} satisfies the normalization condition \eqref{eq:irrep_normalization}.
 For non-identical particles it reads as
\begin{align}
\label{eq:coefficient_Cj}
C_\ell(\myP)^2    = 4 \pi\, (2\ell+1) \times \frac{c}{\myP}.
\end{align}

\paragraph{Identical particles}
The discussion presented above should be slightly modified when the two particle state is composed of identical particles. In the latter case it must satisfy
\begin{equation}
\label{eq:id_symmetry_property}
|\kappa_1,\kappa_2 \>_{id} = (-1)^{2j} |\kappa_2,\kappa_1 \>_{id}.
\end{equation}
We have added the subscript $id$ to explicitly indicate that the state describes a system of two identical particles. 
In order to incorporate \eqref{eq:id_symmetry_property}, we have (instead of simply taking an ordered product) to take either symmetrized (in case of bosons) or anti-symmetrized (in case of fermions) tensor product. Such a  state will thus have the following form 
\begin{equation}
\label{eq:2PS_identical}
|\kappa_1,\kappa_2 \>_{id}\equiv \frac{1}{\sqrt 2}\left(
|m, \vec p_1; j , \lambda_1 \>  \otimes |m, \vec p_2; j, \lambda_2\> + 
(-1)^{2j}|m, \vec p_2; j , \lambda_2 \> \otimes |m, \vec p_1; j, \lambda_1\>\right).
\end{equation}
The normalization of the state \eqref{eq:2PS_identical} follows from \eqref{eq:1PS_normalization}
\begin{equation}
\label{eq:normalization_2PS_id}
{}_{id}\<\kappa_1,\kappa_2 |\kappa_3,\kappa_4 \>_{id} = \delta(\kappa_1 - \kappa_3)\delta(\kappa_2 - \kappa_4)
+ (-1)^{2j} \delta(\kappa_1 - \kappa_4)\delta(\kappa_2 - \kappa_3).
\end{equation}
As before we need to define the identical 2PS in the center of momentum. Adapting \eqref{eq:2PS_com} to the case of identical particles we get
\begin{multline}
\label{eq:2PS_com_id}
|(\myP,\theta,\phi); \lambda_1, \lambda_2 \rangle_{\text{id}} \equiv\\
\frac{1}{\sqrt 2}\left(
|m, +\vec p; j , \lambda_1 \>  \otimes |m, -\vec p; j, \lambda_2\> + 
(-1)^{2j}|m, -\vec p; j , \lambda_2 \> \otimes |m, \vec p; j, \lambda_1\>\right).
\end{multline}
The normalization of the identical COM states \eqref{eq:2PS_com_id} is fixed by \eqref{eq:normalization_2PS_id}. It is still given by \eqref{eq:irrep_normalization} but with
\begin{equation}
\label{eq:convention_gamma_id}
\delta_{\gamma'\gamma} = 
\frac{1}{2}\left(
\delta_{\lambda_{1}\lambda_{1}'}\delta_{\lambda_{2}\lambda_{2}'}
+(-1)^{\ell+\lambda_1-\lambda_2}
\delta_{\lambda_{1}\lambda_{2}'}\delta_{\lambda_{2}\lambda_{1}'}
\right).
\end{equation}
We would now like to decompose the identical two particle state \eqref{eq:2PS_com_id} into irreducible representations
\begin{equation}
\label{eq:irrep_id}
|c,0,\spin,\hel; \lambda_1,\lambda_2 \rangle_{id}\equiv
\frac{1}{2}\left(
|c,0,\spin,\hel; \lambda_1,\lambda_2 \rangle+
(-1)^{\ell+\lambda_1 - \lambda_2}
|c,0,\spin,\hel; \lambda_2,\lambda_1 \rangle
\right).
\end{equation}
The decomposition of identical 2PS is obtained straightforwardly by applying \eqref{eq:2PS_decomposition_com} to both terms in the right-hand side of \eqref{eq:2PS_com_id} which leads to
\begin{equation}
\label{eq:2PS_decomposition_com_id}
|(\myP,\theta,\phi); \lambda_1, \lambda_2 \rangle_{\text{id}} =
\sqrt{2}\sum_{\ell,\lambda}
C_\ell(\myP) e^{i(\lambda_1 + \lambda_2 - \lambda)\phi} d^{(\ell)}_{\lambda \lambda_{12}}(\theta) |c, 0; \spin, \hel; \lambda_1,\lambda_2 \rangle_{id},
\end{equation}
where the coefficient $C_\ell$ is given by \eqref{eq:coefficient_Cj}. For the detailed derivation of these equations see appendix \ref{app:identical_particles}.

\subsection{S-matrix elements}
\label{sec:matrix_elements}

Given a generic state in the reference frame at $t=0$, an observer in another reference frame in the far past ($t=-\infty$) or far future ($t=+\infty$) will see the same state as a (linear combination of) tensor product of one particle states which we refer to as {\it in} or {\it out} asymptotic states respectively. Asymptotic states have a complicated time evolution. One can however establish a formal one to one map between these states and those of some free theory (which evolve trivially with time) by means of a pair of unitary operators $\Omega_{-}$ and $\Omega_{+}$ called the M{\o}ller operators. See section 2.1 in \cite{Karateev:2019ymz} for a recent discussion. In their notation
\begin{equation}
\label{eq:moller_map_1PS}
|\kappa\>_{in}  = \Omega_{-} |\kappa\>_{free},\quad
|\kappa\>_{out} = \Omega_{+} |\kappa\>_{free}.
\end{equation}

Let us now discuss inner products between asymptotic states. Since the M{\o}ller operators are isometric,\footnote{Recall that an isometric operator $O$ on a Hilbert space preserves distances. This implies that $O^\dagger O=1$, however the operator need not be surjective and $O O^\dagger\neq1$ generically. A surjective isometric operator is unitary.} the inner products of only $in$ states or only $out$ states are fixed by the normalization conditions \eqref{eq:irrep_normalization} and \eqref{eq:1PS_normalization}. This means that the only non-trivial matrix elements must include both $in$ and $out$ states.

\paragraph{Four-particle amplitudes}
We start with the most important object for our work
\begin{equation}
\label{eq:2to2_scattering}
{}_{out}\langle\kappa_3,\kappa_4|\kappa_1,\kappa_2\rangle_{in}=
{}_{free}\langle\kappa_3,\kappa_4|S|\kappa_1,\kappa_2\rangle_{free},
\end{equation}
where the scattering operator $S$ is defined via \eqref{eq:moller_map_1PS} as
\begin{equation}
S\equiv \Omega_{+}^\dagger\Omega_{-}.
\end{equation}
M{\o}ller operators are isometric which implies the unitarity of the scattering operator\footnote{This is a non-trivial statement, see \cite{taylor2012scattering} for details.}
\begin{equation}
\label{eq:unitarity}
S^\dag S = 1.
\end{equation}
Poincar\'e invariance implies that\footnote{See the discussion around \eqref{poincare_transformation}.}
\begin{equation}
\label{eq:poincare_invariance}
U(a,\rho)\, S\, U^{-1}(a,\rho) = S,
\end{equation}
where $U$ represents a generic Poincar\'e transformation in the Hilbert space.
Finally it is convenient to split the scattering operator into the trivial part (identity operator) and the interacting part as
\begin{equation}
\label{eq:ST_relation}
S = 1 + i T.
\end{equation}
If $T=0$ we simply recover the free theory.
The matrix element \eqref{eq:2to2_scattering} describes scattering of two particles.
Factoring out the overall delta function due to translation invariance we can define the two to two scattering amplitude as
\begin{equation}
\label{eq:scattering_amplitude}
(2\pi)^4\delta^{(4)}(p_1^\mu + p_2^\mu - p_3^\mu - p_4^\mu) \times
S_{12\rightarrow 34}{}_{\lambda_1, \lambda_2}^{\lambda_3,\lambda_4}(p_1,p_2,p_3,p_4)\equiv
{}_{free}\langle\kappa_3,\kappa_4|S|\kappa_1,\kappa_2\rangle_{free}.
\end{equation}
Equivalently we define the interacting part of the two to two scattering amplitude
\begin{equation}
\label{eq:scattering_amplitude_interacting}
(2\pi)^4\delta^{(4)}(p_1^\mu + p_2^\mu - p_3^\mu - p_4^\mu) \times
T_{12\rightarrow 34}{}_{\lambda_1, \lambda_2}^{\lambda_3,\lambda_4}(p_1,p_2,p_3,p_4)\equiv
{}_{free}\langle\kappa_3,\kappa_4|T|\kappa_1,\kappa_2\rangle_{free}.
\end{equation}
Since the (interacting) scattering amplitude is defined via the one particle states, all the 4-momenta are on-shell
\begin{equation}
\label{eq:on_shell}
p_i^2=-m_i^2.
\end{equation}
We study these amplitudes and their properties in depth in section \ref{sec:scattering_amplitudes}. We sometimes drop the subscript $12\rightarrow 34$ when it is clear from the context which scattering process we describe.
However, it is necessary to keep this subscript when relating amplitudes describing different processes. Using \eqref{eq:ST_relation} we can relate the scattering amplitude \eqref{eq:scattering_amplitude} with its interacting part \eqref{eq:scattering_amplitude_interacting} as
\begin{multline}
S_{12\rightarrow 34}{}_{\lambda_1, \lambda_2}^{\lambda_3,\lambda_4}(p_1,p_2,p_3,p_4) =\\
\left[ \frac{{}_{free}\langle\kappa_3,\kappa_4|\kappa_1,\kappa_2\rangle_{free}}{(2\pi)^4\delta^{(4)}(p_1^\mu + p_2^\mu - p_3^\mu - p_4^\mu)} \right]+
iT_{12\rightarrow 34}{}_{\lambda_1, \lambda_2}^{\lambda_3,\lambda_4}(p_1,p_2,p_3,p_4),
\label{eq:relations_ST}
\end{multline} 
where the first term $[\ldots]$ is a schematic expression which can be evaluated straightforwardly.\footnote{\label{foot:distribution}
For example in the COM frame defined in \eqref{eq:COM_frame} using the normalization condition in spherical coordinates \eqref{eq:normalization_spherical} it is straightforward to write
\begin{equation}
\nn
\left[ \frac{{}_{free}\langle\kappa_3,\kappa_4|\kappa_1,\kappa_2\rangle_{free}}{(2\pi)^4\delta^{(4)}(p_1^\mu + p_2^\mu - p_3^\mu - p_4^\mu)}
\right]_{com} = 
\frac{8\pi \sqrt{s}}{\sqrt{\myP\myP'}}\times
\frac{\delta(\theta)\delta(\phi)}{\sin\theta}
\delta_{m_1 m_3} \delta_{m_2 m_4}
\delta_{j_1 j_3} \delta_{j_2 j_4}
\times
\delta_{\lambda_1\lambda_3} \delta_{\lambda_2\lambda_4}.
\end{equation}
Similarly in case of identical particles with mass $m$ and spin $j$ using \eqref{eq:normalization_spherical_id} we have 
\begin{multline}
\nn
\left[ \frac{{}_{free}\langle\kappa_3,\kappa_4|\kappa_1,\kappa_2\rangle_{free}}{(2\pi)^4\delta^{(4)}(p_1^\mu + p_2^\mu - p_3^\mu - p_4^\mu)}
\right]_{com} =  \frac{32\pi^2 \sqrt{s}}{\sqrt{s-4m^2}}\times
\left(
\frac{\delta(\theta)\delta(\phi)}{\sin\theta}\delta_{\lambda_1\lambda_3}
\delta_{\lambda_2\lambda_4}+(-1)^{2j}
\frac{\delta(\pi-\theta)\delta(\phi+ \pi)}{\sin(\pi-\theta)}
\delta_{\lambda_1\lambda_4}\delta_{\lambda_2\lambda_3}
\right).
\end{multline}
} 
This piece is not a function but a distribution. Due to the relation \eqref{eq:relations_ST}, in practice we will never need to discuss the full amplitude $S$ and instead we will only need the interacting $T$ amplitude.

\paragraph{Partial amplitudes} 

The second matrix element we need is between the $in$ and $out$ states in the irreducible representation
\begin{equation}
{}_{out}\< c', \vec p\,';  \spin', \hel' ; \gamma ' |c, \vec p;  \spin, \hel; \gamma \>_{in}=
{}_{free}\< c', \vec p\,';  \spin', \hel' ; \gamma ' |S|c, \vec p;  \spin, \hel; \gamma \>_{free}.
\end{equation}
Again factoring out the overall delta function due to translational invariance we can define the partial amplitude with a definite spin $\ell$ as
\begin{equation}
\label{eq:partial_amplitude}
(2\pi)^4\delta^{(4)}(p^\mu - {p'}^\mu)\delta_{\ell \ell'} \delta_{\lambda \lambda'}
\times S_\ell{}_\gamma^{\gamma'}(c) =
{}_{free}\< c', \vec p\,';  \spin', \hel' ; \gamma ' |S|c, \vec p;  \spin, \hel; \gamma \>_{free}.
\end{equation}
Equivalently we can define the interacting part of the partial amplitude as
\begin{equation}
\label{eq:partial_amplitude_interacting}
(2\pi)^4\delta^{(4)}(p^\mu - {p'}^\mu)\delta_{\ell \ell'} \delta_{\lambda \lambda'}
\times T_\ell{}_\gamma^{\gamma'}(c) =
{}_{free}\< c', \vec p\,';  \spin', \hel' ; \gamma ' |T|c, \vec p;  \spin, \hel; \gamma \>_{free}.
\end{equation}
We prove that the left-hand side of \eqref{eq:partial_amplitude} and \eqref{eq:partial_amplitude_interacting} must contain $\delta_{\ell\ell'}\delta_{\lambda\lambda'}$ factor in appendix \ref{sec:two_point_amplitudes_COM}.
In a generic situation due to \eqref{eq:irrep_normalization} the two partial amplitudes are simply related as
\begin{equation}
\label{eq:Sj_Tj_relation}
S_\ell{}_\gamma^{\gamma'}(c) = \delta_{\gamma'\gamma} +  iT_\ell{}_\gamma^{\gamma'}(c).
\end{equation} 
In practice we will only need to consider partial amplitudes where the irreps come from the decomposition of COM two particle states \eqref{eq:2PS_decomposition_com}. In that case the additional labels $\gamma$ are multiplicities given by \eqref{eq:gamma_2PS}. The Kronecker delta for distinct particles is given by \eqref{eq:kronecker_gamma} and for identical particles by \eqref{eq:convention_gamma_id}.
We examine partial amplitudes and their properties in detail in section \ref{sec:partial_amplitudes}. In addition, we also derive the relation between partial and scattering amplitudes.

\subsection{Scattering amplitudes and crossing}
\label{sec:scattering_amplitudes}
In this section we carefully study various aspects of the scattering amplitudes \eqref{eq:scattering_amplitude} and its interacting part \eqref{eq:scattering_amplitude_interacting}.

We start with the transformation property under the Poincar\'e group. It directly follows from the transformation property of each state given by \eqref{poincare_transformation}. In the most generic case it reads as
\begin{align}
\label{eq:T_matrix_amplitude_transformation_general}
\resizebox{.9\hsize}{!}{$\displaystyle{
T_{\lambda_1, \lambda_2}^{\lambda_3,\lambda_4}(p_1,p_2,p_3,p_4)=
\sum_{\lambda_i'}
\mathscr{D}^{(  j_1)}_{\lambda'_1\lambda_1}(\vec\omega_1)
\mathscr{D}^{(  j_2)}_{\lambda'_2\lambda_2}(\vec\omega_2)
\mathscr{D}^{*( j_3)}_{\lambda'_3\lambda_3}(\vec\omega_3)
\mathscr{D}^{*(  j_4)}_{\lambda'_4\lambda_4}(\vec\omega_4)
T_{\lambda_1', \lambda_2'}^{\lambda_3',\lambda_4'}(p_1',p_2',p_3',p_4')}$},
\end{align}
where $\vec\omega_i\equiv (\alpha_i,\beta_i,\gamma_i)$ are the Wigner angles for each one particle state defining the (interacting) scattering amplitude.

Let us now introduce the Mandelstam variables which are invariant quantities under Lorentz transformations
\begin{equation}
\label{eq:mandelstam_variables}
s\equiv -(p_1+p_2)^2,\quad
t\equiv -(p_1-p_3)^2,\quad
u\equiv -(p_1-p_4)^2,\quad
s+t+u=\sum_{i=1}^4 m_i^2.
\end{equation}
Using these variables one can split the scattering amplitude 
\eqref{eq:scattering_amplitude_interacting}  into parts   invariant under   Lorentz transformation and parts transforming non-trivially,  
\begin{equation}
\label{eq:tensor_structures}
T_{\lambda_1, \lambda_2}^{\lambda_3,\lambda_4}(p_1,p_2,p_3,p_4) = 
\sum_{I=1}^{(2j_1+1)\ldots(2j_4+1)} T_I(s,t,u)\times \mathbb{T}_I{}_{\lambda_1, \lambda_2}^{\lambda_3,\lambda_4}(p_1,p_2,p_3,p_4).
\end{equation}
We refer to the quantities $T_I(s,t,u)$ as the scalar components of the scattering amplitudes and $\mathbb{T}_I$ are called tensor structures. The latter ensures the correct transformation property of the amplitude as dictated by \eqref{eq:T_matrix_amplitude_transformation_general}. We abuse notation and call both the full amplitude and its scalar components by the same symbol $T$. It should be clear which is which by the presence of indices and arguments. We construct tensor structures explicitly for a particular example in section \ref{sec:example_tensor_structures}. For a general approach see appendix \ref{sec:general_spin_tensor_structures}.

Instead of defining tensor structures and scalar components of the amplitude as in \eqref{eq:tensor_structures} one can evaluate the full amplitude in a particular frame. The standard choice for this frame is the center of mass (COM) defined as
\begin{equation}
\label{eq:COM_frame}
\begin{aligned}
&p_1^{\text{com}}=(E_1,0,0,+\myP),\\
&p_2^{\text{com}}=(E_2,0,0,-\myP),\\
&p_3^{\text{com}}=(E_3,+\myP'\sin \theta ,0,+\myP'\cos \theta),\\
&p_4^{\text{com}}=(E_4,-\myP'\sin \theta ,0,-\myP'\cos \theta).
\end{aligned}
\end{equation}
Here the angle $\theta\in[0,\pi]$. All the parameters in \eqref{eq:COM_frame} can be expressed in terms of the Mandelstam variables. In the simplest case where all four particles have the same mass $m$ we have
\begin{equation}
\label{eq:COM_parameters}
E_i=\frac{\sqrt{s}}{2},\quad
\myP=\myP'=\sqrt{\frac{s}{4}-m^2},\quad
\sin\theta =\frac{2\sqrt{tu}}{s-4m^2},\quad
\cos\theta =\frac{t-u}{s-4m^2}.
\end{equation}
Instead of using $(s,t,u)$ variables to characterize the scattering process one can also use $(s,\theta)$ by using the relations
\begin{equation}
\label{eq:COM_parameters_2}
t = - \frac{s-4m^2}{2}(1-\cos\theta),\quad
u = - \frac{s-4m^2}{2}(1+\cos\theta).
\end{equation}
From these expressions it is clear that the physical range of the Mandelstam variables is
\begin{equation}
\label{eq:physical_range}
s\geq 4m^2,\quad
t\in[4m^2-s,\,0],\quad
u\in[4m^2-s,\,0].
\end{equation}
For the definition of the center of mass frame for the most generic case see appendix \ref{sec:four_point_amplitudes_COM}.

Using the COM frame we can define the scalar components of the interacting scattering amplitude in either of the two equivalent ways\footnote{In order to see the equivalence of two definitions simply compare \eqref{eq:scattering_amplitude_interacting} evaluated in the COM frame \eqref{eq:COM_frame} with \eqref{eq:2PS_com}.}
\begin{equation}
\label{eq:scalar_Tamplitude_definition}
\begin{aligned}
T_{\lambda_1, \lambda_2}^{\lambda_3,\lambda_4}(s,t,u) &\equiv T_{\lambda_1, \lambda_2}^{\lambda_3,\lambda_4}(p^{\text{com}}_1,p^{\text{com}}_2,p^{\text{com}}_3,p^{\text{com}}_4),\\
T_{\lambda_1, \lambda_2}^{\lambda_3,\lambda_4}(s,t,u)\times(2\pi)^4\delta^{(4)}(0) &\equiv
{}_{free}\langle (\myP',\theta,0); \lambda_3, \lambda_4 |T|(\myP,0,0); \lambda_1, \lambda_2 \rangle_{free},
\end{aligned}
\end{equation}
As in \eqref{eq:tensor_structures} we abuse notation and call the full (interacting) amplitude in a generic frame and its particular form in the COM frame by the same symbol. The difference should always be clear from the arguments. 
Given the interacting scattering amplitude in the COM frame one can unambiguously obtain the interacting scattering amplitude in a generic frame by using \eqref{eq:T_matrix_amplitude_transformation_general}. As an example, let us apply a rotation by an angle $\phi$ around the z axis to \eqref{eq:scalar_Tamplitude_definition}. One gets the following relation
\begin{multline}
\label{eq:scalar_Tamplitude_nonzero_phi}
e^{i(\lambda_1-\lambda_2-\lambda_3 -\lambda_4)\phi} T_{\lambda_1, \lambda_2}^{\lambda_3,\lambda_4}(s,t,u)\times(2\pi)^4\delta^{(4)}(0)=\\
{}_{free}\langle (\myP',\theta, \phi); \lambda_3, \lambda_4 |T|(\myP,0,0); \lambda_1, \lambda_2 \rangle_{free}
\end{multline}
This result will be useful when we compute the partial wave decomposition of the scattering amplitude in section \ref{sec:partial_amplitudes}.

We note that the scalar components of the scattering amplitude defined in \eqref{eq:tensor_structures} and the COM frame amplitude defined in \eqref{eq:scalar_Tamplitude_definition} are simply related by a linear transformation which depends only on $s$, $t$ and $u$ variables. This relation can be found by evaluating \eqref{eq:tensor_structures} in the center of mass frame and comparing with \eqref{eq:scalar_Tamplitude_definition}.  We will see an explicit example of this in section \ref{sec:example_tensor_structures}.

\subsubsection*{Parity and time-reversal}
Let us now discuss additional constraints which appear if the system is parity or time-reversal invariant. 
In terms of the $S$  operator (and hence also the $T$ operator), the following must hold
\begin{equation}
\label{eq:pt-constraints}
\parity S \parity^\dag = S,\quad
\time S \time^\dag = S^\dag.
\end{equation}
At the level of COM amplitudes these translate into the following conditions
\begin{align}
\label{eq:amplitudes_parity}
T_{12\rightarrow 34}{}_{\lambda_1,\lambda_2}^{\lambda_3, \lambda_4}(s,t,u) &= \eta_1 \eta_2 \eta_3^* \eta_4^* (-1)^{j_1+j_2+j_3+j_4}(-1)^{\lambda_1 + \lambda_2 +\lambda_3 + \lambda_4} T_{12\rightarrow 34}{}_{-\lambda_1,-\lambda_2}^{-\lambda_3, -\lambda_4}(s,t,u),\\
\label{eq:amplitudes_time_reversal}
T_{12\rightarrow 34}{}_{\lambda_1,\lambda_2}^{\lambda_3, \lambda_4}(s,t,u) &= \varepsilon_1^* \varepsilon_2^* \varepsilon_3 \varepsilon_4 (-1)^{\lambda_1 - \lambda_2 -\lambda_3 + \lambda_4} T_{34\rightarrow 12}{}_{\lambda_3,\lambda_4}^{\lambda_1, \lambda_2}(s,t,u).
\end{align}
We derive them in appendix \ref{app:parity_and_timereversal}.

\subsubsection*{Crossing}
Our goal now is to formulate crossing equations. The case of particles with generic spin was first addressed in  \cite{Trueman:1964zza}. It was further discussed in \cite{Martin:102663,Hara:1970gc,Hara:1971kj}. For a recent discussion see also \cite{deRham:2017zjm}. All the results presented below are carefully derived in appendix \ref{app:crossing_equations}.

\noindent Consider the scattering process of four particles. We denote it schematically by
\begin{equation}
\label{eq:three_scattering_process}
12\rightarrow 34.
\end{equation}
Each particle is characterized by its mass, spin, helicity and 3-momentum, for instance $1 = (m_1,j_1,\lambda_1; \vec p_1)$.
There exist five other related process
\begin{equation}
\label{eq:three_scattering_processes}
\begin{aligned}
\bar 42\rightarrow 3\bar 1,\qquad\qquad
\bar 32\rightarrow \bar 1 4,\\
1 \bar 3\rightarrow \bar 2 4,\qquad\qquad
1 \bar 4\rightarrow 3\bar 2.
\end{aligned}
\end{equation}
together with $\bar 3 \bar 4\rightarrow \bar 1 \bar 2$.
Here if particle $i$ has a charge (or more generally transforms in some representation of a global group) then particle $\bar i$ has the opposite charge (transforms in the conjugate representation). In other words $\bar i$ is the antiparticle of particle $i$. The scattering process \eqref{eq:three_scattering_process} is described by the following interacting part of the amplitude
\begin{equation}
\label{eq:three_scattering_amplitude}
T_{12\rightarrow 34}{\;}_{\lambda_1, \lambda_2}^{\lambda_3,\lambda_4}(p_1, p_2, p_3, p_4),
\end{equation}
whereas the scattering processes in \eqref{eq:three_scattering_processes} are described by
\begin{equation}
\label{eq:three_scattering_amplitudes}
\begin{aligned}
T_{\bar 42\rightarrow 3\bar 1}{\;}_{\lambda_4, \lambda_2}^{\lambda_3,\lambda_1}(p_4, p_2, p_3, p_1),\qquad
T_{\bar{3}2\rightarrow \bar{1}4}{\;}_{\lambda_3, \lambda_2}^{\lambda_1,\lambda_4}(p_3, p_2, p_1, p_4),\\
T_{1 \bar 3\rightarrow \bar 2 4}{\;}_{\lambda_1, \lambda_3}^{\lambda_2,\lambda_4}(p_1, p_3, p_2, p_4),\qquad
T_{1 \bar 4\rightarrow 3\bar 2}{\;}_{\lambda_1, \lambda_4}^{\lambda_3,\lambda_2}(p_1, p_4, p_3, p_2).
\end{aligned}
\end{equation}
Here all 4-momenta have positive energies $p_i^0 > 0$ and are on-shell \eqref{eq:on_shell}.

Under the assumption that the amplitudes in \eqref{eq:three_scattering_amplitude} and \eqref{eq:three_scattering_amplitudes} can be analytically continued in $p_i$ and defined in some common domain of $p_i$ values, one can write a set of crossing equations
\begin{align}
\label{eq:s_t_crossing_general_14}
T_{12\rightarrow 34}{}_{\lambda_1, \lambda_2}^{\lambda_3, \lambda_4}(p_1, p_2, p_3, p_4) &= \epsilon_{14}
T_{\bar 4 2\rightarrow 3\bar 1}{}_{-\lambda_4, +\lambda_2}^{+\lambda_3, -\lambda_1}(-p_4, p_2, p_3, -p_1),\\
\label{eq:s_t_crossing_general_23}
T_{12\rightarrow 34}{}_{\lambda_1, \lambda_2}^{\lambda_3, \lambda_4}(p_1, p_2, p_3, p_4) &=\epsilon_{23} T_{1\bar 3\rightarrow \bar 24}{}_{+\lambda_1, -\lambda_3}^{-\lambda_2, +\lambda_4}(p_1, -p_3, -p_2, p_4).
\end{align}
which are referred to as the $s-t$ crossing equations and
\begin{align}
\label{eq:s_u_crossing_general_13}
T_{12\rightarrow 34}{}_{\lambda_1, \lambda_2}^{\lambda_3, \lambda_4}(p_1, p_2, p_3, p_4) &= \epsilon_{13}
T_{\bar 3 2\rightarrow \bar 1 4}{}_{-\lambda_3, +\lambda_2}^{-\lambda_1, +\lambda_4}(-p_3, p_2, -p_1, p_4),\\
\label{eq:s_u_crossing_general_24}
T_{12\rightarrow 34}{}_{\lambda_1, \lambda_2}^{\lambda_3, \lambda_4}(p_1, p_2, p_3, p_4) &= \epsilon_{24}
T_{1\bar 4\rightarrow 3\bar 2}{}_{+\lambda_1, -\lambda_4}^{+\lambda_3, -\lambda_2}(p_1, -p_4, p_3, -p_2),
\end{align} 
which are referred to as the $s-u$ crossing equations. Here $\epsilon_{14}$, $\epsilon_{23}$, $\epsilon_{13}$ and $\epsilon_{24}$ are some phases which can be fixed in principle using the LSZ procedure. Finally the amplitude for the process $\bar 3 \bar 4\rightarrow \bar 1 \bar 2$ can be related to that of the $12\rightarrow34$ process by using \eqref{eq:s_u_crossing_general_13} and \eqref{eq:s_u_crossing_general_24} one after the other. There are two distinct ways for the analytic continuation of amplitudes discussed in section \ref{app:analytic_continuation}. In writing \eqref{eq:s_t_crossing_general_14} - \eqref{eq:s_u_crossing_general_24} we have made a particular choice, more precisely the one given by \eqref{eq:ac_2}.

Let us now focus on the 23- and 24-crossing equations given by \eqref{eq:s_t_crossing_general_23} and \eqref{eq:s_u_crossing_general_24} respectively and evaluate them in the standard COM \eqref{eq:COM_frame}. We can then use the definition of the center of mass amplitude \eqref{eq:scalar_Tamplitude_definition} in the left-hand side of \eqref{eq:s_t_crossing_general_23} and \eqref{eq:s_u_crossing_general_24} but not in the right-hand side. The right-hand side is not in the center of mass frame of particles 13 and 14 respectively. In order to bring them to this frame we need to perform a Lorentz transformation. The 23-crossing equation then reads
\begin{align}
\nn
T_{12\rightarrow 34}{\;}_{\lambda_1, \lambda_2}^{\lambda_3,\lambda_4}(s,t,u) &=
\epsilon_{23}'\,\sum_{\lambda_i'}e^{i\pi(\lambda_1'+\lambda_4')}\\
&\times d^{(j_1)}_{\lambda'_1\lambda_1}(\alpha_1)
d^{(j_2)}_{\lambda'_2\lambda_2}(\alpha_2)
d^{(j_3)}_{\lambda'_3\lambda_3}(\alpha_3)
d^{(j_4)}_{\lambda'_4\lambda_4}(\alpha_4)
T_{1\bar{3}\rightarrow \bar{2}4}{\;}_{\lambda_1', \lambda_3'}^{\lambda_2',\lambda_4'}(t,s,u),
\label{eq:st_crossing}
\end{align}
where the angles $\alpha_i$ in the equal mass case are given by
\begin{equation}
\label{eq:s-t:equal_mass}
\begin{aligned}
+\cos\alpha_1 = 
-\cos\alpha_2 =
-\cos\alpha_3 =
+\cos\alpha_4 &=
+\frac{s t}{\sqrt{s(s-4m^2)}\sqrt{t(t-4m^2)}},\\
+\sin\alpha_1 = 
-\sin\alpha_2 =
+\sin\alpha_3 =
-\sin\alpha_4 &=
-\frac{2m\,\sqrt{stu}}{\sqrt{s(s-4m^2)}\sqrt{t(t-4m^2)}}.
\end{aligned}
\end{equation}
For the most general case see \eqref{eq:cos_alpha} and \eqref{eq:sin_alpha}.
Similarly the $24$-crossing reads
\begin{align}
\nn
T_{12\rightarrow 34}{\;}_{\lambda_1, \lambda_2}^{\lambda_3,\lambda_4}(s,t,u) &=
\epsilon_{24}'\,\sum_{\lambda_i'}e^{i\pi(\lambda_1'+\lambda_3')}\\
&\times d^{(j_1)}_{\lambda'_1\lambda_1}(\beta_1)
d^{(j_2)}_{\lambda'_2\lambda_2}(\beta_2)
d^{(j_3)}_{\lambda'_3\lambda_3}(\beta_3)
d^{(j_4)}_{\lambda'_4\lambda_4}(\beta_4)
T_{1\bar 4\rightarrow 3 \bar 2}{\;}_{\lambda_1', \lambda_4'}^{\lambda_3',\lambda_2'}(u,t,s),
\label{eq:su_crossing}
\end{align}
where the angles $\beta_i$ in the equal mass case are given by
\begin{equation}
\label{eq:s-u:equal_mass}
\begin{aligned}
+\cos\beta_1 = 
-\cos\beta_2 =
+\cos\beta_3 =
-\cos\beta_4 &=
+\frac{s u}{\sqrt{s(s-4m^2)}\sqrt{u(u-4m^2)}},\\
+\sin\beta_1 = 
-\sin\beta_2 =
-\sin\beta_3 =
+\sin\beta_4 &=
+\frac{2m\,\sqrt{stu}}{\sqrt{s(s-4m^2)}\sqrt{u(u-4m^2)}}.
\end{aligned}
\end{equation}
For the most general case see \eqref{eq:cos_beta} and \eqref{eq:sin_beta}.
Notice that small Wigner d-matrices are $4\pi$ periodic. As a result the angles $\alpha_i$ and $\beta_i$ are not completely fixed by the equations \eqref{eq:s-t:equal_mass} and \eqref{eq:s-u:equal_mass} since sines and cosines are $2\pi$ periodic. Different choices of $\alpha_i$ and $\beta_i$  in the crossing equations satisfying \eqref{eq:s-t:equal_mass} and \eqref{eq:s-u:equal_mass} lead to different additional phases which we combine with $\epsilon_{23}$ and $\epsilon_{24}$ introduced in \eqref{eq:s_t_crossing_general_23} and \eqref{eq:s_u_crossing_general_24} and denote by $\epsilon_{23}'$ and $\epsilon_{24}'$. For each  scattering process these phases can be computed at the leading order in perturbation theory and since the phases are purely kinematic in nature, the result can then be used non-perturbatively.
We also notice that the form of crossing equations in a general frame,  \eqref{eq:s_t_crossing_general_23} and \eqref{eq:s_u_crossing_general_24}, depends on the choice of the analytic continuation, however both analytic continuations lead to the same expressions in the center of mass frame \eqref{eq:st_crossing} and \eqref{eq:su_crossing}.

It is interesting to consider the case when all four particles are massless. Assuming the physical domain \eqref{eq:physical_range} of the Mandelstam variables for the process $12\rightarrow34$ in the limit $m\rightarrow 0$ the expressions \eqref{eq:s-t:equal_mass} and \eqref{eq:s-u:equal_mass} lead to
\begin{align}
&\alpha_1=\pi,\quad
\alpha_2=0,\quad
\alpha_3=0,\quad
\alpha_4=\pi,\\
&\beta_1=\pi,\quad
\beta_2=0,\quad
\beta_3=\pi,\quad
\beta_4=0.
\end{align}
Using the following properties of the small Wigner d-matrices
\begin{equation}
\label{eq:d_propertiese}
d^{(j)}_{\lambda'\lambda}(0)=\delta_{\lambda',\lambda},\qquad
d^{(j)}_{\lambda'\lambda}(\pi)=(-1)^{j-\lambda}\delta_{\lambda',\,-\lambda}
\end{equation}
we get the following crossing equations
\begin{align}
\label{eq:st_crossing_massless}
T_{12\rightarrow 34}{\;}_{\lambda_1, \lambda_2}^{\lambda_3,\lambda_4}(s,t,u) &=
\epsilon_{23}''
T_{1\bar 3\rightarrow \bar 2 4}{\;}_{-\lambda_1, +\lambda_3}^{+\lambda_2,-\lambda_4}(t,s,u),\\
\label{eq:su_crossing_massless}
T_{12\rightarrow 34}{\;}_{\lambda_1, \lambda_2}^{\lambda_3,\lambda_4}(s,t,u) &=
\epsilon_{24}''
T_{1\bar{4}\rightarrow 3\bar{2}}{\;}_{-\lambda_1, +\lambda_4}^{-\lambda_3,+\lambda_2}(u,t,s),
\end{align}
where $\epsilon_{23}''$ and $\epsilon_{24}''$ are some new phases.

\subsubsection*{Neutral identical particles}
\label{sec:idnetical_particles}
In some practical applications one is required to study scattering processes of identical neutral particles with mass $m$ and spin $j$. We discuss this case in great detail in appendix \ref{app:identical_particles}. Here we state only the most important results.

We define the scattering amplitude of identical neutral particles in a generic and in the center of mass frames as
\begin{equation}
\label{eq:scattering_amplitude_interacting_id}
\begin{aligned}
T_{\lambda_1, \lambda_2}^{\lambda_3,\lambda_4}(p_1,p_2,p_3,p_4)
&\times(2\pi)^4\delta^{(4)}(p_1^\mu + p_2^\mu - p_3^\mu - p_4^\mu)\equiv
{}_{id\;free}\langle\kappa_3,\kappa_4|T|\kappa_1,\kappa_2\rangle_{id\;free},\\
T_{\lambda_1, \lambda_2}^{\lambda_3,\lambda_4}(s,t,u)
&\times(2\pi)^4\delta^{(4)}(0)\equiv
{}_{id\;free}\langle (\myP',\theta,0); \lambda_3, \lambda_4 |T|(\myP,0,0); \lambda_1, \lambda_2 \rangle_{id\;free}.
\end{aligned}
\end{equation}
In comparison with \eqref{eq:scattering_amplitude} we drop the subscript $12\rightarrow 34$ since it does not carry any useful information anymore.
The two particle states formed from identical particles were defined in \eqref{eq:2PS_identical}. According to  \eqref{eq:id_symmetry_property} they are (anti)symmetric under the exchange of particles 12 and 34. Due to this condition the following equations hold
\begin{align}
\label{eq:identical_particles_amplitudes1}
T_{\lambda_1,\lambda_2}^{\lambda_3, \lambda_4}(s,t,u) &=(-1)^{-\lambda_1 + \lambda_2 - \lambda_3 +\lambda_4} T_{\lambda_2,\lambda_1}^{\lambda_3, \lambda_4}(s,u,t),\\
\label{eq:identical_particles_amplitudes2}
T_{\lambda_1,\lambda_2}^{\lambda_3, \lambda_4}(s,t,u) &=(-1)^{+\lambda_1 - \lambda_2 - \lambda_3 +\lambda_4} T_{\lambda_1,\lambda_2}^{\lambda_4, \lambda_3}(s,u,t).
\end{align}
See appendix \ref{app:amplitudes_id} for details.
Since the variables $t$ and $u$ are flipped in the left- and right-hand side we refer to them as the t-u crossing equations. Applying them twice we get the following kinematic constraint
\begin{equation}
\label{eq:(12)(34)}
T_{\lambda_1, \lambda_2}^{\lambda_3,\lambda_4}(s,t,u)   =
T_{+\lambda_2, +\lambda_1}^{+\lambda_4,+\lambda_3}(s,t,u).
\end{equation}

In the case of distinct particles the crossing equations \eqref{eq:st_crossing} and \eqref{eq:su_crossing} establish relations between different amplitudes. When particles are identical there is only a single amplitude \eqref{eq:scattering_amplitude_interacting_id} and the crossing equations become constraints on this single amplitude.

One can also combine together \eqref{eq:s_u_crossing_general_13}, \eqref{eq:s_u_crossing_general_24} and \eqref{eq:s_t_crossing_general_14}, \eqref{eq:s_t_crossing_general_23} to obtain
\begin{align}
\label{eq:(13)(24)_gen}
T_{\lambda_1, \lambda_2}^{\lambda_3,\lambda_4}(p_1,p_2,p_3,p_4)  & = 
\,T_{-\lambda_3, -\lambda_4}^{-\lambda_1,-\lambda_2}(-p_3,-p_4,-p_1,-p_2),\\
\label{eq:(14)(23)_gen}
T_{\lambda_1, \lambda_2}^{\lambda_3,\lambda_4}(p_1,p_2,p_3,p_4)  & =
\,T_{-\lambda_4, -\lambda_3}^{-\lambda_2,-\lambda_1}(-p_4,-p_3,-p_2,-p_1).
\end{align}
Using the techniques of appendix \ref{app:crossing_COM_frame} one can bring both sides of these two equations to the center of mass frame and show that
\begin{align}
\label{eq:(13)(24)}
T_{\lambda_1, \lambda_2}^{\lambda_3,\lambda_4}(s,t,u)  & = 
\,T_{-\lambda_3, -\lambda_4}^{-\lambda_1,-\lambda_2}(s,t,u),\\
\label{eq:(14)(23)}
T_{\lambda_1, \lambda_2}^{\lambda_3,\lambda_4}(s,t,u)  & =
\,T_{-\lambda_4, -\lambda_3}^{-\lambda_2,-\lambda_1}(s,t,u).
\end{align}
We refer to the conditions \eqref{eq:(12)(34)}, \eqref{eq:(13)(24)} and \eqref{eq:(14)(23)} as the kinematic constraints associated to the simultaneous permutation of particles (12)(34), (13)(24) and (14)(23) respectively.

As presented, the transition from \eqref{eq:(13)(24)_gen}, \eqref{eq:(14)(23)_gen} to \eqref{eq:(13)(24)}, \eqref{eq:(14)(23)} is very difficult. Strictly speaking the equations \eqref{eq:(13)(24)} and \eqref{eq:(14)(23)} might have some overall helicity independent phase which we have little control over. There is a much simpler way of deriving \eqref{eq:(13)(24)} however. As discussed in appendix \ref{app:PT} one can use the CPT theorem to obtain \eqref{eq:(13)(24)}. The constraint \eqref{eq:(14)(23)} follows from combining \eqref{eq:(13)(24)} with \eqref{eq:(12)(34)}. This discussion indicates (but not proves) that CPT transformation is equivalent to using crossing twice.

\subsection{Counting scattering amplitudes}
\label{sec:counting}
It is useful to count kinematically independent amplitudes in various cases. Looking at the definition of the center of mass amplitude \eqref{eq:scalar_Tamplitude_definition} it is obvious that the number of all possible amplitudes $N_4$ for four different massive particles is
\begin{equation}
N_4=(2j_1 +1)(2j_2 +1)(2j_3 +1)(2j_4 +1),
\end{equation}
since there are no restrictions on the helicity values. Using \eqref{eq:amplitudes_parity} we can further split the amplitudes into parity even and parity odd ones. This is done by taking appropriate linear combinations of
\begin{equation}
\label{eq:amps}
T{}_{\lambda_1,\lambda_2}^{\lambda_3,\lambda_4}
\qquad\text{and}\qquad
T{}_{-\lambda_1,-\lambda_2}^{-\lambda_3,-\lambda_4}.
\end{equation}
Since the scattering amplitude must always contain an even number of fermions and due to \eqref{eq:intrinsic_parities} and the comment below, the product of intrinsic parities
\begin{equation}
\label{eq:product}
\eta_1\eta_2\eta_3^*\eta_4^*
\end{equation}
entering \eqref{eq:amplitudes_parity} is either $+1$ or $-1$\footnote{Parity invariance implies that $\eta_1 \eta_2 = \pm \eta_3 \eta_4$.}. Let us assume for concreteness that the product in \eqref{eq:product} is $+1$. (In the other case when \eqref{eq:product} is $-1$ the role of parity even and odd amplitudes constructed from \eqref{eq:amps} simply flips.)
Having constructed the appropriate linear combination from \eqref{eq:amps} the counting follows straightforwardly. In the case when there are two or four fermions we have the same number of parity even $N_4^+$ and parity odd amplitudes $N_4^-$ which read
\begin{equation}
\label{eq:counting_parity_1}
N_4^\pm=\frac{1}{2}(2j_1 +1)(2j_2 +1)(2j_3 +1)(2j_4 +1).
\end{equation}
In case all particles are bosons we get
\begin{equation}
\label{eq:counting_parity_2}
N_4^\pm=\frac{1}{2}\left((2j_1 +1)(2j_2 +1)(2j_3 +1)(2j_4 +1)\pm 1\right).
\end{equation}
The difference between \eqref{eq:counting_parity_1} and \eqref{eq:counting_parity_1} arises due to the fact that when all particles are bosons there is always a parity even amplitude with all the zero helicities $T{}_{00}^{00}$. The latter is no longer true in the presence of fermions. Clearly, the following is obeyed
\begin{equation}
N_4=N_4^++N_4^-.
\end{equation}
If we impose parity as symmetry of our system only parity even amplitudes will survive.

It is more difficult to perform general counting when particles are identical since we need to take into account the relations \eqref{eq:(12)(34)} - \eqref{eq:(14)(23)}. However it easy to do for any particular case of interest by forming a linear system of all the constraints (due presence of parity, identical and massless particles), solving it and counting the number of independent amplitudes. For example in the case of identical massive Majorana particles and identical massive spin one particles we have
\begin{align}
\label{eq:counting_majorana}
\text{identical Majorana fermions:}\quad &N_4^+=5,\quad\;\;
N_4^-=2,\\
\label{eq:counting_spin=1}
\text{identical spin one bosons:}\quad &N_4^+=17,\quad
N_4^-=10.
\end{align}
Time-reversal does not further reduce these numbers. This can be intuitively understood by noticing that P implies T invariance for neutral particles due to the CPT symmetry.

In case a particle with spin $j$ is massless its helicity can only take two values $+j$ and $-j$. As a result if all four particles in the scattering process are massless and carry a non-zero spin we always have, independently of the precise values of spin, 
\begin{equation}
\text{four different massless particles:}\quad N_4=2^4=16.
\end{equation}
In the case when all particles are identical, massless and carry a non-zero spin we have
\begin{equation}
\label{eq:counting_massless}
\text{identical massless particles:}\quad N_4^+=5,\quad N_4^-=2.
\end{equation}

It was proposed in section 6 of \cite{Costa:2011mg} that the number of scattering amplitudes in $d$ dimensions should be equal to the number of tensor structures of four-point functions in $d-1$ conformal field theories, where massless particles correspond to conserved operators. This correspondence got an explanation in \cite{Kravchuk:2016qvl} where it was noted that the conformal frame analysis of four point function  is equivalent group theoretically to the center of mass analysis of scattering amplitudes. When parity is involved or particles are identical or massless the matching of CFT and amplitude counting is more difficult to confirm. Here we explicitly verify this correspondence on some particular examples. For instance \eqref{eq:counting_parity_1} and \eqref{eq:counting_parity_2} are in a perfect agreement with the formulas (4.47) and (4.49) in \cite{Kravchuk:2016qvl}, results 
\eqref{eq:counting_majorana} and \eqref{eq:counting_spin=1} match (2.40) and (4.58) in \cite{Kravchuk:2016qvl}, finally the very special case \eqref{eq:counting_massless} matches (3.24) in \cite{Kravchuk:2016qvl}.\footnote{The number of parity even conserved  tensor structures in conformal field theories was first computed in \cite{Dymarsky:2013wla}, see table 1. In $d=3$ it is 5 as expected.}

\subsection{Partial amplitudes}
\label{sec:partial_amplitudes}

As we explained in section \ref{sec:two_particle_states}, the two particle states are in a reducible representation of the Poincar\'e group and can be expressed as a direct sum of Poincar\'e irreps according to \eqref{eq:2PS_decomposition_general} or \eqref{eq:2PS_decomposition_com} (in the special case of COM states). This leads to a decomposition (often referred to as the partial wave decomposition) of scattering amplitudes into partial amplitudes.

We start from the definition of the center of mass amplitude \eqref{eq:scalar_Tamplitude_nonzero_phi} and decompose the two particle states there according to \eqref{eq:2PS_decomposition_com}. As a result we get
\begin{multline}
\label{eq:decomposition}
e^{i(\lambda_1-\lambda_2-\lambda_3 -\lambda_4)\phi} T_{\lambda_1, \lambda_2}^{\lambda_3,\lambda_4}(s,t,u) 
\times (2 \pi)^4 \delta^{(4)}(0)=
\sum_{\ell, \ell', \lambda, \lambda'} C_{\ell'}(\myP') C_\ell(\myP)\\
e^{-i(\lambda_3+\lambda_4-\lambda)\phi}
d^{(\ell')}_{\lambda' \lambda_{34}}(\theta) \, d^{(\ell)}_{\lambda \lambda_{12}}(0)
\< c, 0, \spin', \hel'; \lambda_3, \lambda_4  |T|c, 0, \spin, \hel; \lambda_1, \lambda_2  \rangle,
\end{multline}
where we have defined
\begin{equation}
\label{eq:lambdas}
\lambda_{12}\equiv\lambda_1-\lambda_2,\quad
\lambda_{34}\equiv\lambda_3-\lambda_4.
\end{equation}
The coefficient $C_\ell(\myP)$ was computed in \eqref{eq:coefficient_Cj}. Using it we can write
\begin{equation}
C_\ell(\myP')C_\ell(\myP) =  4 \pi (2\ell+1)\frac{\sqrt{s}}{\sqrt{\myP \myP'}}.
\end{equation}
Due to the standard property of the small Wigner d-matrix
\begin{equation}
d^{(\ell)}_{\lambda \lambda_{12}}(0)= \delta_{\lambda \lambda_{12}}
\end{equation}
the dependence on the azimuthal angle $\phi$ in \eqref{eq:decomposition} cancels out on both sides and we obtain
\begin{equation*}
T_{\lambda_1, \lambda_2}^{\lambda_3,\lambda_4}(s,t,u)
\times (2 \pi)^4 \delta^{(4)}(0) =\sum_{\ell, \ell',  \lambda'} C_{\ell'}(\myP') C_\ell(\myP)
d^{(\ell')}_{\lambda' \lambda_{34}}(\theta) 
\< c, 0, \spin', \hel'; \lambda_3, \lambda_4  |T|c, 0, \spin, \lambda_{12}; \lambda_1, \lambda_2  \rangle.
\end{equation*}
Using the definition of the partial amplitudes \eqref{eq:partial_amplitude_interacting} the above can be written in its final form
\begin{align}
\label{eq:partial_wave_decomposition_T}
T_{\lambda_1, \lambda_2}^{\lambda_3,\lambda_4}(s,t,u) &= \sum_{\ell} C_\ell(\myP')C_\ell(\myP) d^{(\ell)}_{\lambda_{12} \lambda_{34}}(\theta) T_\ell{}^{\lambda_3, \lambda_4}_{\lambda_1, \lambda_2}(s).
\end{align}
By using orthogonality of the small Wigner d matrix the decomposition \eqref{eq:partial_wave_decomposition_T} can be inverted and leads to
\begin{align}
\label{eq:partial_T}
T_\ell{}^{\lambda_3, \lambda_4}_{\lambda_1, \lambda_2}(s)  &= \frac{2\ell +1}{2 C_\ell(\myP')C_\ell(\myP)}\times \int_{0}^{\pi} d\theta \sin\theta d^{(\ell)}_{\lambda_{12} \lambda_{34}}(\theta)T_{\lambda_1, \lambda_2}^{\lambda_3,\lambda_4}(s,t,u).
\end{align}
Note that $t$ and $u$ are functions of $s$ and $\theta$. In the equal mass case one has for instance \eqref{eq:COM_parameters_2}.

Analogously to \eqref{eq:partial_T} one can also write the decomposition of the full amplitude
\begin{align}
\label{eq:partial_S}
S_\ell{}^{ \lambda_3, \lambda_4}_{\lambda_1, \lambda_2}(s) &= \frac{2\ell +1}{2 C_\ell(\myP')C_\ell(\myP)} \times
\int_0^{2\pi}\frac{d\phi}{2\pi}
\int_{0}^{\pi} d\theta \sin\theta d^{(\ell)}_{\lambda_{12} \lambda_{34}}(\theta)S_{\lambda_1, \lambda_2}^{\lambda_3,\lambda_4}(s,t,u).
\end{align}
We need to introduce the integration over the azimuthal angle $\phi$ because the disconnected part of the scattering amplitude depends on it, see \eqref{eq:relations_ST} and footnote \ref{foot:distribution}.
A simple relation between \eqref{eq:partial_T} and \eqref{eq:partial_S} partial amplitudes follows from \eqref{eq:Sj_Tj_relation} and \eqref{eq:kronecker_gamma}. It reads\footnote{\label{foot:simple_fact}Here we have used a simple fact that
$\delta_{\lambda_i\lambda_k}\delta_{\lambda_j\lambda_l}
\delta_{\lambda_{ij}\lambda_{kl}}
=\delta_{\lambda_i\lambda_k}\delta_{\lambda_j\lambda_l}.$}
\begin{equation}
\label{eq:pamp_relation}
S_\ell{}^{\lambda_3, \lambda_4}_{\lambda_1, \lambda_2}(s)  =
\delta_{m_1m_3}\delta_{m_2m_4}
\delta_{j_1j_3}\delta_{j_2j_4}
\delta_{\lambda_1\lambda_3}\delta_{\lambda_2\lambda_4}
+ i T_\ell{}^{\lambda_3, \lambda_4}_{\lambda_1, \lambda_2}(s).
\end{equation}
As a consistency check one can obtain this relation in a different way. One can plug \eqref{eq:relations_ST} evaluated in the COM frame together with the very first equation in the footnote  \ref{foot:distribution} into \eqref{eq:partial_S}. The delta functions cancel all the integrals in \eqref{eq:partial_S} and we simply arrive at \eqref{eq:pamp_relation}.

\paragraph{Identical particles}
In case either the incoming particles or the outgoing particles are identical we also have relations between the partial amplitudes due to \eqref{eq:irrep_id}. If the incoming particles 1 and 2 are identical we get
\begin{equation}
\label{eq:partial_id_12}
S_\ell{}^{ \lambda_3, \lambda_4}_{\lambda_1, \lambda_2}(s) = (-1)^{\lambda_2 - \lambda_1 - \ell} S_\ell{}^{\lambda_3, \lambda_4}_{\lambda_2, \lambda_1}(s).
\end{equation}
Similarly if the outgoing particles 3 and 4 are identical we get
\begin{equation}
\label{eq:partial_id_34}
S_\ell{}^{\lambda_3, \lambda_4}_{\lambda_1, \lambda_2}(s) = (-1)^{\lambda_4 - \lambda_3 - \ell} S_\ell{}^{\lambda_4, \lambda_3}_{\lambda_1, \lambda_2}(s).
\end{equation}
For the case of all four particles being identical, we get $\sqrt{\myP\myP'}=\sqrt{s/4-m^2}$ due to \eqref{eq:COM_parameters}, and thus  
\begin{align}
\label{eq:partial_S_id}
S_\ell{}^{ \lambda_3, \lambda_4}_{\lambda_1, \lambda_2}(s)  &=
\frac{\sqrt{s-4m^2}}{32\pi \sqrt{s}} \times
\int_0^{2\pi}\frac{d\phi}{2\pi}
\int_{0}^{\pi} d\theta \sin\theta d^{(\ell)}_{\lambda_{12} \lambda_{34}}(\theta)S_{\lambda_1, \lambda_2}^{\lambda_3,\lambda_4}(s,t,u),\\
\label{eq:partial_T_id}
T_\ell{}^{ \lambda_3, \lambda_4}_{\lambda_1, \lambda_2}(s)  &=
\frac{\sqrt{s-4m^2}}{32\pi \sqrt{s}} \times \int_{0}^{\pi} d\theta \sin\theta d^{(\ell)}_{\lambda_{12} \lambda_{34}}(\theta)T_{\lambda_1, \lambda_2}^{\lambda_3,\lambda_4}(s,t,u).
\end{align}
Notice that we have used here \eqref{eq:2PS_decomposition_com_id} which contains an additional $\sqrt{2}$ factor compared to a non-identical particle case.
Analogously to \eqref{eq:pamp_relation} there is a simple relation between the $S$ and $T$ partial amplitudes that follows from \eqref{eq:Sj_Tj_relation} and \eqref{eq:convention_gamma_id}. It reads\footnote{See footnote \ref{foot:simple_fact}.}
\begin{equation}
\label{eq:partial_S_T_relations}
S_\ell{}^{ \lambda_3, \lambda_4}_{\lambda_1, \lambda_2}(s)  =
\frac{1}{2}\left(
\delta_{\lambda_1\lambda_3}\delta_{\lambda_2\lambda_4}+
(-1)^{\ell-\lambda_{34}}
\delta_{\lambda_1\lambda_4}\delta_{\lambda_2\lambda_3}
\right)
+iT_\ell{}^{ \lambda_3, \lambda_4}_{\lambda_1, \lambda_2}(s).
\end{equation}
This result can also be obtained by plugging \eqref{eq:relations_ST} evaluated in the COM frame together with the second equation in the footnote  \ref{foot:distribution} into \eqref{eq:partial_S_id}. We also notice that in the case of identical scalar particles we recover the standard result, see for example equation (10) in \cite{Paulos:2017fhb}.

\paragraph{Parity and time reversal}
As usual parity and time reversal invariance lead to additional constraints. The parity constraint follows directly from \eqref{eq:parity_pwa_app} and reads
\begin{equation}
\label{eq:parity_pwa}
S_\ell{}^{\lambda_3, \lambda_4}_{\lambda_1, \lambda_2}(s)= \eta_1 \eta_2 \eta_3^* \eta_4^* (-1)^{j_1 - j_2 + j_3 - j_4 - 2\ell} S_\ell{}^{-\lambda_3, -\lambda_4}_{-\lambda_1, -\lambda_2}(s).
\end{equation}
Similarly the time reversal constraint follows from \eqref{eq:time_pwa} and reads
\begin{equation}
\label{eq:time_reversal_pwa}
S_{12\to 34 \;\ell}{}^{\lambda_3, \lambda_4}_{\lambda_1, \lambda_2}(s)= \varepsilon_1^* \varepsilon_2^* \varepsilon_3 \varepsilon_4  S_{34\to 12 \;\ell}{}^{\lambda_1, \lambda_2}_{\lambda_3, \lambda_4}(s).
\end{equation}
Note that we reintroduced the subscripts $12 \rightarrow 34$ and $34 \rightarrow 12$ since Time Reversal is a relation between these two distinct processes.

\subsection{Unitarity}
\label{sec:unitarity}
Unitarity of a quantum theory implies  that the norm of any  state  must be non-negative, \emph{i.e.}  $\forall$ $|\psi\>$ one has $\<\psi|\psi\>\geq0$. Now suppose we are given some set of $N$ states $|\psi_a\>$ 
with $a=1,\ldots,N$. Unitarity then requires that the $N\times N$ hermitian matrix with components $\<\psi_a |\psi_b\>$ is positive semi-definite, namely $\left(\<\psi_a |\psi_b\>\right)\succeq0$. This formulation allows us to impose unitarity constraints on the partial amplitudes straightforwardly.

We start by considering the incoming two particles state (formed from particles 1 and 2) together with the outgoing two particle state (formed from particles 3 and 4).  We decompose each of these states into irreducible representations according to \eqref{eq:2PS_decomposition_general}. 
We then have the following $N=N_\text{in}+N_\text{out}$ independent states transforming in the spin $\ell$ irreducible representation of the Poincar\'e group
\begin{align}
\label{eq:incoming_2ps_wigner}
&|1 \rangle_{\text{in}}  \equiv
|c, \vec p, \spin, \hel; j_1, j_2 \rangle_{\text{in}},\quad
&&|1 \rangle_{\text{out}}  \equiv
|c, \vec p, \spin, \hel; j_3, j_4 \rangle_{\text{out}}, \nonumber\\
&|2 \rangle_{\text{in}}  \equiv
|c, \vec p, \spin, \hel; j_1, j_2 -1 \rangle_{\text{in}},\quad
&&|2 \rangle_{\text{out}}  \equiv
|c, \vec p, \spin, \hel; j_3, j_4 -1 \rangle_{\text{out}}, \nonumber \\
&\qquad\qquad\qquad\vdots &&\qquad\qquad\qquad\vdots \nonumber \\
&|N_{\rm in} \rangle_{\text{in}}  \equiv
|c, \vec p, \spin, \hel; -j_1, -j_2 \rangle_{\text{in}},\quad
&&|N_{\rm out} \rangle_{\text{out}} \equiv
|c, \vec p, \spin, \hel; -j_3, -j_4 \rangle_{\text{out}},
\end{align}
where the number of incoming and outgoing irreducible states is
\begin{equation}
N_\text{in} \equiv (2j_1+1)(2j_2+1),\qquad N_\text{out} \equiv (2j_3+1)(2j_4+1).
\end{equation}
Thus for each spin $\ell$ we can construct the following hermitian $N \times N$ matrix
\begin{equation}
\label{eq:positivity_matrix_1}
\mathcal H_\ell(s) \times(2\pi)^4\delta^4(p'-p)\delta_{\ell'\ell}\delta_{\lambda'\lambda} = \begin{pmatrix}
{}_{\text{in}}\langle a' | b \rangle_{\text{in}} &{}_{\text{in}}\langle a' | b \rangle_{\text{out}}\\
{}_{\text{out}}\langle a' | b \rangle_{\text{in}} & {}_{\text{out}}\langle a' | b \rangle_{\text{out}},
\end{pmatrix}
\end{equation}
where $s=c^2$ is the square of the COM frame total energy and the primes indicate that the conjugated states to \eqref{eq:incoming_2ps_wigner} have all the labels $c$, $p$, $\ell$ and $\lambda$ primed. In defining the matrix $\mathcal H_\ell(s)$ we have also explicitly factored out the overall delta function appearing due to translation invariance. According to the discussion above, unitarity then implies that
\begin{equation}
\label{eq:unitarity_condition}
\mathcal H_\ell(s)\succeq 0,\qquad
\forall \ell\quad\text{ and}\quad s\geq\max(m_1+m_2,m_3+m_4)^2.
\end{equation}

Let us now discuss the components of the matrix \eqref{eq:positivity_matrix_1}. Since the M{\o}ller operators introduced in section \ref{sec:matrix_elements} to define the incoming and outgoing states are isometric, the elements ${}_{\text{in}}\langle a' | b \rangle_{\text{in}}$ and ${}_{\text{out}}\langle a' | b \rangle_{\text{out}}$ are simply fixed by the normalization condition \eqref{eq:irrep_normalization}
\begin{align}
{}_{\text{in}}\langle c_1',\vec{p}^{\,\prime};\spin',\hel'; \lambda_1', \lambda_2'
&|c_1,\vec p;\spin,\hel;\lambda_1, \lambda_2 \rangle_{\text{in}} &=
(2 \pi)^4 
\delta^{(4)}({ p'}^\mu - p^\mu)\,
\delta_{\ell' \ell}\,
\delta_{\lambda' \lambda}\,
\delta_{\lambda_1' \lambda_1}\,
\delta_{\lambda_2' \lambda_2},\\
{}_{\text{out}}\langle c_1',\vec{p}^{\,\prime};\spin',\hel'; \lambda_3', \lambda_4'
&|c_1,\vec p;\spin,\hel;\lambda_3, \lambda_4 \rangle_{\text{out}} &=
(2 \pi)^4 
\delta^{(4)}({ p'}^\mu - p^\mu)\,
\delta_{\ell' \ell}\,
\delta_{\lambda' \lambda}\,
\delta_{\lambda_3' \lambda_3}\,
\delta_{\lambda_4' \lambda_4}.
\end{align}
On the other hand, the matrix elements ${}_{\text{out}}\langle a' | b \rangle_{\text{in}}$ and ${}_{\text{in}}\langle a' | b \rangle_{\text{out}}$ form partial amplitudes according to \eqref{eq:partial_amplitude} 
\begin{align}
{}_{\text{out}}\< c_1', \vec p\,';  \spin', \hel' ; \lambda_3, \lambda_4 &|c_1, \vec p;  \spin, \hel; \lambda_1, \lambda_2 \>_{\text{in}} &= (2\pi)^4\delta^{(4)}(p^\mu - {p'}^\mu)\delta_{\ell \ell'} \delta_{\lambda \lambda'} \, S_\ell{}^{\lambda_3, \lambda_4}_{\lambda_1, \lambda_2}(s),\\
{}_{\text{in}}\< c_1', \vec p';  \spin', \hel' ; \lambda_1, \lambda_2 &|c_1, \vec p;  \spin, \hel; \lambda_3, \lambda_4 \>_{\text{out}} &= (2\pi)^4\delta^{(4)}(p^\mu - {p'}^\mu)\delta_{\ell \ell'} \delta_{\lambda \lambda'} \, S_\ell^*{}^{\lambda_3, \lambda_4}_{\lambda_1, \lambda_2}(s).
\end{align}
Plugging these to \eqref{eq:positivity_matrix_1} we can schematically write the unitarity condition \eqref{eq:unitarity_condition} as
\begin{equation}
\label{eq:positivity_pwa}
\begin{pmatrix}
\delta_{ab} & S_\ell^\dagger{}_{ab} \\
S_\ell{}_{ab} & \delta_{ab} 
\end{pmatrix} \succeq 0.
\end{equation}

\paragraph{Identical particles}
In case either the incoming particles 1 and 2 or outgoing particles 3 and 4 are identical there exist relations \eqref{eq:partial_id_12} and \eqref{eq:partial_id_34} between the partial amplitudes. This introduces redundancies into the condition \eqref{eq:positivity_pwa}. In order to remove these redundancies we restrict our attention to states with $\lambda_1\geq\lambda_2$ (in case of identical incoming particles) and $\lambda_3\geq\lambda_4$ (in case of identical outgoing particles).

\paragraph{Parity invariance}
In the presence of parity invariance various partial amplitudes entering \eqref{eq:positivity_pwa} are related according to \eqref{eq:parity_pwa}. As a consequence the condition \eqref{eq:positivity_pwa} again becomes redundant. One can reformulate the condition \eqref{eq:positivity_pwa} in an equivalent but less redundant way by considering parity eigenstates. One then repeats the procedure above taking into account the fact that Parity invariance forbids transitions between parity even and parity odd states. As a result we get two separate positivity conditions for parity even and parity odd states
\begin{equation}
\mathcal H_\ell^+(s)\succeq 0,\qquad
\mathcal H_\ell^-(s)\succeq 0.
\end{equation}
We will see an explicit example of this in section \ref{sec:example_unitarity}.

\paragraph{Time reversal invariance}
Time reversal invariance relates the scattering amplitudes for the process $12\rightarrow 34$ to the scattering amplitudes for the process $34 \rightarrow 12$. Therefore in general time reversal does not have any implications for the  matrix $\mathcal H_\ell$ since all its elements are scattering amplitudes for the $12 \rightarrow 34$. However in the special case of elastic scattering i.e $12 \rightarrow 12 $, time reversal invariance \eqref{eq:time_reversal_pwa} implies that the sub-matrix $S_{ij}$ is symmetric\footnote{Note that $|\varepsilon_1 \varepsilon_2|^2 = 1$}
\begin{equation}
S_\ell{}^{\lambda_3, \lambda_4}_{\lambda_1, \lambda_2}(s) =
S_\ell{}^{\lambda_1, \lambda_2}_{\lambda_3, \lambda_4}(s)
\end{equation}

\subsection{Kinematic non-analyticities and constraints}
\label{sec:kinematic_non_analiticities}

This section is devoted to the study of the behaviour of COM interacting scattering amplitudes defined in \eqref{eq:scalar_Tamplitude_definition} at some very particular values of the Mandelstam variables $s$, $t$ and $u$. For simplicity we focus on the case of identical particles with mass $m$ and spin $j$.

Using \eqref{eq:COM_parameters} the center of mass frame \eqref{eq:COM_frame} can be written in the following way
\begin{equation}
\label{eq:COM_frame_same_mass}
\begin{aligned}
&p_1^{\text{com}}=\left(\frac{\sqrt s}{2},0,0,+\sqrt{\frac{s}{4}-m^2}\right),\\
&p_2^{\text{com}}=\left(\frac{\sqrt s}{2},0,0,-\sqrt{\frac{s}{4}-m^2}\right),\\
&p_3^{\text{com}}=\left(\frac{\sqrt s}{2},+\frac{\sqrt{tu}}{\sqrt{s-4m^2}} ,0,+\frac{t-u}{2\sqrt{s-4m^2} }\right),\\
&p_4^{\text{com}}=\left(\frac{\sqrt s}{2},-\frac{\sqrt{tu}}{\sqrt{s-4m^2}} ,0,-\frac{t-u}{2\sqrt{s-4m^2}}\right).
\end{aligned}
\end{equation}
The center of mass amplitude \eqref{eq:scalar_Tamplitude_definition} is strictly defined in the physical domain of the Mandelstam variables \eqref{eq:physical_range}. If one attempts however to analytically continue the COM frame amplitudes to arbitrary complex values of $s$, $t$ and $u$, as can be already expected from \eqref{eq:COM_frame_same_mass}, one will encounter non-analyticities (poles and branch points) at 
\begin{equation}
\label{eq:special_points}
s=4m^2,\quad
s=0,\quad
t=0,\quad
u=0.
\end{equation}
Some of these non-analyticities have a purely kinematic nature and have nothing to do with the dynamics of the theory. Our goal here is to isolate them. In what follows we will formulate the problem of kinematic non-analyticities precisely and then discuss each of the special points \eqref{eq:special_points} in detail. (A concrete example will be presented in section \ref{sec:improved_amplitudes}.) For more details on the  subject see chapter 7.3 in \cite{Martin:102663} and references therein.

Recall the definition of helicity states \eqref{poincare_irreps_definition}, two particle center of mass states \eqref{eq:2PS_com} and center of mass amplitudes  \eqref{eq:scalar_Tamplitude_definition}. Using them we can write explicitly the 1PS describing the center of mass scattering process as
\begin{equation}
\label{eq:states_1234}
\begin{aligned}
|m,+\vec p_z; j,\lambda_1 \rangle 
&\equiv e^{-i\eta K_3}
|m,\vec 0; j,\lambda_1 \rangle,\\ 
|m,-\vec p_z; j,\lambda_2 \rangle
&\equiv
e^{-i\pi J_3}e^{-i (\pi-0) J_2} e^{+i\pi J_3}e^{-i\eta K_3}
|m,\vec 0; j, \lambda_2 \rangle, \\
|m,+\vec p_\theta; j,\lambda_3 \rangle 
&\equiv e^{-i \theta J_2} e^{-i\eta K_3} 
|m,\vec 0; j,\lambda_3 \rangle,\\
|m,-\vec p_\theta; j,\lambda_4 \rangle
&\equiv e^{-i\pi J_3}e^{-i (\pi -\theta) J_2} e^{+i\pi J_3} e^{-i\eta K_3} 
|m,\vec 0; j,\lambda_4 \rangle,
\end{aligned}
\end{equation}
where $\vec p_z$ is the 3-momentum in the positive $z$-direction, $\vec p_\theta$ is the 3-momentum in the x-z plane, $J_2$ is the generator of rotations around the $y$-axis, $K_3$ is the boost in the $z$-direction. In the center of mass frame due to \eqref{eq:COM_parameters} the angle $\theta$ and the rapidity $\eta$ defined in \eqref{eq:rapidity} can be related to the Mandelstam variables as follows
\begin{equation}
\label{eq:parameters}
\cos \theta =\frac{t-u}{s-4m^2},\quad
\sin \theta = \frac{2\sqrt{tu}}{s-4m^2},\quad
\sinh \eta =  \frac{\sqrt{s-4m^2}}{2m},\quad
\cosh \eta = \frac{\sqrt{s}}{2m}.
\end{equation} 
The non-analyticities  at $s=4m^2$, $s=0$, $t=0$ and $u=0$ of these expressions 
enter the center of mass amplitude \eqref{eq:scalar_Tamplitude_definition} via \eqref{eq:states_1234}.

The phenomena of kinematic non-analyticities is closely related to the phenomena of kinematic constraints. When defining the COM scattering amplitudes \eqref{eq:scalar_Tamplitude_definition} we have used up all of the Lorentz symmetry to bring the scattering particles to the x-z plane. However at the special points \eqref{eq:special_points} we get an  enhancement of symmetry. For instance at $s=4m^2$ the system is $SO(3)$ symmetric, at $s=0$ the system is $SO(1,1)$ symmetric and at $t=0$ and $u=0$ the system is $SO(2)$ symmetric. This is straightforward to see from \eqref{eq:COM_frame_same_mass}.\footnote{\label{foot:no_poles}Notice that for the physical range of Mandelstam values $t=O(s-4m^2)$ and $u=O(s-4m^2)$ which means that $\cos\theta$ and $\sin\theta$ are finite. This is not the case anymore once we promote $s$ and $t$ to independent complex variables during the analytic continuation process.}
As a consequence of the enhanced symmetry the amplitudes rearrange themselves into irreducible representations of the enhanced symmetry. Only the amplitudes transforming in the trivial representations are allowed to be present, while the rest must vanish. The latter requirement leads to kinematic constraints.\footnote{Kinematic constraints have recently appeared in a similar context in conformal field theories when studying four point correlation functions of local primary operators. They received a proper group theoretic treatment in appendix A of \cite{Kravchuk:2016qvl} (see also appendix D of \cite{Karateev:2019pvw}).}

\paragraph{Special point: $\boldsymbol{s=4m^2}$}
In order to isolate the singular behavior of the COM amplitudes at $s=4m^2$ we perform a simple rewriting of the states \eqref{eq:states_1234} as follows\footnote{The relation
$e^{-i\chi J_2}e^{i \pi J_3}=e^{i \pi J_3}e^{+i\chi J_2}$ is used for the states 2 and 4.}
\begin{equation}
\label{eq:states_1234_modified}
\begin{aligned}
|m,+\vec p_z;j, \lambda_1 \rangle 
&= e^{-i \frac{\theta}{2} J_2} X_+ e^{+i \frac{\theta}{2} J_2}
|m,\vec 0;j, \lambda_1 \rangle, \\
|m,-\vec p_z;j, \lambda_2 \rangle
&= e^{-i \frac{\theta}{2} J_2}e^{+i \pi J_2} X_+ e^{+i \frac{\theta}{2} J_2}
|m,\vec 0;j, \lambda_2 \rangle, \\
|m,+\vec p_\theta;j, \lambda_3 \rangle 
&= e^{-i \frac{\theta}{2} J_2}X_- e^{-i \frac{\theta}{2} J_2}
|m,\vec 0;j, \lambda_3 \rangle, \\
|m,-\vec p_\theta;j, \lambda_4 \rangle
&=e^{-i \frac{\theta}{2} J_2}e^{+i \pi J_2}X_-
e^{-i \frac{\theta}{2} J_2}
|m,\vec 0;j, \lambda_4 \rangle,
\end{aligned}
\end{equation}
where the operators $X_\pm$ are defined as
\begin{equation}
\label{eq:definition_X}
X_\pm\equiv
e^{\pm i \frac{\theta}{2} J_2}e^{-i\eta K_3} e^{\mp i \frac{\theta}{2} J_2} =
e^{-i\left(K_3 \cos \frac{\theta}{2} \mp K_1 \sin \frac{\theta}{2}\right) \eta}.
\end{equation}
In \eqref{eq:definition_X} we have used the commutation properties of the Lorentz generators \eqref{eq:poincare_commutation_4}. Writing
\begin{equation}
\label{eq:half_angle}
\cos \frac{\theta}{2} = \sqrt{\frac{-u}{s-4m^2}},\qquad
\sin \frac{\theta}{2} = \sqrt{\frac{-t}{s-4m^2}},
\end{equation}
we notice that the operators $X_\pm$ are completely regular at $s=4m^2$ since
\begin{equation}
\label{eq:regular}
\eta \cos \frac{\theta}{2}=\frac{\sqrt{-u}}{2m}+O(s-4m^2), \qquad
\eta \sin \frac{\theta}{2}=\frac{\sqrt{-t}}{2m}+O(s-4m^2).
\end{equation}

Plugging the states \eqref{eq:states_1234_modified} into the definition of the COM amplitudes \eqref{eq:scalar_Tamplitude_definition} and using the fact that the scattering operator is invariant under the Poincar\'e transformations\footnote{The same property holds obviously true for the $T$ operators because of \eqref{eq:ST_relation}. The invariance we use here is
$e^{i \frac{\theta}{2} J_2} T e^{-i \frac{\theta}{2} J_2}=T.$} \eqref{eq:poincare_invariance} and the transformation property \eqref{eq:transformation} we can write
\begin{equation}
\label{eq:origin_kinematic}
T_{\lambda_1, \lambda_2}^{\lambda_3, \lambda_4} (s,t,u) =
\sum_{\lambda'}
d^{(j)}_{\lambda'_1 \lambda_1}\left(-\frac{\theta}{2}\right)
d^{(j)}_{\lambda'_2 \lambda_2}\left(-\frac{\theta}{2}\right)
d^{(j)}_{\lambda'_3 \lambda_3}\left(\frac{\theta}{2}\right)
d^{(j)}_{\lambda'_4 \lambda_4}\left(\frac{\theta}{2}\right)
A_{\lambda'_1, \lambda'_2}^{\lambda'_3, \lambda'_4} (s,t,u),
\end{equation}
where the new scattering amplitude $A$ is defined as
\begin{equation}
\label{eq:amplitude_A}
A_{\lambda_1, \lambda_2}^{\lambda_3, \lambda_4} (s,t,u) \equiv
\left(\<\vec 0, \lambda_3|X_-^\dagger\otimes\<\vec 0, \lambda_4|X_-^\dagger e^{-i \pi J_2}\right)
T
\left(X_+|\vec 0, \lambda_1\>\otimes e^{+i \pi J_2} X_+|\vec 0, \lambda_2\>\right).
\end{equation}
From this explicit expression we see that at $s=4m^2$ the amplitude $A$ is completely regular due to \eqref{eq:regular}, more precisely  
\begin{equation}
A_{\lambda_1, \lambda_2}^{\lambda_3, \lambda_4} (s,t,u) =
  O\left( (s-4m^2)^0 \right)\,.
\end{equation}
As a result the non-analytic behavior \eqref{eq:half_angle} at $s=4m^2$ enters the amplitude $T$ only through the Wigner d-matrices in \eqref{eq:origin_kinematic}. Now in order to extract the precise behaviour of the poles in the COM amplitude in practice we simply need to expand \eqref{eq:origin_kinematic} around $s=4m^2$ to the leading order using the explicit expression of the Wigner d matrices \eqref{eq:Wigner_d}, taking into account \eqref{eq:half_angle} and the fact that the functions $A$ are regular at $s=4m^2$.

The expression \eqref{eq:origin_kinematic} together with \eqref{eq:amplitude_A} can also be used to address kinematic constraints. Expanding \eqref{eq:origin_kinematic} to the next to leading order one finds that some linear combinations should vanish as $O\big((s-4m^2)^1\big)$ instead of $O\big((s-4m^2)^0\big)$ or $O\big((s-4m^2)^{-1}\big)$. For simple examples as in section \ref{sec:improved_amplitudes} such linear combinations can be found manually. For more complicated cases one can invoke group theoretic arguments similar to ones in \cite{Kravchuk:2016qvl}.

\paragraph{Special points: $\boldsymbol{t=0}$ and $\boldsymbol{u=0}$}
Kinematic branch points $\sqrt{t}$ and $\sqrt{u}$ enter the center of mass scattering amplitudes via $\sin\theta$ given \eqref{eq:parameters}. Their presence can be deduced by looking at the partial wave decomposition \eqref{eq:partial_wave_decomposition_T} and noticing that the Wigner d matrix there can be written in the form
\begin{equation}
d^{(\ell)}_{\lambda' \lambda}(\theta) = \left(\cos\frac{\theta}{2}\right)^{|\lambda+\lambda'|} \left(\sin \frac{\theta}{2}\right)^{|\lambda-\lambda'|} P^\ell_{\lambda'\lambda}(\cos \theta),
\end{equation}
where $P^\ell_{\lambda'\lambda}$ is a polynomial whose precise definition is irrelevant here but can be deduced from (4.1.19) and (4.1.23) in \cite{Edmonds:1955fi}. The important point is that the polynomial $P^\ell_{\lambda'\lambda}$ depends only on $\cos\theta$ and therefore does not have any branch points as can be seen from \eqref{eq:parameters}.
Using the above, \eqref{eq:half_angle} and \eqref{eq:partial_wave_decomposition_T} we conclude that
\begin{equation}
\label{eq:branch_points_1}
T_{\lambda_1, \lambda_2}^{\lambda_3,\lambda_4}(s,t,u)\sim
d^{(\ell)}_{\lambda_{12} \lambda_{34}}(\theta) \sim
\left(\sqrt{ -u }\right)^{|\lambda_{12}+\lambda_{34}|} \left(\sqrt{ -t }\right)^{|\lambda_{12}-\lambda_{34}|}.
\end{equation}

The $SO(2)$ enhancement of symmetry at $t=0$ ($\theta=0$) and $u=0$ ($\theta=\pi$), see \eqref{eq:COM_frame_same_mass}, leads to kinematic constraints at these two points. This is the simplest case among all the special points and can be easily addressed in full generality. In order to deduce the implications of this $SO(2)$ invariance we inject the identity in the form $1=e^{-i\gamma J_3}e^{+i\gamma J_3}$ to the left and to the right of the $T$ operator in \eqref{eq:scalar_Tamplitude_definition}, where $\gamma$ is some angle. Using the invariance of the $T$ operator \eqref{eq:poincare_invariance} and the fact that the 1PS states along the $z$-direction are the eigenstates of $J_3$, we arrive at
\begin{align}
\label{eq:constraint_1a}
T{}_{\lambda_1, \lambda_2}^{\lambda_3,\lambda_4}(s,t=0) &=
e^{i\gamma(\lambda_1-\lambda_2-\lambda_3+\lambda_4)} T{}_{\lambda_1, \lambda_2}^{\lambda_3,\lambda_4}(s,t=0),\\
\label{eq:constraint_1b}
T{}_{\lambda_1, \lambda_2}^{\lambda_3,\lambda_4}(s,u=0) &=
e^{i\gamma(\lambda_1-\lambda_2+\lambda_3-\lambda_4)} T{}_{\lambda_1, \lambda_2}^{\lambda_3,\lambda_4}(s,u=0).
\end{align}
These constraints should be satisfied for any value of the angle $\gamma$. Thus, the amplitudes must vanish unless $\lambda_1-\lambda_2-\lambda_3+\lambda_4=0$ in the first case and $\lambda_1-\lambda_2+\lambda_3-\lambda_4=0$ in the second case.
This is just conservation of angular momentum along the $z$--axis.

\paragraph{Special point: $\boldsymbol{s=0}$}
Finally let us address the most complicated $s=0$ case. In the vicinity of $s=0$, the rapidity parameter $\eta$ can be written as
\begin{equation}
\label{eq:rapidity_sqrtS}
\eta = \frac{i\pi}{2}-\frac{i\sqrt{s}}{2m}+O(s^{3/2}).
\end{equation}
Let us now define the new rapidity $\xi$ as
\begin{equation}
\label{eq:rapidity_sqrtS_prime}
\xi \equiv \eta  -  \frac{i\pi}{2} = -\frac{i\sqrt{s}}{2m}+O(s^{3/2}).
\end{equation}
By using \eqref{eq:pi_rotate_around_y_boost} - \eqref{eq:pi_rotate_around_x_rotation}
and the following property of the small Wigner d-matrix
\begin{equation}
d^{(j)}_{\lambda' \lambda}(-\pi)=
(-1)^{j+\lambda}\delta_{\lambda', -\lambda},
\end{equation}
we can rewrite the states \eqref{eq:states_1234} in the following way
\begin{equation}
\label{eq:states_1234_s=0}
\begin{aligned}
|m,+\vec p_z; j,\lambda_1 \rangle 
&= e^{+\frac{\pi}{2} K_3} \left(e^{-i\xi K_3}
|m,\vec 0; j,\lambda_1 \rangle\right),\\ 
|m,-\vec p_z; j,\lambda_2 \rangle
&=(-1)^{j+\lambda_2}
e^{-\frac{\pi}{2} K_3} \left(e^{+i\xi K_3}
|m,\vec 0; j, -\lambda_2 \rangle\right), \\
|m,+\vec p_\theta; j,\lambda_3 \rangle 
&= e^{-i\theta J_2}e^{+\frac{\pi}{2} K_3} \left(e^{-i\xi K_3}
|m,\vec 0; j,\lambda_3 \rangle\right),\\
|m,-\vec p_\theta; j,\lambda_4 \rangle
&= (-1)^{j+\lambda_4}e^{-i\theta J_2}
e^{-\frac{\pi}{2} K_3} \left(e^{+i\xi K_3}
|m,\vec 0; j,-\lambda_4 \rangle\right).
\end{aligned}
\end{equation}
These can be further rewritten as\footnote{The steps involved here are as follows. First, we inject the identity operators $\mathbb{I}=e^{\pm i \frac{\pi}{2}J_2}e^{\mp i \frac{\pi}{2}J_2}$ to the left and right of the $e^{\pm i \xi K_3}$ operator and use the following relations
\begin{align*}
e^{-i J_2 \frac{\pi}{2}} e^{\pm i \xi K_3} e^{+i J_2 \frac{\pi}{2}} = e^{\pm i\xi K_1},\qquad
e^{+i J_2 \frac{\pi}{2}} e^{\pm i \xi K_3} e^{-i J_2 \frac{\pi}{2}} = e^{\mp i\xi K_1}.
\end{align*}
Second, we use the following relations
\begin{align*}
e^{\pm \frac{\pi}{2} K_3} e^{+i J_2 \frac{\pi}{2}} = e^{+i J_2 \frac{\pi}{2}}e^{\pm \frac{\pi}{2} K_1},\qquad
e^{\pm \frac{\pi}{2} K_3} e^{-i J_2 \frac{\pi}{2}} = e^{-i J_2 \frac{\pi}{2}} e^{\mp \frac{\pi}{2} K_1}
\end{align*}
to bring all the exponents containing $J_2$ to the left.
Finally, we use
\begin{equation*}
e^{-i \theta J_2} e^{\frac{\pi}{2} K_1} = e^{\frac{\pi}{2} K_1}e^{-\theta K_3}.
\end{equation*}
}
\begin{equation}
\begin{aligned}
|m,+\vec p_z; j,\lambda_1 \rangle 
&=
e^{+\frac{\pi}{2} K_1}e^{+\frac{\pi}{2} K_3}
e^{-i \xi K_1}
\left(e^{-i \frac{\pi}{2}J_2}
|m,\vec 0; j,\lambda_1 \rangle \right),\\ 
|m,-\vec p_z; j,\lambda_2 \rangle
&=
e^{+\frac{\pi}{2} K_1}
e^{-\frac{\pi}{2}K_3}
e^{-i \xi K_1}
\left((-1)^{j+\lambda_2}e^{+i \frac{\pi}{2}J_2}
|m,\vec 0; j, -\lambda_2 \rangle\right), \\
|m,+\vec p_\theta; j,\lambda_3 \rangle 
&=
e^{+ \frac{\pi}{2} K_1}
e^{- (\theta-\frac{\pi}{2}) K_3}
e^{-i \xi K_1}
\left(e^{-i \frac{\pi}{2}J_2}
|m,\vec 0; j,\lambda_3 \rangle\right),\\
|m,-\vec p_\theta; j,\lambda_4 \rangle
&= 
e^{+ \frac{\pi}{2} K_1}
e^{- (\theta+\frac{\pi}{2}) K_3}
e^{-i \xi K_1}
\left((-1)^{j+\lambda_4}e^{+i \frac{\pi}{2}J_2}
|m,\vec 0; j,-\lambda_4 \rangle\right).
\end{aligned}
\end{equation}
We use now invariance of the scattering operator under boosts and the action of rotations on the center of mass states \eqref{eq:transformation}, which for the second and fourth particles become
\begin{equation}
\begin{aligned}
(-1)^{j+\lambda}e^{+i \frac{\pi}{2}J_2}
|m,\vec 0; j, -\lambda \rangle &=
(-1)^{j+\lambda}\sum_{\lambda'}
d^j_{\lambda',-\lambda}(-\frac{\pi}{2})
|m,\vec 0; j, \lambda' \rangle\\
&=
(-1)^{2j}\sum_{\lambda'}
d^j_{\lambda',\lambda}(+\frac{\pi}{2})
|m,\vec 0; j, \lambda' \rangle,
\end{aligned}
\end{equation}
to obtain the final expression
\begin{equation} 
\label{eq:branch_points_2}
T_{\lambda_1, \lambda_2}^{\lambda_3, \lambda_4} (s,t,u) =
\sum_{\lambda'}
d^{(j)}_{\lambda_1' \lambda_1}\left(\frac{\pi}{2}\right)
d^{(j)}_{\lambda_2' \lambda_2}\left(\frac{\pi}{2}\right)
d^{(j)}_{\lambda_3' \lambda_3}\left(\frac{\pi}{2}\right)
d^{(j)}_{\lambda_4' \lambda_4}\left(\frac{\pi}{2}\right)
B_{\lambda'_1, \lambda'_2}^{\lambda'_3, \lambda'_4} (s,t,u),
\end{equation}
where the amplitude $B$ is defined as
\begin{multline}
\label{eq:branch_points}
B_{\lambda_1, \lambda_2}^{\lambda_3, \lambda_4}(s,t,u) \equiv
\left(
\<m,\vec 0; j,\lambda_3 |e^{+i \xi K_1}e^{+\frac{\pi}{2} K_3}\otimes
\<m,\vec 0; j,\lambda_4 |e^{+i \xi K_1}e^{-\frac{\pi}{2} K_3}
\right)e^{-\theta K_3}
T\\
\left(
e^{+\frac{\pi}{2} K_3}e^{-i \xi K_1}|m,\vec 0; j,\lambda_1 \rangle \otimes
e^{-\frac{\pi}{2} K_3}e^{-i \xi K_1}|m,\vec 0; j,\lambda_2 \rangle
\right).
\end{multline}

Let us inspect the structure of this amplitude. We expand it around $\xi=0$ or equivalently around $s=0$ according to \eqref{eq:rapidity_sqrtS_prime}. Schematically speaking, each term in this expansion will contain $(\xi K_1)^n$ with some non-negative integer $n$. We then notice that $(\xi K_1)^n$ are the only operators which change helicities of particles.\footnote{One can define the following operators
\begin{equation*}
K_{\pm} \equiv K_1\pm i K_2
\quad\Rightarrow\quad
K_1=\frac{1}{2}\,(K_++K_-).
\end{equation*}
According to \eqref{eq:poincare_commutation_4} these operators rise and lower helicities of the center of mass states as
\begin{equation*}
J_3 K_\pm |m,\vec 0; j,\lambda \rangle =
(\lambda\pm 1)K_\pm |m,\vec 0; j,\lambda \rangle.
\end{equation*}
}
Now, the only non-zero terms will be the ones with equal total helicity of the states to the left and to the right of the scattering operator $T$. Given a set of helicities $\lambda_i$, the leading term in the $\xi=0$ expansion will contain $(\xi K_1)^n$ with $n=|\lambda_1+\lambda_2-\lambda_3-\lambda_4|$. Using \eqref{eq:rapidity_sqrtS_prime} we conclude that
\begin{equation}
\label{eq:branch_points_3}
B_{\lambda_1, \lambda_2}^{\lambda_3, \lambda_4}(s,t,u) =
\left(\sqrt{s}\right)^{|\lambda_1+\lambda_2-\lambda_3-\lambda_4|}
\times O(s^0).
\end{equation}

From \eqref{eq:branch_points_2} and \eqref{eq:branch_points_3} it follows that the COM amplitudes $T$ get a $\sqrt{s}$ branch point only for odd values of $|\lambda_1+\lambda_2-\lambda_3-\lambda_4|$.
The relations \eqref{eq:branch_points_2} and \eqref{eq:branch_points_3} can be also used to address kinematic singularities. Expanding \eqref{eq:branch_points_2} around $s=0$ at the leading order one can find linear combinations of the amplitudes which behave as $O(s^1)$ instead of $O(s^0)$.

\section{Application: identical Majorana fermions}
\label{sec:application}
We will now use the machinery set up in the previous chapter to study the two to two scattering of identical neutral\footnote{By neutral we mean particles not carrying any $U(1)$ charge and in general not transforming in any non-trivial representation of the global group. In common words it means that the particle is its own antiparticle.} spin $\frac{1}{2}$ fermions also known as Majorana fermions. We will require invariance under parity. As a result we need to specify the intrinsic parity $\eta$ defined in \eqref{eq:parity_1PS_main}. In the two to two scattering of identical  particles we are sensitive only to the value of $\eta^2$. According to \eqref{eq:intrinsic_parities} there are two possibilities
\begin{equation}
\label{eq:eta_options}
\eta^2=-1\quad\text{or}\quad\eta^2=+1.
\end{equation}
For concreteness we assume the former in this section. In the latter situation everything in this section still remains valid except that the meaning of parity even and odd states in section \ref{sec:example_unitarity} is flipped and the role of scalar and pseudoscalar particles is exchanged in section \ref{sec:non-perturbative_couplings}.
Helicity of a spin $\frac{1}{2}$ particle takes only two values: $+\frac{1}{2}$ and $-\frac{1}{2}$. Thus a priori, we have $2^4=16$ helicity amplitudes. However due to the fact that the particles are all identical these amplitudes are related according to \eqref{eq:(12)(34)} - \eqref{eq:(14)(23)}. As a result we can write the following 9 relations
\begin{align}
\nn
&T_{--}^{--}=T_{++}^{++},\qquad
T_{--}^{+-}=T_{--}^{-+},\qquad
T_{-+}^{--}=T_{++}^{+-},\qquad
T_{-+}^{-+}=T_{+-}^{+-},\qquad
T_{-+}^{+-}=T_{+-}^{-+},\\
&T_{-+}^{++}=T_{--}^{-+},\qquad
T_{+-}^{--}=T_{++}^{+-},\qquad
T_{+-}^{++}=T_{--}^{-+},\qquad
T_{++}^{-+}=T_{++}^{+-},
\label{eq:Majorana_identical}
\end{align}
where $+$ and $-$ stand for $+\frac{1}{2}$ and $-\frac{1}{2}$ helicities respectively. Hence out of the 16 amplitudes we are left with 7 independent ones. Requiring parity invariance and noticing that due to \eqref{eq:eta_options} the product of intrinsic parities $\eta_1\eta_2\eta_3^*\eta_4^*=+1$, due to \eqref{eq:amplitudes_parity} we get in addition the following 2 constraints
\begin{equation}
T_{--}^{-+}=-T_{++}^{+-},\qquad
T_{--}^{++}=+T_{++}^{--}.
\label{eq:Majorana_parity}
\end{equation}
As a result out of the 16 helicity amplitudes we are left with only 5 independent ones,  in agreement with \eqref{eq:counting_majorana}, which we denote as
\begin{equation}
\begin{aligned}
\Phi_1 (s,t,u) \equiv T_{++}^{++}(s,t,u), \\
\Phi_2 (s,t,u) \equiv T_{++}^{--}(s,t,u), \\
\Phi_3 (s,t,u) \equiv T_{+-}^{+-}(s,t,u), \\
\Phi_4 (s,t,u) \equiv T_{+-}^{-+}(s,t,u), \\
\Phi_5 (s,t,u) \equiv T_{++}^{+-}(s,t,u).
\end{aligned}
\label{eq:Majorana_independent}
\end{equation}
It is interesting to note that the scattering of identical neutral fermions preserving parity is automatically time-reversal invariant, this can be intuitively understood from the CPT symmetry since charge conjugation is trivial for neutral particles, see appendix \ref{app:PT}.

As discussed in section \ref{sec:idnetical_particles}, in the case of scattering of uncharged identical particles, the crossing equations \eqref{eq:st_crossing} and  \eqref{eq:su_crossing} form highly non-trivial constraints on the scattering amplitudes. For instance, in the case of identical Majorana particles these crossing equations give rise to two sets of $16$ linear equations. Taking into account the relations \eqref{eq:Majorana_identical} and \eqref{eq:Majorana_parity} we simply obtain two sets of 5 linear equations on the independent amplitudes \eqref{eq:Majorana_independent}. They read as
\begin{align}
\label{eq:crossing_st_fermions_amp}
\Phi_I (s,t,u) &=  \sum_{J=1}^5 C_{st}^{IJ}(s,t,u)\Phi_J(t,s,u),\\
\label{eq:crossing_su_fermions_amp}
\Phi_I (s,t,u) &=  \sum_{J=1}^5 C_{su}^{IJ}(s,t,u)\Phi_J(u,t,s),
\end{align}
where the $s-t$ crossing matrix $C_{st}$ is given by
\begin{equation}
\label{eq:crossing_matrix_st}
C_{st}
= -\frac{\epsilon_{23}'}{2}
\begin{pmatrix}
-\sin^2 \alpha & \sin^2 \alpha & -\sin^2 \alpha & 1 + \cos^2 \alpha & 4 \cos \alpha \sin \alpha  \\
\sin^2 \alpha & 1 + \cos^2 \alpha & \sin^2 \alpha & \sin^2 \alpha & -4 \cos \alpha \sin \alpha \\
-\sin^2 \alpha & \sin^2 \alpha & 1 + \cos^2 \alpha   & -\sin^2 \alpha & 4 \cos \alpha \sin \alpha  \\
1+\cos^2 \alpha & \sin^2 \alpha & -\sin^2 \alpha & -\sin^2 \alpha & 4 \cos \alpha \sin \alpha \\
\cos \alpha \sin \alpha \ \ \  & -\cos \alpha \sin \alpha \ \ \  & \cos \alpha \sin \alpha \ \   & \cos \alpha \sin \alpha \ \ \  & 2(\sin^2 \alpha - \cos^2 \alpha)
\end{pmatrix},
\end{equation} 
and the $s-u$ crossing matrix $C_{su}$ is given by 
\begin{equation}
\label{eq:crossing_matrix_su}
C_{su}
= -\frac{\epsilon_{24}'}{2}
\begin{pmatrix}
-\sin^2 \beta & \sin^2 \beta & 1 + \cos^2 \beta & -\sin^2 \beta & 4 \cos \beta \sin \beta  \\
\sin^2 \beta & 1 + \cos^2 \beta & \sin^2 \beta & \sin^2 \beta & -4 \cos \beta \sin \beta \\
1 + \cos^2 \beta & \sin^2 \beta &  -\sin^2 \beta & -\sin^2 \beta & 4 \cos \beta \sin \beta  \\
-\sin^2 \beta & \sin^2 \beta & -\sin^2 \beta & 1 + \cos^2 \beta & 4 \cos \beta \sin \beta \\
\cos \beta \sin \beta\ \    & -\cos \beta \sin \beta\ \ \   & \cos \beta \sin \beta \ \ \  & \cos \beta \sin \beta \ \ \  & 2(\sin^2 \beta - \cos^2 \beta)
\end{pmatrix}.
\end{equation}
The angles $\alpha$ and $\beta$ are defined as follows. Looking at the expressions \eqref{eq:s-t:equal_mass} and \eqref{eq:s-u:equal_mass} we can make the following choice of Wigner angles
\begin{equation}
\label{eq:choice_alpha}
\alpha_1=\alpha,\quad
\alpha_2=\pi+\alpha,\quad
\alpha_3=\pi-\alpha,\quad
\alpha_4=-\alpha.
\end{equation}
\begin{equation}
\label{eq:choice_beta}
\beta_1=\beta,\quad
\beta_2=\pi+\beta,\quad
\beta_3=-\beta,\quad
\beta_4=\pi-\beta,
\end{equation}
where the angles $\alpha$ and $\beta$ obey
\begin{equation}
\label{eq:def_alpha}
\cos\alpha=
\frac{s t}{\sqrt{s(s-4m^2)}\sqrt{t(t-4m^2)}},\quad
\sin\alpha =
-\frac{2m\,\sqrt{stu}}{\sqrt{s(s-4m^2)}\sqrt{t(t-4m^2)}},
\end{equation}
\begin{equation}
\label{eq:def_beta}
\cos\beta=\frac{s u}{\sqrt{s(s-4m^2)}\sqrt{u(u-4m^2)}},\quad
\sin=+\frac{2m\,\sqrt{stu}}{\sqrt{s(s-4m^2)}\sqrt{u(u-4m^2)}}.
\end{equation}
The correct choice of the phases at \eqref{eq:crossing_matrix_st} and \eqref{eq:crossing_matrix_su} will be explained at the end of section \ref{sec:improved_amplitudes}. Here we simply state the correct result, which is
\begin{equation}
\label{eq:choice_phases}
\epsilon_{23}'=\epsilon_{24}'=-1.
\end{equation}
There are two non-trivial consistency checks our matrices \eqref{eq:crossing_matrix_st} and \eqref{eq:crossing_matrix_su} pass.  First, these matrices are involutory, namely they satisfy the following conditions\footnote{More accurately one should write
\begin{equation}
\nn
C_{st}(s,t,u)C_{st}(t,s,u) = 1,\qquad
C_{su}(s,t,u)C_{su}(u,t,s) = 1.
\end{equation}
However, these conditions reduce to \eqref{eq:involutory_condition} by noticing that the matrices $C_{st}(s,t,u)$ and $C_{st}(u,t,s)$ are symmetric in the exchange of $s\leftrightarrow t$ and $s\leftrightarrow u$ respectively. This follows from the fact that the expressions for the angles $\alpha$ and $\beta$ given by \eqref{eq:def_alpha} and \eqref{eq:def_beta} obey the symmetry $s\leftrightarrow t$ and $s\leftrightarrow u$ respectively.
}
\begin{equation}
\label{eq:involutory_condition}
\big(C_{st}(s,t,u)\big)^2 = 1,\qquad
\big(C_{su}(s,t,u)\big)^2 = 1.
\end{equation}
Second, we can obtain the crossing matrix appearing in the $t-u$ crossing equations as
\begin{equation}
\label{eq:crossing_matrix_tu}
C_{tu}(s,t,u) = C_{st}(s,t,u)C_{su}(t,s,u)C_{st}(u,s,t).
\end{equation}
In our case it reads as
\begin{equation}
\label{eq:tu_crossing_matrix}
C_{tu}(s,t,u) = \begin{pmatrix}
                 1 & 0 & 0 & 0 & 0 \\
                 0 & 1 & 0 & 0 & 0 \\
                 0 & 0 & 0 & 1 & 0 \\
                 0 & 0 & 1 & 0 & 0 \\
                 0 & 0 & 0 & 0 & -1 
                 \end{pmatrix}.
\end{equation}
This is in perfect agreement with the result \eqref{eq:identical_particles_amplitudes2}.

\subsection{Improved amplitudes}
\label{sec:improved_amplitudes}
It is important to study the analytic structure (presence of poles and branch cuts) of helicty amplitudes when all the Mandelstam variables $s$, $t$ and $u$ are promoted to the full complex plane. As explained in section \ref{sec:kinematic_non_analiticities} in the case of scattering of spinning particles such amplitudes develop non-analytic behaviour purely due to kinematic reasons. In this section we show how to isolate such kinematic features in the case of Majorana fermions and define improved amplitudes which do not have them.

Due to \eqref{eq:origin_kinematic} all the amplitudes have a pole at $s=4m^2$. Expanding \eqref{eq:origin_kinematic} around this point we get
\begin{equation}
\label{eq:expansion}
\begin{aligned}
\Phi_1  & = \frac{a_1}{s-4m^2} + b_1 + O(s-4m^2),\\
\Phi_2  & = \frac{a_2}{s-4m^2} + b_2 + O(s-4m^2),\\
\Phi_3  & = \frac{a_3}{s-4m^2} + b_3 + O(s-4m^2),\\
\Phi_4  & = \frac{a_4}{s-4m^2} + b_4 + O(s-4m^2),\\
\Phi_5  & = \frac{a_5}{s-4m^2} + b_5 + O(s-4m^2),
\end{aligned}
\end{equation}
where $a_i$ and $b_i$ are some factors which are regular at $s=4m^2$. We do not write them explicitly, their form can be obtained straightforwardly using computer algebra.\footnote{
\label{foot:comment}Notice that identical particles and parity imply constraints on the amplitudes $T$ according to \eqref{eq:Majorana_identical} and \eqref{eq:Majorana_parity}. In order to proceed with the expansion one needs to deduce the analogues of these expressions on the regular $A(s,t,u)$ amplitudes entering \eqref{eq:origin_kinematic} by solving an appropriate system of linear equations.}
We only notice that
\begin{equation}
\label{eq:pole_residues}
a_1=-a_2=a_3=a_4=-i a_5.
\end{equation}
Using \eqref{eq:expansion} we can verify the following kinematic relation at the singular point\footnote{We expand around $s=4m^2$ keeping $t$   independent.
Then in the right-hand side of \eqref{eq:kin_constraint_1} the leading and the next to leading order terms appear to be proportional to the following expression
\begin{equation*}
t-u-2i\sqrt{tu}=2t+(s-4m^2)-2i\sqrt{t(4m^2-s-t)}=0+O(s-4m^2).
\end{equation*}
In the last equality we have used the domain where $t<0$ and $s$ has a small positive imaginary part.}
\begin{equation}
\label{eq:kin_constraint_1}
\Phi_1  - \Phi_2 + \Phi_3 + \Phi_4 + 4 i \Phi_5 = 0 + O(s-4m^2).
\end{equation}

Now due to \eqref{eq:branch_points_2} the amplitude $\Phi_5$ also develops a branch point at $s=0$ as
\begin{equation}
\label{eq:branch_s}
\Phi_5 \sim \sqrt{s},
\end{equation}
whereas all the other amplitudes are regular at $s=0$.
Expanding \eqref{eq:branch_points_2} to the leading order we can verify the following kinematical constraint
\begin{equation}
\label{eq:kin_constraint_3}
\Phi_1  + \Phi_2 - \Phi_3 - \Phi_4 = 0 + O(s).
\end{equation}

Finally we consider the behavior of the amplitudes at $t=0$ and $u=0$ points. Due to \eqref{eq:branch_points_1} the amplitudes $\Phi_1$, $\Phi_2$, $\Phi_3$ and $\Phi_4$ are all analytic at these points. In contrast the amplitude $\Phi_5$ develops a branch point both at $t=0$ and $u=0$ as
\begin{equation}
\label{eq:branch_t_u}
\Phi_5 \sim \sqrt{t u}.
\end{equation}
In addition due to \eqref{eq:constraint_1a} and \eqref{eq:constraint_1b} we have the following constraints
\begin{equation}
\label{eq:kin_constraint_2}
\begin{aligned}
&\Phi_4 =0 + O(t),\,\quad
 \Phi_5 =0 + O(t),\\
&\Phi_3 =0 + O(u),\quad
 \Phi_5 =0 + O(u).
\end{aligned}
\end{equation}

Now that we know precisely the non-analytic behaviour of the amplitudes, we can define new improved amplitudes which are free of the kinematic pole at $s=4m^2$ and kinematic branch points $\sqrt{s}$, $\sqrt{t}$ and $\sqrt{u}$. We denote such improved amplitudes by $H_I(s,t,u)$. The old amplitudes and the new improved amplitudes can be related as
\begin{equation}
\label{eq:ksfa_definition}
\Phi_I(s,t,u) = \sum_{J=1}^5 M_{IJ}^{-1}(s,t,u) H_J(s,t,u),
\end{equation}
where $M(s,t,u)$ is some matrix to be determined. It is constructed by requiring that
\begin{equation}
\Phi_I\sim \frac{1}{s-4m^2},\quad
\Phi_5\sim \sqrt{stu}
\end{equation}
and that the relations \eqref{eq:expansion} along with \eqref{eq:pole_residues}, \eqref{eq:kin_constraint_1}, \eqref{eq:kin_constraint_3} and \eqref{eq:kin_constraint_2} are fulfilled. These requirements do not fix the matrix $M_{IJ}(s,t,u)$ completely. One possible choice is
\begin{equation}
\label{eq:matrix_M}
M_{IJ}(s,t,u) = \left(
\begin{array}{ccccc}
 \frac{4}{s-4m^2} \ \ \ & \frac{-4}{s-4m^2} \ \ \ & \frac{2\, (1-t/u)}{s-4m^2} \ \ \ & \frac{2 (1-u/t)}{s-4m^2} \ \ \ & \frac{s+4m^2}{s-4m^2}\times\frac{2(t-u)}{m\sqrt{s t u}} \\
 0 & 0 & \frac{2}{u} & -\frac{2}{t} & -\frac{8m}{\sqrt{s t u}} \\
 0 & 0 & \frac{2}{u} & -\frac{2}{t} & -\frac{2 s}{m\sqrt{s t u}} \\
 0 & 0 & \frac{2}{u} & \frac{2}{t} & 0 \\
 -\frac{4}{s} & -\frac{4}{s} & \frac{2}{u}+\frac{4}{s} & \frac{2}{t}+\frac{4}{s} & \frac{2 (t-u)}{m\sqrt{s t u}} \\
\end{array}
\right).
\end{equation}
We motivate
this choice in section \ref{sec:example_tensor_structures}.

Having established the relation \eqref{eq:ksfa_definition} we can write the crossing equations \eqref{eq:crossing_st_fermions_amp} and \eqref{eq:crossing_su_fermions_amp} directly in terms of the improved amplitudes as
\begin{align}
\label{eq:crossing_st_fermions_amp_improved}
H_I (s,t,u) &=  \sum_{J=1}^5 \widetilde C_{st}^{IJ}(s,t,u)H_J(t,s,u),\\
\label{eq:crossing_su_fermions_amp_improved}
H_I (s,t,u) &=  \sum_{J=1}^5 \widetilde C_{su}^{IJ}(s,t,u)H_J(u,t,s),
\end{align}
where the crossing matrices $\widetilde C_{st}$ and $\widetilde C_{su}$ read as
\begin{align}
\label{eq:def_C_st_tilde}
\widetilde C_{st} &\equiv  M(s,t,u) C_{st}(s,t,u) M^{-1}(t,s,u),\\
\label{eq:def_C_su_tilde}
\widetilde C_{su} &\equiv  M(s,t,u) C_{su}(s,t,u) M^{-1}(u,t,s).
\end{align}
Plugging here the explicit expressions \eqref{eq:crossing_matrix_st}, \eqref{eq:crossing_matrix_su} and \eqref{eq:matrix_M} we get
\begin{equation}
\label{eq:crossing_matrices_numerical}
\widetilde C_{st} = \left(
\begin{array}{ccccc}
-\frac{1}{4} & -1 & \frac{3}{2} & 1 & -\frac{1}{4} \\
-\frac{1}{4} & \frac{1}{2} & 0 & \frac{1}{2} & \frac{1}{4} \\
\frac{1}{4} & 0 & \frac{1}{2} & 0 & \frac{1}{4} \\
\frac{1}{4} & \frac{1}{2} & 0 & \frac{1}{2} & -\frac{1}{4} \\
-\frac{1}{4} & 1 & \frac{3}{2} & -1 & -\frac{1}{4} \\
\end{array}
\right),\qquad
\widetilde C_{su} = \left(
\begin{array}{ccccc}
-\frac{1}{4} & 1 & -\frac{3}{2} & 1 & -\frac{1}{4} \\
\frac{1}{4} & \frac{1}{2} & 0 & -\frac{1}{2} & -\frac{1}{4} \\
-\frac{1}{4} & 0 & \frac{1}{2} & 0 & -\frac{1}{4} \\
\frac{1}{4} & -\frac{1}{2} & 0 & \frac{1}{2} & -\frac{1}{4} \\
-\frac{1}{4} & -1 & -\frac{3}{2} & -1 & -\frac{1}{4} \\
\end{array}
\right).
\end{equation}
It is remarkable that both matrices turn out to be purely numerical! Just like the original matrices $C_{st}$ and $C_{su}$, the matrices $\widetilde C_{st}$ and $\widetilde C_{su}$ are also involutory, \emph{i.e} $\widetilde C_{st}^2=1$ and $\widetilde C_{su}^2=1$. This follows from the definitions \eqref{eq:def_C_st_tilde}, \eqref{eq:def_C_su_tilde} and the condition \eqref{eq:involutory_condition}. 
Note that similar to \eqref{eq:crossing_matrix_tu} we can compute the $tu$ crossing matrix $\widetilde C_{tu} = \widetilde C_{st}\widetilde C_{su} \widetilde C_{st}$, it reads as
\begin{equation}
\widetilde C_{tu} = \left(
\begin{array}{ccccc}
1 & 0 & 0 & 0 & 0 \\
0 & -1 & 0 & 0 & 0 \\
0 & 0 & -1 & 0 & 0 \\
0 & 0 & 0 & 1 & 0 \\
0 & 0 & 0 & 0 & 1 \\
\end{array}
\right).
\end{equation}
It says that the improved amplitudes defined via \eqref{eq:ksfa_definition} are all eigenfunctions of $tu$ crossing.

The overall sign of the crossing matrices \eqref{eq:crossing_matrices_numerical} depends on the choice of phases in \eqref{eq:choice_phases}. The choice made in \eqref{eq:choice_phases} is the only correct one. In order to see that we can take the Fermi lagrangian and compute the scattering of Majorana fermions to the leading order. We do it in appendix \ref{app:fermi_theory}, the final result for the improved amplitudes is given in \eqref{eq:amplitude_fermi_final}. It automatically satisfies the crossing equations \eqref{eq:crossing_st_fermions_amp_improved} and \eqref{eq:crossing_su_fermions_amp_improved}. Any other choice of phases \eqref{eq:choice_phases} leads to crossing equations which are inconsistent with the perturbative computation of appendix \ref{app:fermi_theory}. This phase choice is independent of any particular model and holds non-perturbatively.

\subsection{Unitarity}
\label{sec:example_unitarity}
The general strategy for imposing unitarity constraints on scattering amplitudes was provided in section \ref{sec:unitarity}. Here we apply that strategy to the case of Majorana fermions. According to section \ref{sec:unitarity} one needs to consider all possible states transforming in the irreducible representations which appear in the decomposition of the two particle state formed from two (identical) Majorana particles. These are 
\begin{equation}
|c,\vec p, \spin, \hel; \lambda_1, \lambda_2 \rangle_{id},
\end{equation}
where $\lambda_i=\pm\frac{1}{2}$. Since the particles are identical we can further restrict our attention to the states with $\lambda_1\geq\lambda_2$. As a result we are left with only three states of the form
\begin{equation}
\label{eq:states_to_consider}
|c,\vec p, \spin, \hel; +, + \rangle_{id},\qquad
|c,\vec p, \spin, \hel; -, - \rangle_{id},\qquad
|c,\vec p, \spin, \hel; +, - \rangle_{id}.
\end{equation}
We further notice that due to \eqref{eq:irrep_id}, the first two states in \eqref{eq:states_to_consider} exist only for even spins $\ell$, whereas the last state in
\eqref{eq:states_to_consider} exists for both even and odd spins $\ell$. Using \eqref{eq:parity_pwa_app} we can form the following three parity eigenstates out of the states \eqref{eq:states_to_consider} 
\begin{align}
\label{eq:state_1}
&|1\rangle \equiv \frac{1}{\sqrt{2}}
\left(
|c,\vec p, \spin, \hel; +, + \rangle_{id} + |c,\vec p, \spin, \hel; -, - \rangle_{id}
\right),\quad \ell\geq 0 \quad (\spin \text{ even}),\\
\label{eq:state_2}
&|2 \rangle \equiv \frac{1}{\sqrt{2}}
\left(
|c,\vec p, \spin, \hel; +, + \rangle_{id} - |c,\vec p, \spin, \hel; -, - \rangle_{id}
\right),\quad  \ell\geq 0 \quad (\spin \text{ even}), \\
\label{eq:state_3}
&|3 \rangle \equiv 
\sqrt{2}\;|c,\vec p, \spin, \hel; +, -  \rangle_{id},\quad \ell\geq 1.
\end{align}
The state $|1\rangle$ is parity odd while the states $|2\rangle$ and $|3 \rangle$ are parity even.
The states \eqref{eq:state_1} - \eqref{eq:state_3} can either be $in$ or $out$ asymptotic states. We now form all possible inner products between such states taking into account that parity eigenstates do not mix since we assumed parity invariance. The states \eqref{eq:state_1} lead to
\begin{equation}
\label{eq:matrix_1}
\ell\geq 0\;\;\text{(even)}:\qquad
 \mathcal H^-_\ell(s)\times\delta_{\ell \ell'}\delta_{\lambda\lambda'}(2\pi)^4\delta^{(4)}(p-p')  \equiv \begin{pmatrix}
                        {}_{in}\langle 1' | 1 \rangle_{in} & {}_{in}\langle 1' | 1 \rangle_{out} \\
                        {}_{out}\langle 1' | 1 \rangle_{in} & {}_{out}\langle 1' | 1 \rangle_{out}
                        \end{pmatrix},                  
\end{equation}
where the primed states have the labels $c'$, $\vec p\,'$, $\ell'$ and $\lambda'$. Analogously the states \eqref{eq:state_2} for $\ell=0$ and the states \eqref{eq:state_3} for odd $\ell\geq 1$ lead to
\begin{align}
\label{eq:matrix_2}
\ell=0:\qquad\mathcal H^+_{\ell}(s)
&\times\delta_{\ell \ell'}\delta_{\lambda\lambda'}(2\pi)^4\delta^{(4)}(p-p')  \equiv \begin{pmatrix}
{}_{in}\langle 2' | 2 \rangle_{in} & {}_{in}\langle 2' | 2 \rangle_{out} \\
{}_{out}\langle 2' | 2 \rangle_{in} & {}_{out}\langle 2' | 2 \rangle_{out}
\end{pmatrix},\\
\label{eq:matrix_3}
\ell\geq 1\;\;\text{(odd)}:\qquad
\mathcal H^+_\ell(s)&\times\delta_{\ell \ell'}\delta_{\lambda\lambda'}(2\pi)^4\delta^{(4)}(p-p')  \equiv \begin{pmatrix}
{}_{in}\langle 3' | 3 \rangle_{in} & {}_{in}\langle 3' | 3 \rangle_{out} \\
{}_{out}\langle 3' | 3 \rangle_{in} & {}_{out}\langle 3' | 3 \rangle_{out}
\end{pmatrix}.    
\end{align}
Finally for even $\ell\geq 2$, the states \eqref{eq:state_2} and \eqref{eq:state_3} can mix. They lead to
\begin{equation}
\label{eq:matrix_4}
H^+_{\ell\geq 2}(s)
\times\delta_{\ell \ell'}\delta_{\lambda\lambda'}(2\pi)^4\delta^{(4)}(p-p')
\equiv \begin{pmatrix}
                        {}_{in}\langle 2' | 2 \rangle_{in} & {}_{in}\langle 2' | 3 \rangle_{in} & {}_{in}\langle 2' | 2 \rangle_{out} & {}_{in}\langle 2' | 3 \rangle_{out} \\
                        {}_{in}\langle 3' | 2 \rangle_{in} & {}_{in}\langle 3' | 3 \rangle_{in} & {}_{in}\langle 3' | 2 \rangle_{out} & {}_{in}\langle 3' | 3 \rangle_{out} \\
                        {}_{out}\langle 2' | 2 \rangle_{in} & {}_{out}\langle 2' | 3 \rangle_{in} & {}_{out}\langle 2' | 2 \rangle_{out} & {}_{out}\langle 2' | 3 \rangle_{out} \\
                        {}_{out}\langle 3' | 2 \rangle_{in} & {}_{out}\langle 3' | 3 \rangle_{in} & {}_{out}\langle 3' | 2 \rangle_{out} & {}_{out}\langle 3' | 3 \rangle_{out} 
                        \end{pmatrix}.
\end{equation}
Let us write explicitly the components of these matrices. The inner product of only $in$ or $out$ states are fixed by our normalization conventions, which read
\begin{equation}
\label{eq:uni_entries_1}
{}_{in}\<a'|b\>_{in} = {}_{out}\<a'|b\>_{out} = \delta_{ab}\times\delta_{\ell \ell'}\delta_{\lambda\lambda'}(2\pi)^4\delta^{(4)}(p-p').
\end{equation}
The inner products between $in$ and $out$ states lead to partial amplitudes, we have
\begin{equation}
\label{eq:uni_entries_2}
\begin{aligned}
{}_{out}\langle 1' | 1 \rangle_{in} &=
\delta_{\ell \ell'}\delta_{\lambda\lambda'}(2\pi)^4\delta^{(4)}(p-p')\times\left(
1 + i \,\left(T_\ell{}_{++}^{++}(s) + T_\ell{}_{++}^{--}(s)\right)\right), \\
{}_{out}\langle 2' | 2 \rangle_{in} &=
\delta_{\ell \ell'}\delta_{\lambda\lambda'}(2\pi)^4\delta^{(4)}(p-p')\times\left(
1 + i \,\left(T_\ell{}_{++}^{++}(s) - T_\ell{}_{++}^{--}(s)\right)\right), \\
{}_{out}\langle 3' |  3 \rangle_{in} &=
\delta_{\ell \ell'}\delta_{\lambda\lambda'}(2\pi)^4\delta^{(4)}(p-p')\times\left(
1 + 2i\, T_\ell{}_{+-}^{+-}(s)\right), \\
{}_{out}\langle 3' |  2 \rangle_{in} &=
{}_{out}\langle 2' |  3 \rangle_{in}=
\delta_{\ell \ell'}\delta_{\lambda\lambda'}(2\pi)^4\delta^{(4)}(p-p')\times\left(
2i\,T_\ell{}_{++}^{+-}(s)\right).
\end{aligned}
\end{equation}
The partial amplitudes entering these expressions are related to scattering amplitudes via \eqref{eq:partial_T_id}. We write here that expression again for the reader's convenience
\begin{equation}
\label{eq:partial_amplitudes}
T_\ell{}^{ \lambda_3, \lambda_4}_{\lambda_1, \lambda_2}(s)  =
\frac{\sqrt{s-4m^2}}{32\pi \sqrt{s}} \times \int_{0}^{\pi} d\theta \sin\theta d^{(\ell)}_{\lambda_{12} \lambda_{34}}(\theta)T_{\lambda_1, \lambda_2}^{\lambda_3,\lambda_4}(s,t,u),\quad
\lambda_{ij}\equiv\lambda_i-\lambda_j.
\end{equation}
In \eqref{eq:uni_entries_2} we used relations \eqref{eq:Majorana_identical}, \eqref{eq:Majorana_parity} and properties of the Wigner d matrices.
As discussed in section \ref{sec:unitarity} unitarity requires the matrices $\mathcal H^-_\ell(s)$ and $\mathcal H^+_\ell(s)$ to be positive semi-definite for all $s\geq 4m^2$ and $\ell$. In what follows we will write these conditions in the final form.

In \eqref{eq:Majorana_independent} we denoted the five independent amplitudes by $\Phi_I(s,t,u)$. In accordance we define the five partial amplitudes as
\begin{equation}
\begin{aligned}
\Phi^\ell_1 (s) \equiv T_\ell{}_{++}^{++}(s), \\
\Phi^\ell_2 (s) \equiv T_\ell{}_{++}^{--}(s), \\
\Phi^\ell_3 (s) \equiv T_\ell{}_{+-}^{+-}(s), \\
\Phi^\ell_4 (s) \equiv T_\ell{}_{+-}^{-+}(s), \\
\Phi^\ell_5 (s) \equiv T_\ell{}_{++}^{+-}(s).
\end{aligned}
\label{eq:Majorana_independent_partial}
\end{equation}
Plugging \eqref{eq:uni_entries_1} and \eqref{eq:uni_entries_2} into \eqref{eq:matrix_1} - \eqref{eq:matrix_4} we can write the semi-definite positivity conditions on the matrices $\mathcal H^-_\ell(s)$ and $\mathcal H^+_\ell(s)$ as
\begin{align}
\label{eq:unitarity_1}
\ell\geq 0\;\;\text{(even)}:\qquad
&\begin{pmatrix}
1 & 1\\
1 & 1
\end{pmatrix}+
i\begin{pmatrix}
0 & -\Phi_1^{\ell*}(s) - \Phi_2^{\ell*}(s) \\
\Phi_1^\ell(s) + \Phi_2^\ell(s) & 0
\end{pmatrix} \succeq 0,\\
\label{eq:unitarity_2}
\ell= 0:\qquad
&\begin{pmatrix}
1 & 1\\
1 & 1
\end{pmatrix}+
i\begin{pmatrix}
0 & -\Phi_1^{\ell*}(s) + \Phi_2^{\ell*}(s) \\
\Phi_1^\ell(s) - \Phi_2^\ell(s) & 0 
\end{pmatrix} \succeq 0,\\
\label{eq:unitarity_3}
\ell\geq 1\;\;\text{(odd)}:\qquad
&\begin{pmatrix}
1 & 1\\
1 & 1
\end{pmatrix}+
2i\begin{pmatrix}
0 & -\Phi_3^{\ell*}(s)  \\
\Phi_3^\ell(s)  & 0 
\end{pmatrix} \succeq 0.
\end{align}
Finally, the matrix \eqref{eq:matrix_4} leads to the following condition
\begin{equation}
\label{eq:unitarity_4}
\ell\geq 2\;\;\text{(even)}:\qquad
\begin{pmatrix}
\mathbb I_{2\times 2} & \mathbb S^{\ell\dagger}_{2\times 2}(s) \\
\mathbb S^\ell_{2\times 2}(s) & \mathbb I_{2\times 2} 
\end{pmatrix} \succeq 0,
\end{equation}
where we have defined
\begin{equation}
\mathbb I_{2\times 2} \equiv
\begin{pmatrix}
1 & 0\\
0 & 1
\end{pmatrix},\qquad
\mathbb S^\ell_{2\times 2}(s) \equiv
\begin{pmatrix}
1 & 0\\
0 & 1
\end{pmatrix}+
i\begin{pmatrix}
\Phi_1^\ell(s) - \Phi_2^\ell(s)  &  \qquad 2 \Phi_5^{\ell}(s) \\
2 \Phi_5^\ell(s) &    \qquad 2\Phi_3^\ell(s)
\end{pmatrix}.
\end{equation}

It is interesting to note that the equations above do not contain the partial amplitude $\Phi_4^\ell(s)$ at all. This is because due to the $t-u$ crossing equations, see \eqref{eq:tu_crossing_matrix}, one has
\begin{equation}
\Phi_4(s,t,u)=\Phi_3(s,u,t)
\end{equation}
Using this inside \eqref{eq:partial_amplitudes} we get the following relation among the partial amplitudes\footnote{In order to show this, we use \eqref{eq:COM_parameters_2} and change the integration variable in \eqref{eq:partial_amplitudes} as $\theta\rightarrow \pi-\theta$. Using the properties of the small Wigner d matrix we get then $T_\ell{}^{-, +}_{+, -}(s) = (-1)^{\ell+1} T_\ell{}^{+, -}_{+, -}(s)$.}
\begin{equation}
\label{eq:relation_partial_34}
\Phi_4^\ell(s) = (-1)^{\ell+1} \Phi_3^\ell(s).
\end{equation}

\subsection{Non-perturbative couplings}
\label{sec:non-perturbative_couplings}

We can use the Majorana fermion scattering amplitude to define several non-perturbative coupling constants. These are useful parameters to describe the allowed space of QFTs.

\paragraph{Quartic coupling}
We begin by considering the value of the amplitude at the crossing symmetric point  
\begin{equation}
\label{eq:crossing_symmetric}
s=t=u=\frac{4m^2}{3}.
\end{equation}
At this point, the improved amplitudes $H_I$ must be invariant  under both the s-t \eqref{eq:crossing_st_fermions_amp_improved} and the s-u \eqref{eq:crossing_su_fermions_amp_improved} crossing equations
\begin{equation}
\label{eq:eigenvector}
\begin{aligned}
H_I\left(\frac{4m^2}{3},\frac{4m^2}{3},\frac{4m^2}{3}\right) = \sum_{J=1}^{5} \tilde C^{IJ}_{st} H_J\left(\frac{4m^2}{3},\frac{4m^2}{3},\frac{4m^2}{3}\right),\\
H_I\left(\frac{4m^2}{3},\frac{4m^2}{3},\frac{4m^2}{3}\right) = \sum_{J=1}^{5} \tilde C^{IJ}_{su} H_J\left(\frac{4m^2}{3},\frac{4m^2}{3},\frac{4m^2}{3}\right).
\end{aligned}
\end{equation}
The most general solution reads
\begin{equation}
\label{eq:csp}
\vec H(4m^2/3,4m^2/3,4m^2/3) = \frac{\lambda}{m^2} \times
\begin{pmatrix}
1 \\
0\\
0\\
1\\
-1
\end{pmatrix},
\end{equation}
where $\vec H$ represents the five amplitudes collectively and $\lambda$ is some parameter. We refer to $\lambda$ as the non-perturbative quartic coupling. 
By comparing \eqref{eq:csp} with the perturbative result \eqref{eq:amplitude_fermi_final} we see that  $\lambda$ can be identified with the coupling in front of the $(\overline \Psi \Psi) (\overline \Psi \Psi)$ interaction term in the Fermi theory.

\paragraph{Cubic (Yukawa) couplings}
Suppose now that our theory is described not only by the Majorana asymptotic state but also by a scalar (spin zero) asymptotic state with mass $M$. We restrict our attention to the values of $M$ in the range $(m,2m)$, where $m$ is the mass of the Majorana asymptotic state. This ensures that the Majorana fermion is the lightest particle in the theory and that the scalar boson is stable.

From general principles we expect such a scalar asymptotic state to manifest itself as a simple pole in the improved scattering amplitudes of Majorana fermions. In full generality one can then write
\begin{equation}
\label{eq:Yukawa}
\vec H_{bound}(s,t,u) = \begin{pmatrix}
\frac{g^2_1}{s-M^2} \\
\frac{g^2_2}{s-M^2}\\
\frac{g^2_3}{s-M^2} \\
\frac{g^2_4}{s-M^2} \\
\frac{g^2_5}{s-M^2} 
\end{pmatrix}
+ \begin{pmatrix}
\frac{g'^2_1}{t-M^2} \\
\frac{g'^2_2}{t-M^2}\\
\frac{g'^2_3}{t-M^2} \\
\frac{g'^2_4}{t-M^2} \\
\frac{g'^2_5}{t-M^2} 
\end{pmatrix}
+ \begin{pmatrix}
\frac{g''^2_1}{u-M^2} \\
\frac{g''^2_2}{u-M^2}\\
\frac{g''^2_3}{u-M^2} \\
\frac{g''^2_4}{u-M^2} \\
\frac{g''^2_5}{u-M^2} 
\end{pmatrix}
+{\rm regular},
\end{equation}
where $g_I$, $g_I'$ and $g_I''$ are 15 arbitrary parameters. In \eqref{eq:Yukawa} we have written only the poles and omitted all the regular terms at $s,t,u=M^2$.
Due to the crossing equations \eqref{eq:crossing_st_fermions_amp_improved} and \eqref{eq:crossing_su_fermions_amp_improved}, the values of $g_I'$ and $g_I''$ get fixed in terms of $g_I$. As a result we are left with 5 undetermined parameters $g_I$. 
We further require that the $s=M^2$ poles contribute only to the zero spin partial amplitudes. This enforces the fact that the particle generating the poles is a scalar. This leads to the following three additional constraints
\begin{equation}
g_2 = g_3 = g_4 = 0.
\end{equation}
Thus we are left with only two parameters $g_1$ and $g_5$.

According to the discussion of section \ref{sec:example_unitarity}, more precisely due to the formulas \eqref{eq:unitarity_1} and \eqref{eq:unitarity_2} one can take combinations of components of partial amplitudes to form  parity odd (-) and parity even (+) partial amplitudes which read as
\begin{equation}
\label{eq:objects_parity}
\Phi_-^\ell(s)\equiv \Phi_1^\ell+\Phi_2^\ell,\qquad
\Phi_+^\ell(s)\equiv \Phi_1^\ell-\Phi_2^\ell.
\end{equation}
The $s=M^2$ poles with the parameter $g_1$ contribute only to the parity even partial amplitude $\Phi_+^{\ell=0}$. We thus interpret $g_1$ as the non-perturbative coupling describing the interaction between two Majorana particles and a scalar parity even particle. The corresponding pole structure of the amplitude reads as
\begin{equation}
\vec H_{scalar}(s,t,u) =
\frac{1}{2}\, g^2 \times\vec P_{scalar}(s,t,u)
+{\rm regular},
\label{eq:object_P_scalar_part}
\end{equation}
where $g\equiv g_1$ and we have defined
\begin{equation}
\label{eq:object_P_scalar}
\vec P_{scalar}(s,t,u)\equiv
\begin{pmatrix}
-\frac{4}{s-M^2} + \frac{1}{t-M^2} + \frac{1}{u-M^2} \\
\frac{1}{t-M^2} - \frac{1}{u-M^2} \\
-\frac{1}{t-M^2} + \frac{1}{u-M^2}\\
-\frac{1}{t-M^2} - \frac{1}{u-M^2}\\
\frac{1}{t-M^2} + \frac{1}{u-M^2}
\end{pmatrix}
\end{equation}
The $s=M^2$ poles with the parameter $g_5$ instead contribute only to the parity odd partial amplitude $\Phi_-^{\ell=0}$.
Thus, the second parameter $g_5$ describes non-perturbatively the interaction between two Majorana particles and a scalar parity odd (pseudoscalar) particle.
The corresponding pole structure of the amplitude reads as
\begin{equation}
\vec H_{pseudoscalar}(s,t,u) =
\frac{1}{2}\,\tilde g^2\times\vec P_{pseudoscalar}(s,t,u)
+{\rm regular},
\label{eq:object_P_pseudoscalar_part}
\end{equation}
where $\tilde g\equiv g_5$ and we have defined
\begin{equation}
\label{eq:object_P_pseudoscalar}
\vec P_{pseudoscalar}(s,t,u)\equiv
\begin{pmatrix}
\frac{1}{t-M^2} + \frac{1}{u-M^2} \\
-\frac{1}{t-M^2}+ \frac{1}{u-M^2} \\
-\frac{1}{t-M^2} +\frac{1}{u-M^2}\\
\frac{1}{t-M^2} + \frac{1}{u-M^2}\\
-\frac{4}{s-M^2} + \frac{1}{t-M^2} + \frac{1}{u-M^2}
\end{pmatrix}.
\end{equation}
The cubic couplings $g\equiv g_1$ and $\tilde g \equiv g_5$ can also be called the non-perturbative Yukawa coupling constants. We also remark that the masses $M$ in \eqref{eq:object_P_scalar} and \eqref{eq:object_P_pseudoscalar} do not have to be the same.

In the discussion above we fixed the overall sign in \eqref{eq:object_P_scalar_part} and \eqref{eq:object_P_pseudoscalar_part} so that the residue at $s=M^2$ has the appropriate sign in the unitarity equations
\eqref{eq:unitarity_1} and \eqref{eq:unitarity_2}.  
Alternatively one can compare \eqref{eq:object_P_scalar_part} and \eqref{eq:object_P_pseudoscalar_part} to the perturbative results \eqref{eq:Yukawa_pert} and \eqref{eq:pseudoYukawa_pert}. This comparison not only fixes the correct signs but also shows that $g$ and $\tilde g$ here can be identified with the couplings appearing in front of the $\varphi \overline \Psi \Psi$ and $\tilde \varphi \overline \Psi \gamma^5 \Psi$ Yukawa interactions respectively in \eqref{eq:Yukawa_lagrangian} and \eqref{eq:pseudoYukawa_lagrangian}.

\subsection{An alternative approach to crossing equations}
\label{sec:crossing_alternative}

We have so far carefully discussed the construction of crossing equations in the COM frame and explicitly showed it in the case of Majorana particle scattering. There is an alternative way of approaching this topic, namely using the fully covariant language based on constructing tensor structures. This relies on the most general representation of a scattering amplitude given in \eqref{eq:tensor_structures}.

In what follows we will construct tensor structures in the particular example of Majorana particle scattering and re-derive the crossing equations. We will describe a general procedure of constructing tensor structures for particles with spin in appendix \ref{sec:general_spin_tensor_structures}.

\subsubsection{Tensor structures}
\label{sec:example_tensor_structures}
As we know from the COM analysis, there are 5 independent amplitudes in the case of identical Majorana particles. As a result there will be 5 linearly independent tensor structures. A particular choice of these tensor structures was made in \cite{Gribov:1963gx}. It reads
\begin{equation}
\label{eq:tensor_structure_compact}
\mathbb{T}_I{}_{\lambda_1, \lambda_2}^{\lambda_3,\lambda_4}(p_1,p_2,p_3,p_4) =
-\frac{1}{2}\,
[\bar u_{\lambda_4} (p_4)\mathcal \cO_I v_{\lambda_3}(p_3)]
\cdot
[\bar v_{\lambda_2}(p_2) \mathcal \cO_I u_{\lambda_1}(p_1)],
\end{equation}
where the five $4\times 4$ matrices $\cO_I$ are given by
\begin{equation}
\cO_1\equiv 1_{4\times 4},\quad
\cO_2\equiv \gamma^\mu,\quad
\cO_3\equiv \sqrt{2}\,\sigma^{\mu\nu},\quad
\cO_4\equiv i\,\gamma^5\gamma^\mu,\quad
\cO_5\equiv \gamma^5.
\end{equation}
The symbol ``$\cdot$'' in \eqref{eq:tensor_structure_compact} means contraction of all the Lorentz indices among $\cO_I$ matrices. Notice also that there is no summation over the repeated index $I$ in \eqref{eq:tensor_structure_compact}.
For the readers convenience we also write \eqref{eq:tensor_structure_compact} explicitly
\begin{equation}
\label{tensor_structures_4_majorana}
\begin{aligned}
\mathbb{T}_1{}_{\lambda_1, \lambda_2}^{\lambda_3,\lambda_4}(p_1,p_2,p_3,p_4) &= -\frac{1}{2}[\bar{u}_{\lambda_4} (p_4) v_{\lambda_3}(p_3)][\bar v_{\lambda_2}(p_2) u_{\lambda_1}(p_1)],\\
\mathbb{T}_2{}_{\lambda_1, \lambda_2}^{\lambda_3,\lambda_4}(p_1,p_2,p_3,p_4) &= -\frac{1}{2}[\bar u_{\lambda_4} (p_4)\gamma_\mu v_{\lambda_3}(p_3)][\bar v_{\lambda_2}(p_2) \gamma^\mu u_{\lambda_1}(p_1)],\\
\mathbb{T}_3{}_{\lambda_1, \lambda_2}^{\lambda_3,\lambda_4}(p_1,p_2,p_3,p_4) &= \;- \;\,[ \bar u_{\lambda_4} (p_4)\sigma_{\mu \nu} v_{\lambda_3}(p_3)][\bar v_{\lambda_2}(p_2)\sigma^{\mu \nu} u_{\lambda_1}(p_1)],\\
\mathbb{T}_4{}_{\lambda_1, \lambda_2}^{\lambda_3,\lambda_4}(p_1,p_2,p_3,p_4) &= + \frac{1}{2}[ \bar u_{\lambda_4} (p_4)\gamma_5 \gamma_\mu v_{\lambda_3}(p_3)][\bar v_{\lambda_2}(p_2)\gamma_5 \gamma^\mu u_{\lambda_1}(p_1)],\\
\mathbb{T}_5{}_{\lambda_1, \lambda_2}^{\lambda_3,\lambda_4}(p_1,p_2,p_3,p_4) &= -\frac{1}{2}[\bar u_{\lambda_4} (p_4) \gamma_5 v_{\lambda_3}(p_3)][\bar v_{\lambda_2}(p_2) \gamma_5 u_{\lambda_1}(p_1)].
\end{aligned}
\end{equation}
The objects $u_\lambda(p)$ and $v_\lambda(p)$ are the usual 4-spinor solutions to the Dirac equation and
\begin{equation}
\bar u_\lambda(p) \equiv  u_\lambda^\dagger(p)\gamma^0,\quad
\bar v_\lambda(p) \equiv  v_\lambda^\dagger(p)\gamma^0.
\end{equation}
We use the Weyl (also known as chiral) basis for the Dirac $\gamma^\mu$ matrices given in \eqref{eq:Dirac_matrices} and the helicity basis of states. With these conventions the spinor solutions to the Dirac equation read
\begin{equation}
\label{eq:Dirac_solutions}
\begin{aligned}
u_{\frac{1}{2}}(p) &= \frac{1}{\sqrt{2}}\,
\begin{pmatrix}
\;\;\,\sqrt{p^0-\myP}\,\cos \frac{\theta}{2} &\\
\;\;\,\sqrt{p^0-\myP}\,\sin \frac{\theta}{2} &  e^{+i \phi}\\
\;\;\,\sqrt{p^0+\myP}\,\cos \frac{\theta}{2} &\\
\;\;\,\sqrt{p^0+\myP}\,\sin \frac{\theta}{2} & e^{+i \phi}
\end{pmatrix},\;
u_{-\frac{1}{2}}(p) = \frac{1}{\sqrt{2}}\,
\begin{pmatrix}
-\sqrt{p^0+\myP}\,\sin \frac{\theta}{2} &  e^{-i \phi}\\
\,\;\;\sqrt{p^0+\myP}\,\cos \frac{\theta}{2} &\\
-\sqrt{p^0-\myP}\,\sin \frac{\theta}{2} & e^{-i \phi}\\
\,\;\;\sqrt{p^0-\myP}\,\cos \frac{\theta}{2} &
\end{pmatrix},\\
v_{\frac{1}{2}}(p) &= \frac{1}{\sqrt{2}}\,
\begin{pmatrix}
-\sqrt{p^0+\myP}\,\sin \frac{\theta}{2} &  e^{-i \phi}\\
\;\;\,\sqrt{p^0+\myP}\,\cos \frac{\theta}{2} &\\
\;\;\,\sqrt{p^0-\myP}\,\sin \frac{\theta}{2} & e^{-i \phi}\\
-\sqrt{p^0-\myP}\,\cos \frac{\theta}{2} &
\end{pmatrix},\;
v_{-\frac{1}{2}}(p) = \frac{1}{\sqrt{2}}\,
\begin{pmatrix}
-\sqrt{p^0-\myP}\,\cos \frac{\theta}{2} &\\
-\sqrt{p^0-\myP}\,\sin \frac{\theta}{2} & e^{+i \phi}\\
\;\;\,\sqrt{p^0+\myP}\,\cos \frac{\theta}{2} &\\
\;\;\,\sqrt{p^0+\myP}\,\sin \frac{\theta}{2} &e^{+i \phi}
\end{pmatrix}.
\end{aligned}
\end{equation}
We notice that the objects $u_\lambda(p)$ and $\bar v_\lambda (p)$ transform in the spin-1/2 representation, namely they get rotated by $D^{(1/2)}_{\lambda' \lambda}$, whereas the objects $v_\lambda (p)$ and $\bar u_\lambda (p)$ transform in the conjugate spin-1/2 representation, namely they get rotated by $D^{*(1/2)}_{\lambda' \lambda}$. The tensor structures \eqref{tensor_structures_4_majorana} are constructed in such a way that all the Lorentz indices are contracted. They depend only on the helicity labels and thus transform only in the Little group induced representation.

The choice of tensor structures \eqref{tensor_structures_4_majorana} is very convenient because of the following reason. Plugging \eqref{tensor_structures_4_majorana} into \eqref{eq:tensor_structures} and evaluating the amplitudes in the COM frame we get
\begin{equation}
\label{eq:ts_com}
T_{\lambda_1, \lambda_2}^{\lambda_3, \lambda_4}(s,t,u) = \sum_{I=1}^{5} H_I(s,t,u) \mathbb{T}_I{}_{\lambda_1, \lambda_2}^{\lambda_3,\lambda_4}(p_1^{com},p_2^{com},p_3^{com},p_4^{com}).
\end{equation}
In the left-hand side of \eqref{eq:ts_com} we get the 16 COM amplitudes. They are related  by \eqref{eq:Majorana_identical} and \eqref{eq:Majorana_parity}. The 5 independent amplitudes were chosen in \eqref{eq:Majorana_independent} and given special names $\Phi_I(s,t,u)$. For the five independent amplitudes  $\Phi_I(s,t,u)$ the right-hand side of \eqref{eq:ts_com} simply becomes \eqref{eq:ksfa_definition}.
In other words the functions $H_I(s,t,u)$ appearing in \eqref{eq:ts_com} are precisely the kinematic singularity free amplitudes found in section \ref{sec:improved_amplitudes}. This means that all the kinematic singularities and constraints are taken care of by the tensor structures!
Having established this we can write \eqref{eq:tensor_structures} explicitly in the case of identical Majorana fermions. It reads
\begin{equation}
\label{eq:amplitude_with_tensor_structures_repeated}
T_{\lambda_1, \lambda_2}^{\lambda_3,\lambda_4}(p_1,p_2,p_3,p_4) = 
\sum_{I=1}^{5} H_I(s,t,u)\; \mathbb{T}_I{}_{\lambda_1, \lambda_2}^{\lambda_3,\lambda_4}(p_1,p_2,p_3,p_4),
\end{equation}
where the basis of tensor structures is given by \eqref{tensor_structures_4_majorana} and the functions $H_I(s,t,u)$ in this basis are precisely the improved amplitudes defined in section \ref{sec:improved_amplitudes}.

\subsubsection{Verification of crossing matrices}
In this section we will re-derive the crossing equations \eqref{eq:crossing_st_fermions_amp_improved} and \eqref{eq:crossing_su_fermions_amp_improved}.

The amplitude \eqref{eq:amplitude_with_tensor_structures_repeated} is defined for $p_i^0\geq 0$ $(i=1,2,3,4)$ as usual. It can however be analytically continued to $p_i^0< 0$ domain. There are several options for such an analytic continuation. Throughout this paper we use option \eqref{eq:ac_2}. 
Let us now analytically continue both sides of \eqref{eq:amplitude_with_tensor_structures_repeated} in $p_2$ and $p_3$ to their negative values, we obtain
\begin{equation}
T_{\lambda_1, \lambda_2}^{\lambda_3,\lambda_4}(p_1,-p_2,-p_3,p_4) = 
\sum_{I=1}^{5} H_I(s,t,u)\; \mathbb{T}_I{}_{\lambda_1, \lambda_2}^{\lambda_3,\lambda_4}(p_1,-p_2,-p_3,p_4),
\end{equation}
Exchanging the role of labels 2 and 3 we get then
\begin{equation}
\label{st_crossed_amplitude_with_tensor_structure}
T_{\lambda_1, \lambda_3}^{\lambda_2,\lambda_1}(p_1,-p_3,-p_2,p_4) = 
\sum_{I=1}^{5} H_I(t,s,u)\; \mathbb{T}_I{}_{\lambda_1, \lambda_3}^{\lambda_2,\lambda_4}(p_1,-p_3,-p_2,p_4).
\end{equation}
Plugging \eqref{eq:amplitude_with_tensor_structures_repeated} and \eqref{st_crossed_amplitude_with_tensor_structure} into the crossing equation \eqref{eq:s_t_crossing_general_23} taken with the positive sign for fermions, see \eqref{eq:crossing_LSZ_ac2} for details, one gets
\begin{equation}
\label{eq:crossing equation_structures}
\sum_{I=1}^{5} H_I(s,t,u)\; \mathbb{T}_I{}_{\lambda_1, \lambda_2}^{\lambda_3,\lambda_4}(p_1,p_2,p_3,p_4)=
\sum_{I=1}^{5} H_I(t,s,u)\; \mathbb{T}_I{}_{+\lambda_1, -\lambda_3}^{-\lambda_2,+\lambda_4}(p_1,-p_3,-p_2,p_4).
\end{equation}
The analytic continuations of $u$ and $v$ objects are given in \eqref{eq:uv_continuation_2}. Using these and the explicit form of tensor structures
\eqref{eq:tensor_structure_compact} one gets
\begin{equation}
\label{eq:negative_structure_to_positive_structure}
\begin{aligned}
\mathbb{T}_I{}_{+\lambda_1, -\lambda_3}^{-\lambda_2,+\lambda_4}(p_1,-p_3,-p_2,p_4) &= [\bar u_{\lambda_4} (p_4)\mathcal O_I v_{-\lambda_2}(-p_2)]\cdot[\bar v_{-\lambda_3}(-p_3) \mathcal O_I u_{\lambda_1}(p_1)]  \\
&= -\;
[\bar u_{\lambda_4} (p_4)\mathcal O_I u_{\lambda_2}(p_2)]
\cdot
[\bar u_{\lambda_3}(p_3) \mathcal O_I u_{\lambda_1}(p_1)]\\
&= \sum_{I=1}^5 \widetilde C^{st}_{IJ}\mathbb{T}_I{}_{\lambda_1, \lambda_2}^{\lambda_3,\lambda_4}(p_1,p_2,p_3,p_4),
\end{aligned}
\end{equation}
where the matrix $\widetilde C^{st}_{IJ}$ is precisely the crossing matrix \eqref{eq:crossing_matrices_numerical}. In the third line of \eqref{eq:negative_structure_to_positive_structure} we have used the Fierz identities, see for example \cite{Dreiner:2008tw}. Plugging \eqref{eq:negative_structure_to_positive_structure} into \eqref{eq:crossing equation_structures} and using the fact that the structures \eqref{eq:tensor_structure_compact} form a basis, we obtain the final crossing equations
\begin{equation}
H_I(s,t,u) = \sum_{J=1}^5 \widetilde C^{st}_{IJ} H_J(t,s,u),
\end{equation}
which coincide with \eqref{eq:crossing_st_fermions_amp_improved}. Using   identical arguments one can obtain the $s-u$ crossing equations \eqref{eq:crossing_su_fermions_amp_improved}.

At first glance, it might seem that the way of deriving crossing equations using tensor structures is much simpler than the COM frame approach. This is not necessarily the case especially if one works with higher spin particles. The main issue here is the construction of a linearly independent basis of tensor structures. In practice there are many different looking tensor structures one can write. They are however related via a complicated set of Fierz-like identities. Luckily in the case of identical Majorana particles the problem was already thoroughly studied and  the set of linearly independent tensor structures \eqref{tensor_structures_4_majorana} was well known.

Once the linearly independent basis of structures is chosen the troubles are unfortunately not over. In the process of deriving the crossing equations all allowed tensor structures reappear and they need to be expressed back in terms of the chosen basis of structures via the Fierz-like identities. In the case of identical Majorana fermions this is precisely the step done in the last line of \eqref{eq:negative_structure_to_positive_structure} which can be  quite tedious for particles with general spin.

\section{Numerical bounds}
\label{sec:results}
In this section we  numerically estimate non-perturbative bounds on quantum field theories where the scattering of Majorana particles can be defined.

In sections \ref{sec:improved_amplitudes} and \ref{sec:example_unitarity} we carefully derived the crossing equations and the unitarity constraints which any scattering amplitude of Majorana particles must satisfy.
Our precise goal here is to derive various bounds on the non-perturbative coupling constants that we defined in 
 section \ref{sec:non-perturbative_couplings}.
In section \ref{sec:setup} we explain the numerical setup which allows for this. We present the numerical results in sections \ref{sec:quartic} and \ref{sec:qubic}.

\subsection{Setup}
\label{sec:setup}
We use the numerical approach of \cite{Paulos:2017fhb}.
The first step of this approach is to write the most general ansatz for the scattering amplitude. Before addressing Majorana fermions let us quickly recap the scalar case. The non-trivial part of the scattering of identical scalars with mass $m$ is described by a function of three Mandelstam variables $T(s,t,u)$. To proceed it is crucial to assume maximal analyticity, namely that the amplitude is an analytic function of $s$, $t$ and $u$ complex variables independently modulo the standard branch cuts at
\begin{equation}
s\in [4m^2,+\infty],\qquad
t\in [4m^2,+\infty],\qquad
u\in [4m^2,+\infty],
\end{equation}
where $m$ is the mass of the scalar particle. Given a $z$ complex plane, to mimic the above situation, one can define a function which is analytic in the whole complex plane modulo $z\in [4m^2,+\infty]$ branch cut. We choose\footnote{The physical domain is defined via $
s+i\epsilon$ with $\epsilon>0$.	We can thus rotate the cuts using the identity
\begin{equation}
\nn
\sqrt{4m^2-s} = -i\,\sqrt{s-4m^2}.
\end{equation}
}
\begin{equation}
\label{eq:rho}
\myRho(z;z_0) \equiv \frac{\sqrt{4m^2-z_0}-\sqrt{4m^2-z}}{\sqrt{4m^2-z_0}+\sqrt{4m^2-z}}.
\end{equation}
Here $z_0<4m^2$ is a free parameter which can be chosen at our will. We can then represent the interacting part of the scalar amplitude as a simple power series in terms of the functions \eqref{eq:rho} in the following way
\begin{equation}
\label{eq:scalar_ansatz}
T(s,t,u) = \sum_{a=0}^\infty\sum_{b=0}^\infty\sum_{c=0}^\infty
\alpha_{abc}\;\myRho(s,s_0)^a\myRho(t,t_0)^b\myRho(u,u_0)^c,
\end{equation}
where $\alpha_{abc}$ are some real parameters. The ansatz \eqref{eq:scalar_ansatz} has a lot of redundancies due to the condition $s+t+u=4m^2$. One can attempt to remove them in various ways. In this paper we do it in a slightly drastic manner by imposing
\begin{equation} 
\alpha_{abc} = 0\qquad
{\rm if} \qquad  abc\neq 0\,.
\end{equation}
This choice is motivated by the Mandelstam representation \cite{Mandelstam:1959bc}, see appendix C in \cite{Paulos:2017fhb} for further details.
It is convenient to choose the values of $s_0$, $t_0$ and $u_0$ to all be the crossing symmetric point
\begin{equation}
\label{eq:crossin_symmetric_point}
s_0=t_0=u_0 =\frac{4}{3}\,m^2.
\end{equation}
The $s-t$ and $s-u$ crossing equations for the scalar amplitude are very simple. They read
\begin{equation}
\label{eq:crossin_scalars}
T(s,t,u) =
T(t,s,u)=
T(u,t,s).
\end{equation}
Plugging here the ansatz \eqref{eq:scalar_ansatz} we see that the coefficients $\alpha_{abc}$ must be fully symmetric under the permutation of indices $a$, $b$ and $c$.
The scalar amplitude $T(s,t,u)$ can also have poles when other particles exist in the theory or if the asymptotic state is allowed to have self-interactions. In this case one should extend the ansatz \eqref{eq:scalar_ansatz} to include such poles. We will see how it works in the case of Majorana fermions and skip further discussion of the scalar case.

In the case of Majorana fermions we defined five amplitudes $\vec\Phi$, where the vector denotes the five components $\Phi_1$, $\Phi_2$, $\Phi_3$, $\Phi_4$ and $\Phi_5$ collectively. These amplitudes contain kinematic non-analyticities. In order to remove them in section \ref{sec:improved_amplitudes} we introduced a new set of improved amplitudes denoted by $\vec H$. Again the vector here denotes the five components $H_1$, $H_2$, $H_3$, $H_4$ and $H_5$ collectively. We assume now that these five improved amplitudes are maximally analytic. Analogously to the scalar case, this allows us to write the following most general ansatz
\begin{align}
\nn
\vec H(s,t,u) &= 
\frac{1}{2}\, g^2 \times\vec P_{scalar}(s,t,u) +
\frac{1}{2}\, \tilde g^2 \times\vec P_{pseudoscalar}(s,t,u) \\
&+\sum_{a=0}^\infty\sum_{b=0}^\infty\sum_{c=0}^\infty
\vec\alpha_{abc}\;\myRho(s,s_0)^a\myRho(t,t_0)^b\myRho(u,u_0)^c,
\label{eq:fermions_ansatz}
\end{align}
where $\vec P_{scalar}$ and $\vec P_{pseudoscalar}$ are the terms containing poles defined in \eqref{eq:object_P_scalar} and \eqref{eq:object_P_pseudoscalar} and $\vec\alpha_{abc}$ are some real parameters. In order to remove the redundancies in the ansatz \eqref{eq:fermions_ansatz} as in the scalar case we require
\begin{equation}
\vec \alpha_{abc} = 0
\qquad {\rm if}
\qquad
abc\neq 0\,.
\end{equation}
The ansatz \eqref{eq:fermions_ansatz} has an infinite number of terms. In order to work with it one has to introduce the following truncation
\begin{equation}
\sum_{a=0}^\infty\sum_{b=0}^\infty\sum_{c=0}^\infty \longrightarrow
\sum_{a+b+c\leq N_{max}},
\end{equation}
where $N_{max}$ is some cut-off parameter. In practice its value is taken to be around 20.

We now require the ansatz \eqref{eq:fermions_ansatz} to satisfy the crossing equations \eqref{eq:crossing_st_fermions_amp_improved} and \eqref{eq:crossing_su_fermions_amp_improved}. The pole terms have been already constructed to obey the crossing equations. However, for the parameters $\vec\alpha_{abc}$ we get a non-trivial system of linear algebraic equations with constant coefficients (independent of $s$, $t$ and $u$ variables). For a chosen $N_{max}$ we solve this system with computer algebra and plug the solution into \eqref{eq:fermions_ansatz}. As a result we get a fully crossing invariant expression. From now on when we refer to  \eqref{eq:fermions_ansatz} we assume that the above procedure has been done and that \eqref{eq:fermions_ansatz} is fully crossing symmetric.

The second step is to compute the partial amplitudes. In order to do that we need to obtain the amplitudes $\vec \Phi(s,t,u)$ (containing all the kinematic non-analyticities) by plugging the ansatz \eqref{eq:fermions_ansatz} into \eqref{eq:ksfa_definition}. We then compute the partial amplitudes $\vec \Phi^\ell(s)$ using \eqref{eq:partial_amplitudes}, see also \eqref{eq:Majorana_independent} and \eqref{eq:Majorana_independent_partial}. In doing this one needs to perform a set of integrals which have the following form
\begin{equation}
\label{eq:core integral}
\int_{0}^{\pi} d\theta \sin\theta\, d^{(\ell)}_{ij}(\theta)\;
\myRho\big(t(s,\theta),t_0\big)^a\myRho\big(u(s,\theta),u_0\big)^b
\end{equation}
for the following set of indices in the d-matrix  
\begin{equation}
(i,j)=\{ (0,0),\,(1,1),\,(1,-1),\,(0,1) \}.
\end{equation}
The integral \eqref{eq:core integral} is hard  to compute analytically. Hence, we perform the integration numerically   for some tabulated values $a$, $b$, $\ell$ and $s$. We do it with Mathematica requiring between 20 and 30 digits of precision. The computed partial amplitudes can then be plugged into the unitarity constraints \eqref{eq:unitarity_1} - \eqref{eq:unitarity_4}. These become a set of numerical semi-definite positivity conditions for different values of spin $\ell$ and $s$.

It is important to explain how we choose the values of   $\ell$ and $s$. Let us start with the former. In principle one needs to consider unitarity conditions for all the spins up to $\ell=\infty$. However, realistically this is not possible and one needs to introduce another truncation, namely we impose unitarity only for a finite set of spins
\begin{equation}
\ell=0,1,2,\ldots,L_{max}.
\end{equation}
In the majority of computations we take the values of $L_{max}$ to be as follows
\begin{equation}
\label{eq:spin_cut-off}
L_{max} = N_{max} + 20.
\end{equation}
In the next section we justify our choice by varying $L_{max}$ and $N_{max}$ separately. Let us address the choice of $s$ values now. The unitarity constraints \eqref{eq:unitarity_1} - \eqref{eq:unitarity_4} are imposed in the region
\begin{equation}
\label{eq:region}
s\in[4m^2,+\infty]
\end{equation}
slightly above the branch cut. We take 300 different values of $s$ in the region \eqref{eq:region}. One can spread these points differently, in practice we use the Chebyshev distribution.\footnote{More precisely, we define a variable $\phi(s)$ by $\myRho(s,s_0)=e^{i \phi(s)}$. Notice that $s \in [4m^2,+\infty]$ corresponds to
$\phi \in [0,\pi]$. Then, we pick a grid $\phi_k = \frac{\pi}{2} \left[1-\cos\frac{\pi (k-1/2)}{n}\right]$ with $n=300$ and $k=1,\dots, n$.}

To summarize, we wrote the unitarity constraints in terms of the unknown real coefficients $\{g^2 ,\tilde g ^2, \vec \alpha_{abc}\}$ originally appearing in the (crossing symmetric) ansatz \eqref{eq:fermions_ansatz}. These conditions were written in a positive semi-definite form. We can now look for these coefficients numerically using   semi-definite programming. For this we employ SDPB \cite{Simmons-Duffin:2015qma,Landry:2019qug}.\footnote{SDPB works only with real matrices. The unitarity conditions \eqref{eq:unitarity_1} - \eqref{eq:unitarity_4} are formulated in terms of the hermitian matrices. In order to recast those conditions into the form used by SDPB one needs to use the equivalence
\begin{equation}
H\succeq 0 \quad\Leftrightarrow\quad
\begin{pmatrix}
Re(H) & -Im(H) \\
Im(H) &  Re(H)
\end{pmatrix}\succeq 0,
\end{equation}
where $H$ is some hermitian matrix.
}
In the following sections we define two different optimization problems and provide the numerical results. All the optimization problems we consider below use the following normalization
\begin{equation}
m=1.
\end{equation}
This simply means that all the dimensionful quantities are measured in terms of the mass of the Majorana asymptotic state.

A word about choosing various parameters of the setup such as the number of $s$ values or the precision of the numerical approximation for the integrals \eqref{eq:core integral}. When performing the numerical analysis we made sure that our results do not depend on these parameters. This is simply done by performing the same computation with two different sets of parameters and confirming that the outcome is stable under such a change. The choice we made in this work guarantees at least two digits of precision in the final answer.

In order to obtain the numerical results presented below we  consumed 0.4 million CPU Hours on the EPFL SCITAS cluster.

\subsection{Quartic coupling}
\label{sec:quartic}
We now apply the strategy described in the previous section to bound the quartic coupling defined in \eqref{eq:csp} in the absence of poles
\begin{equation}
g=\tilde g =0.
\end{equation}
At the crossing symmetric point \eqref{eq:crossin_symmetric_point} the $\myRho$ function vanishes and thus the ansatz \eqref{eq:fermions_ansatz} depends only on the five coefficients $\vec \alpha_{000}$. Crossing implies that only one of those coefficients is really independent. By comparing the ansatz at the crossing symmetric point with \eqref{eq:csp} we conclude that
\begin{equation}
\alpha^2_{000} = \alpha^3_{000} = 0,\qquad
\alpha^1_{000} = \alpha^4_{000} = -\alpha^5_{000} =\lambda,
\end{equation}
where $\lambda$ is the non-perturbative quartic coupling.

\emph{Optimization problem:} we search for the coefficients $\vec \alpha_{abc}$ such that the non-perturbative quartic coupling $\lambda$ has the smallest/largest value and the unitarity constraints \eqref{eq:unitarity_1} - \eqref{eq:unitarity_4} are satisfied. By solving this problem numerically we conclude that the quartic coupling must be in the following interval
\begin{equation}
\label{eq:bound_quartic}
\frac{\lambda}{32\pi} \in [-3.25,\; +1.74].
\end{equation}
This is reasonably compatible with the expectation $\lambda \lesssim (4\pi)^2$  from \emph{naive dimensional analysis}  \cite{Manohar:1983md,Jenkins:2013sda}.

\begin{figure}[!htb]
	\begin{center}
		\includegraphics[scale=0.90]{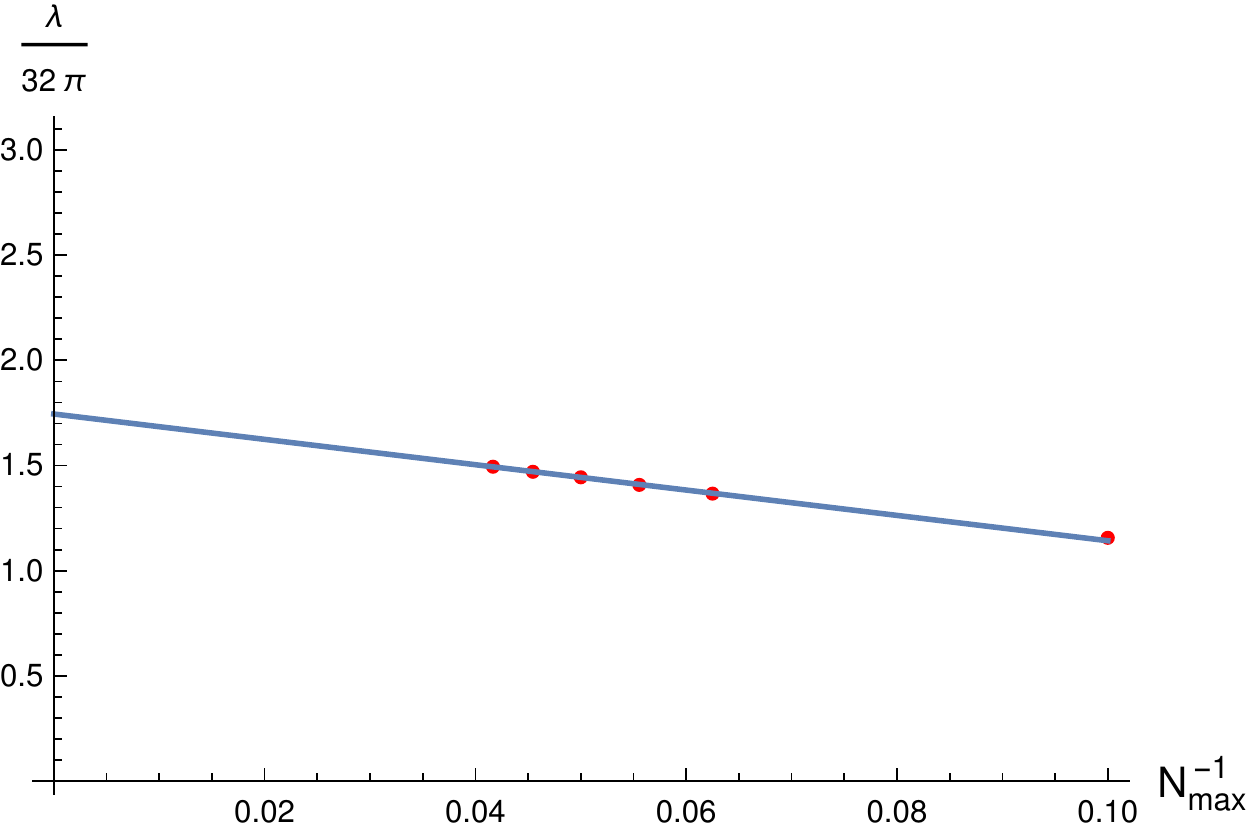}
		\caption{Upper bound on the quartic coupling $\lambda$ as a function of $N_{max}^{-1}$. The numerical results are indicated by the red dots. They correspond to $N_{max}=10, 16,18,20,22$ and $24$. The blue line represents the linear fit of the three points $N_{max}=20,\,22$ and $24$. It is described by $\lambda/(32\pi) = 1.74 - 6.02\, N_{max}^{-1}$ equation. The spin cut-off parameter used here is $L_{max}=N_{max}+20$.}
		\label{fig:quartic_upper}
	\end{center}
\end{figure}

\begin{figure}[!htb]
	\begin{center}											\includegraphics[scale=0.90]{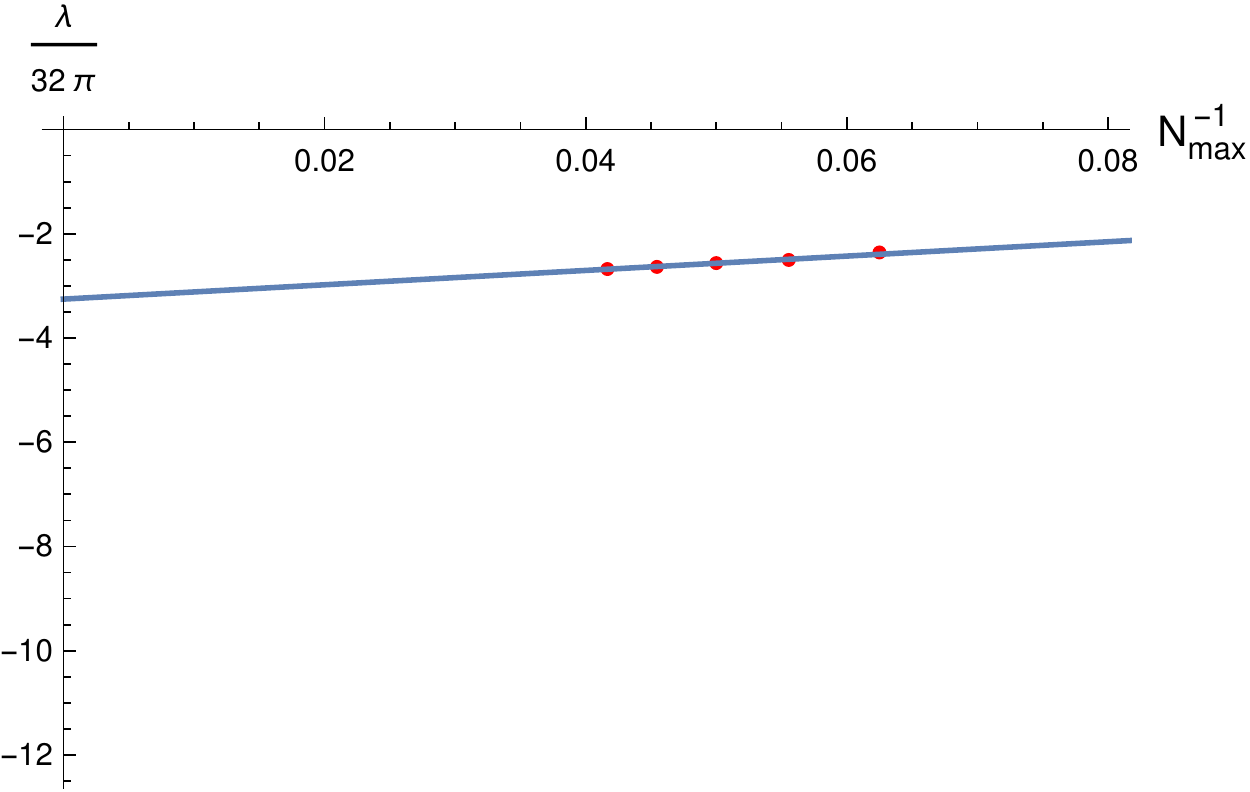}
		\caption{Lower bound on the quartic coupling $\lambda$ as a function of $N_{max}^{-1}$. The numerical results are indicated by the red dots. They correspond to $N_{max}=10, 16,18,20,22$ and $24$. The blue line represents the linear fit of the three points $N_{max}=20,\,22$ and $24$. It is described by $\lambda/(32\pi) = -3.25 + 13.76\, N_{max}^{-1}$ equation. The spin cut-off parameter used here is $L_{max}=N_{max}+20$.}
		\label{fig:quartic_lower}
	\end{center}
\end{figure}

Let us now discuss the details of this result. Among other parameters our numerical setup depends on $N_{max}$. In figures \ref{fig:quartic_upper} and \ref{fig:quartic_lower} we present the upper and lower bound on $\lambda$ as a function of $N_{max}^{-1}$. The highest value we probe is $N_{max}=24$. One can see that the bounds get weaker as we increase $N_{max}$. Intuitively this is easy to understand: upon increasing $N_{max}$ the ansatz becomes more general and thus a larger/smaller coupling is attainable.
We then perform an extrapolation of our results to $N_{max}=\infty$. The bound \eqref{eq:bound_quartic} already includes this extrapolation.

\begin{figure}[!htb]
	\begin{center}
		\includegraphics[scale=0.85]{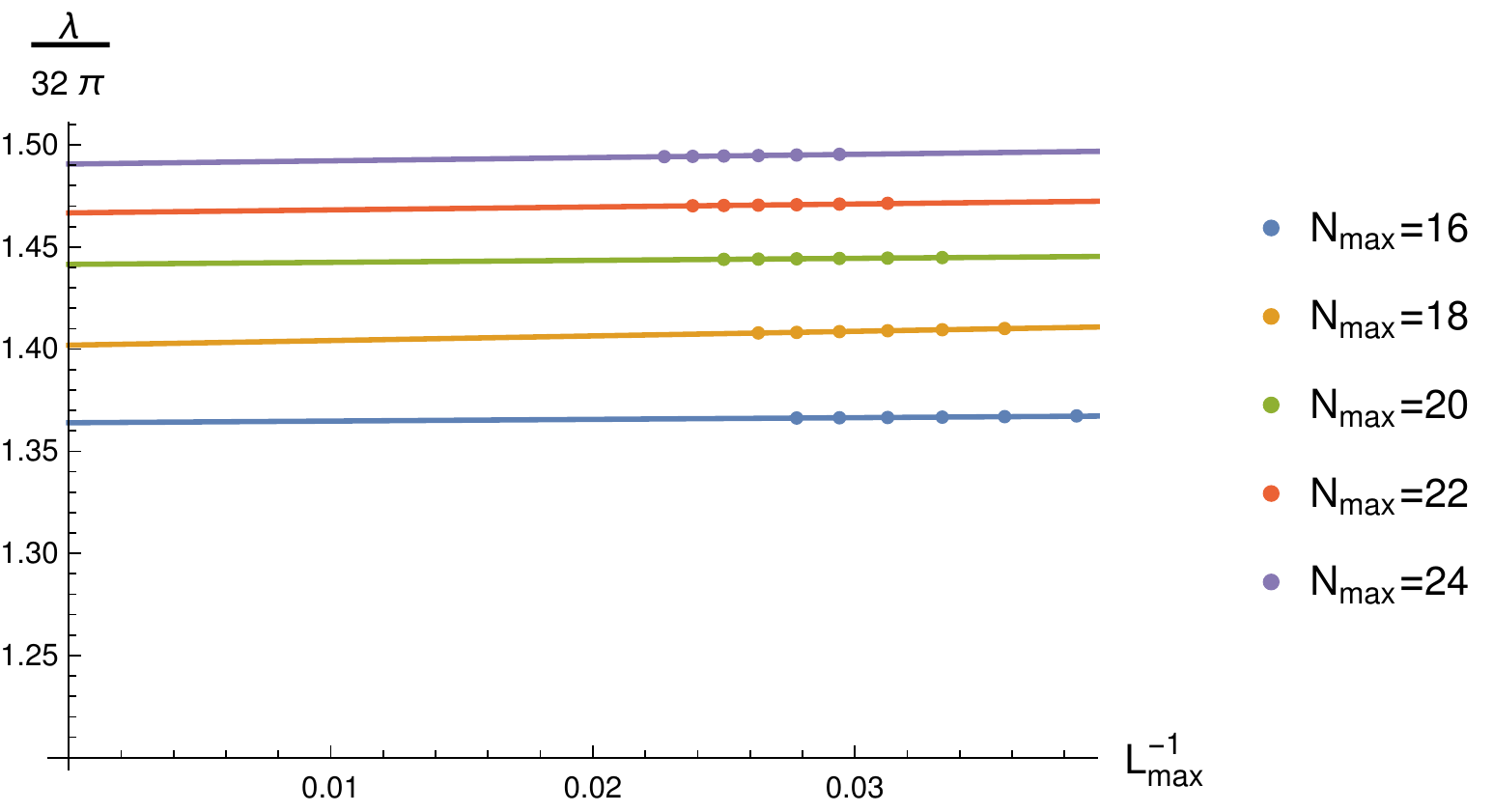}
		\caption{Upper bound on the quartic coupling $\lambda$ as a function of $L_{max}^{-1}$. The dots represent the numerical results. The solid lines represent the linear extrapolation in $L_{max}^{-1}$ based on the last four points for each $N_{max}$. Different colours correspond to different values of $N_{max}$ indicated in the right-hand side of the plot.}
		\label{fig:spin_upper}
	\end{center}
\end{figure}
\begin{figure}[!htb]
	\begin{center}
		\includegraphics[scale=0.85]{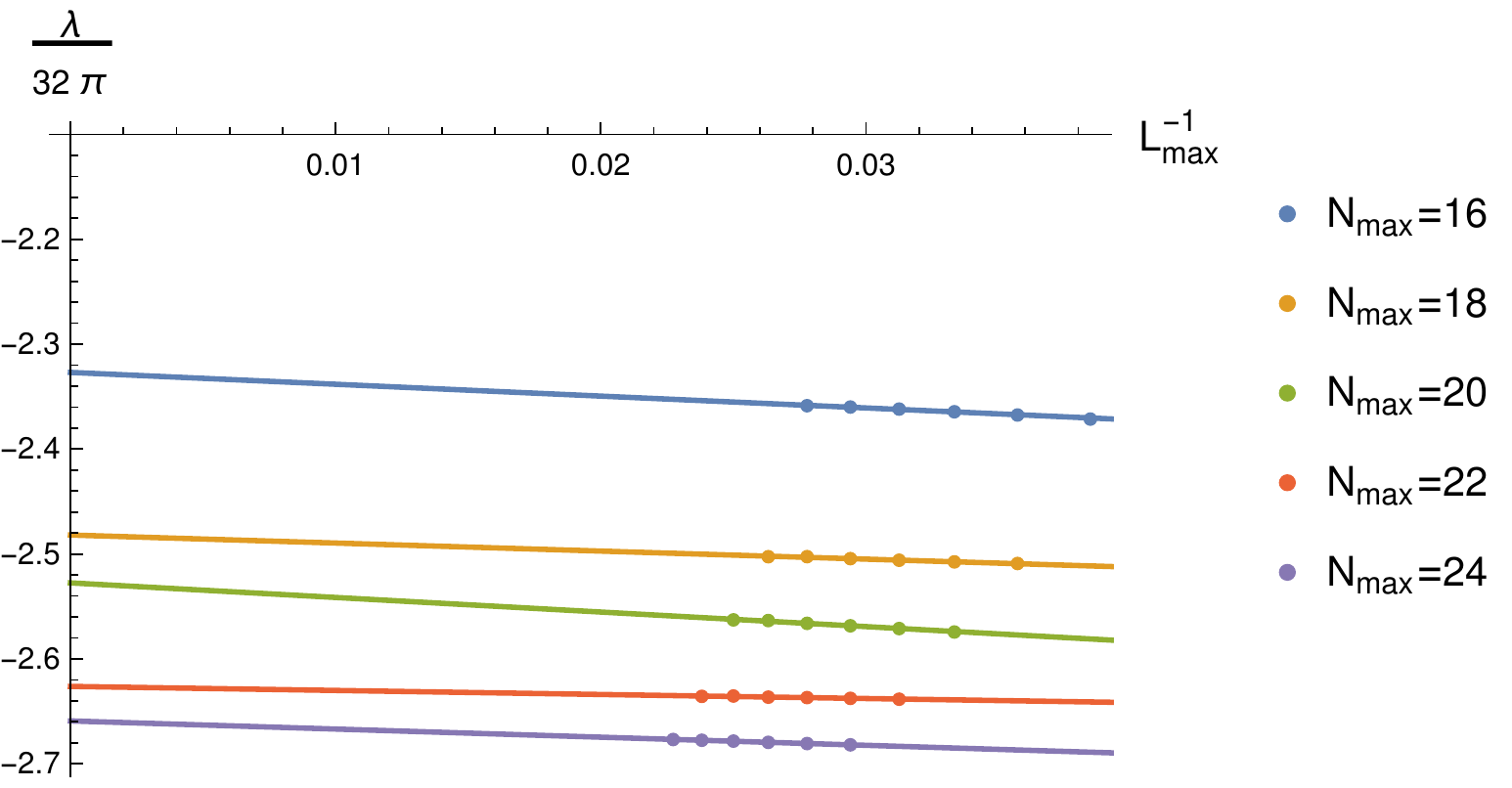}
		\caption{Lower bound on the quartic coupling $\lambda$ as a function of $L_{max}^{-1}$. The dots represent the numerical results. The solid lines represent the linear extrapolation in $L_{max}^{-1}$ based on the last four points for each $N_{max}$. Different colours correspond to different values of $N_{max}$ indicated in the right-hand side of the plot.}
		\label{fig:spin_lower}
	\end{center}
\end{figure}

In making figures \ref{fig:quartic_upper} and \ref{fig:quartic_lower} we have used the spin cut-off value  $L_{max}=N_{max}+20$ as indicated in \eqref{eq:spin_cut-off}. Let us now relax that condition and see the dependence of the bounds also on $L_{max}$. In figures \ref{fig:spin_upper} and \ref{fig:spin_lower} we present the upper and lower bound on the quartic coupling $\lambda$ as a function of $L_{max}^{-1}$ for various values of $N_{max}$. We also perform a linear extrapolation to $L_{max}=\infty$.

When constructing figures \ref{fig:quartic_upper} and \ref{fig:quartic_lower} one could have   used the points extrapolated to $L_{max}=\infty$  and only then perform the extrapolation to $N_{max}=\infty$. We have done this exercise and found  the bound $[-3.34,+1.73]$ which is not too different from \eqref{eq:bound_quartic}.\footnote{In the case of the lower bound the $L_{max}=\infty$ results have a less clear pattern for the subsequent extrapolation in $N_{max}$.  Figure \ref{fig:spin_lower} suggests that one should extrapolate in $N_{max}$ in steps of 4.  In order to do that more numerical data with higher $N_{max}$ is needed.}
The reader may use the difference between these results as an indicator of the precision with which we have estimated the optimal bounds for the quartic coupling $\lambda$.

\subsection{Cubic Yukawa couplings}
\label{sec:qubic}
Let us now present the upper bound on the cubic (Yukawa) couplings. All the plots below are made using \eqref{eq:spin_cut-off} spin cut-off value. One should in principle perform the extrapolation to $L_{max}=\infty$. However, we expect that this procedure will only make the bounds stronger analogously to the previous section and thus skipping such an extrapolation makes our bounds only more conservative.

\begin{figure}[!htb]
	\begin{center}
		\includegraphics[scale=0.85]{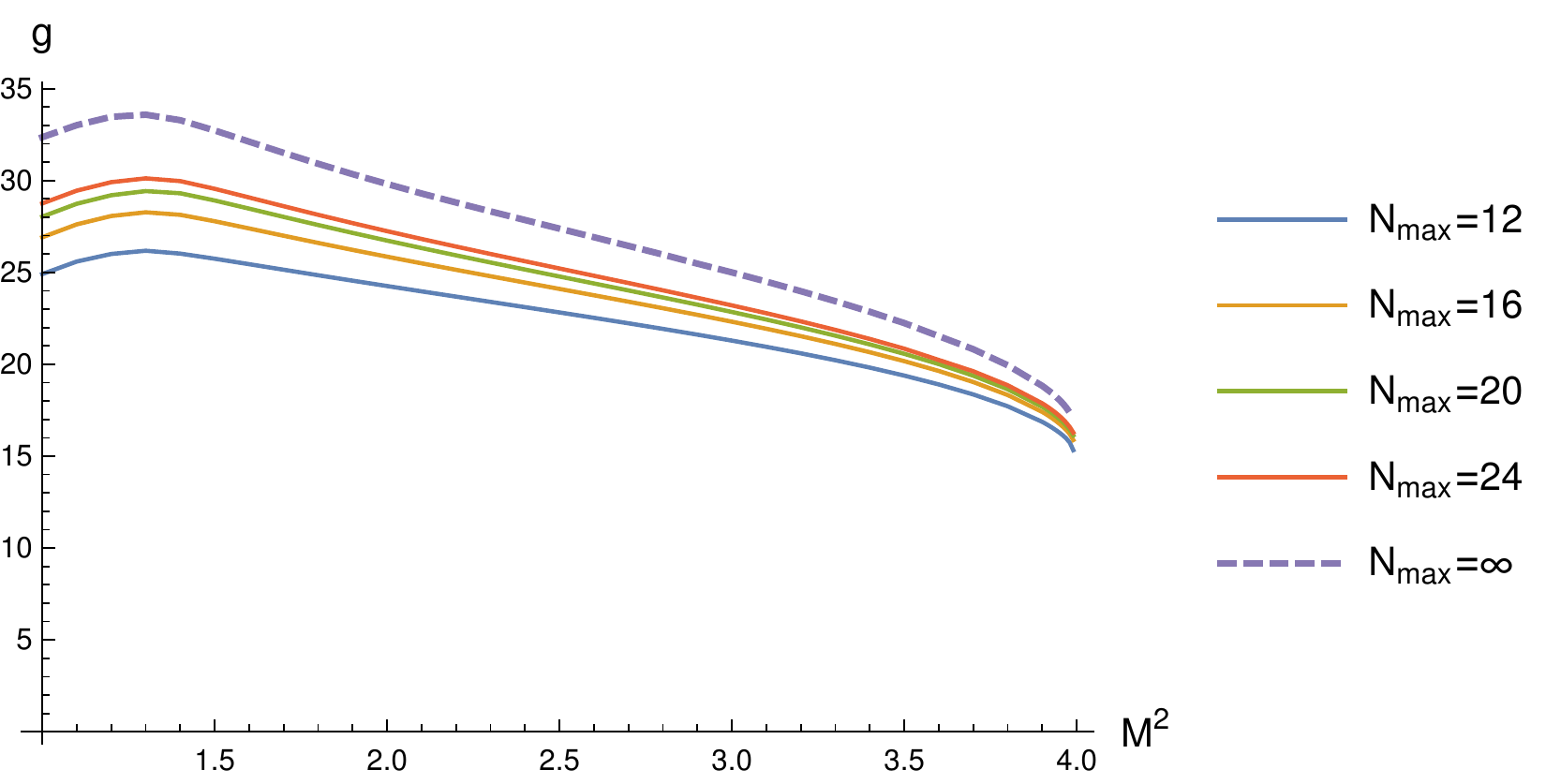}
		\caption{Upper bound on the cubic Yukawa coupling $g$ as a function of the scalar particle mass $M$. The bound is constructed for $N_{max}=12,\,16,\,20,\,24$ and $L_{max}=N_{max}+20$. Using $N_{max}=20$ and $N_{max}=24$ we also perform a linear extrapolation of the bound to $N_{max}=\infty$. The latter is indicated by the dashed line.}
		\label{fig:Yukawa}
	\end{center}
\end{figure}

We start with the situation where we have a scalar particle with mass $M$ and no pseudoscalar particles. In other words $\tilde g=0$. The bound on $g$ as a function of $M$ for various values of $N_{max}$ is given in figure \ref{fig:Yukawa}. As in the previous section the bound gets weaker when the value of $N_{max}$ increases. For each $M$ we perform a linear extrapolation to $N_{max}\rightarrow \infty$ analogously to the previous section. The final extrapolated bound is also shown in figure \ref{fig:Yukawa}.

\begin{figure}[!htb]
	\begin{center}
		\includegraphics[scale=0.85]{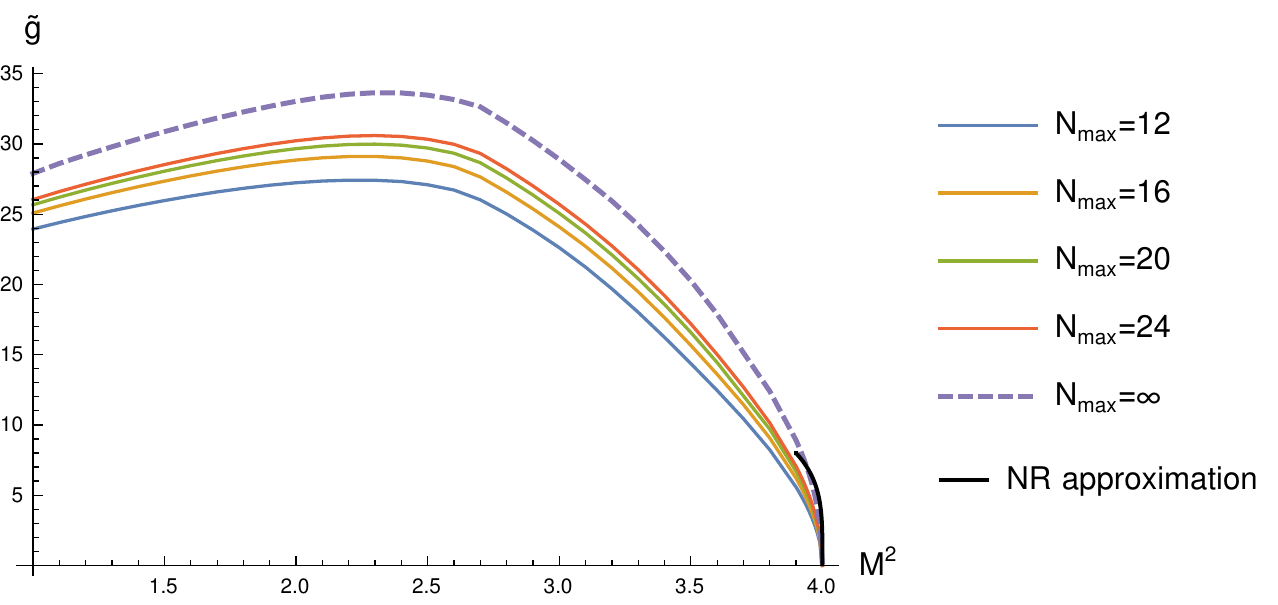}
		\caption{Upper bound on the cubic pseudo-Yukawa coupling $\tilde g$ as a function of the scalar particle mass $M$. The bound is constructed for $N_{max}=12,\,16,\,20,\,24$ and $L_{max}=N_{max}+20$. Using $N_{max}=20$ and $N_{max}=24$ we also perform a linear extrapolation of the bound to $N_{max}=\infty$. The latter is indicated by the dashed line. The black line indicates an analytic prediction for the upper bound in the vicinity of $M^2=4$ computed in \eqref{eq:analytic_bound}.}
		\label{fig:pseudoYukawa}
	\end{center}
\end{figure}

Now consider the case where there is a pseudoscalar particle in the theory and no scalar particle, namely $g=0$. We can construct an upper bound on the $\tilde g$ coupling as a function of the pseudoscalar mass $M$. The result for different $N_{max}$ is given in figure \ref{fig:pseudoYukawa}. In the figure we also present the extrapolated bound to $N_{max}\rightarrow\infty$. It is interesting to note that the bound gets stronger when we approach $M^2=4$ point. At $M^2=4$ we are forced to have $\tilde g=0$. This situation is very different from figure \ref{fig:Yukawa}.  
As a consistency check we compute in appendix \ref{app:close_to_threshold} an analytic expression for the upper bound in the vicinity of the threshold $M^2=4$. It is given by  \eqref{eq:analytic_bound}. In figure \ref{fig:pseudoYukawa} it is indicated by the black solid line. We see that our numerical result is in agreement with the analytic one.

Similarly to the case of the quartic coupling $\lambda$, the order of magnitude of our bounds on Yukawa and pseudo-Yukawa couplings is compatible with the expectation $ g \sim  \tilde{g} \lesssim  4\pi$  from \emph{naive dimensional analysis}  \cite{Manohar:1983md,Jenkins:2013sda}.


\newpage
\section{Conclusion}
\label{sec:conclusions}

In this article, we setup the formalism for the numerical S-matrix bootstrap approach to scattering amplitudes of spinning particles in 4d QFTs.
We explain the general case and perform the numerical analysis for the particular case of 2 to 2 scattering of identical massive neutral spin $\frac{1}{2}$ fermions, \emph{i.e.} Majorana fermions.
In principle, our nonperturbative bound \eqref{eq:bound_quartic} on the quartic coupling applies to neutrinos but this is purely academic because neutrinos are very weakly coupled with $  \lambda  \sim G_F  m_\nu^2 \lesssim  10^{-24}$.
Our bound also applies to  pseudo Goldstinos. Goldstinos are massless Majorana fermions in QFTs with spontaneously broken supersymmetry. In the presence of a small explicit supersymmetry breaking they acquire a small mass and are referred to as pseudo Goldstinos. 
In the future it would be also interesting to study massless Majorana fermions and derive bounds on the  low energy effective field theories for massless Goldstinos  \cite{Volkov:1973ix, Komargodski:2009rz}.  

In the future, we would like to consider more realistic applications of our methods. For example, the mixed system of 2 to 2 scattering amplitudes involving pions and nucleons seems to be a feasible target. This would be a significant extension of the pion S-matrix bootstrap analysis of \cite{Guerrieri:2018uew} and it would give access to many more physical observables, including the mass of the deuteron. 
It would be interesting to perform this analysis both in the theoretically cleaner case of massless pions and the more realistic case of massive pions. In the first case, one should be able to obtain bounds on the Wilson coefficients of the chiral lagrangian including nucleons \cite{Weinberg:1990rz}.
Notice that the results of this paper do not apply to nucleons because we assumed that the fermion was the lightest particle. The existence of a lighter particle (the pion) changes the analytic structure of the fermion-fermion scattering amplitude, creating an extended unitarity region \cite{Homrich:2019cbt}. 
 
Another interesting application is pure Yang-Mills theory. In practice, one would bootstrap 2 to 2 scattering amplitudes involving  the stable glueballs (that can have nonzero spin).
One may hope that the $SU(2)$ theory has a special place near the boundary of the allowed space of (strongly coupled) scattering amplitudes of this type.

It is also interesting to consider massless spinning particles like photons and gravitons.
In the case of photons, one can use the S-matrix bootstrap to bound the leading Wilson coefficients in the Euler-Heisenberg lagrangian \cite{PhotonBootstrap}.
The case of gravitons is more interesting but also more difficult. The first obstacle is that the usual S-matrix does not exist in four spacetime dimensions due to IR divergences. Pragmatically, one may bypass this difficulty by working in higher dimensions. However, this means that we will have to redo the general analysis of this paper for spacetime dimension $d\ge 5$.\footnote{It should not be  difficult to write crossing equations for the graviton scattering using tensor structures, see \cite{Chowdhury:2019kaq}. However, one will have to  compute the analogous of Wigner d-matrices for the $SO(d-1)$ Little group for $d\ge 5$.}
We leave this endeavour for the future.

The use of S-matrix consistency conditions to bound the space of low energy Effective Field Theories (EFT) has a long history  \cite{Adams:2006sv,  Vecchi:2007na, Manohar:2008tc, Nicolis:2009qm, Bellazzini:2015cra, Cheung:2016yqr, Bellazzini:2016xrt,  deRham:2017avq, deRham:2017zjm, deRham:2018qqo, NimaCERN, Bellazzini:2020cot,Tolley:2020gtv, Caron-Huot:2020cmc}.
Our numerical approach can be thought of as a systematic algorithm to optimize such bounds. However, we use a \emph{primal} formulation, where we rule in amplitudes and approach the boundary of the allowed space from within as we increase the numerical truncation. On the other hand, the recent methods of \cite{Bellazzini:2020cot, Tolley:2020gtv, Caron-Huot:2020cmc} solve the \emph{dual} problem and therefore rule out theories. It would be very interesting to develop a dual formulation that includes the full set of consistency conditions that we impose in our primal formulation.

In 2+1 dimensions, the dichotomy between bosons and fermions breaks down and particles can be anyons \cite{Wilczek:1982wy}. It would be very interesting to study their scattering amplitudes. Explicit computations in Chern-Simons-matter theories \cite{Jain:2014nza} show that the crossing equations need to be modified to accommodate the non-trivial phases of anyon statistics. Nevertheless, a systematic relativistic S-matrix theory of anyons is still lacking. In addition, one could consider the coupled system of  photons in 3+1 dimensions interacting with anyons living in a  2+1 dimensional subspace, as a reasonable model  for topological quantum computation \cite{Kitaev:1997wr}.

\section*{Acknowledgments}
The authors are grateful to the organizers of the $S$-matrix bootstrap project meeting in Azores and the participants for useful discussions. We especially thank Petr Kravchuk and Andrea Guerrieri for giving some key insights for the project. We also thank Gabriel Cuomo, Kelian H\"{a}ring, Riccardo Rattazzi, Lorenzo Ricci and Mikhail Shaposhnikov for discussions. In addition we thank Brando Bellazzini for carefully reading the paper and giving his valuable feedback.

The authors are supported by the
Simons Foundation grant 488649 (Simons Collaboration on the Nonperturbative Bootstrap) and
by the Swiss National Science Foundation through the project 200021-169132 and through the
National Centre of Competence in Research SwissMAP.
The computations in this paper were run on the EPFL SCITAS cluster.

\appendix
\section{Details of working with spin}
\label{sec:particles_and_states}
In this appendix we provide many technical details which support the discussion in the main text of the paper. We start in appendix \ref{app:euclidean_rotations} by reviewing 3d Euclidean rotations and then move to the discussion of the Poincar\'e group in appendix \ref{app:poincare}. These cover most of the basics required in section \ref{sec:states}. We define the vector and spinor representations in appendix
\ref{app:representations}. Finally we derive the Wigner angles in a particular situation (crucial for appendix \ref{app:crossing_COM_frame}) in appendix \ref{app:wigner_angles}.

\subsection{Euclidean rotations in 3d}
\label{app:euclidean_rotations}
Rotations in $3d$ are generated by three generators $J_1$, $J_2$ and $J_3$ which satisfy the algebra
\begin{equation}
\label{eq:so(3)_algebra}
[J_i, J_j] = i\epsilon_{ijk}J_k\,.
\end{equation}
Physically, these three generators correspond to infinitesimal rotations about the 3 axes. The Casimir operator which commutes with all the generators is
\begin{equation}
\label{eq:Jsq}
J^2\equiv J_1^2 + J_2^2 + J_3^2,\qquad
[J^2,J_i]=0.
\end{equation}
Any generic rotation can be written as 
\begin{equation}
\label{eq:rotation_parameterization_theta}
R(\theta_i) = \exp\left(-\sum_{k=1}^3 i \theta_k J_k\right).
\end{equation}
For our purposes a more useful way to write a rotation is to use  the Euler angles $(\alpha, \beta, \gamma)$ instead of angles $\theta_i$. A generic rotation in the Euler form reads as
\begin{equation}
\label{eq:rotation_Euler}
R(\alpha,\beta,\gamma)\equiv
\exp(-i\alpha J_3)\exp(-i\beta J_2)\exp(-i\gamma J_3).
\end{equation}
The Euler angles $\alpha$, $\beta$ and $\gamma$ can be related to the $\theta_i$ angles in~\eqref{eq:rotation_parameterization_theta}. We do not write this relation explicitly since it is complicated and not very illuminating. 
 
In quantum mechanical theories, the classical group of symmetries gets extended and therefore in the case at hand, we need to consider the central extension of $SO(3)$ which is the $SU(2)$ group. Henceforth when we talk about rotations we will mean the $SU(2)$ group. 
The unitary representations of the $SU(2)$ are finite-dimensional and are classified by the eigenvalue of the Casimir operator $J^2$.
The usual basis for these representations is formed by choosing eigenvectors of the $J_3$ operator and thus these vectors are labelled by two parameters $\spin$ and $\lambda$:
\begin{equation}
\label{eq:basis_SO(3)_irreps}
\begin{aligned}
J^2|\spin,
\lambda \rangle =& \, {\spin}({\spin} + 1) |\spin,\lambda\rangle, \\
J_3   |\spin,
\lambda \rangle =& \, \lambda |\spin, \lambda \rangle,
\end{aligned}
\end{equation}
where $\spin$ is a non-negative integer or half-integer and $\lambda = -\spin, -\spin+1, \ldots, \spin-1, \spin$.

Given such a spin $\spin$ representation, a generic rotation parametrized using Euler angles as in~\eqref{eq:rotation_Euler} acts on it in the following way
\begin{equation}
\label{eq:generic_rotation}
R(\alpha,\beta,\gamma) |\spin,\lambda\rangle = \sum_{\lambda}
|\spin,\lambda' \rangle\langle \spin,\lambda' | R |\spin,\lambda\rangle = \sum_{\lambda'}
 \mathscr{D}^{\spin}_{\lambda'\lambda} (\alpha,\beta,\gamma)|\spin,\lambda'\rangle,
\end{equation}
where in the first equality we inject an identity operator as a sum over all the states and in the second equality we have defined the Wigner $\mathscr{D}$-matrix
\begin{equation}
\label{big_D_matrix}
\mathscr{D}^{\spin}_{\lambda'\lambda} (\alpha,\beta,\gamma) \equiv
\langle \spin,\lambda' |
R(\alpha,\beta,\gamma)
|{\spin},\lambda\rangle=
\exp\left(-i(\alpha \lambda'+\gamma \lambda)\right)\times
d^{\spin}_{\lambda'\lambda} (\beta),
\end{equation}
and the (small) Wigner d-matrix
\begin{equation}
\label{eq:small_d_matrix}
d^\spin_{\lambda'\lambda} (\beta) \equiv
\langle \spin,\lambda' |
\exp(-i\beta J_2)
|\spin,\lambda\rangle.
\end{equation}
Since the rotation operator is unitary, the inverse of a rotation can be written in terms of the complex conjugate of a $\mathscr{D}$-matrix as
\begin{equation}
\label{eq:conjugated_D}
\langle \spin , \lambda^\prime | R^{-1}(\alpha, \beta, \gamma) | \spin, \lambda \rangle = \langle \spin,\lambda |
R(\alpha,\beta,\gamma)
|\spin,\lambda' \rangle^* = \mathscr{D}^{\spin \,*}_{\lambda \lambda^\prime}(\alpha, \beta, \gamma).
\end{equation}
The general form of~\eqref{eq:small_d_matrix} has the following simple expression
\begin{equation}
\label{eq:def_d}
\begin{aligned}
d^{\spin}_{\lambda' \lambda} (\beta) &=
\sqrt{(j+\lambda')!(j-\lambda')!(j+\lambda)!(j-\lambda)!}\\
&\times
\sum_{\nu}
(-1)^\nu
\frac{\left(\cos(\beta/2)\right)^{2j+\lambda-\lambda'-2\nu}
	\left(-\sin(\beta/2)\right)^{\lambda'-\lambda+2\nu}}
{\nu!(j-\lambda'-\nu)!(j+\lambda-\nu)!(\nu+\lambda'-\lambda)!}.
\end{aligned}
\end{equation}
Note that setting the $\lambda'$ and $\lambda$ indices to 0 gives the familiar Legendre polynomials
\begin{equation}
d^{\spin}_{0 0}(\beta) = P_\spin (\cos \beta).
\end{equation}
The small Wigner $d$-matrix is real. From its explicit expression one can conclude
\begin{equation}
d^{\spin}_{\lambda' \lambda} (\beta)
= (-1)^{\lambda'-\lambda}d^{\spin}_{-\lambda',-\lambda} (\beta)
= (-1)^{\lambda'-\lambda}d^{\spin}_{\lambda \lambda'} (\beta).
\end{equation}
As a consequence we also have
\begin{equation}
\label{eq:wignerd_wignerdstar}
\mathscr{D}^{\spin *}_{\lambda' \lambda} (\alpha,\beta,\gamma) =
(-1)^{\lambda'-\lambda}\mathscr{D}^{\spin}_{-\lambda',-\lambda} (\alpha,\beta,\gamma).
\end{equation}
The Wigner D-matrix satisfies the following important orthogonality relations
\begin{align}
\label{eq:D_matrix_property_1}
\sum_{\lambda'}\mathscr{D}^{\spin*}_{\lambda'\lambda_2} (\alpha,\beta,\gamma)
\mathscr{D}^{\spin}_{\lambda'\lambda_1} (\alpha,\beta,\gamma)
&=\delta_{\lambda_1\lambda_2},\\
\label{eq:D_matrix_property_2}
\int_0^{2\pi}d\alpha\int_{-1}^{+1}d\cos\beta\int_0^{2\pi}d\gamma\;
\mathscr{D}^{\spin_1*}_{\lambda_1'\lambda_1} (\alpha,\beta,\gamma)
\mathscr{D}^{\spin_2}_{\lambda_2'\lambda_2} (\alpha,\beta,\gamma)
&=
\frac{8\pi^2}{2\spin_1 + 1}\,
\delta_{\spin_1 \spin_2}\delta_{\lambda_1'\lambda_2'}\delta_{\lambda_1\lambda_2}.
\end{align}
The small Wigner $d$ matrix satisfies the following orthogonality condition instead
\begin{equation}
\label{eq:D_matrix_property_3}
\int_0^\pi d\beta \sin \beta\, d^{\,\spin_1}_{\lambda' \lambda}(\beta) d^{\,\spin_2}_{\lambda' \lambda} (\beta) = \frac{2}{2\spin_1 + 1} \delta_{\spin_1 \spin_2}
\end{equation}

Since the spin $\spin$ representations described above are unitary, the dual representation is the same as the complex conjugate representation and moreover, as the spin $\spin$ representations are irreducible, the complex conjugate representations are also irreducible.  We denote the basis of states in the dual spin $\ell$ representation by
\begin{equation*}
|\ell,\lambda'\rangle^\text{dual}.
\end{equation*}
Under rotations they transform as
\begin{equation}
\label{eq:transformation_dual}
R(\alpha,\beta,\gamma)|\ell,\lambda\rangle^\text{dual}
=\sum_{\lambda'}\mathscr{D}^{\spin *}_{\lambda' \lambda} (\alpha,\beta,\gamma)
|\ell,\lambda'\rangle^\text{dual}.
\end{equation}
The dual representations are actually equivalent to the standard spin $\spin$ representations. In order to show that let us rewrite \eqref{eq:wignerd_wignerdstar} in the following form
\begin{equation}
\label{eq:identity_D}
\mathscr{D}^{\spin *}_{\lambda' \lambda} (\alpha,\beta,\gamma)=
\sum_{\lambda_1,\lambda_2}\left(U^{-1}\right)_{\lambda' \lambda_1}  \mathscr{D}^{\spin}_{\lambda_1 \lambda_2} (\alpha,\beta,\gamma) U_{\lambda_2 \lambda},
\end{equation} 
where we have defined
\begin{equation}
\label{eq:conjugate_spin_representation}
\begin{aligned}
U_{\lambda' \lambda} &\equiv d^\spin_{\lambda' \lambda} (+\pi) = (-1)^{\spin - \lambda} \delta_{\lambda', -\lambda},\\
\left(U^{-1}\right)_{\lambda' \lambda}&= (-1)^{\spin + \lambda} \delta_{\lambda', -\lambda}.
\end{aligned}
\end{equation}
In order to confirm that $U^{-1}U=UU^{-1}=1$ and to show the results below, notice the following identity
\begin{equation}
1=(-1)^{2(\ell\pm\lambda)},
\end{equation}
which holds true since $\ell\pm \lambda$ is always an integer. Using \eqref{eq:conjugate_spin_representation} we can then relate the basis states in two representation as follows
\begin{equation}
\label{eq:dual}
|\ell,\lambda\rangle^\text{dual}=\sum_{\lambda'}
U_{\lambda' \lambda}|\ell,\lambda'\rangle=(-1)^{\ell-\lambda}|\ell,-\lambda\rangle.
\end{equation}
In order to show this, we simply rotate both sides of \eqref{eq:dual}. It then follows that
\begin{equation}
\label{eq:dual_trnasformation}
\begin{aligned}
R(\alpha,\beta,\gamma)|\ell,\lambda\rangle^\text{dual}&=
\sum_{\lambda',\lambda''}
\mathscr{D}^{\spin}_{\lambda'' \lambda'} (\alpha,\beta,\gamma)U_{\lambda' \lambda}
|\ell,\lambda''\rangle\\
&=\sum_{\mu,\lambda',\lambda'',\lambda'''}
U_{\lambda'' \mu}
\left(U^{-1}\right)_{\mu \lambda'''}
\mathscr{D}^{\spin}_{\lambda''' \lambda'} (\alpha,\beta,\gamma)U_{\lambda' \lambda}
|\ell,\lambda''\rangle\\
&=\sum_{\mu,\lambda''}
U_{\lambda'' \mu}
\mathscr{D}^{\spin*}_{\mu \lambda} (\alpha,\beta,\gamma)
|\ell,\lambda''\rangle\\
&=\sum_{\mu}
\mathscr{D}^{\spin*}_{\mu \lambda} (\alpha,\beta,\gamma)
|\ell,\mu\rangle^\text{dual}.
\end{aligned}
\end{equation}
Where in the first line we used \eqref{eq:generic_rotation}, in the second line we inserted the identity, we used \eqref{eq:identity_D} in the third line and finally we used \eqref{eq:dual} in the fourth line. Thus we see that the identification \eqref{eq:dual} leads consistently to \eqref{eq:transformation_dual}.

\subsection{Poincar\'e group}
\label{app:poincare}
We now consider the group of symmetries of Minkowski space \emph{i.e.} the Poincar\'e group. We begin by recalling its defining representation, mainly to set the notation, and then we recall its algebra and unitary representations.

\subsubsection{Defining representation}
Given a 4-vector
\begin{equation}
x^\mu\equiv\{t,\vec x\},\quad
\mu=0,1,2,3,
\end{equation}
one can define the following transformation
\begin{equation}
\label{eq:poincare_transformation_vectors}
x^\mu\longrightarrow x^{\prime\mu}=a^\mu+\Lambda(\omega)^\mu{}_\nu x^\nu,
\end{equation} 
where $a^\mu$ and $\omega_{\rho\sigma}$ are Lie parameters of the transformation. The transformation matrix $\Lambda$ obeys the constraint
\begin{equation}
\label{eq:constraint_lambda}
\eta^{\mu\nu}=\Lambda^\mu{}_\rho\Lambda^\nu{}_\sigma\eta^{\rho\sigma},\quad
\eta_{\mu\nu}=\{-+++\},
\end{equation}
where $\eta_{\mu\nu}$ is the metric. This implies that $\omega_{\rho\sigma} = -\omega_{\sigma\rho}$.

The transformations~\eqref{eq:poincare_transformation_vectors} form the Poincar\'e group, which is also known as the inhomogeneous Lorentz group. It is denoted by 
\begin{equation}
ISO(1,3) \equiv R^{1,3} \rtimes O(1,3),\qquad
O(1,3) = SO^+(1,3)  \rtimes P  \rtimes T,
\end{equation}
where $R^{1,3}$ is the group of 4d Minkowski translations, $ SO^+(1,3)$ is the proper orthochronous Lorentz group and $P$ and $T$ are discrete transformations called parity and time reversal which act on the coordinates as follows
\begin{equation}
\label{eq:two_discrete_symmetries}
x^\mu\longrightarrow (t,-\vec x),\quad
x^\mu \longrightarrow  (-t,\vec x).
\end{equation}
We restrict the $O(1,3)$ group to its $ SO^+(1,3)$ subgroup by requiring that the generic Lorentz transformation $\Lambda(a,\omega)$ obeys
\begin{equation}
\det \Lambda = +1,\quad
\Lambda^0{}_0\geq +1.
\end{equation}

We require our quantum system to be invariant only under the restricted Poincar\'e group denoted by
\begin{equation}
\label{eq:iso_group}
ISO^+(1,3) \equiv R^{1,3} \rtimes SO^+(1,3).
\end{equation}
Parity or time reversal symmetry may or may not be present. 
The discussion above was about the classical group of symmetries. Once again, in quantum mechanical theories the Lorentz group $SO(1,3)$ is centrally extended to its double-cover, the $SL(2,\mathbb C)$ group.

\subsubsection{Poincar\'e Algebra}
\label{app:poincare_algebra}
A generic Poincar\'e transformation can be written in terms of infinitesimal generators
\begin{equation}
\label{eq:generic_poincare_transformation_definition}
U(a,\omega)=\exp(-ia_\mu P^\mu)\Lambda(\omega),\quad
\Lambda(\omega)\equiv\exp\left(-\frac{i}{2}\,\omega_{\rho\sigma} M^{\rho\sigma}\right).
\end{equation}
Here $P^\mu$ and $M^{\rho\sigma}$ are the generators of translations and $4d$ Lorentz tranformations respectively.
The generators satisfy the following algebra
\begin{align}
\label{eq:poincare_commutation_1}
[P_\mu,P_\nu] &=0,\\
\label{eq:poincare_commutation_2}
[M_{\mu\nu},P_\lambda] &=i\,(\eta_{\mu\lambda}P_\nu-\eta_{\nu\lambda}P_\mu),\\
\label{eq:poincare_commutation_3}
[M_{\mu\nu},M_{\lambda\sigma}] &=i\,(
\eta_{\mu\lambda}M_{\nu\sigma}-\eta_{\nu\lambda}M_{\mu\sigma}+
\eta_{\mu\sigma}M_{\lambda\nu}-\eta_{\nu\sigma}M_{\lambda\mu}).
\end{align}
There are two Casimir operators which commute with all the generators, they are
\begin{equation}
\label{eq:casimirs}
C_1\equiv -P^2,\quad
C_{2}\equiv W^2,\quad
W^\mu\equiv\epsilon^{\mu\nu\rho\sigma}M_{\nu\rho}P_{\sigma}.
\end{equation}
where $W^\mu$ is called the Pauli-Lubanski pseudovector. Using the definitions~\eqref{eq:casimirs} and the commutation relations~\eqref{eq:poincare_commutation_1} and \eqref{eq:poincare_commutation_2} we can write\footnote{The Lorentzian 4d epsilon symbol $\epsilon^{\mu\nu\lambda\sigma}$ is fully antisymmetric. It is defined by $\epsilon^{0123}=-\epsilon_{0123}=+1$. Instead the Euclidean 4d epsilon symbol obeys instead $\epsilon^{1234}=\epsilon_{1234}=+1.$ It has the following property
	\begin{equation*}
	\sum_{a=1}^4\epsilon^{abcd}\epsilon_{ab'c'd'}=
	\delta^b_{b'}\delta^c_{c'}\delta^d_{d'}
	-\delta^b_{b'}\delta^c_{d'}\delta^d_{c'}
	-\delta^b_{c'}\delta^c_{b'}\delta^d_{d'}
	+\delta^b_{c'}\delta^c_{d'}\delta^d_{b'}
	-\delta^b_{d'}\delta^c_{c'}\delta^d_{b'}
	+\delta^b_{d'}\delta^c_{b'}\delta^d_{c'}.
	\end{equation*}}
\begin{equation}
W^2 = -2\,M_{\mu \nu} M^{\nu \mu}  \, C_1 - 4\, P_\mu M^{\mu \nu} M_{\nu \sigma}  P^\sigma.
\end{equation}

Let us consider the purely Lorentz part $\Lambda(\omega)$ of the generic Poincar\'e transformation~\eqref{eq:generic_poincare_transformation_definition}. It is convenient to split it into two parts. First, we define boosts
\begin{equation}
\label{eq:boost}
B(\vec\eta)\equiv\exp\left(-i\eta_i K^i\right),\quad
K^i\equiv M^{0i},\quad
\eta_i\equiv \omega_{0i},
\end{equation}
where $K^i$ are the three boost generators. Second, we define rotations
\begin{equation}
\label{eq:rotation}
R(\vec\theta)\equiv\exp\left(-i\theta_i J^i\right),\quad
J^i\equiv \frac{1}{2} \epsilon^{ijk} M^{jk},\quad\theta_i\equiv\epsilon_{ijk} \omega_{jk},
\end{equation}
where $J^i$ are the generators of rotations around $i$th axis and $\theta_i$ are the angles of rotations around the $i$th axis. For completeness we also write explicitly
\begin{equation}
\label{eq:generators_3d_rotation}
\vec J = \{M^{23},\,M^{31},\,M^{12}\},\quad
\vec \theta =\{\omega^{23},\,\omega^{31},\,\omega^{12}\}.
\end{equation}
Pure rotations form an $SO(3)$ subgroup of the Lorentz group which one can verify by computing the algebra of the operators $\vec J =\{M^{23},\,M^{31},\,M^{12}\}$ and seeing that it matches \eqref{eq:so(3)_algebra}.
In terms of boost and rotation generator the Lorentz algebra \eqref{eq:poincare_commutation_3} can be rewritten as
\begin{equation}
\label{eq:poincare_commutation_4}
[J_i,J_j]=+i\epsilon_{ijk} J_k,\quad
[J_i,K_j]=+i\epsilon_{ijk} K_k,\quad
[K_i,K_j]=-i\epsilon_{ijk} K_k.
\end{equation}
We can use the above commutation relations along with the Baker-Campbell-Hausdorff formula
\begin{equation}
e^{\xi A} B e^{-\xi A} = B + \xi [A,B] + \frac{\xi^2}{2}[A,[A,B]] + \ldots
\end{equation}
to get commutation relations between finite boosts and rotations. We list here three such relations that will turn out to be useful later
\begin{align}
\label{eq:y_rotate_boost}
e^{-i J_2 \theta} e^{-i K_3 \eta} e^{i J_2 \theta} &= e^{-i(K_3 \cos \theta + K_1 \sin \theta) \eta},\\
\label{eq:y_rotate_z}
e^{-i J_2 \theta} e^{-i J_3 \phi} e^{i J_2 \theta} &= e^{-i(J_3 \cos \theta + J_1 \sin \theta) \phi},\\
\label{eq:z_rotate_y}
e^{-i J_3 \phi} e^{-i J_2 \theta} e^{i J_3 \phi} &= e^{-i(J_2 \cos \phi - J_1 \sin \phi) \theta}.
\end{align}
In particular, we will use the following special cases of the above equations repeatedly
\begin{align}
\label{eq:pi_rotate_around_y_boost}
e^{\pm i \pi J_2} e^{-i K_3 \eta} = e^{i K_3 \eta}e^{\pm i \pi J_2},\\
\label{eq:pi_rotate_around_y_rotation}
e^{\pm i \pi J_2} e^{-i J_3 \phi} = e^{i J_3 \phi}e^{\pm i \pi J_2},\\
\label{eq:pi_rotate_around_x_rotation}
e^{\pm i \pi J_3} e^{-i J_2 \theta} = e^{i J_2 \theta}e^{\pm i \pi J_3}.
\end{align}

\subsubsection{Unitary representation}
We now review the unitary representation of the restricted Poincar\'e group $ISO^+(1,3)$. We refer to vectors in this representation as states. The unitary representation is characterized by the eigenvalues of two Casimirs $-P^2$ and $W^2$ as defined in \eqref{eq:casimirs}.

Let us denote the eigenvalue of the first Casimir $-P^2$ by
\begin{equation}
c^2=-P^2=-p^2.
\end{equation}
We focus on the case when $c^2>0$ only. We can chose the basis of states to be eigenvalues of $\vec P$. We denote such a basis by
\begin{equation}
\label{eq:states_A}
|c, \vec p ;\ldots\rangle,
\end{equation}
where the dots stand for other labels yet to be discussed.
Notice that the energy $p^0$ is related to $c$ as
\begin{equation}
c^2 = -p^2  \quad \Rightarrow \quad p^0 = + \sqrt{c^2 + \vec p^2}.
\end{equation}
Thus, we can also use $p^0$ instead of $c$ to label the representation.
There are two disconnected but equivalent regions for $p^0$ which can be related by time-reversal. We consider only positive energy states i.e those with $p^0>0$.

We now focus on the center of mass states, namely the states with $\vec p = 0$. 
We notice that 3d spatial rotations leave the $\vec p = 0$ condition invariant. This means that the set of states \eqref{eq:states_A} with $\vec p = 0$  must furnish a representation of the $SU(2)$ group.\footnote{This follows for example from the equality $W^2|c,\vec 0; \ldots \rangle =-4\,P^2 J^2 \,|c,\vec 0;\ldots\rangle$, which can be deduced using the results of appendix \ref{app:poincare_algebra}. We see that the second Casimir $W^2$ for the center of mass states simply reduces to the $J^2$ Casimir of the $SU(2)$ group defined in \eqref{eq:Jsq}.} We often refer to this group as the Little group. $SU(2)$ representations were already discussed in section \ref{app:euclidean_rotations}. They are labeled by the (half)integer $\ell$. The basis of states is labeled by the eigenvalues of the $J_3$ generator. We can thus fill the dots in \eqref{eq:states_A} when $\vec p =0$
\begin{equation}
\label{eq:com_frame_poincare_irreps_appendix}
|c, \vec 0; \spin, \lambda \rangle.
\end{equation} 
Under 3d rotation the state \eqref{eq:com_frame_poincare_irreps_appendix} transforms according to \eqref{eq:rotation_Euler}, 
\begin{equation}
\label{eq:com_frame_rotation_appendix}
R(\alpha,\beta,\gamma) |c,\vec 0; \spin,\lambda\rangle = \sum_{\lambda'}
 \mathscr{D}^{\spin}_{\lambda'\lambda} (\alpha,\beta,\gamma)|c,\vec 0; \spin,\lambda'\rangle.
\end{equation}

We now need to define a basis of states with generic values of $\vec p$ from \eqref{eq:com_frame_poincare_irreps_appendix} by applying an appropriate Lorentz transformation which we denote by $U_h(\vec p)$. In other words
\begin{equation}
\label{eq:helicity_basis}
|c ,\vec p; \spin,\lambda\rangle =
U_{\text{h}}(\vec p)
|c,\vec 0;\spin,\lambda\rangle.
\end{equation}
The most convenient choice of the transformation $U_h(\vec p)$ is as follows
\begin{equation}
\label{eq:unitary_transformation_helicity}
U_{\text{h}}(\vec p) =
R(\phi,\theta,-\phi)\exp(-i \eta K_3),
\end{equation}
where $(\phi,\theta,-\phi)$ are the three Wigner angles and $\eta$ is the rapidity related to the four-momentum by \eqref{eq:rapidity}. Here the boost generates a non-zero 3-momentum along the z-axis. The rotation then brings this 3-momentum to the required direction $\vec p$, where $(\phi,\theta)$ are the spherical angles of $\vec p$.\footnote{\label{foot:periodicity_R}
Notice that a rotation $R(\phi,\theta,\gamma)$ with any value of $\gamma$ would do the job. Such a rotation is $4\pi$ periodic in $\phi$. With the particular choice $\gamma=-\phi$, the rotation becomes instead $2\pi$ periodic in $\phi$, namely $R(\phi+2\pi,\theta,-(\phi+2\pi))=R(\phi,\theta,-\phi)$.}
 
The choice \eqref{eq:unitary_transformation_helicity} is known as the helicity boost and the basis \eqref{eq:helicity_basis} is known as the helicity basis. This name comes from the fact that the states are the eigenstates of the helicity operator defined as
\begin{equation}
\label{eq:helicity}
\mathbb{H}\equiv(\vec J\cdot \vec P).
\end{equation}
In other words one has
\begin{equation}
\mathbb{H}|c,\vec p; \spin,\lambda \rangle=
\lambda\, \myP\;
|c,\vec p;\spin,\lambda \rangle,
\end{equation}
where $\myP$ is the length of $\vec p$. Moreover the  helicity label $\lambda$ remains invariant under any 3d rotation as can be seen from
\begin{equation}
\label{eq:rotation_helicity_basis}
\begin{aligned}
R(\alpha,\beta,\gamma) |c,\vec p;\spin,\lambda ;\gamma\rangle
&=R(\alpha,\beta,\gamma) R(\phi,\theta,-\phi)B_3( \eta)
|c,\vec 0;\spin,\lambda ;\gamma\rangle \\
&=R(\alpha',\beta',\gamma') B_3( \eta)
|c,\vec 0;\spin,\lambda ;\gamma\rangle \\
&=R(\alpha',\beta',-\alpha') \exp(-i(\alpha'+\gamma')J_3)B_3( \eta)
|c,\vec 0;\spin,\lambda ;\gamma\rangle \\
&=\exp(-i\lambda(\alpha'+\gamma'))R(\alpha',\beta',-\alpha') B_3( \eta)
|c,\vec 0;\spin,\lambda ;\gamma\rangle\\
&= \exp(-i\lambda(\alpha'+\gamma')) U_h(\vec p\,')|c,\vec 0;\spin,\lambda ;\gamma\rangle,\\
&= \exp(-i\lambda(\alpha'+\gamma')) |c,\vec p\,';\spin,\lambda ;\gamma \rangle.
\end{aligned}
\end{equation}
Here in the second line we use the fact two rotations give another rotation, and the parameters $(\alpha',\beta',\gamma')$ can be expressed in terms of $(\alpha,\beta,\gamma,\phi,\theta)$. In the fourth line we use the fact that $J_3$ commutes with $K_3$. Finally we obtain the three-momentum $\vec p\,'$ which has $(\alpha',\beta')$ spherical angles and $|\vec p\,'|=|\vec p\,|$.
Thus, contrary to the rotations of the center of mass states~\eqref{eq:com_frame_rotation_appendix}, the rotation of a state in a generic frame with non-zero momentum $\vec p$ only changes  the direction of its three-momentum but not its helicity. This can be understood intuitively since the helicities in the helicity eigenstates are always aligned with the three-momentum.

Finally let us discuss transformation properties of the states \eqref{eq:helicity_basis} under a generic Poincar\'e transformation $U(a, \omega)$, where $a$ and $\omega$ are its Lie parameters as discussed in appendix \eqref{app:poincare_algebra}. One has
\begin{equation}
\label{eq:generic_poincare_transformation}
\begin{aligned}
U(a,\omega)
|c,\vec p;\spin,\lambda;\gamma \rangle &=
\exp(-ia_\mu P^\mu)\Lambda(\omega)|c,\vec p;\spin,\lambda;\gamma \rangle\\
&=\exp(-ia_\mu P^\mu)
U_{\text{h}}(\vec p\,')U_{\text{h}}(\vec p\,')^{-1}
\Lambda(\omega)
U_{\text{h}}(\vec p)
|c,\vec 0;\spin,\lambda;\gamma\rangle\\
&=\exp(-ia_\mu P^\mu)
U_{\text{h}}(\vec p\,')R(\alpha,\beta,\gamma)
|c,\vec 0;\spin,\lambda;\gamma \rangle\\
&=\exp(-ia_{\mu}p^{\prime\mu})
\sum_{\lambda'} \mathscr{D}^{\spin}_{\lambda'\lambda}(\alpha,\beta,\gamma)|c,\vec p\,';\spin,\lambda';\gamma\rangle.
\end{aligned}
\end{equation}
Here in the second line we inserted the identity operator in the form
\begin{equation}
\label{eq:identity_app}
\mathbb{I}=U_{\text{h}}(\vec p\,')U_{\text{h}}(\vec p\,')^{-1}.
\end{equation}
The key point lies in the third line where we notice that the following product of Lorentz group elements is a pure rotation
\begin{equation}
\label{eq:Wigners_rotation}
R(\alpha(p,\omega), \beta(p,\omega), \gamma(p,\omega))=
U_{\text{h}}(\vec p\,')^{-1}
\Lambda(\omega)
U_{\text{h}}(\vec p),\quad
p^{\prime\mu}\equiv\Lambda^\mu{}_\nu(w) p^\nu.
\end{equation}
This can be seen as follows. The transformation \eqref{eq:Wigners_rotation} takes the rest frame states to the rest frames states in the following way: $\vec 0 \rightarrow \vec p \rightarrow \vec p\,' \rightarrow \vec 0$. The rotation \eqref{eq:Wigners_rotation} is known as a Wigner rotation. In the third line of \eqref{eq:generic_poincare_transformation} we use \eqref{eq:com_frame_rotation_appendix}. Finally  the action of $U_h (\vec p \, ')$ just sends the COM frame state to the helicity state with final momentum $\vec p\, '$. 

The Wigner angles $(\alpha,\beta,\gamma)$ in the left-hand side of \eqref{eq:Wigners_rotation} are determined in terms of $(p,\omega)$. They can be computed for example by choosing a particular finite dimensional representation and comparing the final matrices in the left- and right-hand side of~\eqref{eq:Wigners_rotation}. However these expressions are too cumbersome to be presented in the most general case. In practice we only need to  consider a few special cases. The most important one for our paper is discussed in appendix \ref{app:wigner_angles}.

\subsubsection{Clebsch-Gordan coefficients}
\label{app:clebsch_gordan_coefficients}

In this appendix we compute in detail the Clebsch-Gordan coefficient $C^\ell_\lambda$ defined in \eqref{eq:CG}. For convenience let us recall its definition here
\begin{equation}
\label{eq:CB_coefficients_def}
(2\pi)^4\delta^{(4)}(p^\mu-p_1^\mu-p_2^\mu)\delta_{\alpha\gamma}\times
C^{\spin}_{\lambda}(\vec p_1, \vec p_2, \alpha)\equiv
\langle c,\vec p;\spin ,\lambda ;\gamma|\kappa_1,\kappa_2\rangle,
\end{equation}
where $\kappa_1$ and $\kappa_2$ are the one-particle states (with masses $m_1$, $m_2$, spins $j_1$,  $j_2$, helicities $\lambda_1$, $\lambda_2$ and three-momenta $\vec p_1$, $\vec p_2$) and $\alpha$ is the multiplicity label of the two-particle states which reads as
\begin{equation}
\label{eq:alpha_multipl}
\alpha=(m_1,m_2,j_1,j_2,\lambda_1,\lambda_2).
\end{equation}
Using it the decomposition of generic two particle states can be written as
\begin{equation}
\label{eq:2PS_decomposition_elaborated}
|\kappa_1,\kappa_2\rangle = \sum_{\spin ,\lambda}
C^{\spin}_{\lambda}(\vec p_1, \vec p_2, \alpha)|c,\vec p;\spin,\lambda;\alpha\rangle.
\end{equation}

We start by bringing the states in the right-hand side of \eqref{eq:CB_coefficients_def} to the center of mass frame. This is done by injecting an identity operator
\begin{equation}
\label{eq:identity_app_2}
\mathbb{I}=U_{\text{h}}(\vec p) U^{-1}_{\text{h}}(\vec p)
\end{equation}
composed out of the helicity boosts \eqref{eq:unitary_transformation_helicity} into the definition of the Clebsch-Gordan coefficient \eqref{eq:CB_coefficients_def}, we then have
  \begin{equation}
\label{eq:determine_C_first_step}
C^{\spin}_{\lambda}(\vec p_1, \vec p_2)=
C^{\spin}_{\lambda}(\vec p_1^{\;\prime}, -\vec p_1^{\;\prime}, \alpha)\times \mathscr{D}^{j_1}_{\lambda_1' \lambda_1}(\vec \omega_1)\mathscr{D}^{j_2}_{\lambda_2' \lambda_2}(\vec \omega_2).
\end{equation}
The value of $\vec p$ in \eqref{eq:identity_app_2} is chosen in such a way that
the inverse helicity boost $U^{-1}_{\text{h}}(\vec p)$ brings the pair of vectors $(\vec p_1,\vec p_2)$ to $(\vec p_1^{\;\prime}, -\vec p_1^{\;\prime})$ which are in the center of mass frame. The Wigner angles $\vec \omega_1$ and $\vec \omega_2$ correspond to the Wigner rotations $W_1$ and $W_2$ defined as
\begin{equation}
W_i \equiv U^{-1}_h(\Lambda \vec p_i)\Lambda U_h(\vec p_i), \quad \Lambda = U^{-1}_h(\vec p).
\end{equation}

In practice we never need the general expression \eqref{eq:determine_C_first_step}. We will therefore not attempt to derive the Wigner angles $\vec\omega_1$ and $\vec\omega_2$. What we will need instead is the Clebsch-Gordan coefficient in the center of mass frame
\begin{equation}
\label{eq:CB_COM}
C^{\spin}_{\lambda}(\vec p, -\vec p, \alpha),
\end{equation}
which enters in the right-hand side  of \eqref{eq:determine_C_first_step}. Notice that we dropped the primes and the subscripts compared to \eqref{eq:determine_C_first_step}. Recalling the definition of the two-particle center of mass states \eqref{eq:2PS_com}
\begin{equation}
\label{eq:2PS_spherical}
|(\myP,\theta,\phi); \lambda_1, \lambda_2 \rangle
\equiv
|m_1,\vec p;j_1,\lambda_1\rangle\otimes
|m_2,-\vec p;j_2,\lambda_2\rangle,
\end{equation}
where $(\myP,\theta,\phi)$ are the spherical coordinates of $\vec p$, and using the definition \eqref{eq:CB_coefficients_def} we can write the Clebsch-Gordan coefficient \eqref{eq:CB_COM} as
\begin{equation}
\label{eq:CB_COM_another_writing}
(2 \pi)^4 \delta^{(4)}(0)\times
C^{\spin}_{\lambda}(\vec p, -\vec p, \alpha)\equiv
\langle c,\vec 0;\spin,\lambda;\lambda_1, \lambda_2 |(\myP,\theta,\phi); \lambda_1, \lambda_2 \rangle.
\end{equation}
We also notice that the state \eqref{eq:2PS_spherical} obeys the following relation
\begin{equation}
\label{eq:relation_rotation_2PS_spherical}
R(\phi, \theta, -\phi)|(\myP,0,0); \lambda_1, \lambda_2 \rangle=  e^{-2i\phi\lambda_2}|(\myP,\theta,\phi); \lambda_1, \lambda_2 \rangle.
\end{equation}
We prove it later in this section.

In order to compute the Clebsch-Gordan coefficient \eqref{eq:CB_COM_another_writing} we inject the identity operator
\begin{equation}
\mathbb{I}=R(\phi, \theta, -\phi)R^{-1}(\phi, \theta, -\phi)
\end{equation}
in the right-hand side of \eqref{eq:CB_COM_another_writing}. Due to \eqref{eq:com_frame_rotation_appendix} and \eqref{eq:relation_rotation_2PS_spherical} the matrix element in the right-hand side of \eqref{eq:CB_COM_another_writing} becomes
\begin{equation}
\label{eq:mel}
\begin{aligned}
\langle c, 0,  \spin, \lambda |(\myP,\theta,\phi); \lambda_1, \lambda_2 \rangle
&= e^{+2i\phi\lambda_2}\sum_{\lambda'}\mathscr D^{\spin}_{\lambda\lambda'}(\phi, \theta, -\phi)\;
\langle c, 0,  \spin, \lambda' |(\myP,0,0); \lambda_1, \lambda_2 \rangle\\
&=  e^{+2i\phi\lambda_2}\mathscr D^{\spin}_{\lambda\,\lambda_{12}}(\phi, \theta, -\phi)\;
\langle c, 0,  \spin, \lambda_{12} |(\myP,0,0); \lambda_1, \lambda_2 \rangle.
\end{aligned}
\end{equation}
In the second line we have used the fact that the states here are eigenvectors of the $J_3$ generator and we have defined
\begin{equation}
\lambda_{12}\equiv\lambda_1-\lambda_2.
\end{equation}
Finally, we denote the matrix element in the right-hand side of the second line in \eqref{eq:mel} by
\begin{equation}
\label{eq:coefficient_C}
(2 \pi)^4 \delta^{(4)}(0)\times
C_\ell(\myP)\equiv \langle c_1, 0, \spin, \lambda_{12} |(\myP,0,0); \lambda_1, \lambda_2 \rangle.
\end{equation}
As indicated, the coefficient $C_\ell(\myP)$ can only depend on the spin label $\ell$ and the length $\myP$. Its value is fixed by the choice of normalization. We will derive it shortly. Plugging \eqref{eq:mel} and \eqref{eq:coefficient_C} into \eqref{eq:CB_COM_another_writing} we derive the final expression of the Clebsch-Gordan coefficient
\begin{equation}
\label{eq:CB_final}
C^{\spin}_{\lambda}(\vec p, -\vec p, \alpha) =
C_\ell(\myP)e^{+2i\phi\lambda_2}\mathscr D^{\spin}_{\lambda\,\lambda_{12}}(\phi, \theta, -\phi).
\end{equation}

Using \eqref{eq:CB_final}, see also \eqref{eq:2PS_spherical}, we can write the decomposition \eqref{eq:2PS_decomposition_elaborated} in the center of mass frame. It reads
\begin{align}
\nn
|(\myP,\theta,\phi); \lambda_1, \lambda_2 \rangle
&= e^{+2i\phi\lambda_2}
\sum_{\spin,\lambda}C_{ \spin}(\myP)\;\mathscr{D}^{ \spin}_{\lambda \lambda_{12}} (\phi,\theta,-\phi)|c,\vec 0; \spin,\lambda;\alpha\rangle \nonumber \\
&= \sum_{\spin,\lambda} C_{ \spin}(\myP)\;e^{i(\lambda_1+\lambda_2 -\lambda)\phi}d^{ \spin}_{\lambda \lambda_{12}} (\theta)|c,\vec 0; \spin,\lambda;\alpha\rangle.
\label{eq:COM_state_decomposition_appendix}
\end{align}
We can invert the above equation using \eqref{eq:D_matrix_property_2} and the orthogonality of the exponential function
\begin{equation}
\label{eq:COM_state_decomposition_inverse_appendix}
|c,\vec 0; \spin,\lambda;\alpha\rangle = \frac{2\spin +1}{4 \pi C_\spin(\myP)} \int_0^{2\pi}d\phi\int_{-1}^{+1}d\cos\theta
e^{-i(\lambda_1+\lambda_2 -\lambda)\phi} d^{ \spin}_{\lambda \lambda_{12}} (\theta)|(\myP,\theta,\phi); \lambda_1, \lambda_2 \rangle .
\end{equation}

\paragraph{Derivation of \eqref{eq:relation_rotation_2PS_spherical}}
Let us denote the vector $\vec p$ aligned with the direction of the z-axis by $\vec p_z$. Notice that as defined both $\vec p$ and $\vec p_z$ have the same length $\myP$. Using the definition \eqref{eq:2PS_spherical} we can write
\begin{multline}
R(\phi, \theta, -\phi)
|(\myP,0,0); \lambda_1, \lambda_2 \rangle
=\\
\left(R(\phi, \theta, -\phi)|m_1,\vec p_z;j_1,\lambda_1\rangle\right)
\otimes
\left(R(\phi, \theta, -\phi)|m_2,-\vec p_z;j_2,\lambda_2\rangle\right).
\end{multline}
From \eqref{eq:rotation_helicity_basis} it is clear that the rotation operators bring the two one-particle states from $\vec p_z$ and $-\vec p_z$ configuration to the $\vec p$ and $-\vec p$ configuration with some additional phases $\xi_1$ and $\xi_2$. In other words
\begin{equation}
\label{eq:result_xi}
R(\phi, \theta, -\phi)
|(\myP,0,0); \lambda_1, \lambda_2 \rangle = 
e^{i(\xi_1+\xi_2)}|(\myP,\theta,\phi); \lambda_1, \lambda_2 \rangle
\end{equation}
The goal of this section is to compute the phases $\xi_1$ and $\xi_2$.

Let us start from the state with $\vec p_z$. Its spherical angles are $(0,0)$. Using the definition of the helicity basis \eqref{eq:helicity_basis} we can simply write
\begin{equation}
\begin{aligned}
R(\phi, \theta, -\phi)|m_1,\vec p_z;j_1,\lambda_1\rangle &=
R(\phi, \theta, -\phi)R(0, 0, 0)e^{-i\eta K_3}|m_1,\vec 0;j_1,\lambda_1\rangle\\
&=|m_1,\vec p;j_1,\lambda_1\rangle.
\end{aligned}
\end{equation}
Thus, we conclude that
\begin{equation}
\label{eq:xi_1}
\xi_1=0.
\end{equation}

Let us now address the state with $-\vec p_z$. According to \eqref{eq:negative_spherical} its spherical angles are $(\pi,\pi)$ instead. Using again the definition of the helicity basis \eqref{eq:helicity_basis} we can write
\begin{equation}
\label{eq:derivation_xi2}
\begin{aligned}
R(\phi, \theta, -\phi)|m_2,\vec p_z;j_2,\lambda_2\rangle &=
R(\phi, \theta, -\phi)R(\pi, \pi, -\pi)e^{-i\eta K_3}|m_2,\vec 0;j_2,\lambda_2\rangle\\
&=R(\pi + \phi,\pi- \theta, -\pi-\phi)e^{-2i\phi J_3}e^{-i\eta K_3}|m_2,\vec 0;j_2,\lambda_2\rangle\\
&=e^{-2i\phi\lambda_2}|m_2,-\vec p;j_2,\lambda_2\rangle.
\end{aligned}
\end{equation}
In the second line of \eqref{eq:derivation_xi2} we have  repeatedly used the identities \eqref{eq:pi_rotate_around_y_rotation} and \eqref{eq:pi_rotate_around_x_rotation}. In the last line we have used the fact that $J_3$ and $K_3$ commute and that the center of mass state is the eigenstate of the $J_3$ generator. Thus, we conclude that
\begin{equation}
\label{eq:xi_2}
\xi_2=-2\lambda_2\phi.
\end{equation}
Combining \eqref{eq:result_xi} with \eqref{eq:xi_1} and \eqref{eq:xi_2} we arrive at the desired property \eqref{eq:relation_rotation_2PS_spherical}.

\paragraph{Computation of \eqref{eq:coefficient_C}}
The coefficient $C_{ \spin}(\myP)$ in \eqref{eq:coefficient_C} is fixed by the normalization condition \eqref{eq:1PS_normalization}. In what follows we carefully compute it.
Using \eqref{eq:normalization_2PS} and performing the change of variables, see appendix A of \cite{Karateev:2019ymz} for details, we get
\begin{align}
\langle\kappa_1',\kappa_2'|\kappa_1,\kappa_2\rangle 
&= (2\pi)^6 4p_1^{0}p_2^{0}\,
\delta^{(3)}(\vec p_1^{\,\prime}-\vec p_1)\delta^{(3)}(\vec p_2^{\,\prime}- \vec p_2)\delta_{\alpha'\alpha}\\
&= (2\pi)^6 4\,\sqrt{\frac{-p^2}{\myP^2}}\,
\delta^{(4)}(p^{\prime\mu}-p^\mu)\delta(\cos\theta'-\cos\theta)\delta(\phi'-\phi)\delta_{\alpha'\alpha},
\label{eq:normalization_2}
\end{align}
where $\alpha$ is given by \eqref{eq:alpha_multipl} and
\begin{equation}
\delta_{\alpha'\alpha} \equiv
\delta_{m_1'm_1}\delta_{m_2'm_2}
\delta_{j_1'j_1}\delta_{j_2'j_2}
\delta_{\lambda'_1\lambda_1}\delta_{\lambda'_2\lambda_2}.
\end{equation}
Now let us take the norm of both sides of \eqref{eq:COM_state_decomposition_inverse_appendix} and use \eqref{eq:irrep_normalization} and \eqref{eq:normalization_2} to get
\begin{equation}
\begin{aligned}
\delta_{ \spin' \spin}\,
\delta_{\lambda'\lambda}\,
\delta_{\gamma'\gamma} =  \left(\frac{2\spin +1}{4\pi|C_\spin(\myP)|}\right)^2 &(2\pi)^2 4\,\sqrt{\frac{-p^2}{\myP^2}}   \\
& \times \int_0^{2\pi} d\phi \int_{-1}^{+1} d\cos\theta e^{-i(\lambda_1+\lambda_2 -\lambda)\phi} d^{ \spin}_{\lambda \lambda_{12}} (\theta)  \\
&\times \int_0^{2\pi} d\phi' \int_{-1}^{+1} d\cos\theta' 
e^{i(\lambda_1'+\lambda_2' -\lambda')\phi'} d^{ \spin}_{\lambda' \lambda_{12}'} (\theta') \\ 
& \qquad \qquad \times \delta (\cos \theta' - \cos \theta) \delta (\phi' - \phi) \delta_{\alpha'\alpha}.
\end{aligned}
\end{equation}
The delta functions over the angular coordinates removes two of the integrals and sets $\phi' = \phi$ and $\cos \theta' = \cos \theta$. We can then use the orthogonality of the small Wigner $d$ matrix \eqref{eq:D_matrix_property_3} along with the orthogonality of the exponential function to obtain
\begin{equation}
1 = \left(\frac{2\spin +1}{4\pi |C_\spin(\myP)|}\right)^2
(2\pi)^2 4\,\sqrt{\frac{-p^2}{\myP^2}}\,
\frac{4\pi}{2\spin+1}.
\end{equation}
Noticing that in the center of mass frame $-p^2=c^2$, we immediately get
\begin{equation}
\label{eq:C_l}
|C_{\spin}(\myP)|^2 = 4\pi (2\spin +1)\times\frac{c}{\myP}.
\end{equation}
The phase of the $C_{\spin}(\myP)$ coefficient is unobservable. In all the final formulas it will enter in the form \eqref{eq:C_l}. Thus we can simply set this phase to zero and obtain the final expression
\begin{equation}
\label{eq:C_l_final}
C_{\spin}(\myP) = \sqrt{4\pi (2\spin +1)\times\frac{c}{\myP}}.
\end{equation}

\subsection{Finite dimensional Lorentz representations}
\label{app:representations}
Let us discuss two particular finite-dimensional representations on the Lorentz group, namely the vector and spinor representations.

\subsubsection*{Vector representation}
The generators of the Lorentz transformation \eqref{eq:generic_poincare_transformation_definition} obey the algebra \eqref{eq:poincare_commutation_3}. In the vector representation of the Lorentz group, the generators satisfying \eqref{eq:poincare_commutation_3} can be written as
\begin{equation}
\big[M^{\mu\nu}\big]_{\rho\sigma} = -i\,(\delta^\mu{}_\rho\delta^\nu{}_\sigma-\delta^\mu{}_\sigma\delta^\nu{}_\rho)
\quad\Rightarrow\quad
\big[M^{\mu\nu}\big]^\rho{}_\sigma = -i\,(\eta^{\mu\rho}\delta^\nu_\sigma-\delta^\mu_\sigma\eta^{\nu\rho}).
\end{equation}
According to \eqref{eq:boost} and \eqref{eq:rotation} they split into generators of boosts and rotations as
\begin{align}
\nn
&\big[K^1\big]^\rho{}_\sigma = i\begin{pmatrix}
0\;\, & +1\;\, & 0\;\, & 0\\
+1\;\, & 0\;\, & 0\;\, & 0\\
0\;\, & 0\;\, & 0\;\, & 0\\
0\;\, & 0\;\, & 0\;\, & 0
\end{pmatrix},\;
&&\big[K^2\big]^\rho{}_\sigma =  i\begin{pmatrix}
0\;\, & 0\;\, & +1\;\, & 0\\
0\;\, & 0\;\, & 0\;\, & 0\\
+1\;\, & 0\;\, & 0\;\, & 0\\
0\;\, & 0\;\, & 0\;\, & 0
\end{pmatrix},\;
&&&\big[K^3\big]^\rho{}_\sigma =  i\begin{pmatrix}
0\;\, & 0\;\, & 0\;\, & +1\\
0\;\, & 0\;\, & 0\;\, & 0\\
0\;\, & 0\;\, & 0\;\, & 0\\
+1\;\, & 0\;\, & 0\;\, & 0
\end{pmatrix},\\
\nn
&\big[J^1\big]^\rho{}_\sigma = -i\begin{pmatrix}
0\;\, & 0\;\, & 0\;\, & 0\\
0\;\, & 0\;\, & 0\;\, & 0\\
0\;\, & 0\;\, & 0\;\, & +1\\
0\;\, & 0\;\, & -1\;\, & 0
\end{pmatrix},
&&\big[J^2\big]^\rho{}_\sigma =  -i\begin{pmatrix}
0\;\, & 0\;\, & 0\;\, & 0\\
0\;\, & 0\;\, & 0\;\, & -1\\
0\;\, & 0\;\, & 0\;\, & 0\\
0\;\, & +1\;\, & 0\;\, & 0
\end{pmatrix},
&&&\big[J^3\big]^\rho{}_\sigma =  -i\begin{pmatrix}
0\;\, & 0\;\, & 0\;\, & 0\\
0\;\, & 0\;\, & +1\;\, & 0\\
0\;\, & -1\;\, & 0\;\, & 0\\
0\;\, & 0\;\, & 0\;\, & 0
\end{pmatrix}.
\end{align}
Using \eqref{eq:boost} the matrices of finite transformations follow straightforwardly, for instance for the boost along the $z$-axis we get
\begin{equation}
\label{eq:boost_fundamental}
B_3(\eta)^\mu{}_\nu = \begin{pmatrix}
\cosh\eta & \;\;0 & \;\;\;0\;\; & \sinh\eta\\
0            & \;\;1 & \;\;\;0\;\; & 0\\
0            & \;\;0 & \;\;\;1\;\; & 0\\
\sinh\eta & \;\;0 & \;\;\;0\;\; & \cosh\eta\\
\end{pmatrix}.
\end{equation}
Similarly using \eqref{eq:rotation} for the rotation around the $y$-axis and $z$-axis we get respectively
\begin{equation}
\label{eq:rotation_fundamental}
R_2(\beta)^\mu{}_\nu = \begin{pmatrix}
1\;\; & 0             & \;\;\;0\;\;\; & 0\\
0\;\; & \cos\beta  & \;\;\;0\;\;\; & \sin\beta\\
0\;\; & 0             & \;\;\;1\;\;\; & 0\\
0\;\; & -\sin\beta & \;\;\;0\;\;\; & \cos\beta\\
\end{pmatrix},\quad
R_3(\gamma)^\mu{}_\nu = \begin{pmatrix}
1\;\; & 0             & \;\;\;0\;\;\; & 0\\
0\;\; & \cos\gamma   & -\sin\gamma & \;\;\;0\;\;\;\\
0\;\; & \sin\gamma   & \cos\gamma & \;\;\;0\;\;\;\\
0\;\; & 0            & 0 & \;\;\;1\;\;\;\\
\end{pmatrix}.
\end{equation}

In defining a 1PS we apply a boost along the positive direction of the $z$-axis to the particle at rest. The boost parameter can be found from
\begin{equation}
\begin{pmatrix}
p^0\\
0\\
0\\
\myP
\end{pmatrix}
= B_3(\eta)
\begin{pmatrix}
m\\
0\\
0\\
0
\end{pmatrix},\qquad\eta\geq 0.
\end{equation}
Using \eqref{eq:boost_fundamental} we get
\begin{equation}
\label{eq:boost_parameter}
\cosh\eta = \frac{p^0}{m},\quad
\sinh\eta = \frac{\myP}{m}.
\end{equation}
Consider now a state with the three-momentum $\vec p$ constrained to the $xz$-plane
\begin{equation}
\label{eq:vector_xz_cart}
p^\mu = \{p^0, p_x, 0, p_z\}.
\end{equation}
In terms of rapidity and spherical coordinates it is described by the following parameters
\begin{equation}
\label{eq:vector_xz}
p^\mu:\quad(\eta, \theta).
\end{equation}
The components of the vector \eqref{eq:vector_xz_cart} can be expressed in terms of the components \eqref{eq:vector_xz} as
\begin{equation}
\label{eq:momentum_xz_plane}
p^0=m\cosh \eta,\quad
\myP=m\sinh\eta,\quad
p_x=\myP\sin\theta,\qquad
p_z=\myP\cos\theta.
\end{equation}
By definition \eqref{poincare_irreps_definition} the helicity state is constructed by applying \eqref{eq:unitary_transformation_helicity} to the center of mass states. We have
\begin{equation}
\label{eq:4vector_p}
\begin{pmatrix}
p^0\\
p_x\\
0\\
p_z
\end{pmatrix}
= U_{\text{h}}(p)
\begin{pmatrix}
m\\
0\\
0\\
0
\end{pmatrix},
\end{equation}
where in the vector representation the matrix $U_{\text{h}}(p)$ reads as
\begin{equation}
\label{eq:helicit_boost}
U_{\text{h}}(p) = R_2(\theta)B_3(\eta)=
\begin{pmatrix}
\cosh\eta & 0 & \;\;\;0\;\;\;\;\; & \sinh\eta\\
\sinh\eta \sin\theta & \;\;\cos\theta\;\; & \;\;\;0\;\;\;\;\; & \cosh\eta \sin\theta\\
0            & \;\;0 & \;\;\;1\;\;\;\;\; & 0\\
\sinh\eta \cos\theta & \;\;-\sin\theta\;\; & \;\;\;0\;\;\;\;\; & \cosh\eta \cos\theta\\
\end{pmatrix}.
\end{equation}

\subsubsection*{Spinor representation}
In order to define the spinor representation of the Lorentz group, we first define the $4 
\times 4 $ gamma matrices in our conventions\footnote{Note that we work in the Weyl (also known as chiral) basis for the gamma matrices.}
\begin{equation}
\label{eq:Dirac_matrices}
\gamma^\mu \equiv \begin{pmatrix}
               0 & \sigma^\mu \\
                 \bar \sigma^\mu & 0   
                  \end{pmatrix},
\end{equation}
where 
\begin{equation}
\label{4Pauli}
\sigma^\mu = (I, \vec \sigma) \quad \text{and} \quad \bar \sigma^\mu = (I, -\vec \sigma)
\end{equation}
and $\vec \sigma$ are the usual $2 \times 2$ Pauli matrices:
\begin{equation}
\sigma_1 = \begin{pmatrix}
0 & 1 \\
1 & 0   
\end{pmatrix}
, \quad
\sigma_2 = \begin{pmatrix}
0 & -i \\
i & 0   
\end{pmatrix}
, \quad
\sigma_3 = \begin{pmatrix}
1 & 0 \\
0 & -1   
\end{pmatrix}.
\end{equation}
From the explicit form of the gamma matrices it is easy to verify that they satisfy
\begin{equation}
\{\gamma^\mu, \gamma^\nu\} = -2 \eta^{\mu \nu},
\end{equation}
where $\{A,B\}\equiv AB + BA$ is the anti-commutator. 
We can now define the generators of the spinorial representation of the Lorentz group:
\begin{equation}
S^{\mu \nu} \equiv \frac{i}{4}[\gamma^\mu,\gamma^\nu].
\end{equation}
These generators satisfy the Lorentz algebra \eqref{eq:poincare_commutation_3} and we can split them into boost generators $K^i = S^{0i}$ and rotation generators $J^i = \frac{1}{2}\epsilon^{ijk}S^{jk}$. For the reader's convenience, we write out these matrices explicitly
\begin{align}
\nn
& K^1\ = \left(
\begin{array}{cccc}
0 & -\frac{i}{2} & 0 & 0 \\
-\frac{i}{2} & 0 & 0 & 0 \\
0 & 0 & 0 & \frac{i}{2} \\
0 & 0 & \frac{i}{2} & 0 \\
\end{array}
\right),
&& K^2 =  \left(
\begin{array}{cccc}
0 & -\frac{1}{2} & 0 & 0 \\
\frac{1}{2} & 0 & 0 & 0 \\
0 & 0 & 0 & \frac{1}{2} \\
0 & 0 & -\frac{1}{2} & 0 \\
\end{array}
\right),
&&& K^3 =  \left(
\begin{array}{cccc}
-\frac{i}{2} & 0 & 0 & 0 \\
0 & \frac{i}{2} & 0 & 0 \\
0 & 0 & \frac{i}{2} & 0 \\
0 & 0 & 0 & -\frac{i}{2} \\
\end{array}
\right),\\
\nn
& J^1 = \left(
\begin{array}{cccc}
0 & \frac{1}{2} & 0 & 0 \\
\frac{1}{2} & 0 & 0 & 0 \\
0 & 0 & 0 & \frac{1}{2} \\
0 & 0 & \frac{1}{2} & 0 \\
\end{array}
\right),
&&J^2 =  \left(
\begin{array}{cccc}
0 & -\frac{i}{2} & 0 & 0 \\
\frac{i}{2} & 0 & 0 & 0 \\
0 & 0 & 0 & -\frac{i}{2} \\
0 & 0 & \frac{i}{2} & 0 \\
\end{array}
\right),
&&&J^3 = \left(
\begin{array}{cccc}
\frac{1}{2} & 0 & 0 & 0 \\
0 & -\frac{1}{2} & 0 & 0 \\
0 & 0 & \frac{1}{2} & 0 \\
0 & 0 & 0 & -\frac{1}{2} \\
\end{array}
\right).
\end{align}
Using \eqref{eq:boost} and \eqref{eq:rotation} the matrices of finite transformations follow straightforwardly, for instance for a boost along the $z$-axis by rapidity $\eta$ we get
\begin{equation}
\label{eq:boost_fundamental_spinor}
B_3(\eta) = \left(
\begin{array}{cccc}
e^{-\frac{\eta }{2}} & 0 & 0 & 0 \\
0 & e^{\eta /2} & 0 & 0 \\
0 & 0 & e^{\eta /2} & 0 \\
0 & 0 & 0 & e^{-\frac{\eta }{2}} \\
\end{array}
\right),
\end{equation}
while for rotations about the $y$-axis by an angle $\theta$ we get 
\begin{equation}
\label{eq:rotation_fundamental_spino}
R_2(\theta) = \left(
\begin{array}{cccc}
\cos \left(\frac{\theta }{2}\right) & -\sin \left(\frac{\theta }{2}\right) & 0 & 0 \\
\sin \left(\frac{\theta }{2}\right) & \cos \left(\frac{\theta }{2}\right) & 0 & 0 \\
0 & 0 & \cos \left(\frac{\theta }{2}\right) & -\sin \left(\frac{\theta }{2}\right) \\
0 & 0 & \sin \left(\frac{\theta }{2}\right) & \cos \left(\frac{\theta }{2}\right) \\
\end{array}
\right).
\end{equation}
In case of a rotation about the $z$-axis by an angle $\phi$ we have
\begin{equation}
R_3(\phi)=\left(
\begin{array}{cccc}
	e^{-\frac{i \phi}{2} } & 0 & 0 & 0 \\
	0 & e^{\frac{i \phi }{2}} & 0 & 0 \\
	0 & 0 & e^{-\frac{i \phi}{2} } & 0 \\
	0 & 0 & 0 & e^{\frac{i \phi }{2}} \\
\end{array}
\right).
\end{equation}

\subsection{An example of the Wigner rotation}
\label{app:wigner_angles}
Consider now the vector \eqref{eq:vector_xz} and the following Lorentz transformation applied to it
\begin{equation}
\label{eq:lorentz_transformation_rbr}
\Lambda = R_2(\psi_2)B_3(\chi)R_2(\psi_1),\quad \psi_i\in[0,\pi],
\end{equation}
which implements a rotation around the $y$-axis by an angle $\psi_1$ followed by a boost along the positive $z$-axis with the rapidity parameter $\chi$ and another rotation around the $y$-axis by an angle $\psi_2$. As a result we get the following 4-momentum
\begin{equation}
\label{eq:pprime}
p^{\prime \mu} = \Lambda^\mu{}_\nu p^\nu
\end{equation}
which is described by the parameters
\begin{equation}
\label{eq:vector_xz_prime}
p^{\prime\mu}:\quad(\eta', \theta').
\end{equation}
and also lies in the $xz$-plane.
The four-vector $p^{\prime \mu}$ can be generated using the helicity boost $U_{\text{h}}(p')=U_{\text{h}}(\eta',\theta')$ analogously to \eqref{eq:4vector_p}.
The components of $p^{\prime \mu}$ can be found from \eqref{eq:pprime}. The values of $(\eta',\theta')$ in the helicity boost matrix then follow straightforwardly.

The Wigner rotation associated to generic Lorentz transformations are defined in \eqref{eq:Wigners_rotation}. In case of the Lorentz transformation \eqref{eq:lorentz_transformation_rbr}, upon plugging the above results in the definition \eqref{eq:Wigners_rotation}, we get the following explicit result
\begin{equation}
\label{eq:wigner_rotation}
R_{\text{wigner}} =
U_{\text{h}}^{-1}(p')
\,\Lambda\,
U_{\text{h}}(p)=
\begin{pmatrix}
1 & 0 & \;\;\;0\;\;\;\;\; & 0\\
0 & \;\;\cos\omega\;\; & \;\;\;0\;\;\;\;\; & \sin\omega\\
0            & \;\;0 & \;\;\;1\;\;\;\;\; & 0\\
0 & \;\;-\sin\omega\;\; & \;\;\;0\;\;\;\;\; & \cos\omega\\
\end{pmatrix},
\end{equation}
where the Wigner angle written in compact form read as
\begin{equation}
\label{eq:wigner_angles}
\cos\omega = \frac{p^0 p^{\prime 0}-m^2\cosh\chi}
{\myP\, \myP'},\quad
\sin\omega = \frac{m\sinh\chi}
{\myP'}\,\sin(\theta+\psi_1).
\end{equation}
The Wigner angle $\omega$ depends on five parameters $(\eta,\theta,\chi,\psi_1,\psi_2)$. The full form of the angle $\omega$ reads as
\begin{equation}
\cos \omega = \frac{A}{\sqrt{B^2-1}},\quad
\sin \omega = \frac{\sin(\theta+\psi_1)\sinh\chi}{\sqrt{B^2-1}},
\end{equation}
where we have defined
\begin{align}
A &\equiv \sinh \eta \cosh \chi + \cosh \eta \sinh \chi \cos(\theta+\psi_1),\\
B &\equiv \cosh \eta \cosh \chi + \sinh \eta \sinh \chi \cos(\theta+\psi_1).
\end{align}

\section{Parity and time-reversal}
\label{app:parity_and_timereversal}
In this section we will discuss the discrete symmetries of the full Poincar\'e group, namely parity $\parity$ and time-reversal $\time$.

\subsection{Parity}
\label{app:parity}
Parity in the defining vector representation of the Lorentz group is given by the following matrix
\begin{equation}
\label{eq:parity_defining}
\parity =
\begin{pmatrix}
1\;\, & 0\;\, & 0\;\, & 0\\
0\;\, & -1\;\, & 0\;\, & 0\\
0\;\, & 0\;\, & -1\;\, & 0\\
0\;\, & 0\;\, & 0\;\, & -1
\end{pmatrix}.
\end{equation}
We denote the parity operator in the (infinite-dimensional) unitary representation by the same symbol $\parity$. It obeys the following
commutation relations with the generators of the Poincar\'e group
\begin{equation}
\label{eq:parity_commutation}
\parity P^\mu \parity^\dag = (P^0, -\vec P),\quad
\parity K^i \parity^\dag = -K^i,\quad
\parity J^i \parity^\dag = J^i.
\end{equation}

In what follows we will use \eqref{eq:parity_commutation} to derive the action of parity on one- and two-particle states. We will then derive constraints on the scattering amplitudes due to parity. Let us begin with the following preliminary computation of the parity transformation property of the helicity boost operator \eqref{eq:unitary_transformation_helicity},
\begin{equation}
\label{parity_action_on_helicity_boost}
\begin{aligned}
\parity U_{\text{h}}(\vec p) \parity^\dag =&
\parity R(\phi,\theta,-\phi) \parity^\dagger\parity B_3(\eta) \parity^\dag\\
=&R(\phi,\theta,-\phi) B_3(-\eta) \\
=&R(\phi,\theta,-\phi) R(0,\pi,0) B_3(\eta) R^\dag(0,\pi,0).
\end{aligned}
\end{equation}
Here in the first line we injected the identity $\mathbb{I}=\parity^\dagger\parity$, we then used the commutation properties \eqref{eq:parity_commutation} in the second line. Finally, in the third line we used \eqref{eq:pi_rotate_around_y_boost}.

\paragraph{One-particle states}
Consider the action of parity on a one-particle state \eqref{eq:shorthand_1PS} in the rest frame. Since parity commutes with all of the rotation generators $J_i$, it must leave the helicity of the particle invariant.\footnote{This can be seen by applying parity to the eigenvector conditions \eqref{eq:basis_SO(3)_irreps}.} Therefore the most general possible action is a simple multiplication by a phase which we denote by $\eta$. In other words
\begin{equation}
\label{eq:parity_action_rest_frame}
\parity |m,\vec 0;j,\lambda \rangle =
\eta|m,\vec 0;j, \lambda \rangle.
\end{equation}
This phase $\eta$ is called the intrinsic parity of the particle. Due to the discussion of section 3.3 in \cite{Weinberg:1995mt}, one can always define parity operator $\parity$ in such a way that either $\parity^2=+1$ or $\parity^2=-1$. As a result, applying \eqref{eq:parity_action_rest_frame} consecutively we conclude that
\begin{equation}
\eta^2=+1\quad\text{or}\quad\eta^2=-1.
\end{equation}

We can now deduce the action of parity on a generic one-particle state \eqref{eq:shorthand_1PS}, see also \eqref{eq:helicity_basis}. One has
\begin{equation}
\label{eq:1PSstep1}
\begin{aligned}
\parity |m,\vec p;j,\lambda\rangle &= \parity U_h(\vec p)|m, \vec 0; j, \lambda \rangle\\
&=\parity U_h(\vec p) \parity^\dagger \parity |m, \vec 0; j, \lambda \rangle \\
&= \eta R(\phi,\theta,-\phi) R(0,\pi,0) B_3(\eta) R^\dag(0,\pi,0)|m, \vec 0; j, \lambda \rangle \\
&= \eta (-1)^{j+\lambda} R(\phi,\theta,-\phi) R(0,\pi,0)B_3(\eta)|m, \vec 0; j, -\lambda \rangle.
\end{aligned}
\end{equation}
Here in the third line we used \eqref{parity_action_on_helicity_boost} and \eqref{eq:parity_action_rest_frame}, instead in the fourth line we used \eqref{eq:com_frame_rotation_appendix}, \eqref{eq:generic_rotation} and the following property of the small Wigner d-matrix
\begin{equation}
\label{eq:property_d_par}
d^j_{\lambda'\lambda}(-\pi)=(-1)^{j+\lambda'}\delta_{-\lambda,\lambda'}.
\end{equation}
Next, by repeatedly using \eqref{eq:pi_rotate_around_y_rotation} and \eqref{eq:pi_rotate_around_x_rotation} one can show that
\begin{equation}
\label{RRpi}
R(\phi,\theta,-\phi) R(0,\pi,0) =
R(\phi+\pi,\pi -\theta,-(\phi+\pi))e^{-i(2\pi +2\phi)J_3}.
\end{equation}
Inserting this relation into \eqref{eq:1PSstep1} and using the fact that $J_3$ commutes with $K_3$ and that the $\vec p=0$ state is the eigenstate of $J_3$ we conclude that
\begin{equation}
\begin{aligned}
\parity |m,\vec p;j,\lambda\rangle &= 
\eta (-1)^{j+\lambda}
e^{+i(2\pi +2\phi)\lambda}
R(\phi+\pi,\pi -\theta,-(\phi+\pi))
B_3(\eta)|m, \vec 0; j, -\lambda \rangle\\
&=\eta (-1)^{j+3\lambda}e^{2 i \lambda \phi}U_h(-\vec p) |m, \vec 0; j,- \lambda \rangle\\
&=\eta (-1)^{j-\lambda}e^{2 i \lambda \phi}|m, -\vec p; j, -\lambda \rangle.
\end{aligned}
\end{equation}
Here in the second line we used \eqref{eq:negative_spherical} and \eqref{eq:unitary_transformation_helicity}. In the third line we used \eqref{eq:helicity_basis} and the fact that
\begin{equation}
\label{eq:fact_par}
e^{4i\pi\lambda}=e^{4i\pi j}=+1
\end{equation}
for any $\lambda$ and $j$ which are integer or half-integer.
Summarizing, the final expression for the action of the parity operator on a one-particle states reads as
\begin{equation}
\label{eq:parity_action_1PS}
\parity |m,\vec p;j,\lambda \rangle = \eta (-1)^{j-\lambda}e^{2 i \lambda \phi}|m,-\vec p;j, -\lambda \rangle.
\end{equation}

\paragraph{Two-particle COM states}
From the action of parity on one-particle states \eqref{eq:parity_action_1PS}, one can conclude the action of parity on two-particle center of mass states defined in \eqref{eq:2PS_com}. One has
\begin{eqnarray}
\mathcal P |(\myP, \theta, \phi), \lambda_1 , \lambda_2 \rangle &=&  \mathcal P \left( |m_1, \vec p; j_1, \lambda_1 \rangle \otimes |m_2, -\vec p; j_2, \lambda_2 \rangle \right) \nonumber \\
        &=& \eta_1 \eta_2 (-1)^{j_1-\lambda_1}(-1)^{j_2-\lambda_2}e^{2 i \phi \lambda_1} e^{2 i (\phi+\pi) \lambda_2} |m_1, -\vec p; j_1, -\lambda_1 \rangle \otimes |m_2, \vec p; j_2, -\lambda_2 \rangle \nonumber \\
        &=& \eta_1 \eta_2 (-1)^{j_1 - j_2 - \lambda_1 - \lambda_2} e^{2i\phi(\lambda_1 + \lambda_2)}|(\myP, \pi - \theta, \phi + \pi), -\lambda_1, -\lambda_2 \rangle,
        \nn
\end{eqnarray}
where $\eta_1$ and $\eta_2$ are the intrinsic parities of the first and the second particle respectively. Notice also that in the third line we used for the second particle the identity
\begin{equation}
\label{eq:identity_hel}
1=e^{2\pi i(\lambda\pm j)},
\end{equation}
which holds true since $\lambda\pm j$ is always an integer. To summarize, we have
\begin{eqnarray}
\label{parity_action_COM_generic}
\mathcal P |(\myP, \theta, \phi), \lambda_1 , \lambda_2 \rangle = \eta_1 \eta_2 (-1)^{j_1 - j_2 - \lambda_1 - \lambda_2} e^{2i\phi(\lambda_1 + \lambda_2)}
|(\myP, \pi - \theta, \phi + \pi), -\lambda_1, -\lambda_2 \rangle.
\end{eqnarray}

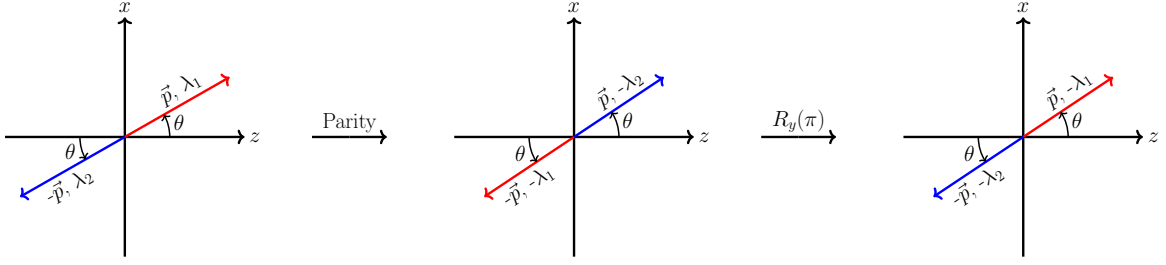
\begin{figure}
\centering
\resizebox{\textwidth}{!}{
\begin{tikzpicture}
    \coordinate (origin) at (0,0);

    \draw[line width=0.8mm,black,->] (origin) -- ++(4,0) node (z) [black,right] {\huge $z$};
    \draw[line width=0.8mm,black,->] (origin) -- ++(0,4) node (x)  [black, above] { \huge $x$};
\draw[line width=0.8mm,black,-] (origin) -- ++(-4,0) node (zminus) {};
    \draw[line width=0.8mm,black,-] (origin) -- ++(0,-4) node (xminus) {}  ;
     
    \draw[line width=0.8mm,red,->] (origin) -- ++(3.5,2) coordinate (p) node[pos = 0.6,above, black, sloped]  {\huge $\vec p$, $\lambda_1$};
    \draw[line width=0.8mm,blue,->] (origin) -- ++(-3.5,-2) coordinate (pminus) node[pos = 0.6,below, black, sloped] { \huge -$\vec p$, $\lambda_2$};

    \pic [draw,line width=0.5mm, ->,  "\huge $\theta$", angle eccentricity=1.25, angle radius = 1.5cm] {angle = z--origin--p};
    \pic [draw,line width=0.5mm, ->,"\huge $\theta$", angle eccentricity=1.25, angle radius = 1.5cm] {angle = zminus--origin--pminus};
    
    \draw[line width=0.8mm,black, -]  (6.25,0) --  (7.5,0) node  [black, above] {\huge Parity};
   \draw[line width=0.8mm,black, ->] (7.5,0) -- (8.75,0);

\coordinate (origin2) at (15,0);
\coordinate (pprime) at (18,2);
\coordinate (pprimeminus) at (12,-2);
  
    \draw[line width=0.8mm,black,->] (origin2) -- (19,0) node (zprime) [black,right] {\huge $z$};
    \draw[line width=0.8mm,black,->] (origin2) -- (15,4) node (xprime)  [black, above] {\huge $x$};
\draw[line width=0.8mm,black,-] (origin2) -- (11,0) node (zprimeminus) {};
   \draw[line width=0.8mm,black,-] (origin2) --(15,-4) node (xprimeminus) {}  ;

    \draw[line width=0.8mm,blue,->] (origin2) -- (pprime) node[pos = 0.6,above, black, sloped] {\huge $\vec p$, -$\lambda_2$};
    \draw[line width=0.8mm,red,->] (origin2) -- (pprimeminus) node[pos = 0.6,below, black, sloped] {\huge -$\vec p$, -$\lambda_1$};

    \pic [draw,line width=0.5mm, ->, "\huge $\theta$", angle eccentricity=1.25, angle radius = 1.5cm] {angle = zprime--origin2--pprime};
    \pic [draw,line width=0.5mm, ->, "\huge $\theta$", angle eccentricity=1.25, angle radius = 1.5cm] {angle = zprimeminus--origin2--pprimeminus};

\draw[line width=0.8mm,black, -]  (21.25,0) --  (22.5,0) node  [black, above] {\huge $R_y(\pi)$};
   \draw[line width=0.8mm,black, ->] (22.5,0) -- (23.75,0);

\coordinate (origin3) at (30,0);
\coordinate (pprimeprime) at (33,2);
\coordinate (pprimeprimeminus) at (27,-2);
%
    \draw[line width=0.8mm,black,->] (origin3) -- (34,0) node (zprimeprime) [black,right] {\huge $z$};
    \draw[line width=0.8mm,black,->] (origin3) -- (30,4) node (xprimeprime)  [line width=0.8mm,black, above] {\huge $x$};
\draw[line width=0.8mm,black,-] (origin3) -- (26,0) node (zprimeprimeminus) {};
   \draw[line width=0.8mm,black,-] (origin3) --(30,-4) node (xprimeprimeminus) {}  ;
    \draw[line width=0.8mm,red,->] (origin3) -- (pprimeprime) node[pos = 0.6,above, black, sloped] {\huge $\vec p$, -$\lambda_1$};
    \draw[line width=0.8mm,blue,->] (origin3) -- (pprimeprimeminus) node[pos = 0.6, below, black, sloped] {\huge -$\vec p$, -$\lambda_2$};

    \pic [draw,line width=0.5mm, ->, "\huge $\theta$", angle eccentricity=1.25, angle radius = 1.5cm] {angle = zprimeprime--origin3--pprimeprime};
    \pic [draw,line width=0.5mm, ->, "\huge $\theta$", angle eccentricity=1.25, angle radius = 1.5cm] {angle = zprimeprimeminus--origin3--pprimeprimeminus};

  \end{tikzpicture} 
 }
 \caption{The geometric picture behind \eqref{parity_action_COM} and \eqref{eq:property_COM}. {\color{red}Particle 1} is red and {\color{blue}particle 2} is blue. \ }
 \label{fig:parity_picture}
 \end{figure}

In principle \eqref{parity_action_COM_generic} is our final answer. However, for applications to scattering amplitudes we need to bring \eqref{parity_action_COM_generic} to a different form. We focus on the case where $\phi=0$, when \eqref{parity_action_COM_generic} simplifies to
\begin{eqnarray}
\label{parity_action_COM}
\mathcal P |(\myP, \theta, 0), \lambda_1 , \lambda_2 \rangle = \eta_1 \eta_2 (-1)^{j_1 - j_2 - \lambda_1 - \lambda_2}
|(\myP, \pi - \theta, \pi), -\lambda_1, -\lambda_2 \rangle.
\end{eqnarray}
Here the two-particle state in the right-hand side by definition reads as
\begin{equation}
\label{eq:2PS_par}
|(\myP,\pi - \theta, \pi), -\lambda_1, -\lambda_2 \rangle
= |m_1, -\vec p, j_1, -\lambda_1 \rangle \otimes |m_2, \vec p, j_2, -\lambda_2 \rangle,
\end{equation}
where the three-vector $\vec p$ has $(\theta,0)$ spherical angles and the three-vector $-\vec p$ has $(\pi-\theta,\pi)$ spherical angles.
Using \eqref{eq:pi_rotate_around_y_rotation} and \eqref{eq:pi_rotate_around_x_rotation} one can derive the following relations
\begin{align}
\label{eq:rel_par_1}
R(\pi,\pi-\theta,-\pi) &= 
e^{-i\pi J_2}R(0,\theta,-2\pi),\\
\label{eq:rel_par_2}
R(0,\theta,0) &= 
e^{-i\pi J_2}R(\pi,\pi-\theta,-\pi).
\end{align}
Using the definition of helicity states \eqref{eq:helicity_basis} and the relation \eqref{eq:rel_par_1} we conclude that the first one-particle state in \eqref{eq:2PS_par} can be written as
\begin{equation}
\label{eq:par_eq1}
\begin{aligned}
|m_1, -\vec p; j_1, -\lambda_1 \rangle &=
R(\pi, \pi-\theta, -\pi) B(\eta)|m_1, \vec 0; j_1, -\lambda_1 \rangle  \\
&= e^{-i\pi J_2}R(0, \theta, 0) e^{2\pi i J_3} B_3(\eta)|m_1, \vec 0; j_1, -\lambda_1 \rangle  \\
&= e^{-2\pi i \lambda_1}e^{-i\pi J_2}R(0, \theta, 0) B_3(\eta)|m_1, \vec 0; j_1, -\lambda_1 \rangle  \\
&= (-1)^{-2j_1}e^{-i\pi J_2}|m_1, \vec p; j_1, -\lambda_1 \rangle.
\end{aligned}
\end{equation}
In the third line we used the fact that $J_3$ commutes with $K_3$ and that the state with $\vec p=0$ is the eigenstate of $J_3$. In the last equality we used \eqref{eq:identity_hel}. Analogously using \eqref{eq:rel_par_2} for the second one-particle state in \eqref{eq:2PS_par} we conclude that
\begin{equation}
\label{eq:par_eq2}
|m_2, \vec p; j_2, -\lambda_2 \rangle=
e^{-i\pi J_2}|m_2, -\vec p; j_2, -\lambda_2 \rangle.
\end{equation}
Plugging \eqref{eq:par_eq1} and \eqref{eq:par_eq2} into \eqref{eq:2PS_par} we obtain the following relation
\begin{equation}
\label{eq:property_COM}
|(\myP,\pi - \theta, \pi), -\lambda_1, -\lambda_2 \rangle =
(-1)^{-2j_1}e^{-i\pi J_2}|(\myP,\theta, 0), -\lambda_1, -\lambda_2 \rangle.
\end{equation}
Finally, plugging \eqref{eq:property_COM} into \eqref{parity_action_COM} and  
using an obvious identity $(-1)^{-j-\lambda}=(-1)^{j+\lambda}$ which holds true since $j+\lambda$ is always an integer, we obtain the desired expression
\begin{equation}
\label{eq:parity_com}
\mathcal P |(\myP,\theta,0); \lambda_1, \lambda_2 \rangle = \eta_1 \eta_2 (-1)^{j_1+j_2+\lambda_1+\lambda_2}e^{-i\pi J_2} |(\myP,\theta,0); -\lambda_1, -\lambda_2 \rangle.
\end{equation}
The benefit of this equation is that the states in the left- and right-hand side are in the same configuration contrary to \eqref{parity_action_COM}. For a pictorial representation of the above formulas see  figure \ref{fig:parity_picture}.
 
\paragraph{Two-particle irreps}
The two-particle states can be decomposed into states in the irreducible representation of the Poincar\'e group. We refer to them as the two-particle irreps. In the center of mass frame such a decomposition and its inverse are given by \eqref{eq:COM_state_decomposition_appendix} and \eqref{eq:COM_state_decomposition_inverse_appendix} respectively. Applying parity to \eqref{eq:COM_state_decomposition_inverse_appendix} and using \eqref{parity_action_COM_generic} we get
\begin{align}
\label{eq:irrep_par}
\parity |c, 0; \spin, \hel; \lambda_1, \lambda_2 \rangle &= \mathcal N \int_0^{2 \pi} d\phi \int_{0}^{\pi} d\theta \sin\theta \, e^{-i\phi(\lambda_1+\lambda_2-\lambda)} d^{(\ell)}_{\lambda \lambda_{12}}(\theta)\parity |(\myP,\theta,\phi); \lambda_1, \lambda_2 \rangle\\
&= \eta^\text{com}_{12}\mathcal N \int_0^{2 \pi} d\phi \int_{0}^{\pi} d\theta \sin\theta \, e^{i(\lambda_1+\lambda_2+\lambda)\phi} d^{(\ell)}_{\lambda \lambda_{12}}(\theta)|(\myP, \pi - \theta, \phi + \pi), -\lambda_1, -\lambda_2 \rangle,
\nn
\end{align}
where we have defined
\begin{equation}
\label{eq:defs_par}
\mathcal N \equiv \frac{2\spin +1}{4 \pi C_\spin(\myP)},\qquad
\eta_{12}^\text{com} \equiv \eta_1 \eta_2 (-1)^{j_1 - j_2 - \lambda_1 - \lambda_2}.
\end{equation}
Changing the integration variables from $\theta$ and $\phi$ to $\theta'\equiv\pi-\theta$ and $\phi'\equiv \phi+\pi$, using the following property of the small Wigner d-matrix
\begin{equation}
\label{eq:d_relation_par}
d^{(\ell)}_{\lambda' \lambda} (\pi-\theta)=(-1)^{\ell+\lambda'}
d^{(\ell)}_{\lambda', -\lambda}(\theta),
\end{equation}
the second line of \eqref{eq:irrep_par} can be written as
\begin{multline}
\parity |c, 0; \spin, \hel; \lambda_1, \lambda_2 \rangle = 
(-1)^{-\lambda_1-\lambda_2-\lambda}(-1)^{\ell+\lambda}\eta^{com}_{12}\mathcal N \\
\quad \quad \quad \times \int_{\pi}^{ 3\pi} d\phi' \int_{0}^{\pi} d\theta' \sin\theta' e^{i(\lambda_1+\lambda_2+\lambda)\phi'}d^{(\ell)}_{\lambda, -\lambda_{12}}(\theta')|(p, \theta', \phi'), -\lambda_1, -\lambda_2 \rangle.
\end{multline}

Consider now a function $f(\phi)$ which is $2\pi$ periodic in $\phi$. The following property then holds
\begin{equation}
\label{eq:integration_periodic}
\int_0^{2\pi} d\phi f(\phi) = \int_{\phi_0}^{\phi_0+2\pi} d\phi f(\phi)
\end{equation}
for any real $\phi_0$. We notice that both $e^{i(\lambda_1+\lambda_2+\lambda)\phi'}$ and $|(p, \theta', \phi'), -\lambda_1, -\lambda_2 \rangle$ are $2 \pi$ periodic in $\phi'$. The former follows from the fact that $\lambda_1 + \lambda_2 + \lambda$ is always an integer. The latter follows from our definition of the helicity basis \eqref{eq:unitary_transformation_helicity}, see in particular footnote \ref{foot:periodicity_R}.
We can then use the definition \eqref{eq:COM_state_decomposition_inverse_appendix} one more time to conclude that
\begin{equation}
\parity |c, 0; \spin, \hel; \lambda_1, \lambda_2 \rangle = 
(-1)^{-\lambda_1-\lambda_2-\lambda}(-1)^{\ell+\lambda}\eta^\text{com}_{12}
|c, 0; \spin, \hel; -\lambda_1, -\lambda_2 \rangle.
\end{equation}
Plugging in \eqref{eq:defs_par} and using \eqref{eq:identity_hel} the above can be brought to the following final form
\begin{equation}
\label{eq:parity_pwa_app}
\parity |c, 0; \spin, \hel; \lambda_1, \lambda_2 \rangle = \eta_1 \eta_2 (-1)^{\spin - j_1 + j_2} |c, 0; \spin, \hel; -\lambda_1, -\lambda_2 \rangle.
\end{equation}

\paragraph{Constraints on scattering amplitudes}
In parity invariant theories the scattering operators $S$ and $T$ obey
\begin{equation}
\label{eq:par_S}
S=\parity S \parity^\dagger=\parity^\dagger S \parity,\qquad
T=\parity T \parity^\dagger=\parity^\dagger T \parity.
\end{equation}
Using \eqref{eq:par_S} in the definition of the center of mass amplitude \eqref{eq:scalar_Tamplitude_definition} we obtain the following constraint on the COM scattering amplitudes
\begin{equation}
\label{eq:par_cond}
\langle (\myP',\theta,0) ; \lambda_3 , \lambda_4 |T|(\myP,0,0); \lambda_1 , \lambda_2 \rangle=
\langle (\myP',\theta,0) ; \lambda_3 , \lambda_4 | \mathcal P^\dagger T \mathcal P |(\myP,0,0); \lambda_1 , \lambda_2 \rangle.
\end{equation}
Using \eqref{eq:parity_com} the right-hand side of this equation can  be written as
\begin{multline}
\langle (\myP',\theta,0) ; \lambda_3 , \lambda_4 | \mathcal P^\dagger T \mathcal P |(\myP,0,0); \lambda_1 , \lambda_2 \rangle = \eta_1\eta_2 \eta_3^* \eta_4^* (-1)^{j_1+j_2+\lambda_1+\lambda_2}(-1)^{j_3+j_4+\lambda_3+\lambda_4}\\
\langle (\myP',\theta,0) ; -\lambda_3 , -\lambda_4 | e^{i\pi J_2} T e^{-i\pi J_2}    |(\myP,0,0); -\lambda_1 , -\lambda_2 \rangle.
\end{multline}
Plugging this into \eqref{eq:par_cond}, using the fact the scattering operator $T$ is invariant under rotations and invoking the definition of the COM amplitudes we obtain the final constraint
\begin{equation}
\label{eq:amplitudes_parity_app}
T_{12\rightarrow 34}{}_{\lambda_1,\lambda_2}^{\lambda_3, \lambda_4}(s,t,u) = \eta_1 \eta_2 \eta_3^* \eta_4^* (-1)^{j_1+j_2+j_3+j_4}(-1)^{\lambda_1 + \lambda_2 +\lambda_3 + \lambda_4} T_{12\rightarrow 34}{}_{-\lambda_1,-\lambda_2}^{-\lambda_3, -\lambda_4}(s,t,u).
\end{equation}

\subsection{Time-reversal}
\label{app:time-reversal}
In the defining vector representation time-reversal is given by the following matrix
\begin{equation}
\label{eq:time_defining}
\time =
\begin{pmatrix}
-1\;\, & 0\;\, & 0\;\, & 0\\
0\;\, & 1\;\, & 0\;\, & 0\\
0\;\, & 0\;\, & 1\;\, & 0\\
0\;\, & 0\;\, & 0\;\, & 1
\end{pmatrix}.
\end{equation}
We use the same symbol $\time$ to denote the time-reversal operator in the (infinite-dimensional) unitary representation.
Using \eqref{eq:time_defining} one can deduce the following commutation properties of the time-reversal operator with finite rotations and boosts in the unitary representation
\begin{equation}
\time e^{-i \theta_i J_i} \time^\dagger = e^{-i \theta_i J_i} \quad \text{and} \quad \time e^{-i \eta_i K_i} \time^\dagger = e^{i \eta_i K_i}. 
\end{equation}
Similarly, the action of time-reversal on the translation operators is given by
\begin{equation}
\time e^{-i P^0 t} \time^\dagger = e^{+i P^0 t} \quad \text{and}\quad \time e^{+i \vec P\cdot \vec x} \time^\dagger = e^{+i \vec P \cdot \vec x}.
\end{equation}
We  recall now that $\time$ is anti-unitary, namely it obeys the following condition
\begin{equation}
\label{eq:antiunitary}
\time i \time^\dagger = - i.
\end{equation}
Using these facts we deduce the following commutation relations of $\time$ with the generators of the Poincar\'e group
\begin{equation}
\label{eq:time-reversal_generators}
\time P^\mu  \time^\dag = (P^0, -\vec P),\qquad
\time K^i \time^\dag = K^i,\qquad
\time J^i \time^\dag =  -J^i.
\end{equation}

In what follows we will use \eqref{eq:time-reversal_generators} to derive the action of time-reversal on one- and two-particle states. Then (as in the previous section) we will derive constraints on the scattering amplitudes.
As before, we begin by computing the transformation property of the helicity eigenstate boost \eqref{eq:unitary_transformation_helicity} under time-reversal,
\begin{equation}
\label{eq:helicity_boost_timereversal}
\begin{aligned}
\time U_{\text{h}}(\vec p) \time^\dag
=&\time R(\phi,\theta,-\phi) \time^\dagger\time B_3(\eta) \time^\dag\\
=&R(\phi,\theta,-\phi) B_3(-\eta) \\
=&R(\phi,\theta,-\phi) R(0,\pi,0) B_3(\eta) R^\dag(0,\pi,0).
\end{aligned}
\end{equation}
Here in the first line we injected the identity $\mathbb{I}=\time^\dagger\time$, we then used the commutation properties \eqref{eq:time-reversal_generators} together with \eqref{eq:antiunitary} in the second line. Finally, in the third line we used \eqref{eq:pi_rotate_around_y_boost}. Interestingly enough, \eqref{eq:helicity_boost_timereversal} is the same as \eqref{parity_action_on_helicity_boost}. Thus, we expect that in what follows we will be able to utilize many intermediate results from the previous section.

\paragraph{One-particle states}
Let us first deduce the action of time-reversal on a one-particle state in the rest frame. Consider the following relation
\begin{equation}
\begin{aligned}
\time R(\alpha,\beta,\gamma) |m,\vec 0;j,\lambda\rangle&=
\time R(\alpha,\beta,\gamma)\time^\dagger \time |m,\vec 0;j,\lambda\rangle\\
&=
R(\alpha,\beta,\gamma)\time |m,\vec 0;j,\lambda\rangle,
\end{aligned}
\end{equation}
which holds true due to \eqref{eq:antiunitary} and \eqref{eq:time-reversal_generators}. On the other hand we also have
\begin{equation}
\begin{aligned}
\time R(\alpha,\beta,\gamma) |m,\vec 0;j,\lambda\rangle&=
\time \mathscr{D}^{\spin}_{\lambda'\lambda}(\alpha,\beta,\gamma) |m,\vec 0;j,\lambda\rangle\\
&=
\mathscr{D}^{\spin*}_{\lambda'\lambda}(\alpha,\beta,\gamma) \time|m,\vec 0;j,\lambda'\rangle.
\end{aligned}
\end{equation}
Here in the first line we simply used \eqref{eq:com_frame_rotation_appendix}. Instead the second line follows from \eqref{eq:antiunitary} and the explicit form of the large Wigner D-matrix, see \eqref{big_D_matrix} and \eqref{eq:def_d}.
Comparing the above two expressions we conclude that
\begin{equation}
\label{eq:transf_time_reversed_1PS}
R(\alpha,\beta,\gamma)\time |m,\vec 0;j,\lambda\rangle
=\mathscr{D}^{\spin*}_{\lambda'\lambda}(\alpha,\beta,\gamma) \time|m,\vec 0;j,\lambda'\rangle.
\end{equation}
By comparing \eqref{eq:transf_time_reversed_1PS} with \eqref{eq:transformation_dual}, we see that the time reversal transformed one-particle states (in the center of mass) are in the dual spin $\ell$ representation. Using \eqref{eq:dual} we conclude that\footnote{From $J_3 \time = -\time J_3$ and \eqref{eq:basis_SO(3)_irreps} we could have only concluded that helicity flips under-time reversal.}
\begin{equation}
\time |m,\vec 0;j,\lambda\rangle =
\varepsilon\, (-1)^{j-\lambda}|m,\vec 0;j,-\lambda\rangle,
\end{equation}
where $\varepsilon$ is the proportionality coefficient obeying $|\varepsilon|^2=1$. One can always define $\time$ in such a way that $\time^2=+1$ or $\time^2=-1$. Thus, we have
\begin{equation}
\varepsilon^2=+1\quad\text{or}\quad\varepsilon^2=-1.
\end{equation}

In order to obtain the action of time-reversal on a one-particle state, we use the definition of the helicity basis \eqref{eq:helicity_basis} and \eqref{eq:helicity_boost_timereversal}. We have
\begin{equation}
\begin{aligned}
\time |m,\vec p;j,\lambda\rangle &= \time U_h(\vec p)|m, \vec 0, j, \lambda \rangle\\
&=\time U_h(\vec p) \time^\dagger \time |m, \vec 0, j, \lambda \rangle\\
&= R(\phi,\theta,-\phi) R(0,\pi,0) B(+\vec \eta) R^\dag(0,\pi,0)\time |m, \vec 0, j, \lambda \rangle \\
&= \varepsilon (-1)^{j-\lambda} R(\phi,\theta,-\phi) R(0,\pi,0)B(+\eta)R^\dagger(0,\pi,0)|m, \vec 0, j, -\lambda \rangle \\
&=\varepsilon R(\phi,\theta,-\phi) R(0,\pi,0)B(+\vec \eta)|m, \vec 0, j, \lambda \rangle.
\end{aligned}
\end{equation}
In going from the fourth to the fifth line we used \eqref{eq:property_d_par} and \eqref{eq:identity_hel}. Utilizing \eqref{RRpi} and \eqref{eq:identity_hel} the above result can be brought to the following final form
\begin{equation}
\label{eq:time_reversal_1PS}
\time |m,\vec p;j,\lambda\rangle = \varepsilon (-1)^{2j} e^{-2i\lambda \phi}|m,-\vec p, j; \lambda \rangle.
\end{equation}

\paragraph{Two-particle COM states}
From the action of time-reversal on one-particle states \eqref{eq:parity_action_1PS}, one concludes the action of time-reversal on two-particle center of mass states defined in \eqref{eq:2PS_com}. One has
\begin{eqnarray}
\time |(\myP, \theta, \phi); \lambda_1 , \lambda_2 \rangle &=&  \time \left( |m_1, \vec p, j_1, \lambda_1 \rangle \otimes |m_2, -\vec p, j_2, \lambda_2 \rangle \right) \\
        &=& \varepsilon_1 \varepsilon_2 (-1)^{2j_1}(-1)^{2j_2}e^{-2 i \phi \lambda_1} e^{-2 i (\phi+\pi) \lambda_2} |m_1, -\vec p, j_1, \lambda_1 \rangle \otimes |m_2, \vec p, j_2, \lambda_2 \rangle \nonumber.
\end{eqnarray}
Using \eqref{eq:identity_hel} we can bring the above to the final form
\begin{equation}
\label{time_action_2_particle_general_com}
\time |(\myP, \theta, \phi); \lambda_1 , \lambda_2 \rangle =
\varepsilon_1 \varepsilon_2 (-1)^{2j_1}  e^{-2i(\lambda_1 + \lambda_2)\phi}|(\myP, \pi - \theta, \phi + \pi); \lambda_1, \lambda_2 \rangle.
\end{equation}

\paragraph{Two-particle irreps}
Let us repeat that the two-particle states can be decomposed into states in the irreducible representation of the Poincar\'e group. We refer to them as the two-particle irreps. In the center of mass frame such a decomposition and its inverse were given by \eqref{eq:COM_state_decomposition_appendix} and \eqref{eq:COM_state_decomposition_inverse_appendix} respectively. Applying time-reversal to \eqref{eq:COM_state_decomposition_inverse_appendix} and using \eqref{time_action_2_particle_general_com} we get
\begin{align}
\nn
\time |c, 0; \spin, \hel; \lambda_1, \lambda_2 \rangle &= \mathcal N \int_0^{2 \pi} d\phi \int_{0}^{\pi} d\theta \sin\theta \,\time e^{-i(\lambda_1+\lambda_2-\lambda)\phi} d^{(\ell)}_{\lambda \lambda_{12}}(\theta) |(\myP,\theta,\phi); \lambda_1, \lambda_2 \rangle\\
\label{eq:irrep_time}
&= \mathcal N \int_0^{2 \pi} d\phi \int_{0}^{\pi} d\theta \sin\theta \, e^{+i(\lambda_1+\lambda_2-\lambda)\phi} d^{(\ell)}_{\lambda \lambda_{12}}(\theta)\time |(\myP,\theta,\phi); \lambda_1, \lambda_2 \rangle\\
&= \varepsilon^\text{com}_{12}\mathcal N \int_0^{2 \pi} d\phi \int_{0}^{\pi} d\theta \sin\theta \, e^{-i(\lambda_1+\lambda_2+\lambda)\phi} d^{(\ell)}_{\lambda \lambda_{12}}(\theta)|(\myP, \pi - \theta, \phi + \pi), \lambda_1, \lambda_2 \rangle.
\nn
\end{align}
Notice that in going from the first to the second line we used \eqref{eq:antiunitary}. The constant $\mathcal N$ was defined in \eqref{eq:defs_par} and we have introduced for brevity
\begin{equation}
\label{eq:brevity}
\varepsilon^\text{com}_{12} \equiv \varepsilon_1 \varepsilon_2 (-1)^{2j_1}.
\end{equation}

Changing the integration variables from $\theta$ and $\phi$ to $\theta'\equiv\pi-\theta$ and $\phi'\equiv \phi+\pi$, using the following property of the small Wigner d-matrix
\begin{equation}
d^{(\ell)}_{\lambda' \lambda} (\pi-\theta)=(-1)^{\lambda-\ell}
d^{(\ell)}_{-\lambda' \lambda}(\theta),
\end{equation}
we can bring \eqref{eq:irrep_time} to the following form
\begin{multline}
\time |c, 0; \spin, \hel; \lambda_1, \lambda_2 \rangle =
\varepsilon^\text{com}_{12}\mathcal N
e^{i(\lambda_1+\lambda_2+\lambda)\pi}
(-1)^{\lambda_1-\lambda_2-\ell}
\\ \times\int_\pi^{3 \pi} d\phi' \int_{0}^{\pi} d\theta' \sin\theta' \, e^{-i(\lambda_1+\lambda_2+\lambda)\phi} d^{(\ell)}_{-\lambda \lambda_{12}}(\theta')|(\myP, \theta', \pi'), \lambda_1, \lambda_2 \rangle.
\end{multline}
Using \eqref{eq:COM_state_decomposition_inverse_appendix} and \eqref{eq:integration_periodic} we conclude that
\begin{equation}
\time |c, 0; \spin, \hel; \lambda_1, \lambda_2 \rangle =
\varepsilon^\text{com}_{12}
e^{i(\lambda_1+\lambda_2+\lambda)\pi}
(-1)^{\lambda_1-\lambda_2-\ell}
|c, 0; \spin, -\hel; \lambda_1, \lambda_2 \rangle.
\end{equation}
Using \eqref{eq:brevity} and  
the obvious identity $(-1)^{\ell-\lambda}=(-1)^{\lambda-\ell}$ which holds true since $\ell-\lambda$ is always an integer, we obtain our final expression
\begin{equation}
\label{eq:time_pwa}
\mathcal T |c, 0, \spin, \hel; \lambda_1, \lambda_2 \rangle = \varepsilon_1 \varepsilon_2 (-1)^{\ell-\lambda}|c, 0, \spin, -\hel; \lambda_1, \lambda_2 \rangle.
\end{equation}

\paragraph{Constraints on scattering amplitudes}
In time-reversal invariant theories the scattering operators $S$ and $T$ obey \begin{equation}
\label{eq:time_S}
\time S \time^\dagger = S^\dagger,\qquad
\time T \time^\dagger = T^\dagger.
\end{equation}
Consider the states
\begin{equation}
\label{eq:state_psi_phi_definition}
|\psi\rangle = | (\myP',\theta,0); \lambda_3, \lambda_4 \rangle,
\qquad
|\phi \rangle = T| (\myP,0,0); \lambda_1, \lambda_2 \rangle.
\end{equation}
By definition the anti-unitary time-reversal operator $\time$ satisfies
\begin{equation}
\label{eq:time_on_inner_product}
 \langle \time \psi |\time \phi \rangle^* = \langle \psi | \phi \rangle ,
\end{equation}
where 
\begin{equation}
|\time \phi \rangle \equiv \time |\phi \rangle \quad \text{and} \quad |\time \psi \rangle \equiv \time |\psi \rangle.
\end{equation}
Using \eqref{time_action_2_particle_general_com} and time-reversal invariance of the $T$ operator \eqref{eq:time_S}, we have
\begin{equation}
\label{eq:time_on_state_phi}
\begin{aligned}
|\time \phi \rangle &= T \time | (\myP,0,0); \lambda_1, \lambda_2 \rangle \\ 
&= \varepsilon_1 \varepsilon_2 (-1)^{2j_1} T^\dagger |(\myP, \pi ,  \pi), \lambda_1, \lambda_2 \rangle
\end{aligned}
\end{equation}
and 
\begin{equation}
\label{eq:time_on_state_psi}
\langle \time \psi | = \varepsilon_3^* \varepsilon_4^* (-1)^{2j_3}  \langle (\myP, \pi - \theta ,  \pi), \lambda_3, \lambda_4 |.
\end{equation}
Plugging \eqref{eq:time_on_state_phi} and \eqref{eq:time_on_state_psi} along with the definitions \eqref{eq:state_psi_phi_definition} into \eqref{eq:time_on_inner_product}, we arrive at
\begin{equation}
\label{eq:time_condition}
\begin{aligned}
\langle (\myP',\theta,0) ; \lambda_3 , \lambda_4 |T&|(\myP,0,0); \lambda_1 , \lambda_2 \rangle=\\
&\varepsilon_1^*\varepsilon_2^*\varepsilon_3\varepsilon_4
(-1)^{2j_1}(-1)^{2j_3}
\langle (\myP',\pi-\theta,\pi) ; \lambda_3 , \lambda_4 | T^\dagger |(\myP,\pi,\pi); \lambda_1 , \lambda_2 \rangle^*\\
&\varepsilon_1^*\varepsilon_2^*\varepsilon_3\varepsilon_4
(-1)^{2j_1}(-1)^{2j_3}
\langle (\myP,\pi,\pi) ; \lambda_1 , \lambda_2 | T |(\myP',\pi-\theta,\pi); \lambda_3, \lambda_4 \rangle.
\end{aligned}
\end{equation}
We are now left with bringing the matrix element in the last line of \eqref{eq:time_condition} to the standard COM frame. For that we will use the following identity
\begin{equation}
\label{eq:reorganization}
R( \pi, \pi+\theta'-\theta, -\pi)=
R(\pi,\pi-\theta,0)
R(0,\theta',0)e^{i \pi J_3} ,
\end{equation}
which is a simple reorganization of exponents in the definition of the Euler rotation \eqref{eq:rotation_Euler}.

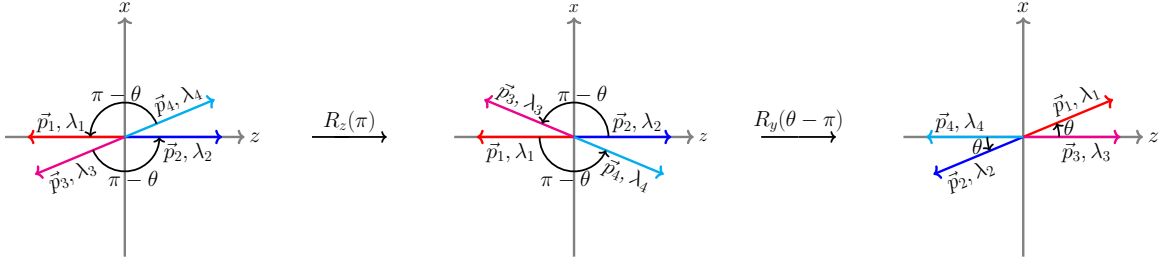
\begin{figure}
\resizebox{\textwidth}{!}{
\begin{tikzpicture}
    \coordinate (origin) at (0,0);

    \draw[line width=0.8mm,gray,->] (origin) -- ++(4,0) node (z) [black,right] {\huge $z$};
    \draw[line width=0.8mm,gray,->] (origin) -- ++(0,4) node (x)  [black, above] { \huge $x$};
\draw[line width=0.8mm,gray,-] (origin) -- ++(-4,0) node (zminus) {};
    \draw[line width=0.8mm,gray,-] (origin) -- ++(0,-4) node (xminus) {}  ;
    \draw[line width=0.8mm,blue,->] (origin) -- ++(3.25,0) coordinate (p1) node[pos = 0.65,below, black, sloped]  {\huge $\vec p_2, \lambda_2$};
    \draw[line width=0.8mm,red,->] (origin) -- ++(-3.25,-0) coordinate (p2) node[pos = 0.65,above, black, sloped] { \huge $\vec p_1, \lambda_1$};
     
    \draw[line width=0.8mm,magenta,->] (origin) -- ++(-3,-1.25) coordinate (p3) node[pos = 0.65,below, black, sloped]  {\huge $\vec p_3, \lambda_3$};
    \draw[line width=0.8mm,cyan,->] (origin) -- ++(3,1.25) coordinate (p4) node[pos = 0.65,above, black, sloped] { \huge $\vec p_4, \lambda_4$};

    \pic [draw,line width=0.6mm, ->,  "\huge $\pi - \theta$", angle eccentricity=1.25, angle radius = 1.15cm] {angle = p3--origin--z};
    \pic [draw,line width=0.6mm, ->,"\huge $\pi - \theta$", angle eccentricity=1.25, angle radius = 1.15cm] {angle = p4--origin--zminus};
    
    
    \draw[thick,line width=0.6mm,black, -]  (6.25,0) --  (7.5,0) node  [black, above] {\huge $R_z(\pi)$};
   \draw[thick,line width=0.6mm,black, ->] (7,0) -- (8.75,0);

\coordinate (origin2) at (15,0);
\coordinate (pprime) at (18,2);
\coordinate (pprimeminus) at (12,-2);
  
    \draw[line width=0.8mm,gray,->] (origin2) -- (19,0) node (zprime) [black,right] {\LARGE $z$};
    \draw[line width=0.8mm,gray,->] (origin2) -- (15,4) node (xprime)  [black, above] {\LARGE $x$};
\draw[line width=0.8mm,gray,-] (origin2) -- (11,0) node (zminusprime) {};
   \draw[line width=0.8mm,gray,-] (origin2) --(15,-4) node (xminusprime) {}  ;

    \draw[line width=0.8mm,blue,->] (origin2) -- ++(3.25,0) coordinate (p1prime) node[pos = 0.65,above, black, sloped]  {\huge $\vec p_2, \lambda_2$};
    \draw[line width=0.8mm,red,->] (origin2) -- ++(-3.25,-0) coordinate (p2prime) node[pos = 0.65,below, black, sloped] { \huge $\vec p_1, \lambda_1$};
     
    \draw[line width=0.8mm,magenta,->] (origin2) -- ++(-3,1.25) coordinate (p3prime) node[pos = 0.65,above, black, sloped]  {\huge $\vec p_3, \lambda_3$};
    \draw[line width=0.8mm,cyan,->] (origin2) -- ++(3,-1.25) coordinate (p4prime) node[pos = 0.65,below, black, sloped] { \huge $\vec p_4, \lambda_4$};

    \pic [draw,line width=0.6mm, ->,  "\huge $\pi - \theta$", angle eccentricity=1.25, angle radius = 1.15cm] {angle = zprime--origin2--p3prime};
    \pic [draw,line width=0.6mm, ->,"\huge $\pi - \theta$", angle eccentricity=1.25, angle radius = 1.15cm] {angle = zminusprime--origin2--p4prime};

\draw[thick, line width = 0.6mm, black, -]  (21.25,0) --  (22.5,0) node  [black, above] {\huge $R_y(\theta-\pi)$};
   \draw[thick, line width = 0.6mm,black, ->] (22.5,0) -- (23.75,0);

\coordinate (origin3) at (30,0);
\coordinate (pprimeprime) at (33,2);
\coordinate (pprimeprimeminus) at (27,-2);
%
    \draw[line width=0.8mm,gray,->] (origin3) -- (34,0) node (zprimeprime) [black,right] {\huge $z$};
    \draw[line width=0.8mm,gray,->] (origin3) -- (30,4) node (xprimeprime)  [black, above] {\huge $x$};
\draw[line width=0.8mm,gray,-] (origin3) -- (26,0) node (zminusprimeprime) {};
   \draw[line width=0.8mm,gray,-] (origin3) --(30,-4) node (xminusprimeprime) {}  ;
%
%
   
\draw[line width=0.8mm,red,->] (origin3) -- ++(3,1.25) coordinate (p1primeprime) node[pos = 0.65,above, black, sloped] { \huge $\vec p_1, \lambda_1$};   
   
    \draw[line width=0.8mm,blue,->] (origin3) -- ++(-3,-1.25) coordinate (p2primeprime) node[pos = 0.65,below, black, sloped]  {\huge $\vec p_2, \lambda_2$};

    \draw[line width=0.8mm,magenta,->] (origin3) -- ++(3.25,0) coordinate (p3primeprime) node[pos = 0.65,below, black, sloped]  {\huge $\vec p_3, \lambda_3$};
    \draw[line width=0.8mm,cyan,->] (origin3) -- ++(-3.25,0) coordinate (p4primeprime) node[pos = 0.65,above, black, sloped] { \huge $\vec p_4, \lambda_4$};

    \pic [draw,line width=0.6mm, ->,  "\huge $\theta$", angle eccentricity=1.3, angle radius = 1.2cm] {angle = zprimeprime--origin3--p1primeprime};
    \pic [draw,line width=0.6mm, ->,"\huge $\theta$", angle eccentricity=1.3, angle radius = 1.2cm] {angle = zminusprimeprime--origin3--p2primeprime};

  \end{tikzpicture}
  
}
\caption{The geometric picture involved in going from \eqref{eq:time_condition} to \eqref{eq:time_pre_result}. {\color{red}Particle 1} is red, {\color{blue}particle 2} is blue, {\color{magenta}particle 3} is magenta and {\color{cyan}particle 4} is cyan. }
\label{fig:time_picture_1}
\end{figure}

Let us start by the following rewriting of the 34 two-particle state
\begin{equation}
\label{eq:rewriting_1}
\begin{aligned}
|(\myP',\pi-\theta,\pi); \lambda_3, \lambda_4 \rangle &=
e^{2i\pi \lambda_4}R( \pi, \pi-\theta, -\pi)|(\myP',0,0); \lambda_3, \lambda_4 \rangle\\
&=e^{2i\pi \lambda_4}
R(\pi,\pi-\theta,0)
R(0,0,0)e^{i \pi J_3}|(\myP',0,0); \lambda_3, \lambda_4 \rangle\\
&=
e^{i\pi (\lambda_3+\lambda_4)}R(\pi,\pi-\theta,0)
|(\myP',0,0); \lambda_3, \lambda_4 \rangle.
\end{aligned}
\end{equation}
Here in the first line we used \eqref{eq:relation_rotation_2PS_spherical}. We used \eqref{eq:reorganization} with $\theta'=0$ in the second line. Finally we used the fact that the states with the momentum along the z-axis are eigenstates of $J_3$ generators. Analogously we can write the 12 two-particle state as
\begin{equation}
\label{eq:rewriting_2}
\begin{aligned}
|(\myP,\pi,\pi); \lambda_1, \lambda_2 \rangle &=e^{2i\pi\lambda_2}
R( \pi, \pi, -\pi)|(\myP,0,0); \lambda_1, \lambda_2 \rangle\\
&=
e^{2i\pi\lambda_2}R(\pi,\pi-\theta,0)
R(0,\theta,0)e^{i \pi J_3}|(\myP,0,0); \lambda_1, \lambda_2 \rangle\\
&=
e^{i\pi (\lambda_1+\lambda_2)}R(\pi,\pi-\theta,0)
|(\myP,\theta,0); \lambda_1, \lambda_2 \rangle,
\end{aligned}
\end{equation}
where in the second line we used \eqref{eq:reorganization} with $\theta'=\theta$.
Plugging both \eqref{eq:rewriting_1} and \eqref{eq:rewriting_2} into the last line of \eqref{eq:time_condition} we conclude that
\begin{multline}
\label{eq:time_pre_result}
\langle (\myP',\theta,0) ; \lambda_3 , \lambda_4 |T|(\myP,0,0); \lambda_1 , \lambda_2 \rangle=
\varepsilon_1^*\varepsilon_2^*\varepsilon_3\varepsilon_4
(-1)^{2j_1}(-1)^{2j_3}
e^{i\pi (-\lambda_1-\lambda_2+\lambda_3+\lambda_4)}\\
\langle (\myP,\theta,0) ; \lambda_1 , \lambda_2 |\left(R(\pi,\pi-\theta,0)\right)^\dagger T R(\pi,\pi-\theta,0) |(\myP',0,0); \lambda_3, \lambda_4 \rangle.
\end{multline}
The series of steps from \eqref{eq:time_condition} to \eqref{eq:time_pre_result} is depicted in figure \ref{fig:time_picture_1}.

Using the fact that the scattering operator is invariant under rotations and that according to \eqref{eq:s_chan} - \eqref{eq:u_chan} the Mandelstam variables remain invariant under the exchange $\myP \leftrightarrow \myP'$, we can use the definition of the center of mass amplitudes \eqref{eq:scalar_Tamplitude_definition} and write \eqref{eq:time_pre_result} in its final form
\begin{equation}
\label{eq:amplitudes_time_reversal_app}
T_{12\rightarrow 34}{}_{\lambda_1,\lambda_2}^{\lambda_3, \lambda_4}(s,t,u) = \varepsilon_1^* \varepsilon_2^* \varepsilon_3 \varepsilon_4 (-1)^{\lambda_1 - \lambda_2 -\lambda_3 + \lambda_4} T_{34\rightarrow 12}{}_{\lambda_3,\lambda_4}^{\lambda_1, \lambda_2}(s,t,u).
\end{equation}
Here we also simplified the phases according to \eqref{eq:identity_hel}.
Notice that whereas parity \eqref{eq:amplitudes_parity} imposes a constraint on the same amplitude, time reversal relates the process $12 \rightarrow 34$ to the process $34 \rightarrow 12$, which are in general different.

\subsection{$\parity\time$}
\label{app:PT}
Let us conclude this appendix by discussing the situation when our physical system is symmetric under simultaneous application of parity and time-reversal, in other words under the $\parity\time$ transformation. 
Due to \eqref{eq:parity_action_1PS} and \eqref{eq:time_reversal_1PS}, the one-particle states transform under $\parity\time$ as follows
\begin{equation}
\label{eq:1PS_PT}
\parity\time |m,\vec p;j,\lambda \rangle = \eta\varepsilon (-1)^{j-\lambda}|m,\vec p;j, -\lambda \rangle.
\end{equation}
One can then derive constraints posed by $\parity\time$ on scattering amplitudes. The simplest way to obtain them is to combine \eqref{eq:amplitudes_parity_app} and \eqref{eq:amplitudes_time_reversal_app}. One then has
\begin{equation}
\label{eq:amplitudes_PT}
T_{12\rightarrow 34}{}_{\lambda_1,\lambda_2}^{\lambda_3, \lambda_4}(s,t,u) = \zeta_1^* \zeta_2^* \zeta_3 \zeta_4 (-1)^{j_1 - j_2 -j_3 + j_4} T_{34\rightarrow 12}{}_{-\lambda_3,-\lambda_4}^{-\lambda_1, -\lambda_2}(s,t,u),
\end{equation}
where we have defined the phases as
\begin{equation}
\zeta_i\equiv \varepsilon_i \eta_i.
\end{equation}

It is interesting to consider the case of identical neutral particles with spin $j$ and the phase $\zeta$. Then the equation \eqref{eq:amplitudes_PT} becomes
\begin{equation}
\label{eq:amplitudes_PT_neutral}
T{}_{\lambda_1,\lambda_2}^{\lambda_3, \lambda_4}(s,t,u) = T{}_{-\lambda_3,-\lambda_4}^{-\lambda_1, -\lambda_2}(s,t,u),
\end{equation}
where we have removed the subscript $12\rightarrow 34$ since all the particles are identical and used the fact that $|\zeta|^2=1$.

Any consistent quantum field theory must be CPT invariant \cite{Streater:1989vi}. This means that one can always introduce the so called CPT operator which we denote by $\Sigma$. In the system of neutral particles one can choose
\begin{equation}
\Sigma= \parity \time.
\end{equation}
As a result the constraint \eqref{eq:amplitudes_PT_neutral} on the scattering amplitudes of identical particles is not an additional assumption but rather a consequence of the CPT theorem.

\section{Identical particles}
\label{app:identical_particles}
In this appendix we consider the special situation where a two-particle state (2PS) describes a system of two identical particles with mass $m$ and spin $j$. Such a system possesses  Bose or Fermi (anti-)symmetry. In other words, the two-particle state must satisfy
\begin{equation}
\label{eq:id_symmetry_property_app}
|\kappa_1,\kappa_2 \>_{id} = (-1)^{2j} |\kappa_2,\kappa_1 \>_{id}.
\end{equation}
Here we have added a subscript $id$ to explicitly indicate that the state describes a system of two identical particles.
In order to satisfy \eqref{eq:id_symmetry_property_app}, we have  to take the symmetrized, in case of bosons, and the anti-symmetrized, in case of fermions, tensor product, \emph{i.e}
\begin{equation}
\label{eq:2PS_identical_app}
|\kappa_1,\kappa_2 \>_{id}\equiv \frac{1}{\sqrt 2}\left(
|m, \vec p_1; j , \lambda_1 \>  \otimes |m, \vec p_2; j, \lambda_2\> + 
(-1)^{2j}|m, \vec p_2; j , \lambda_2 \> \otimes |m, \vec p_1; j, \lambda_1\>\right).
\end{equation}
We remind the reader that $\otimes$ denotes the ordered tensor product that was used in \eqref{eq:2PS_tensor_product} to define generic two-particle states.
The normalization of the state \eqref{eq:2PS_identical_app} follows from \eqref{eq:1PS_normalization} and reads
\begin{equation}
\label{eq:normalization_2PS_id_app}
{}_{id}\<\kappa_1,\kappa_2 |\kappa_3,\kappa_4 \>_{id} = \delta(\kappa_1 - \kappa_3)\delta(\kappa_2 - \kappa_4)
+ (-1)^{2j} \delta(\kappa_1 - \kappa_4)\delta(\kappa_2 - \kappa_3).
\end{equation}

\subsection{Two-particle COM states}
\begin{figure}
\resizebox{\textwidth}{!}{
\begin{tikzpicture}
    \coordinate (origin) at (0,0);

    \draw[thick,gray,->] (origin) -- ++(4,0) node (z) [black,right] {\huge $z$};
    \draw[thick,gray,->] (origin) -- ++(0,4) node (x)  [black, above] {\huge $x$};
\draw[thick,gray,-] (origin) -- ++(-4,0) node (zminus) {};
    \draw[thick,gray,-] (origin) -- ++(0,-4) node (xminus) {}  ;
     
    \draw[thick,red,->] (origin) -- ++(3,2) coordinate (p) node[pos = 0.6,above, black, sloped] {\LARGE $\vec p$, $\lambda_1$};
    \draw[thick,red,->] (origin) -- ++(-3,-2) coordinate (pminus) node[pos = 0.6,below, black, sloped] {\LARGE -$\vec p$, $\lambda_2$};

    \pic [draw, ->, "\LARGE $\theta$", angle eccentricity=1.5,angle radius = 1cm] {angle = z--origin--p};
    \pic [draw, ->, "\LARGE $\theta$", angle eccentricity=1.5,angle radius = 1cm] {angle = zminus--origin--pminus};
    
    \draw[thick,black, -]  (6,0) --  (7.5,0) node  [black, above] {\LARGE $R_z(\pi)$};
   \draw[thick,black, ->] (7.5,0) -- (9,0);

\coordinate (origin2) at (15,0);
\coordinate (pprime) at (18,-2);
\coordinate (pprimeminus) at (12,2);
  
    \draw[thick,gray,->] (origin2) -- (19,0) node (zprime) [black,right] {\huge $z$};
    \draw[thick,gray,->] (origin2) -- (15,4) node (xprime)  [black, above] {\huge $x$};
\draw[thick,gray,-] (origin2) -- (11,0) node (zprimeminus) {};
   \draw[thick,gray,-] (origin2) --(15,-4) node (xprimeminus) {}  ;

    \draw[thick,red,->] (origin2) -- (pprime) node[pos = 0.6,above, black, sloped] {\LARGE -$\vec p\,'$, $\lambda_1$};
    \draw[thick,red,->] (origin2) -- (pprimeminus) node[pos = 0.6,above, black, sloped] {\LARGE $\vec p\,'$, $\lambda_2$};

    \pic [draw, ->, "\LARGE $\pi - \theta$", angle eccentricity=1.3,angle radius = 1cm] {angle = zprime--origin2--pprimeminus};
    \pic [draw, ->, "\LARGE $\pi - \theta$", angle eccentricity=1.3,angle radius = 1cm] {angle = zprimeminus--origin2--pprime};

  \end{tikzpicture}  
 }
 \caption{The geometric picture behind \eqref{eq:identical_particles_com_2}. The particles are identical and are therefore represented by the same colour.}
 \label{fig:pi_rotation_z}
\end{figure}
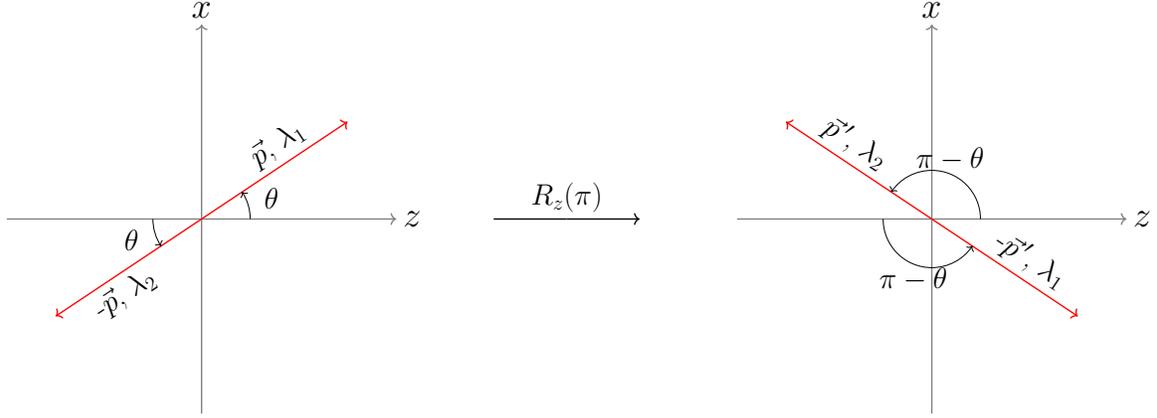

As before we need to define the identical 2PS in the center of momentum. Adapting \eqref{eq:2PS_com} to the case of identical particles we get
\begin{multline}
\label{eq:2PS_com_id_app}
|(\myP,\theta,\phi); \lambda_1, \lambda_2 \rangle_{id} \equiv\\
\frac{1}{\sqrt 2}\left(
|m, +\vec p; j , \lambda_1 \>  \otimes |m, -\vec p; j, \lambda_2\> + 
(-1)^{2j}|m, -\vec p; j , \lambda_2 \> \otimes |m, \vec p; j, \lambda_1\>\right).
\end{multline}
In the notation \eqref{eq:2PS_com} this can be equivalently written as
\begin{equation}
\label{eq:2PS_com_id_app_alt}
|(\myP,\theta,\phi); \lambda_1, \lambda_2 \rangle_{id} =
\frac{1}{\sqrt 2}\left(
|(\myP,\theta,\phi); \lambda_1, \lambda_2 \rangle + 
(-1)^{2j}
|(\myP,\pi-\theta,\pi+\phi); \lambda_2, \lambda_1 \rangle\right).
\end{equation}
The normalization of these states is fixed by \eqref{eq:normalization_2PS_id}. Analogously to \eqref{eq:normalization_spherical} we can write it in spherical coordinates as
\begin{multline}
\label{eq:normalization_spherical_id}
{}_{id}\<(\myP,\theta,\phi); \lambda_1, \lambda_2
|(\myP', 0, 0); \lambda_1', \lambda_2' \rangle_{id} =
(2\pi)^4\delta^4(0) \times
\frac{16\pi^2 \sqrt{s}}{\sqrt{\myP\myP'}}\times\\
\left(
\frac{\delta(\theta)\delta(\phi)}{\sin\theta}\delta_{\lambda_1\lambda_1'}
\delta_{\lambda_2\lambda_2'}+(-1)^{2j}
\frac{\delta(\pi-\theta)\delta(\phi+\pi)}{\sin(\pi-\theta)}
\delta_{\lambda_2\lambda_1'}\delta_{\lambda_1\lambda_2'}
\right).
\end{multline}
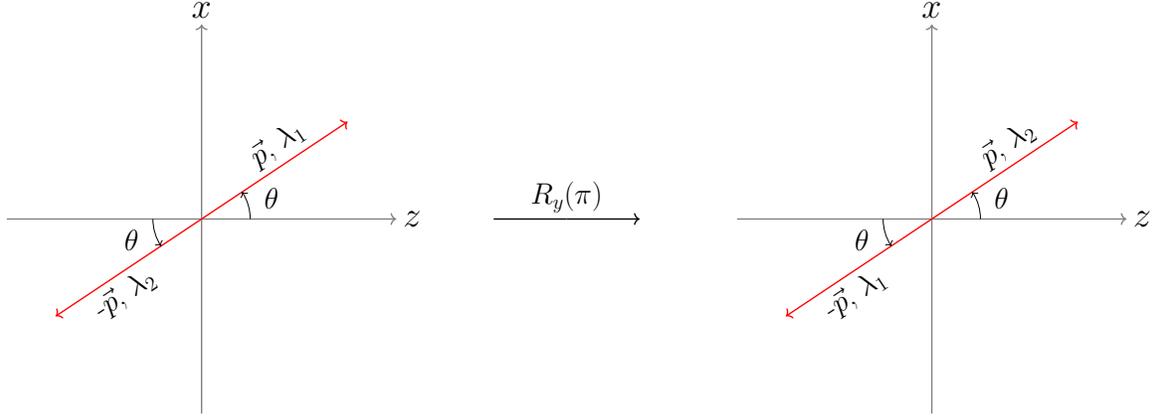
\begin{figure}
\resizebox{\textwidth}{!}{
\begin{tikzpicture}
    \coordinate (origin) at (0,0);

    \draw[thick,gray,->] (origin) -- ++(4,0) node (z) [black,right] {\huge $z$};
    \draw[thick,gray,->] (origin) -- ++(0,4) node (x)  [black, above] {\huge $x$};
\draw[thick,gray,-] (origin) -- ++(-4,0) node (zminus) {};
    \draw[thick,gray,-] (origin) -- ++(0,-4) node (xminus) {}  ;
     
    \draw[thick,red,->] (origin) -- ++(3,2) coordinate (p) node[pos = 0.6,above, black, sloped] {\LARGE $\vec p$, $\lambda_1$};
    \draw[thick,red,->] (origin) -- ++(-3,-2) coordinate (pminus) node[pos = 0.6,below, black, sloped] {\LARGE -$\vec p$, $\lambda_2$};

    \pic [draw, ->, "\LARGE $\theta$", angle eccentricity=1.5,angle radius = 1cm] {angle = z--origin--p};
    \pic [draw, ->, "\LARGE $\theta$", angle eccentricity=1.5,angle radius = 1cm] {angle = zminus--origin--pminus};
    
    \draw[thick,black, -]  (6,0) --  (7.5,0) node  [black, above] {\LARGE $R_y(\pi)$};
   \draw[thick,black, ->] (7.5,0) -- (9,0);

\coordinate (origin2) at (15,0);
\coordinate (pprime) at (18,2);
\coordinate (pprimeminus) at (12,-2);
  
    \draw[thick,gray,->] (origin2) -- (19,0) node (zprime) [black,right] {\huge $z$};
    \draw[thick,gray,->] (origin2) -- (15,4) node (xprime)  [black, above] {\huge $x$};
\draw[thick,gray,-] (origin2) -- (11,0) node (zprimeminus) {};
   \draw[thick,gray,-] (origin2) --(15,-4) node (xprimeminus) {}  ;

    \draw[thick,red,->] (origin2) -- (pprime) node[pos = 0.6,above, black, sloped] {\LARGE $\vec p$, $\lambda_2$};
    \draw[thick,red,->] (origin2) -- (pprimeminus) node[pos = 0.6,below, black, sloped] {\LARGE -$\vec p$, $\lambda_1$};

    \pic [draw, ->, "\LARGE $\theta$", angle eccentricity=1.5,angle radius = 1cm] {angle = zprime--origin2--pprime};
    \pic [draw, ->, "\LARGE $\theta$", angle eccentricity=1.5,angle radius = 1cm] {angle = zprimeminus--origin2--pprimeminus};

  \end{tikzpicture}
}
\caption{The geometric picture behind \eqref{eq:identical_particles_com_1}. The particles are identical and are therefore represented by the same colour.}
\label{fig_pi_rotation_y}
\end{figure}
The symmetry \eqref{eq:id_symmetry_property_app} for the two-particle states in the center of mass reads as
\begin{equation}
\label{eq:2PS_com_identical_app}
|(\myP,\theta,\phi); \lambda_1, \lambda_2 \rangle_{id}  
=(-1)^{2j}|(\myP,\pi - \theta,\phi+\pi); \lambda_2, \lambda_1 \rangle_{id}.
\end{equation}
Let us now restrict our attention on the special case $\phi=0$ and derive the following two relations
\begin{align}
\label{eq:identical_particles_com_1}
|(\myP,\theta,0); \lambda_1, \lambda_2 \rangle_{id} &= e^{-i\pi J_2} |(\myP,\theta,0); \lambda_2, \lambda_1 \rangle_{id},\\
\label{eq:identical_particles_com_2}
|(\myP,\theta,0) ; \lambda_1 , \lambda_2 \rangle_{id} &=(-1)^{\lambda_1-\lambda_2}e^{-i\pi J_3} |(\myP,\pi - \theta,0) ; \lambda_2 , \lambda_1 \rangle_{id}.
\end{align}
The first one simply follows from \eqref{eq:2PS_com_identical_app} with $\phi=0$ and \eqref{eq:property_COM} and is shown in figure \ref{fig_pi_rotation_y}. The second one follows from \eqref{eq:2PS_com_identical_app} with $\phi=0$ and \eqref{eq:relation_rotation_2PS_spherical} and is shown in figure \ref{fig:pi_rotation_z}. 

More precisely
\begin{equation}
\label{eq:inter_id}
\begin{aligned}
|(\myP,\theta,\phi); \lambda_1, \lambda_2 \rangle_{id}  
&=(-1)^{2j}e^{2i\pi \lambda_1}R(\pi,\pi-\theta,-\pi)
|(\myP,0,0); \lambda_2, \lambda_1 \rangle_{id}\\
&=(-1)^{2j}e^{2i\pi \lambda_1}e^{-i\pi J_3}R(0,\pi-\theta,0)e^{+i\pi J_3}
|(\myP,0,0); \lambda_2, \lambda_1 \rangle_{id}\\
&=(-1)^{2j}e^{i\pi(\lambda_2+\lambda_1)}e^{-i\pi J_3}R(0,\pi-\theta,0)
|(\myP,0,0); \lambda_2, \lambda_1 \rangle_{id}\\
&=(-1)^{2j}e^{i\pi(\lambda_1+\lambda_2)}e^{-i\pi J_3}
|(\myP,\pi-\theta,0); \lambda_2, \lambda_1 \rangle_{id}.
\end{aligned}
\end{equation}
In the first and the fourth line we used \eqref{eq:relation_rotation_2PS_spherical}. In the second line we simply used the definition of Euler rotations \eqref{eq:rotation_Euler}. Finally in the third line we used the fact that the states with the three-momentum along the z-axis are eigenvector of $J_3$. Using \eqref{eq:identity_hel} in the last line of \eqref{eq:inter_id} we obtain \eqref{eq:identical_particles_com_2}.

\subsection{Two-particle irreps}
We would now like to decompose identical two particle states \eqref{eq:2PS_identical_app} into irreducible representations analogously to the generic two-particle state decomposition \eqref{eq:2PS_decomposition_com}. The two-particle states in the irreducible representation are denoted by
\begin{equation}
\label{eq:irrep_app}
|c,0,\spin,\hel; \lambda_1,\lambda_2 \rangle_{id}.
\end{equation}
As before the subscript $id$ emphasizes the fact that we do not deal with a generic situation but with the identical particle case. In what follows we need to define the states \eqref{eq:irrep_app} precisely and fix their normalization.

In a generic situation, the two-particle states in the reducible and irreducible representations are related by \eqref{eq:COM_state_decomposition_inverse_appendix}. Since the Lorentz transformation property of the states are the same for identical and distinct particles, we can just use the decomposition formula \eqref{eq:COM_state_decomposition_inverse_appendix} but with an unspecified normalization $\mathcal N_{id}$ which we fix later in the section. One has
\begin{equation}
\label{eq:decomposition_id}
|c,\vec 0; \spin,\lambda;\lambda_1,\lambda_2\rangle_{id} = \mathcal N_{id} \int_0^{2\pi}d\phi\int_{-1}^{+1}d\cos\theta
e^{-i(\lambda_1+\lambda_2 -\lambda)\phi} d^{ \spin}_{\lambda \lambda_{12}} (\theta)|(\myP,\theta,\phi); \lambda_1, \lambda_2 \rangle_{id}.
\end{equation}
We can use this formula to deduce the symmetry property of the state \eqref{eq:irrep_app} under the exchange of two particles. Plugging \eqref{eq:2PS_com_identical_app} into the right-hand side of \eqref{eq:decomposition_id} we get 
\begin{equation}
\label{eq:relations_id_irrep}
\begin{aligned}
&(-1)^{2j}\mathcal N_{id} \int_0^{2 \pi} d\phi \int_{0}^{\pi} d\theta \sin\theta \, e^{-i(\lambda_1+\lambda_2-\lambda)\phi} d^{(\ell)}_{\lambda \lambda_{12}}(\theta) |(\myP,\pi-\theta,\phi+\pi); \lambda_2, \lambda_1 \rangle_{id}= \\
& (-1)^{2j}\mathcal N_{id} \int_\pi^{3 \pi} d\phi' \int_{0}^{\pi} d\theta' \sin\theta'e^{-i(\lambda_1+\lambda_2-\lambda)(\phi'-\pi)}d^{(\ell)}_{\lambda \lambda_{12}}(\pi-\theta')|(\myP,\theta',\phi'); \lambda_2, \lambda_1 \rangle_{id}= \\
& (-1)^{2j}e^{i\pi(\lambda_1+\lambda_2-\lambda)}(-1)^{\spin + \hel}\times \\
& \qquad \quad \quad \mathcal N_{id} \int_\pi^{3 \pi} d\phi' \int_{0}^{\pi} d\theta' \sin\theta'e^{-i(\lambda_1+\lambda_2-\lambda)\phi'}d^{(\ell)}_{\lambda \lambda_{21}}(\theta')|(\myP,\theta',\phi'); \lambda_2, \lambda_1 \rangle_{id}.
\end{aligned}
\end{equation}
In the second line we changed the integration variables from $\theta$ and $\phi$ to $\theta'\equiv\pi-\theta$ and $\phi'\equiv \phi+\pi$. In the third line we used \eqref{eq:d_relation_par} to rewrite the small Wigner d-matrix. Also recall the definition
\begin{equation}
\lambda_{12}\equiv \lambda_1-\lambda_2.
\end{equation}
The last line in \eqref{eq:relations_id_irrep} simply contains the two-particle irrep \eqref{eq:decomposition_id}. In order to see this, we refer the reader to \eqref{eq:integration_periodic} and the  discussion below. Combining \eqref{eq:decomposition_id} and \eqref{eq:relations_id_irrep}, and taking into account \eqref{eq:identity_hel} we finally get\footnote{Notice that due to \eqref{eq:irreps_prop_id} the two-particle irreps with $\lambda_1=\lambda_2$ exist only for even spins $\spin$. When $\lambda_1\neq\lambda_2$ we can form two linear combinations from two-particle irreps, one of which exists only for even spins $\spin$ and the other one only for odd $\spin$.}
\begin{equation}
\label{eq:irreps_prop_id}
|c,\vec 0; \spin,\lambda;\lambda_1,\lambda_2\rangle_{id} =
(-1)^{\spin + \lambda_1 - \lambda_2}|c, 0; \spin, \hel; \lambda_2, \lambda_1 \rangle_{id}.
\end{equation}

We can now define two-particle irreps \eqref{eq:irrep_app} describing identical particles in terms of generic two-particle irreps by requiring that \eqref{eq:irrep_app} automatically satisfies  the condition \eqref{eq:irreps_prop_id}. Our choice here is as follows
\begin{equation}
\label{eq:irrep_id_app}
|c,0,\spin,\hel; \lambda_1,\lambda_2 \rangle_{id}\equiv
\frac{1}{2}\left(
|c,0,\spin,\hel; \lambda_1,\lambda_2 \rangle+
(-1)^{\ell+\lambda_1 - \lambda_2}
|c,0,\spin,\hel; \lambda_2,\lambda_1 \rangle
\right).
\end{equation}
The normalization of the states \eqref{eq:irrep_id_app} follows from the normalization of each of the two terms fixed by \eqref{eq:irrep_normalization} together with \eqref{eq:kronecker_gamma}. As a result the normalization of the states \eqref{eq:irrep_id_app} is given by \eqref{eq:irrep_normalization} along with 
\begin{equation}
\label{eq:convention_gamma_id_app}
\delta_{\gamma'\gamma} = 
\frac{1}{2}\left(
\delta_{\lambda_{1}\lambda_{1}'}\delta_{\lambda_{2}\lambda_{2}'}
+(-1)^{\ell+\lambda_1-\lambda_2}
\delta_{\lambda_{1}\lambda_{2}'}\delta_{\lambda_{2}\lambda_{1}'}
\right).
\end{equation}

Having defined \eqref{eq:irrep_id_app}, we can apply \eqref{eq:2PS_decomposition_com} to both terms in the right-hand side of \eqref{eq:2PS_com_id_app_alt}.
Using the property of the small Wigner d-matrix \eqref{eq:d_relation_par} we get
\begin{equation}
\label{eq:2PS_decomposition_com_id_app}
|(\myP,\theta,\phi); \lambda_1, \lambda_2 \rangle_{\text{id}} =
\sqrt{2}\sum_{\ell,\lambda}
C_\ell(\myP) e^{i(\lambda_1 + \lambda_2 - \lambda)\phi} d^{(\ell)}_{\lambda \lambda_{12}}(\theta) |c, 0; \spin, \hel; \lambda_1,\lambda_2 \rangle_{id},
\end{equation}
where the coefficient $C_\ell(\myP)$ is given by \eqref{eq:coefficient_Cj}.
As before we can invert the above equation to get precisely \eqref{eq:decomposition_id} with
\begin{equation}
\mathcal N_{id} = \frac{2\ell +1}{4 \pi \sqrt{2} C_\ell(\myP)}.
\end{equation}

\subsection{Constraints on scattering amplitudes}
\label{app:amplitudes_id}
Consider now the scattering of identical particles.\footnote{More generally one can consider scattering processes with only incoming or outgoing particles being identical. As presented, most of the results in this section still apply to these situations.} The amplitude describing such a situation is defined as
\begin{align}
\label{sec:idnetical_particles_app}
T_{\lambda_1, \lambda_2}^{\lambda_3,\lambda_4}(s,t,u)\times(2\pi)^4\delta^{(4)}(0)\equiv
{}_{id}\langle (\myP',\theta,0); \lambda_3, \lambda_4 |T|(\myP,0,0); \lambda_1, \lambda_2 \rangle_{id},
\end{align}
where the identical two-particle states were defined in \eqref{eq:2PS_com_id_app_alt}. We now deduce constraints the amplitude \eqref{sec:idnetical_particles_app} must obey in order to incorporate the symmetry property \eqref{eq:2PS_com_identical_app}. In practice we will use \eqref{eq:2PS_com_identical_app} rewritten in the form \eqref{eq:identical_particles_com_2}.

Let us start with the outgoing particles 3 and 4. Using \eqref{eq:identical_particles_com_2} one can rewrite the right-hand side of \eqref{sec:idnetical_particles_app} as
\begin{equation}
\label{eq:outgoing_tu}
\begin{aligned}
{}_{id}\langle (\myP',\theta,0); &\lambda_3, \lambda_4 |T|(\myP,0,0); \lambda_1, \lambda_2 \rangle_{id}\\
&=(-1)^{\lambda_3-\lambda_4}
{}_{id}\langle (\myP',\pi-\theta,0); \lambda_4, \lambda_3 |e^{i\pi J_3}T|(\myP,0,0); \lambda_1, \lambda_2 \rangle_{id}\\
&=(-1)^{\lambda_3-\lambda_4}
{}_{id}\langle (\myP',\pi-\theta,0); \lambda_4, \lambda_3 |e^{i\pi J_3}T
e^{-i\pi J_3}e^{i\pi J_3}|(\myP,0,0); \lambda_1, \lambda_2 \rangle_{id}\\
&=(-1)^{\lambda_3-\lambda_4+\lambda_1-\lambda_2}
{}_{id}\langle (\myP',\pi-\theta,0); \lambda_4, \lambda_3 |T|(\myP,0,0); \lambda_1, \lambda_2 \rangle_{id}.
\end{aligned}
\end{equation}
In the third line we simply injected the identity operator made out of $z$-rotations. In the fourth line we used the fact that the states with the momentum along the z-axis are eigenstates of $J_3$ generators. Looking at the definition of the Mandelstam variables \eqref{eq:t_chan} and \eqref{eq:u_chan} we see that the exchange $\theta \leftrightarrow\pi-\theta$ simply corresponds to $t\leftrightarrow u$. Combining \eqref{eq:outgoing_tu} with \eqref{sec:idnetical_particles_app} we obtain
\begin{align}
\label{eq:identical_particles_amplitudes1_app}
T_{\lambda_1,\lambda_2}^{\lambda_3, \lambda_4}(s,t,u) &=(-1)^{\lambda_1 - \lambda_2 - \lambda_3 +\lambda_4} T_{\lambda_1,\lambda_2}^{\lambda_4, \lambda_3}(s,u,t),
\end{align}
which is nothing but the (34) $t-u$ crossing equation.

Let now address the incoming particles 1 and 2. First, we need the following relation which  holds true for generic two-particle states (and therefore also for identical two-particle states)
\begin{equation}
\label{eq:rel_1_id}
|(\myP,\theta,0) ; \lambda_3 , \lambda_4 \rangle =(-1)^{\lambda_3-\lambda_4}e^{-i\pi J_3}e^{-i\pi J_2} |(\myP,\pi-\theta,0) ; \lambda_3 , \lambda_4 \rangle.
\end{equation}
It simply follows from \eqref{eq:relation_rotation_2PS_spherical} and \eqref{eq:pi_rotate_around_x_rotation}, see also \eqref{eq:rotation_Euler} and \eqref{eq:unitary_transformation_helicity}. Due to \eqref{eq:identical_particles_com_2} and \eqref{eq:relation_rotation_2PS_spherical} we also have the following relation for identical particles
\begin{align}
\label{eq:rel_2_id}
|(\myP,0,0) ; \lambda_1 , \lambda_2 \rangle_{id} &=(-1)^{\lambda_1-\lambda_2}e^{-i\pi J_3}e^{-i\pi J_2} |(\myP,0,0) ; \lambda_2 , \lambda_1 \rangle_{id}.
\end{align}
Using both \eqref{eq:rel_1_id} and \eqref{eq:rel_1_id} in the right-hand side of \eqref{sec:idnetical_particles_app}, and the fact that the scattering operator is invariant under rotations we conclude that
\begin{align}
\label{eq:identical_particles_amplitudes2_app}
T_{\lambda_1,\lambda_2}^{\lambda_3, \lambda_4}(s,t,u) &=(-1)^{\lambda_1 - \lambda_2 + \lambda_3 -\lambda_4} T_{\lambda_2,\lambda_1}^{\lambda_3, \lambda_4}(s,u,t),
\end{align}
which is nothing but the (12) $t-u$ crossing equation.

Combining both crossing equations \eqref{eq:identical_particles_amplitudes1_app} and \eqref{eq:identical_particles_amplitudes2_app} we obtain the following purely kinematic constraint
\begin{align}
\label{eq:identical_particles_amplitudes3_app}
T_{\lambda_1,\lambda_2}^{\lambda_3, \lambda_4}(s,t,u) &= T_{\lambda_2,\lambda_1}^{\lambda_4, \lambda_3}(s,t,u).
\end{align}
Here we used the fact that $\lambda_1 - \lambda_2 + \lambda_3 -\lambda_4$ is always an integer.

\section{Center of mass frame}
\label{app:center_mass_frame}

In this section we define the center of mass frame describing two-, three- and four-point amplitudes.

\subsection{Two-point amplitudes}
\label{sec:two_point_amplitudes_COM}

We start with the two-point amplitude defined in \eqref{eq:partial_amplitude_interacting}. It reads
\begin{equation}
\label{eq:2pa}
{}_{free}\< c_1, \vec p_1;  \spin_1, \hel_1 ; \gamma_1 |T|c_2, \vec p_2;  \spin_2, \hel_2; \gamma_2 \>_{free}.
\end{equation}
It is non-zero only for $\vec p_1=\vec p_2$ due to translation symmetry. By using the three boost generators one can set
\begin{equation}
\label{eq:2pa_configuration}
\vec p_1=\vec p_2 =0.
\end{equation}
The remaining symmetry is $SO(3)$. The matrix element \eqref{eq:2pa} at the point \eqref{eq:2pa_configuration} must be invariant under this $SO(3)$ symmetry. Applying an $SO(3)$ rotation to \eqref{eq:2pa} we get
\begin{multline}
\label{eq:constraint_2pa}
{}_{free}\< c_1, \vec 0;  \spin_1, \hel_1 ; \gamma_1 |T|c_2, \vec 0;  \spin_2, \hel_2; \gamma_2 \>_{free}=\\
\sum_{\lambda_1',\lambda_2'}
{}_{free}\< c_1, \vec 0;  \spin_1, \hel_1' ; \gamma_1 |T|c_2, \vec 0;  \spin_2, \hel_2'; \gamma_2 \>_{free}\times
\mathscr{D}^{*(\spin_1)}_{\lambda'_1\lambda_1} (\vec{\omega_1})
\mathscr{D}^{(\spin_2)}_{\lambda'_2\lambda_2}  (\vec{\omega_2}).
\end{multline}
In order to make the amplitude invariant one has to demand that
\begin{equation}
\label{eq:2pa_condition}
{}_{free}\< c_1, \vec 0;  \spin_1, \hel_1 ; \gamma_1 |T|c_2, \vec 0;  \spin_2, \hel_2; \gamma_2 \>_{free} \propto \delta_{\ell_1 \ell_2}\delta_{\lambda_1 \lambda_2}.
\end{equation}
Then due to the orthogonality \eqref{eq:D_matrix_property_1} the Wigner D-matrices disappear completely and one recovers the invariance of the two-point amplitude. The condition \eqref{eq:2pa_condition} means that there is only one independent two-point amplitude
\begin{equation}
N_2=1.
\end{equation}

\subsection{Three-point amplitudes}
\label{sec:three_point_amplitudes_COM}

Consider the following matrix element
\begin{equation}
\label{eq:3pt}
{}_{out}\langle\kappa_1,\kappa_2|c, \vec p;  \spin, \hel; \gamma \>_{in}=
{}_{free}\langle\kappa_1,\kappa_2|S|c, \vec p;  \spin, \hel; \gamma \>_{free}.
\end{equation}
It appears in the discussion of poles, see for example section 2.5.1 in \cite{Karateev:2019ymz}. It also appears in computations of partial amplitudes, see the end of appendix \ref{sec:general_spin_tensor_structures}.  It is interesting to ask what happens if the ket state in \eqref{eq:3pt} is actually a one particle state, namely when we deal with the following object
\begin{equation}
\label{eq:3pt_decay}
{}_{out}\langle\kappa_1,\kappa_2|\kappa_3 \>_{in}=
{}_{free}\langle\kappa_1,\kappa_2|S|\kappa_3 \>_{free}.
\end{equation}
It describes the decay process of the asymptotic state 3 into two asymptotic states 1 and 2. Strictly speaking such a matrix element must be zero, since asymptotic states by definition cannot decay.\footnote{If a particle is unstable an observer after waiting long enough will see the decay product which is described by true asymptotic states.} In some circumstances when a particle is unstable but lives long enough it might be useful however to treat it as an approximate asymptotic state and prescribe physical meaning to \eqref{eq:3pt_decay}.

Let us discuss the spin structure of \eqref{eq:3pt_decay} in the COM frame. 
By using the three boosts we can set $\vec p_3 =0$. By using two rotations we can move to the following final frame
\begin{equation}
\label{eq:three_point_cmf}
\begin{aligned}
p_1^\mu &=\{E_1,\; 0,\; 0,\; +\myP\},\\
p_2^\mu &=\{E_2,\; 0,\; 0,\; -\myP\},\\
p_3^\mu &=\{E_3,\; 0,\; 0,\;0\},
\end{aligned}
\end{equation}
where the energies read as 
\begin{equation}
E_1 = \frac{\big|m_3^2+m_1^2-m_2^2\big|}{2\,m_3},\quad
E_2 = \frac{\big|m_3^2-m_1^2+m_2^2\big|}{2\,m_3},\quad
E_3 = m_3
\end{equation}
and we have
\begin{equation}
\myP = \frac{1}{2m_3}\;
\sqrt{(m_3+m_1+m_2)(m_3-m_1-m_2)(m_3-m_1+m_2)(m_3+m_1-m_2)}.
\end{equation}
The value $\myP$ should be real, this enforces the condition
\begin{equation}
m_3\geq m_1+m_2.
\end{equation}

Having chosen the frame \eqref{eq:three_point_cmf} one is left with a single generator of rotations around the z-axis. This means that one has a remaining $SO(2)$ symmetry and the matrix element \eqref{eq:3pt_decay} in the frame \eqref{eq:three_point_cmf} must be invariant under it. To find the consequence of this symmetry one can
inject the identity operator composed out of z-rotations into \eqref{eq:3pt_decay} twice and requiring that the matrix element remains invariant. It leads to the following constraint
\begin{equation}
\label{eq:three_point_constraint}
\lambda_3 = \lambda_1 - \lambda_2.
\end{equation}

Given the condition \eqref{eq:three_point_constraint} one can easily count the number of different three-point amplitudes $N_3$. It is given by
\begin{equation}
N_3 = \sum_{\lambda_1=-j_1}^{+j_1} \sum_{\lambda_2=-j_2}^{+j_2} \sum_{\lambda_3=-j_3}^{+j_3} \delta_{\lambda_1+\lambda_2,\lambda_3}.
\end{equation}
This expression was rewritten in a compact form in \cite{Kravchuk:2016qvl}, it reads as
\begin{equation}\label{eq:counting_3}
N_3=(2j_1+1)(2j_2+1)-r(r+1),\quad
r\equiv\max(j_1+j_2-j_3,\,0),\quad
j_1\leq j_2\leq j_3.
\end{equation}

\subsection{Four-point amplitudes}
\label{sec:four_point_amplitudes_COM}
The four-point amplitude was defined in \eqref{eq:2to2_scattering}. Using all the generators of the Lorentz group, we can bring this amplitude to the following frame \begin{equation}
\label{eq:four_point_cmf}
\begin{aligned}
p_1^\mu &=\{E_1,0,0,+\myP \},\\
p_2^\mu &=\{E_2,0,0,-\myP\},\\
p_3^\mu &=\{E_3,+\myP'\sin\theta,0,+\myP'\cos\theta\},\\
p_4^\mu &=\{E_4,-\myP'\sin\theta,0,-\myP'\cos\theta\},
\end{aligned}
\end{equation}
where $\myP\geq 0$, $\myP'\geq 0$  and $\theta \in [0,\pi]$ and the energies are given by
\begin{equation}
\label{eq:energies}
E_1 = \sqrt{m_1^2+\myP^2},\quad
E_2 = \sqrt{m_2^2+\myP^2},\quad
E_3 = \sqrt{m_3^2+\myP^{\prime 2}},\quad
E_4 = \sqrt{m_4^2+\myP^{\prime 2}}.
\end{equation}
The Mandelstam variables in the COM frame \eqref{eq:four_point_cmf} then read as
\begin{align}
\label{eq:s_chan}
s &= (E_1+E_2)^2 = (E_3+E_4)^2,\\
\label{eq:t_chan}
t &= m_1^2 + m_3^2 -2E_1E_3 + 2 \myP\myP' \cos\theta,\\
\label{eq:u_chan}
u &= m_1^2 + m_4^2 -2E_1E_4 - 2 \myP\myP' \cos\theta.
\end{align}
In the case of four-point COM amplitudes there is no additional symmetry 
left. Thus, the total number of amplitudes is obtained by simply counting all possible helicity configurations
\begin{equation}
N_4 = (2\ell_1+1)(2\ell_2+1)(2\ell_3+1)(2\ell_4+1).
\end{equation}

The relations \eqref{eq:s_chan} - \eqref{eq:u_chan} express the Mandelstam variables $(s,t,u)$ in terms of $(\myP,\myP',\theta)$. We can also invert these relations as follows. From \eqref{eq:s_chan} we get
\begin{equation}
\label{eq:p_pp}
\myP  = \frac{\mathcal{L}_{12}(s)}{2\sqrt{s}},\qquad
\myP' = \frac{\mathcal{L}_{34}(s)}{2\sqrt{s}},
\end{equation}
where we have defined
\begin{equation}
\label{eq:aux_1}
\mathcal{L}_{ij}(s)\equiv\sqrt{\big(s-(m_i-m_j)^2\big)\big(s-(m_i+m_j)^2\big)}.
\end{equation}
Plugging \eqref{eq:p_pp} into \eqref{eq:energies} we get the energies
\begin{align}
\label{eq:energies_CMF}
E_1 = \frac{s+m_1^2-m_2^2}{2\sqrt{s}},\;\;
E_2 = \frac{s-m_1^2+m_2^2}{2\sqrt{s}},\;\;
E_3 = \frac{s+m_3^2-m_4^2}{2\sqrt{s}},\;\;
E_4 = \frac{s-m_3^2+m_4^2}{2\sqrt{s}}.
\end{align}
Subtracting \eqref{eq:u_chan} from \eqref{eq:t_chan} and using \eqref{eq:p_pp} and \eqref{eq:energies_CMF} we get
\begin{align}
\label{eq:cos}
\cos\theta &= \frac{s(t-u)+(m_1^2-m_2^2)(m_3^2-m_4^2)}{\mathcal{L}_{12}(s)\mathcal{L}_{34}(s)}.
\end{align}
In the range $\theta \in [0,\pi]$ we can also write unambiguously 
\begin{equation}
\label{eq:sin}
\sin\theta = \sqrt{1-\cos^2\theta} = 
\frac{2\sqrt{s}\,\sqrt{\Phi}}{\mathcal{L}_{12}(s)\mathcal{L}_{34}(s)},
\end{equation}
where we have defined
\begin{align}
\nn
\Phi &\equiv stu
-s (m_2^2-m_4^2)(m_1^2-m_3^2)
-t\, (m_1^2-m_2^2)(m_3^2-m_4^2)
+ \Delta_t\,(m_1^2m_4^2-m_2^2m_3^2)\\
&=stu
-s (m_2^2-m_3^2)(m_1^2-m_4^2)
+u (m_1^2-m_2^2)(m_3^2-m_4^2)
+ \Delta_u(m_1^2m_3^2-m_2^2m_4^2)
\label{eq:definition_Phi}
\end{align}
together with
\begin{equation}
\label{eq:definition_delta}
\Delta_t    \equiv -m_1^2+m_2^2+m_3^2-m_4^2,\qquad
\Delta_u    \equiv -m_1^2+m_2^2-m_3^2+m_4^2.
\end{equation}

Let us study the physical ranges of the Mandelstam variables. From \eqref{eq:energies} and  \eqref{eq:s_chan} the following inequalities follow
\begin{equation}
\label{eq:inequalities}
s\geq
\max\Big((m_1+m_2)^2,\,(m_3+m_4)^2\Big).
\end{equation}
Notice, that due to \eqref{eq:inequalities} all the energies $E_i$ in \eqref{eq:energies_CMF} are positive and $\mathcal{L}_{12}(s)$, $\mathcal{L}_{34}(s)$ are real as they should be. 
Since the value of $\cos\theta$ is bounded to be in the $[-1,+1]$ interval, we can derive the following constraints on the values of the variable $t$ from \eqref{eq:cos}
\begin{equation}
\label{eq:range_t}
\begin{aligned}
t &\in[v+
\frac{\mathcal{L}_{12}(s)\mathcal{L}_{34}(s)}{2s},v-
\frac{\mathcal{L}_{12}(s)\mathcal{L}_{34}(s)}{2s}],\\
v&\equiv \frac{1}{2}\,(m_1^2+m_2^2+m_3^2+m_4^2-s)-
\frac{(m_1^2-m_2^2)(m_3^2-m_4^2)}{2s}.
\end{aligned}
\end{equation}
From \eqref{eq:inequalities} and \eqref{eq:range_t} in the equal mass case we recover the familiar result
\begin{equation}
\label{eq:range_t_eqmass}
s\in[4m^2,\,\infty),\quad
t\in[-(s-4m^2),\, 0].
\end{equation}

There is a very special situation when
\begin{equation}
\label{eq:special_case}
\mathcal{L}_{12}(s) =0\quad{\text{\bf or}}\quad\mathcal{L}_{34}(s) =0.
\end{equation}
From \eqref{eq:p_pp} we see that this corresponds to when either incoming or outgoing particles are at rest. In such a situation we cannot define the angle between incoming and outgoing particles which can be see from \eqref{eq:cos} which is singular at \eqref{eq:special_case}. The values of $s$ which lead to \eqref{eq:special_case} are 
\begin{equation}
s = (m_1\pm m_2)^2,\quad
s = (m_3\pm m_4)^2\,.
\end{equation}

\section{Crossing equations}
\label{app:crossing_equations}
The goal of this appendix is to prove the crossing relations in a general frame  \eqref{eq:s_t_crossing_general_14} - \eqref{eq:s_u_crossing_general_24} and then derive the crossing equations in the COM frame \eqref{eq:st_crossing} and \eqref{eq:su_crossing} together with the Wigner angles $\alpha_i$ and $\beta_i$.

\subsection{Analytic continuation of four-momenta}
\label{app:analytic_continuation}
Consider the interacting part of the scattering amplitude
\begin{equation}
\label{eq:int_amp}
T_{\lambda_1,\lambda_2}^{\lambda_3,\lambda_4}(p_1^{\mu_1},p_2^{\mu_2},p_3^{\mu_3},p_4^{\mu_4}).
\end{equation}
This function is defined only for non-negative energies $p_i^0\geq 0$, provided that the following constraint is satisfied
\begin{equation}
\label{eq:constraint}
p_i^2=-m^2_i,
\end{equation}
where $m_i\geq0$ are the masses of particles.
However crossing requires us to evaluate the amplitude \eqref{eq:int_amp} at $-p_i^\mu$ points which means that one must extend the definition of \eqref{eq:int_amp} also to negative values of energies $p_i^0$. 
This can be done by analytically continuing the amplitude \eqref{eq:int_amp} in each component of four 4-vectors $p_i^\mu$ to the full complex plane while keeping the constraints \eqref{eq:constraint} satisfied.\footnote{In other words by using the analytic continuation, the amplitude \eqref{eq:int_amp} can be defined as a function of $4 \times 4 = 16$ complex variables which satisfy the four constraints \eqref{eq:constraint}.} To perform this analytic continuation, 
one needs to choose the path in (complexified) momentum space.

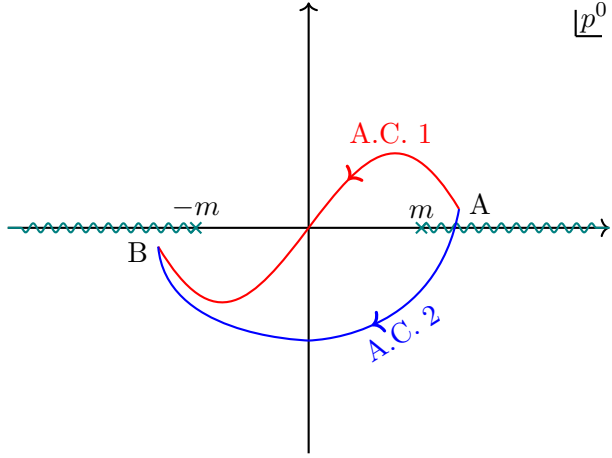
\begin{figure}[t]
	\begin{center}
		\begin{tikzpicture}
    \coordinate (origin) at (0,0);

      \draw[thick,black] (3.8,2.8) node[] {$p^0$};
      \draw[thick,black,-](3.55,2.55) -- ++ (0,0.35);
      \draw[thick,black,-](3.55,2.55) -- ++ (0.35,0);
    \draw[thick,black,->] (origin) -- ++(4,0) node (z) [black,right]{} ;
    \draw[thick,black,->] (origin) -- ++(0,3) node (x)  [black, above] {};
\draw[thick,black,-] (origin) -- ++(-4,0) node (zminus) {};
    \draw[thick,black,-] (origin) -- ++(0,-3) node (xminus) {}  ;
     
    \draw[thick] (1.5,0) node[branch point,draw=teal,thick] {};
    \draw (1.5,0) node[above] {$m$};
    \draw [-,snake=snake,thick, teal, segment amplitude=.6mm,segment length=2mm,line after snake=1mm] (1.5,0) to (4,0);
     \draw[thick] (-1.5,0) node[branch point,draw=teal,thick] {};
     \draw (-1.5,0) node[above] {$-m$};
     \draw [-,snake=snake,thick, teal, segment amplitude=.6mm,segment length=2mm,line after snake=1mm] (-1.5,0) to (-4,0);
   
   \draw[thick,red,-, decoration={ markings,mark = at position 0.7 with {\arrow[line width=1.2pt]{>}}},
  postaction={decorate}] (2,0.25) .. controls (1.25,1.5) and (0.75,1) .. (0,0) node[pos = 0.45,above, red] {A.C. 1};
   \draw[thick,red,-] (-2,-0.25) .. controls (-1.25,-1.5) and (-0.75,-1) .. (0,0);
  
\draw[thick,blue,-, decoration={ markings,mark = at position 0.7 with {\arrow[line width=1.2pt]{>}}},
  postaction={decorate}] (2,0.25) .. controls (1.7,-1.4) and (0.2,-1.49) .. (0,-1.5) node[pos = 0.45,below, blue,sloped] {A.C. 2};
   \draw[thick,blue,-] (0,-1.5) .. controls (-0.2,-1.49) and (-1.9,-1.4).. (-2,-0.25);   
   
 \draw (2,0.05) node[above right, black] {A};
   \draw (-2,-0.05) node[below left, black] {B};   
   
  \end{tikzpicture}
  
		\caption{The complex $p^0$ plane. We depict the analytic structure of the function $\myP(p^0)$ defined in \eqref{eq:fatb}. It has two branch cuts. The original domain of the function $\myP$ is given by positive values $p^0$ slightly above the right cut. We define $\myP$ for negative values of $p^0$ by an analytic continuation. The two different options are depicted in red and blue. Two paths together encircle the $+m$ branch point and thus differ by a monodromy around that point.}
		\label{fig:ac}
	\end{center}
\end{figure}

To continue the discussion in more detail let us focus for simplicity on a function of a single 4-momentum
\begin{equation}
\label{eq:f}
f(p^\mu)
\end{equation} 
defined for $p^0\geq 0$ and $p^2=-m^2$ with the non-negative mass $m$. The case of four 4-momenta \eqref{eq:int_amp} follows straightforwardly by the repeated use of the steps here for each $i=1,2,3,4$. Using the spherical coordinates $\myP$, $\theta$ and $\phi$ one can write
\begin{equation}
\label{eq:spherical_coordinates}
p^0 = \sqrt{\myP^2+m^2},\qquad
\begin{cases}
p^1= \myP \sin\theta\cos\phi,\\
p^2= \myP \sin\theta\sin\phi,\\
p^3= \myP \cos\theta,
\end{cases}
\end{equation}
where $\myP\geq 0$ is the length of the 3-vector $\vec p$. The relation $p^2=-m^2$ can be rewritten as
\begin{equation}
\label{eq:fatb}
\myP =i \sqrt{(m-p^0)(p^0+m)}.
\end{equation}
If we promote $p^0$ to the full complex plane, the function \eqref{eq:fatb} will have an analytic structure as depicted on figure \ref{fig:ac} with two branch points $\pm m$ and two branch cuts ending in these points.

In order to study crossing we need to defined \eqref{eq:f} at the following point
\begin{equation}
\label{eq:f_an}
f(-p^\mu).
\end{equation} 
This can be achieved by performing the following analytic continuation
\begin{equation}
\label{eq:continuation}
p^0 +i\epsilon \rightarrow \text{complex value} \rightarrow -p^0-i\epsilon,\quad
\vec p \rightarrow \text{complex value} \rightarrow -\vec p.
\end{equation}
In the $p^0$ complex plane two different options for such an analytic continuation are  depicted in figure \ref{fig:ac} \footnote{To be precise, the two different paths for analytic continuation take us to two different points on the Riemann surface.}. Note that the original domain of physical energy $p^0\geq m$ by convention lies slightly above the right-hand branch cut.

Now let us investigate the behavior of the function \eqref{eq:fatb} depending on the chosen path of the analytic continuation. If the path does not cross any branch cuts we simply get
\begin{equation}
\label{eq:case_1}
\myP \rightarrow \text{complex value} \rightarrow +\myP.
\end{equation}
In case we cross once one of the branch cuts we get an extra phase due to the monodromy around the associated branch point which leads to
\begin{equation}
\nn
p^0\pm m \rightarrow (p^0\pm m)e^{2\pi i}.
\end{equation}
As a result for this path we get the following
\begin{equation}
\label{eq:case_2}
\myP \rightarrow \text{complex value} \rightarrow -\myP.
\end{equation}
To summarize, the analytic continuation \eqref{eq:continuation} can be implemented in two different ways depending on the path chosen on figure \ref{fig:ac}. The two distinct options \eqref{eq:case_1} and \eqref{eq:case_2} due to \eqref{eq:spherical_coordinates} read as
\begin{align}
\label{eq:ac_1}
&p^0 \rightarrow -p^0,\quad
\myP \rightarrow +\myP,\quad
\theta \rightarrow \pi -\theta,\quad
\phi \rightarrow \pi + \phi,\\
\label{eq:ac_2}
&p^0 \rightarrow -p^0,\quad
\myP \rightarrow -\myP,\quad
\theta \rightarrow \theta,\quad
\quad\;\;\;\phi \rightarrow \phi.
\end{align}
The first option \eqref{eq:ac_1} is commonly used in the literature. The second option \eqref{eq:ac_2} is more suitable for massless particles since two branch cuts in figure \ref{fig:ac} unite and one cannot choose a path for the analytic continuation without crossing any of the two branch cuts. For treating massless particles one can also use the first option \eqref{eq:ac_1} and take the limit $m\rightarrow 0$ at the very end.

The helicity states are defined via \eqref{poincare_irreps_definition} and at a practical level depend only on the rapidity and two angles $(\eta,\theta,\phi)$. Using the definition of rapidity \eqref{eq:rapidity} we can write \eqref{eq:ac_1} and \eqref{eq:ac_2} as
\begin{align}
\label{eq:ac_1f}
&\eta\rightarrow i \pi -\eta,\quad
\theta \rightarrow \pi -\theta,\quad
\phi \rightarrow \pi + \phi,\\
\label{eq:ac_2f}
&\eta\rightarrow i \pi +\eta,\quad
\theta \rightarrow \theta,\quad
\quad\;\;\;\phi \rightarrow \phi.
\end{align}

\subsection{Crossing equations in a general frame}
\label{app:analytically_continued_transformations}
Given an amplitude of the $12\rightarrow 34$ process in a generic frame we would like to relate it in this appendix to the four ``crossed'' amplitudes associated to the following processes
\begin{equation}
\bar 4 2\rightarrow 3 \bar 1,\qquad
1 \bar 3\rightarrow \bar 2 4,\qquad
\bar 3 2 \rightarrow \bar 1 4,\qquad
1 \bar 4\rightarrow 3 \bar 2.
\end{equation}
We refer to these relations as the crossing equations. To be concrete we will focus on writing the crossing equations for $12\rightarrow 34$ and $\bar 4 2\rightarrow 3 \bar 1$ amplitudes, the rest will follow by analogy.  

Given  an amplitude
\begin{equation}
\label{eq:amplitude}
T_{12\rightarrow 34}{\;}_{\lambda_1, \lambda_2}^{\lambda_3,\lambda_4}(p_1,p_2,p_3,p_4),
\end{equation}
one can obtain the amplitude $\bar 4 2\rightarrow 3 \bar 1$ by crossing particles 1 and 4. The ingoing particle 1 becomes then the outgoing one and the outgoing particle 4 becomes the ingoing one. As a result we need to make the following replacements: $p_1^\mu\rightarrow -p_1^\mu$ and $p_4^\mu\rightarrow -p_4^\mu$.  Moreover if a particle $1$ has a charge (or more generally transforms in some representation of a global group) the particle $\bar i$ has the opposite charge (transforms in the conjugate representation).
As a result \eqref{eq:amplitude} under crossing 1-4 becomes
\begin{equation}
\label{eq:amplitude_crossed}
T_{\bar 42\rightarrow 3\bar 1}{\;}_{\lambda_4, \lambda_2}^{\lambda_3,\lambda_1}(-p_4,p_2,p_3,-p_1).
\end{equation}
In both \eqref{eq:amplitude} and \eqref{eq:amplitude_crossed} all 4-momenta have positive energies $p_i^0 > 0$ and are on-shell \eqref{eq:on_shell}.

Without using the LSZ reduction formula one can only postulate crossing. It states that one can define a single ``mother amplitude'' with complex values of 4-momenta such that all the amplitudes in \eqref{eq:amplitude} and \eqref{eq:amplitude_crossed} are its boundary values. This is known as the Mandelstam hypothesis. One cannot however simply equate \eqref{eq:amplitude} and \eqref{eq:amplitude_crossed}. At the very least these amplitudes should have the same Lorentz transformation properties. We will use this requirement to fix the crossing equations up to an overall phase. This is the original way by which general spin crossing equations were derived in \cite{Trueman:1964zza}.

Consider the transformation property of the amplitude describing the $12\rightarrow 34$ process under generic Lorentz transformations $\Lambda$. According to \eqref{eq:T_matrix_amplitude_transformation_general} it reads as
\begin{align}
\label{eq:transformation_1}
&T_{12\rightarrow 34}{\;}_{\lambda_1, \lambda_2}^{\lambda_3,\lambda_4}(p_1,p_2,p_3,p_4)=\\
\nn
&\sum_{\lambda_i'}
		\mathscr{D}^{(j_1)}_{\lambda'_1\lambda_1} (p_1,\Lambda)
		\mathscr{D}^{(j_2)}_{\lambda'_2\lambda_2} (p_2,\Lambda)
		\mathscr{D}^{*(j_3)}_{\lambda'_3\lambda_3}(p_3,\Lambda)
		\mathscr{D}^{*(j_4)}_{\lambda'_4\lambda_4}(p_4,\Lambda)
T_{12\rightarrow 34}{\;}_{\lambda_1', \lambda_2'}^{\lambda_3',\lambda_4'}(p_1',p_2',p_3',p_4').
\end{align}
Similarly the crossed amplitude describing the $\bar 4 2\rightarrow 3 \bar 1$ process transforms as
\begin{align}
\label{eq:transformation_2}
&T_{\bar 42\rightarrow 3\bar 1}{\;}_{\lambda_4, \lambda_2}^{\lambda_3,\lambda_1}(-p_4,p_2,p_3,-p_1)=\\
\nn
&\sum_{\lambda_i'}
\mathscr{D}^{*(j_1)}_{\lambda'_1\lambda_1}(-p_1,\Lambda)
\mathscr{D}^{(j_2)}_{\lambda'_2\lambda_2} (p_2,\Lambda)
\mathscr{D}^{*(j_3)}_{\lambda'_3\lambda_3}(p_3,\Lambda)
\mathscr{D}^{(j_4)}_{\lambda'_4\lambda_4} (-p_4,\Lambda)
T_{\bar 42\rightarrow 3\bar 1}{\;}_{\lambda_4', \lambda_2'}^{\lambda_3',\lambda_1'}(-p_4',p_2',p_3',-p_1').
\end{align}
In the equations \eqref{eq:transformation_1} and \eqref{eq:transformation_2} we have schematically denoted the arguments of the Wigner D-matrices by $(p_i,\Lambda)$. In practice they depend only on three Wigner angles which correspond to the $p_i\rightarrow \Lambda p_i$ Lorentz transformation. We denote these angles as
\begin{align}
\label{eq:wigner}
\mathscr{D}^{(j_i)}_{\lambda'_i\lambda_i} (\alpha_i,\beta_i,\gamma_i) &\equiv \mathscr{D}^{(j_i)}_{\lambda'_i\lambda_i} (+p,\Lambda),\\
\mathscr{D}^{(j_i)}_{\lambda'_i\lambda_i} (\bar\alpha_i,\bar\beta_i,\bar\gamma_i) &\equiv \mathscr{D}^{(j)}_{\lambda'\lambda_i} (-p_i,\Lambda),
\label{eq:wigner_bar}
\end{align}
where $i=1,2,3,4$. The Wigner angles $(\alpha_i,\beta_i,\gamma_i)$ generically differ from $(\bar\alpha_i,\bar\beta_i,\bar\gamma_i)$. They are however closely related. The main technical task of this appendix is to understand precisely how.

We would like to equate the amplitudes which transform in the same way under  Lorentz transformations. Thus, in practice we need to compare the transformation properties of \eqref{eq:transformation_1} and \eqref{eq:transformation_2}. This in turn boils down to comparing the Wigner D-matrices
\begin{equation}
\nn
\mathscr{D}^{(j_1)}_{\lambda'_1\lambda_1} (+p_1,\Lambda)
\quad\text{vs.}\quad
\mathscr{D}^{*(j_1)}_{\lambda'_1\lambda_1} (-p_1,\Lambda)
\qquad\text{\bf and}\qquad
\mathscr{D}^{*(j_4)}_{\lambda'_4\lambda_4} (+p_4,\Lambda)
\quad\text{vs.}\quad
\mathscr{D}^{(j_4)}_{\lambda'_4\lambda_4} (-p_4,\Lambda).
\end{equation}
In order to do this recall that according to \eqref{eq:generic_poincare_transformation} and \eqref{eq:Wigners_rotation} the Wigner rotation matrices are defined as
\begin{align}
\label{eq:original}
\mathscr{D}^{(j)}_{\lambda'\lambda} (\alpha,\beta,\gamma)=\mathscr{D}^{(j)}_{\lambda'\lambda} (+p,\Lambda):\quad
R(\alpha,\beta,\gamma) &=
U_{\text{h}}(+p')^{-1}
\Lambda
U_{\text{h}}(+p),\\
\label{eq:after_continuation}
\mathscr{D}^{(j)}_{\lambda'\lambda}(\bar\alpha,\bar\beta,\bar\gamma)=\mathscr{D}^{(j)}_{\lambda'\lambda} (-p,\Lambda):\quad
R(\bar\alpha,\bar\beta,\bar\gamma) &=
U_{\text{h}}(-p')^{-1}
\Lambda
U_{\text{h}}(-p),
\end{align}
where the helicity transformation $U_{\text{h}}(+p)$ is defined via  \eqref{eq:unitary_transformation_helicity}. We repeat this definition here for convenience
\begin{align}
\label{eq:hel_mat}
U_{\text{h}}(+p) &= R(\phi,\theta,-\phi) \times B_3(\eta).
\end{align}
The quantity $U_{\text{h}}(-p)$ is defined from $U_{\text{h}}(+p)$ by  analytic continuation. As discussed in the previous section there are two distinct ways to do it, using \eqref{eq:ac_1f} or \eqref{eq:ac_2f}. Below we address the two options separately.

\subsubsection{Analytic continuation 1}
Using \eqref{eq:hel_mat} and \eqref{eq:ac_1f} we get
\begin{align}
\label{eq:continuation_1}
U_{\text{h}}(-p) &= R(\pi+\phi,\,\pi-\theta,\,-\pi-\phi) \times B_3(i\pi-\eta).
\end{align}
We can then focus on the vector representation defined in appendix \ref{app:representations} and compute the Wigner angles in \eqref{eq:original} and \eqref{eq:after_continuation} brute force using computer algebra.\footnote{Given a generic $3\times3$ rotation matrix it is straightforward to determine $\tan\alpha$, $\cos\beta$ and $\tan\gamma$ without any ambiguity. In order to determine the rest of trigonometric functions it is necessary to choose the region of $\beta$ angle. One can either have $\beta\in[0,\pi]$ or $\beta\in[0, -\pi]$.}${}^,$\footnote{By convention $\beta\in[0,\pi]$, however we are free to choose either $\bar \beta\in[0,\pi]$ or $\bar \beta\in[0,-\pi]$. We make the latter choice, however both options lead to the same conclusion in the very end.}${}^,$\footnote{We first perform this computation for a generic infinitesimal Lorentz transformation. We then focus on some simple finite transformations like boosts along the $x$, $y$ and $z$ axes separately.}
Comparing the results we conclude that
\begin{equation}
\bar\alpha=-\alpha+2\phi',\quad
\bar\beta =-\beta,\quad
\bar\gamma= -\gamma-2\phi.
\end{equation}
From the properties of the Wigner $D$-matrix it then follows that
\begin{equation}
\begin{aligned}
\mathscr{D}^{(j)}_{\lambda',\lambda} (\bar\alpha,\bar\beta,\bar\gamma) &=
e^{-2i\phi'\lambda'}\mathscr{D}^{(j)}_{\lambda',\lambda} (-\alpha,-\beta,-\gamma) e^{+2i\phi\lambda}\\
&=e^{i(\pi-2\phi')\lambda'}e^{-i(\pi-2\phi)\lambda}
\mathscr{D}^{*(j)}_{\lambda'\lambda} (\alpha,\beta,\gamma),
\end{aligned}
\end{equation}
see appendix A.2 of \cite{Martin:102663} for the summary of properties of D matrices. In other words
\begin{equation}
\label{eq:relation_D-matrices}
\mathscr{D}^{(j)}_{\lambda'\lambda} (-p,\Lambda)=
e^{i(\pi-2\phi')\lambda'}e^{-i(\pi-2\phi)\lambda}
\mathscr{D}^{*(j)}_{\lambda'\lambda} (+p,\Lambda).
\end{equation}

With the help of \eqref{eq:relation_D-matrices} one can rewrite the transformation property \eqref{eq:transformation_2} in the following form\footnote{\label{foot:remark_ac}In writing this formula we assumed that the  relation between $\mathscr{D}^{*(j)}_{\lambda'\lambda} (-p,\Lambda)$ and $\mathscr{D}^{(j)}_{\lambda'\lambda} (+p,\Lambda)$ is obtained from \eqref{eq:relation_D-matrices} by taking complex conjugation. This is not quite correct because we are interested in the analytic continuation along the same path in the $p^0$ complex plane and not along the complex conjugated path.  However, the two continuations are related by a monodromy around  square root branch points that can only give rise to an   overall helicity-independent phase which is irrelevant for the discussion below.  }
\begin{align}
&T_{\bar 42\rightarrow 3\bar 1}{\;}_{\lambda_4, \lambda_2}^{\lambda_3,\lambda_1}(-p_4,p_2,p_3,-p_1)=
\sum_{\lambda_i'}
e^{-i(\pi-2\phi'_1)\lambda'_1}e^{+i(\pi-2\phi_1)\lambda_1}
e^{i(\pi-2\phi'_4)\lambda'_4}e^{-i(\pi-2\phi_4)\lambda_4}
\\\nn
&\mathscr{D}^{(j_1)}_{\lambda'_1\lambda_1}(+p_1,\Lambda)
\mathscr{D}^{(j_2)}_{\lambda'_2\lambda_2} (+p_1,\Lambda)
\mathscr{D}^{*(j_3)}_{\lambda'_3\lambda_3}(+p_1,\Lambda)
\mathscr{D}^{*(j_4)}_{\lambda'_4\lambda_4}(+p_4,\Lambda)
T_{\bar 42\rightarrow 3\bar 1}{\;}_{\lambda_4', \lambda_2'}^{\lambda_3',\lambda_1'}(-p_4',p_2',p_3',-p_1'),,
\end{align} 
which can be rewritten as
\begin{multline}
\label{eq:transformation_final}
\Bigg(
e^{-i(\pi-2\phi_1)\lambda_1}
e^{+i(\pi-2\phi_4)\lambda_4}
T_{\bar 42\rightarrow 3\bar 1}{\;}_{\lambda_4, \lambda_2}^{\lambda_3,\lambda_1}(-p_4,p_2,p_3,-p_1)\Bigg)=
\sum_{\lambda_i'}
\mathscr{D}^{(j_1)}_{\lambda'_1\lambda_1}(+p_1,\Lambda)
\mathscr{D}^{(j_2)}_{\lambda'_2\lambda_2} (+p_1,\Lambda)
\\
\mathscr{D}^{*(j_3)}_{\lambda'_3\lambda_3}(+p_1,\Lambda)
\mathscr{D}^{*(j_4)}_{\lambda'_4\lambda_4}(+p_4,\Lambda)
\Bigg(
e^{-i(\pi-2\phi'_1)\lambda'_1}
e^{i(\pi-2\phi'_4)\lambda'_4}
T_{\bar 42\rightarrow 3\bar 1}{\;}_{\lambda_4', \lambda_2'}^{\lambda_3',\lambda_1'}(-p_4',p_2',p_3',-p_1')\Bigg).
\end{multline}
Comparing \eqref{eq:transformation_1} and \eqref{eq:transformation_final} we see that the following two objects transform in the same way under generic Lorentz transformations
\begin{equation}
\label{eq:two_obects}
T_{12\rightarrow 34}{\;}_{\lambda_1, \lambda_2}^{\lambda_3,\lambda_4}(p_1,p_2,p_3,p_4)
\;\;\text{vs.}\;\;
e^{-i(\pi-2\phi_1)\lambda_1}
e^{i(\pi-2\phi_4)\lambda_4}
T_{\bar 42\rightarrow 3\bar 1}{\;}_{\lambda_4, \lambda_2}^{\lambda_3,\lambda_1}(-p_4,p_2,p_3,-p_1).
\end{equation}
As a result, the only way to write crossing equations which involve objects transforming in the same way under generic Lorentz tranformation is to equate the objects in \eqref{eq:two_obects}. This procedure leaves undetermined an overall helicity independent phase. The other crossing equations follow by simply re-labeling the indices. 

Our final answer reads as
\begin{equation}
\label{eq:crossing_ac_1}
\begin{aligned}
T_{12\rightarrow 34}{\;}_{\lambda_1, \lambda_2}^{\lambda_3,\lambda_4}(p_1,p_2,p_3,p_4)
&=\epsilon_{14}^{(1)}
e^{-i(\pi-2\phi_1)\lambda_1}
e^{i(\pi-2\phi_4)\lambda_4}
T_{\bar 42\rightarrow 3\bar 1}{\;}_{\lambda_4, \lambda_2}^{\lambda_3,\lambda_1}(-p_4,p_2,p_3,-p_1),\\
T_{12\rightarrow 34}{\;}_{\lambda_1, \lambda_2}^{\lambda_3,\lambda_4}(p_1,p_2,p_3,p_4)
&=\epsilon_{23}^{(1)}
e^{-i(\pi-2\phi_2)\lambda_2}
e^{i(\pi-2\phi_3)\lambda_3}
T_{1\bar3\rightarrow \bar 2 4}{\;}_{\lambda_1, \lambda_3}^{\lambda_2,\lambda_4}(p_1,-p_3,-p_2,p_4),\\
T_{12\rightarrow 34}{\;}_{\lambda_1, \lambda_2}^{\lambda_3,\lambda_4}(p_1,p_2,p_3,p_4)
&=\epsilon_{13}^{(1)}
e^{-i(\pi-2\phi_1)\lambda_1}
e^{i(\pi-2\phi_3)\lambda_3}
T_{\bar 32\rightarrow 1\bar 4}{\;}_{\lambda_3, \lambda_2}^{\lambda_1,\lambda_4}(-p_3,p_2,-p_1,p_4),\\
T_{12\rightarrow 34}{\;}_{\lambda_1, \lambda_2}^{\lambda_3,\lambda_4}(p_1,p_2,p_3,p_4)
&=\epsilon_{24}^{(1)}
e^{-i(\pi-2\phi_2)\lambda_2}
e^{i(\pi-2\phi_4)\lambda_4}
T_{1\bar 4\rightarrow 3\bar 2}{\;}_{\lambda_1, \lambda_4}^{\lambda_3,\lambda_2}(p_1,-p_4,p_3,-p_2),
\end{aligned}
\end{equation}
where $\epsilon_{14}^{(1)}$, $\epsilon_{23}^{(1)}$, $\epsilon_{13}^{(1)}$ and $\epsilon_{24}^{(1)}$ are the helicity independent phases unfixed by this procedure.

\subsubsection{Analytic continuation 2}
Using \eqref{eq:hel_mat} and \eqref{eq:ac_2f} we get
\begin{equation}
\label{eq;helicity_boost_ac_2}
U_{\text{h}}(-p) = R(\phi,\,\theta,\,-\phi) \times B_3(i\pi+\eta)=
-U_{\text{h}}(+p)R(\pi,0,0).
\end{equation}
Plugging it into \eqref{eq:after_continuation} one gets
\begin{align}
\nn
\mathscr{D}^{(j)}_{\lambda'\lambda} (\bar\alpha,\bar\beta,\bar\gamma)=
\mathscr{D}^{(j)}_{\lambda'\lambda} (-p,\Lambda):\quad
R(\bar\alpha,\bar\beta,\bar\gamma) &=
R^{-1}(\pi,0,0)\Big(
U_{\text{h}}(+p')^{-1}
\Lambda
U_{\text{h}}(+p)\Big)R(\pi,0,0)\\
&=R^{-1}(\pi,0,0)R(\alpha,\beta,\gamma)R(\pi,0,0).
\end{align}
Using the properties of the small Wigner d-matrices we can write then
\begin{align}
\mathscr{D}^{(j)}_{\lambda'\lambda} (\bar\alpha,\bar\beta,\bar\gamma)
=e^{+i\pi \lambda'}\mathscr{D}^{(j)}_{\lambda'\lambda}(\alpha,\beta,\gamma)e^{-i\pi \lambda}
=\mathscr{D}^{*(j)}_{-\lambda',-\lambda}(\alpha,\beta,\gamma).
\end{align}
In other words we get
\begin{equation}
\mathscr{D}^{(j)}_{\lambda'\lambda} (-p,\Lambda)=
\mathscr{D}^{*(j)}_{-\lambda',-\lambda} (+p,\Lambda).
\end{equation}
Using this we can write the transformation property \eqref{eq:transformation_2} as\footnote{See footnote \ref{foot:remark_ac}.}
\begin{align}
&T_{\bar 42\rightarrow 3\bar 1}{\;}_{\lambda_4, \lambda_2}^{\lambda_3,\lambda_1}(-p_4,p_2,p_3,-p_1)=\\
\nn
&\sum_{\lambda_i'}
\mathscr{D}^{(j_1)}_{-\lambda'_1,-\lambda_1} (+p_1,\Lambda)
\mathscr{D}^{(j_2)}_{\lambda'_2\lambda_2}    (+p_2,\Lambda)
\mathscr{D}^{*(j_3)}_{\lambda'_3\lambda_3}   (+p_3,\Lambda)
\mathscr{D}^{*(j_4)}_{-\lambda'_4,-\lambda_4}(+p_4,\Lambda)
T_{\bar 42\rightarrow 3\bar 1}{\;}_{\lambda_4', \lambda_2'}^{\lambda_3',\lambda_1'}(-p_4',p_2',p_3',-p_1').
\end{align}
Let us rename $\lambda_1$ and $\lambda_4$ and call them $-\lambda_1$ and $-\lambda_4$. We can also do the same for $\lambda_1'$ and $\lambda_4'$ since they are dummy indices and the summation covers all the options. We get then
\begin{align}
\label{eq:transformation_2_final}
&T_{\bar 42\rightarrow 3\bar 1}{\;}_{-\lambda_4, +\lambda_2}^{+\lambda_3,-\lambda_1}(-p_4,p_2,p_3,-p_1)=\\
\nn
&\sum_{\lambda_i'}
\mathscr{D}^{(j_1)}_{\lambda'_1,\lambda_1} (+p_1,\Lambda)
\mathscr{D}^{(j_2)}_{\lambda'_2\lambda_2}  (+p_2,\Lambda)
\mathscr{D}^{*(j_3)}_{\lambda'_3\lambda_3} (+p_3,\Lambda)
\mathscr{D}^{*(j_4)}_{\lambda'_4,\lambda_4}(+p_4,\Lambda)
T_{\bar 42\rightarrow 3\bar 1}{\;}_{-\lambda_4', +\lambda_2'}^{+\lambda_3',-\lambda_1'}(-p_4',p_2',p_3',-p_1').
\end{align}

Comparing \eqref{eq:transformation_1} and \eqref{eq:transformation_2_final} we conclude that the following two objects transform in the same way under the generic Lorentz transformations
\begin{equation}
\label{eq:two_obects_2}
T_{12\rightarrow 34}{\;}_{\lambda_1, \lambda_2}^{\lambda_3,\lambda_4}(+p_1,p_2,p_3,+p_4)
\quad\text{vs.}\quad
T_{\bar 42\rightarrow 3\bar 1}{\;}_{-\lambda_4, +\lambda_2}^{+\lambda_3,-\lambda_1}(-p_4,p_2,p_3,-p_1).
\end{equation}
Analogous discussion holds for other crossings we can thus write the following equations
\begin{equation}
\label{eq:crossing_ac_2}
\begin{aligned}
T_{12\rightarrow 34}{\;}_{\lambda_1, \lambda_2}^{\lambda_3,\lambda_4}(p_1,p_2,p_3,p_4)
&=\epsilon_{14}^{(2)}\;
T_{\bar 42\rightarrow 3\bar 1}{\;}_{-\lambda_4, +\lambda_2}^{+\lambda_3,-\lambda_1}(-p_4,+p_2,+p_3,-p_1),\\
T_{12\rightarrow 34}{\;}_{\lambda_1, \lambda_2}^{\lambda_3,\lambda_4}(p_1,p_2,p_3,p_4)
&=\epsilon_{23}^{(2)}\;
T_{1\bar3\rightarrow \bar 2 4}{\;}_{+\lambda_1, -\lambda_3}^{-\lambda_2,+\lambda_4}(+p_1,-p_3,-p_2,+p_4),\\
T_{12\rightarrow 34}{\;}_{\lambda_1, \lambda_2}^{\lambda_3,\lambda_4}(p_1,p_2,p_3,p_4)
&=\epsilon_{13}^{(2)}\;
T_{\bar 32\rightarrow 1\bar 4}{\;}_{-\lambda_3, +\lambda_2}^{-\lambda_1,+\lambda_4}(-p_3,+p_2,-p_1,+p_4),\\
T_{12\rightarrow 34}{\;}_{\lambda_1, \lambda_2}^{\lambda_3,\lambda_4}(p_1,p_2,p_3,p_4)
&=\epsilon_{24}^{(2)}\;
T_{1\bar 4\rightarrow 3\bar 2}{\;}_{+\lambda_1, -\lambda_4}^{+\lambda_3,-\lambda_2}(+p_1,-p_4,+p_3,-p_2),
\end{aligned}
\end{equation}
where as before $\epsilon_{14}^{(2)}$, $\epsilon_{23}^{(2)}$, $\epsilon_{13}^{(2)}$ and $\epsilon_{24}^{(2)}$ are the helicity independent phases unfixed by this procedure.

\subsection{Crossing equations in a general frame: LSZ derivation}
\label{app:LSZ}
In this section we derive the crossing equation \eqref{eq:s_t_crossing_general_23} from the LSZ reduction formula in the case of spin $1/2$ fermion scattering. The latter is carefully discussed in section 41 of \cite{Srednicki:2007qs}.  The LSZ reduction formula in this case reads as
\begin{eqnarray}
\label{eq:lsz_reduction}
T_{12\rightarrow 34}{}_{\lambda_1, \lambda_2}^{\lambda_3, \lambda_4}(p_1, p_2, p_3, p_4) &=& \int d^4 x_1 \, d^4 x_2\, d^4 x_3 \,d^4 x_4 \nonumber \\
&& \quad \times 
\quad e^{-i p_3 x_3}\,[ \bar{u}_{\lambda_3} (p_3) (-i \slashed{\partial}_3 + m_3)]_{\alpha_3} \nonumber \\
&& \quad \times 
\quad e^{-i p_4 x_4}\,[ \bar{u}_{\lambda_4} (p_4) (-i \slashed{\partial}_4 + m_4)]_{\alpha_4} \nonumber \\
&&  \quad \times \quad \langle \Omega | T\{\Psi_{\alpha_4}(x_4)\Psi_{\alpha_3}(x_3)\overline \Psi_{\alpha_1}(x_1) \overline\Psi_{\alpha_2}(x_2)\}| \Omega \rangle_{connected} \nonumber \\
&&  \quad \times 
\quad [(i \overleftarrow{\slashed{\partial}}_1 + m_1) u_{\lambda_1}(p_1)]_{\alpha_1}\, e^{i p_1 x_1}\nonumber \\
&&\quad \times 
\quad [(i \overleftarrow{\slashed{\partial}}_2 + m_2) u_{\lambda_2}(p_2)]_{\alpha_2}\, e^{i p_2 x_2},
\end{eqnarray}
where $|\Omega\>$ denotes the vacuum state and $\Psi_i(x)$ are 4-component Majorana or Dirac fields with masses $m_i$. Analogously one can write
\begin{eqnarray}
\label{eq:lsz_crossed}
T_{1\bar 3\rightarrow \bar 24}{}_{\lambda_1, \lambda_3}^{\lambda_2, \lambda_4}(p_1, p_3, p_2, p_4) &=& \int d^4 x_1 \, d^4 x_2\, d^4 x_3 \,d^4 x_4 \nonumber \\
&& \quad \times 
\quad e^{i p_3 x_3}\,[ \bar{v}_{\lambda_3} (p_3) (-i \slashed{\partial}_3 + m_3)]_{\alpha_3} \nonumber \\
&& \quad \times 
\quad e^{-i p_4 x_4}\,[ \bar{u}_{\lambda_4} (p_4) (-i \slashed{\partial}_4 + m_4)]_{\alpha_4} \nonumber \\
&&  \quad \times \quad \langle \Omega | T\{\Psi_{\alpha_4}(x_4)\overline \Psi_{\alpha_2}(x_2)\overline \Psi_{\alpha_1}(x_1) \Psi_{\alpha_3}(x_3)\}| \Omega \rangle_{connected} \nonumber \\
&&  \quad \times 
\quad [(i \overleftarrow{\slashed{\partial}}_1 + m_1) u_{\lambda_1}(p_1)]_{\alpha_1}\, e^{i p_1 x_1}\nonumber \\
&&\quad \times 
\quad [(i \overleftarrow{\slashed{\partial}}_2 + m_2) v_{\lambda_2}(p_2)]_{\alpha_2}\, e^{-i p_2 x_2}.
\end{eqnarray}
In the above equations all the momenta have positive energy, namely $p_i^0>0$ .  In order to relate the two processes \eqref{eq:lsz_reduction} and \eqref{eq:lsz_crossed} one can analytically continue the latter process in $p_2$ and $p_3$ to allow for negative energies. Assuming such an analytic continuation exists one gets
\begin{eqnarray}
T_{1\bar 3\rightarrow \bar 24}{}_{\lambda_1, \lambda_3}^{\lambda_2, \lambda_4}(p_1, -p_3, -p_2, p_4) &=& \int d^4 x_1 \, d^4 x_2\, d^4 x_3 \,d^4 x_4 \nonumber \\
&& \quad \times 
\quad e^{-i p_3 x_3}\,[ \bar{v}_{\lambda_3} (-p_3) (-i \slashed{\partial}_3 + m_3)]_{\alpha_3} \nonumber \\
&& \quad \times 
\quad e^{-i p_4 x_4}\,[ \bar{u}_{\lambda_4} (p_4) (-i \slashed{\partial}_4 + m_4)]_{\alpha_4} \nonumber \\
&&  \quad \times \quad \langle \Omega | T\{\Psi_{\alpha_4}(x_4)\overline \Psi_{\alpha_2}(x_2)\overline \Psi_{\alpha_1}(x_1) \Psi_{\alpha_3}(x_3)\}| \Omega \rangle_{connected} \nonumber \\
&&  \quad \times 
\quad [(i \overleftarrow{\slashed{\partial}}_1 + m_1) u_{\lambda_1}(p_1)]_{\alpha_1}\, e^{i p_1 x_1}\nonumber \\
&&\quad \times 
\quad [(i \overleftarrow{\slashed{\partial}}_2 + m_2) v_{\lambda_2}(-p_2)]_{\alpha_2}\, e^{i p_2 x_2}.
\label{eq:LSZ_crossing_final}
\end{eqnarray}
Notice also that the anticommutation of fermionic operators manifests itself in the following relation
\begin{multline}
\label{eq:permutations}
\langle \Omega | T\{\Psi_{\alpha_4}(x_4)\overline \Psi_{\alpha_2}(x_2)\overline \Psi_{\alpha_1}(x_1) \Psi_{\alpha_3}(x_3)\}| \Omega \rangle=\\
- \langle \Omega | T\{\Psi_{\alpha_4}(x_4)\Psi_{\alpha_3}(x_3)\overline \Psi_{\alpha_1}(x_1) \overline\Psi_{\alpha_2}(x_2)\}| \Omega \rangle.
\end{multline}
According to appendix \ref{app:analytic_continuation} there are two ways of choosing the analytic continuation when writing \eqref{eq:LSZ_crossing_final}. Below we discuss both of them in order. 

\section*{Analytic continuation 1}
Using the analytic continuation \eqref{eq:ac_1} and the definition \eqref{eq:Dirac_solutions} it follows straightforwardly
\begin{equation}
\begin{aligned}
\label{eq:uv_continuation}
u_\lambda (-p) &= e^{i \lambda (\pi + 2\phi)} v_\lambda(p),\qquad
\bar u_\lambda(-p) = e^{i \lambda (\pi - 2\phi)} \bar v_\lambda(p), \\
v_\lambda (-p) &= e^{i \lambda (\pi - 2\phi)} u_\lambda(p),\qquad
\bar v_\lambda(-p) = e^{i \lambda (\pi + 2\phi)} \bar u_\lambda(p).
\end{aligned}
\end{equation} 
Plugging \eqref{eq:uv_continuation} and  \eqref{eq:permutations} into \eqref{eq:LSZ_crossing_final} and comparing the result with \eqref{eq:lsz_reduction}
we arrive at the following crossing equation
\begin{equation}
T_{12\rightarrow 34}{}_{\lambda_1, \lambda_2}^{\lambda_3, \lambda_4}(p_1, p_2, p_3, p_4) = - e^{-i \lambda_2 (\pi - 2\phi_2)}e^{-i \lambda_3 (\pi + 2\phi_3)}T_{1\bar 3\rightarrow \bar 24}{}_{\lambda_1, \lambda_3}^{\lambda_2, \lambda_4}(p_1, -p_3, -p_2, p_4).
\end{equation}
Analogously one derives the other three crossing equations. The complete summary of crossing equations reads
\begin{equation}
\label{eq:crossing_LSZ_ac1}
\begin{aligned}
T_{12\rightarrow 34}{}_{\lambda_1, \lambda_2}^{\lambda_3, \lambda_4}(p_1, p_2, p_3, p_4) = - e^{-i \lambda_1 (\pi - 2\phi_1)}e^{-i \lambda_4 (\pi + 2\phi_4)}
T_{\bar 4 2\rightarrow 3\bar 1}{}_{\lambda_4, \lambda_2}^{\lambda_3, \lambda_1}(-p_4, p_2, p_3, -p_1),\\
T_{12\rightarrow 34}{}_{\lambda_1, \lambda_2}^{\lambda_3, \lambda_4}(p_1, p_2, p_3, p_4) = - e^{-i \lambda_2 (\pi - 2\phi_2)}e^{-i \lambda_3 (\pi + 2\phi_3)}T_{1\bar 3\rightarrow \bar 24}{}_{\lambda_1, \lambda_3}^{\lambda_2, \lambda_4}(p_1, -p_3, -p_2, p_4),\\
T_{12\rightarrow 34}{}_{\lambda_1, \lambda_2}^{\lambda_3, \lambda_4}(p_1, p_2, p_3, p_4) = - e^{-i \lambda_1 (\pi - 2\phi_1)}e^{-i \lambda_3 (\pi + 2\phi_3)}
T_{\bar 3 2\rightarrow \bar 1 4}{}_{\lambda_3, \lambda_2}^{\lambda_1, \lambda_4}(-p_3, p_2, -p_1, p_4),\\
T_{12\rightarrow 34}{}_{\lambda_1, \lambda_2}^{\lambda_3, \lambda_4}(p_1, p_2, p_3, p_4) = - e^{-i \lambda_2 (\pi - 2\phi_2)}e^{-i \lambda_4 (\pi + 2\phi_4)}
T_{1\bar 4\rightarrow 3\bar 2}{}_{\lambda_1, \lambda_4}^{\lambda_3, \lambda_2}(p_1, -p_4, p_3, -p_2).
\end{aligned}
\end{equation}
We can now compare the above to \eqref{eq:crossing_ac_1} and deduce the undetermined phases\footnote{Here we use the fact that $e^{2i\pi \lambda}=-1$ since $\lambda$ is half-integer.
}
\begin{equation}
\epsilon^{(1)}_{14} = \epsilon^{(1)}_{23} =
\epsilon^{(1)}_{13} =
\epsilon^{(1)}_{24} = +1.
\end{equation}

\section*{Analytic continuation 2}
Using the analytic continuation \eqref{eq:ac_2} and the definition \eqref{eq:Dirac_solutions} it follows straightforwardly
\begin{equation}
\begin{aligned}
\label{eq:uv_continuation_2}
u_\lambda (-p) = i\,v_{-\lambda}(p),\qquad
\bar u_\lambda(-p) = i\,\bar v_{-\lambda}(p),\\
v_\lambda (-p) = i\,u_{-\lambda}(p),\qquad
\bar v_\lambda(-p) = i\,\bar u_{-\lambda}(p).
\end{aligned}
\end{equation} 
Plugging \eqref{eq:uv_continuation_2} and  \eqref{eq:permutations} into \eqref{eq:LSZ_crossing_final} and comparing the result with \eqref{eq:lsz_reduction}
we arrive at the following crossing equation
\begin{equation}
T_{12\rightarrow 34}{}_{\lambda_1, \lambda_2}^{\lambda_3, \lambda_4}(p_1, p_2, p_3, p_4) = T_{1\bar 3\rightarrow \bar 24}{}_{+\lambda_1, -\lambda_3}^{-\lambda_2, +\lambda_4}(p_1, -p_3, -p_2, p_4).
\end{equation}
Analogously one derives the other three crossing equations. The complete summary of crossing equations reads
\begin{equation}
\label{eq:crossing_LSZ_ac2}
\begin{aligned}
T_{12\rightarrow 34}{}_{\lambda_1, \lambda_2}^{\lambda_3, \lambda_4}(p_1, p_2, p_3, p_4) = 
T_{\bar 4 2\rightarrow 3\bar 1}{}_{-\lambda_4, +\lambda_2}^{+\lambda_3, -\lambda_1}(-p_4, p_2, p_3, -p_1),\\
T_{12\rightarrow 34}{}_{\lambda_1, \lambda_2}^{\lambda_3, \lambda_4}(p_1, p_2, p_3, p_4) = T_{1\bar 3\rightarrow \bar 24}{}_{+\lambda_1, -\lambda_3}^{-\lambda_2, +\lambda_4}(p_1, -p_3, -p_2, p_4),\\
T_{12\rightarrow 34}{}_{\lambda_1, \lambda_2}^{\lambda_3, \lambda_4}(p_1, p_2, p_3, p_4) = 
T_{\bar 3 2\rightarrow \bar 1 4}{}_{-\lambda_3, +\lambda_2}^{-\lambda_1, +\lambda_4}(-p_3, p_2, -p_1, p_4),\\
T_{12\rightarrow 34}{}_{\lambda_1, \lambda_2}^{\lambda_3, \lambda_4}(p_1, p_2, p_3, p_4) = 
T_{1\bar 4\rightarrow 3\bar 2}{}_{+\lambda_1, -\lambda_4}^{+\lambda_3, -\lambda_2}(p_1, -p_4, p_3, -p_2).
\end{aligned}
\end{equation}
Comparing the above with \eqref{eq:crossing_ac_2}, we see that all the previously undetermined phases are
\begin{equation}
\epsilon^{(2)}_{14} =
\epsilon^{(2)}_{23} =
\epsilon^{(2)}_{13} =
\epsilon^{(2)}_{24} = +1.
\end{equation}

\section*{Concluding remarks}
We have derived in this appendix the crossing equations for generic spin $1/2$ particles in a general frame using the LSZ reduction formula. These depend on the analytic continuation. For the analytic continuation \eqref{eq:ac_1} our crossing equations are given by \eqref{eq:crossing_LSZ_ac1}. For the analytic continuation \eqref{eq:ac_2} our crossing equations are given by \eqref{eq:crossing_LSZ_ac2}. All these formulas remain (almost) the same even if some of the particles have  spin different from 1/2. This follows from the fact that the spin structures of a generic spin particle can be represented by products of $u$ and $v$ objects. The only change in the crossing equations comes in the overall sign since some particles can now commute instead.

\subsection{Crossing equations in the center of mass frame}
\label{app:crossing_COM_frame}
The goal of this appendix is to write the crossing equations \eqref{eq:crossing_ac_1} and \eqref{eq:crossing_ac_2} in the center of mass frame. We will see that both analytic continuations lead to the same center of mass equations. The crossing equations 1-4 and 2-3 are called the $s-t$ equations. Since they carry identical information, we focus only on the 2-3 crossing equation. Instead the crossing equations 1-3 and 2-4 are called the $s-u$ equations. Since they also carry the same information we focus only the 2-4 crossing equation. The discussion of the $s-u$ center of mass equations is identical to the $s-t$ one, we will therefore only provide the final results without any intermediate steps.

\subsubsection{$s-t$ crossing equation}
According to appendix \ref{app:analytically_continued_transformations} the 2-3 crossing equation depending on the analytic continuation can take either of the two forms
\begin{align}
\label{eq_1}
T_{12\rightarrow 34}{\;}_{\lambda_1, \lambda_2}^{\lambda_3,\lambda_4}(p_1,p_2,p_3,p_4)
&=\epsilon_{23}^{(1)}
e^{-i(\pi-2\phi_2)\lambda_2}
e^{i(\pi-2\phi_3)\lambda_3}
T_{1\bar3\rightarrow \bar 2 4}{\;}_{\lambda_1, \lambda_3}^{\lambda_2,\lambda_4}(P_1, P_2, P_3, P_4),\\
\label{eq_2}
T_{12\rightarrow 34}{\;}_{\lambda_1, \lambda_2}^{\lambda_3,\lambda_4}(p_1,p_2,p_3,p_4)
&=\epsilon_{23}^{(2)}\;
T_{1\bar3\rightarrow \bar 2 4}{\;}_{+\lambda_1, -\lambda_3}^{-\lambda_2,+\lambda_4}(P_1, P_2, P_3, P_4),
\end{align}
where we have defined
\begin{equation}
P_1\equiv p_1,\quad
P_2\equiv-p_3,\quad
P_3\equiv-p_2,\quad
P_4\equiv p_4.
\end{equation}
In the right hand-side of \eqref{eq_1} and \eqref{eq_2} $P_1$ and $P_2$ describe the incoming particles 1 and $\bar 3$ respectively, whereas $P_3$ and $P_4$ describe the outgoing particles $\bar 2$ and 4 respectively. The Mandelstam variables associated to the left-hand side are as usual
\begin{equation}
s\equiv -(p_1+p_2)^2,\quad
t\equiv -(p_1-p_3)^2,\quad
u\equiv -(p_1-p_4)^2.
\end{equation} 
Instead the Mandelstam variables associated to the right-hand side are
\begin{equation}
\begin{aligned}
S&\equiv - (P_1+P_2)^2= - (p_1-p_3)^2= t,\\
T&\equiv - (P_1-P_3)^2= - (p_1+p_2)^2= s,\\
U&\equiv - (P_1-P_4)^2= - (p_1-p_4)^2= u.
\end{aligned}
\end{equation}
The Mandelstam variables remain invariant by definition under any Lorentz transformation.

Let us now evaluate the left-hand side of \eqref{eq_1} and \eqref{eq_2} in the center of mass frame of the $12\rightarrow 34$ process denoted by $p^\text{com}_i$. It is defined in \eqref{eq:COM_frame}, we write it here again for convenience
\begin{equation}
\begin{aligned}
&p_1^{\text{com}}\equiv(E_1,0,0,+\myP),\\
&p_3^{\text{com}}\equiv(E_3,+\myP'\sin \theta ,0,+\myP'\cos \theta),\\
&p_2^{\text{com}}\equiv(E_2,0,0,-\myP),\\
&p_4^{\text{com}}\equiv(E_4,-\myP'\sin \theta ,0,-\myP'\cos \theta).
\end{aligned}
\end{equation}
The main feature of this frame is that it respects the center of mass conditions
\begin{equation}
\label{eq:COM_property}
\begin{aligned}
p_1^\text{com}+p_2^\text{com} = (E_1+E_2,0,0,0),\\
p_3^\text{com}+p_4^\text{com} = (E_3+E_4,0,0,0).
\end{aligned}
\end{equation}
Then the right-hand side of \eqref{eq_1} and \eqref{eq_2} will depend on
\begin{equation}
\label{eq:crossed_frame}
\begin{aligned}
P_1^\text{com}&\equiv(E_1,0,0,\myP),\\
P_2^\text{com}&\equiv(-E_3,-\myP'\sin\theta,0,-\myP'\cos\theta),\\
P_3^\text{com}&\equiv(-E_2,0,0,\myP),\\
P_4^\text{com}&\equiv(E_4,-\myP'\sin\theta,0,-\myP'\cos\theta).
\end{aligned}
\end{equation}
The latter obey
\begin{equation}
\label{eq:COM_property_crossed}
\begin{aligned}
P_1^\text{com}-P_3^\text{com} = (E_1+E_2,0,0,0),\\
P_4^\text{com}-P_2^\text{com} = (E_3+E_4,0,0,0).
\end{aligned}
\end{equation}
However, this is not the standard COM frame of the process $12\rightarrow 34$ since it differs from \eqref{eq:COM_property}. We refer to \eqref{eq:crossed_frame} as the (23) crossed COM frame.

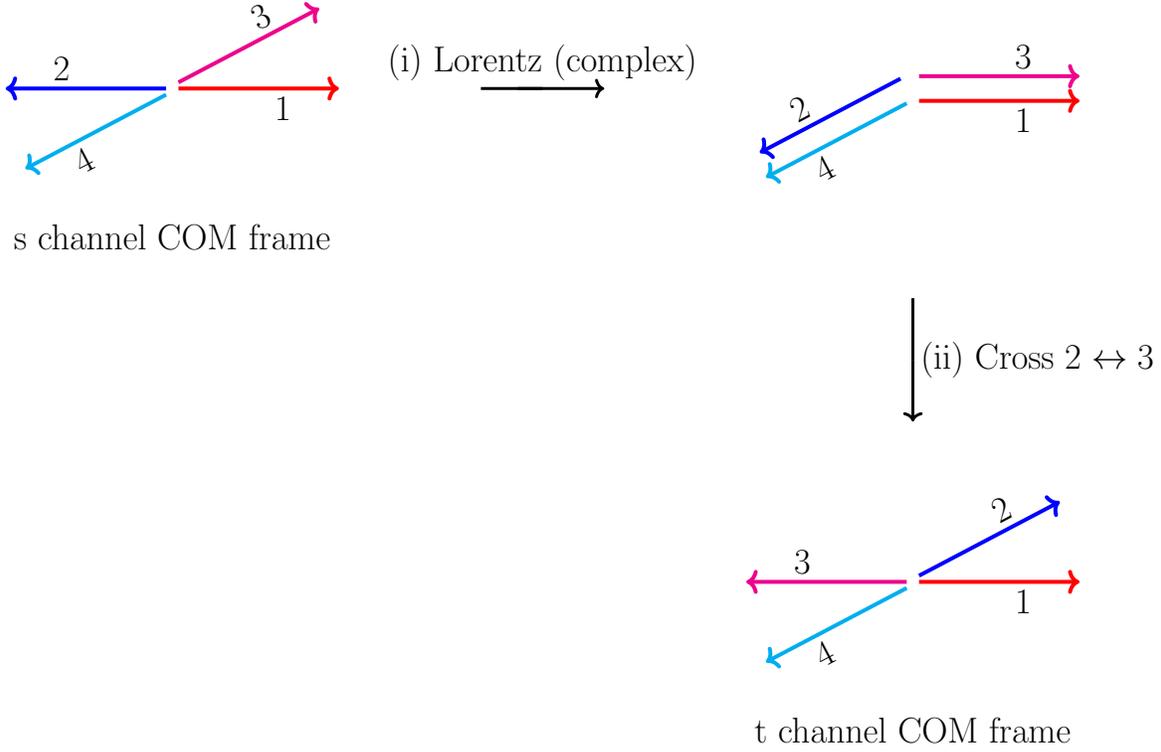
\begin{figure}
	\centering
	\resizebox{\textwidth}{!}{
		\begin{tikzpicture}
			\coordinate (origin) at (0,0);
			
			\draw[] (0,-3) node[] {\huge s channel COM frame};
			\draw[line width=0.8mm,red,->] (0.125,0) -- ++(3.25,0) coordinate (p1) node[pos = 0.65,below, black, sloped]  {\huge $1$};
			\draw[line width=0.8mm,blue,->] (-0.125,0) -- ++(-3.25,-0) coordinate (p2) node[pos = 0.65,above, black, sloped] { \huge $2$};
			
			\draw[line width=0.8mm,cyan,->] (-0.125,-0.125) -- ++(-2.85,-1.5) coordinate (p4) node[pos = 0.65,below, black, sloped]  {\huge $4$};
			\draw[line width=0.8mm,magenta,->] (0.125,0.125) -- ++(2.85,1.5) coordinate (p3) node[pos = 0.65,above, black, sloped] { \huge $3$};
			
			
			
			\draw[thick,line width=0.6mm,black, -]  (6.25,0) --  (7.5,0) node  [black, above] {\huge (i) Lorentz (complex)};
			\draw[thick,line width=0.6mm,black, ->] (7,0) -- (8.75,0);

			\coordinate (origin2) at (15,0);
			\coordinate (pprime) at (18,2);
			\coordinate (pprimeminus) at (12,-2);
			%

			\draw[line width=0.8mm,red,->] (15.125,-0.25) -- ++(3.25,0) coordinate (p1prime) node[pos = 0.65,below, black, sloped]  {\huge $1$};
			\draw[line width=0.8mm,blue,->] (14.75,0.2) -- ++(-2.85,-1.5) coordinate (p2prime) node[pos = 0.65,above, black, sloped] { \huge $2$};
			
			\draw[line width=0.8mm,magenta,->] (15.125,0.25) -- ++(3.25,0) coordinate (p3prime) node[pos = 0.65,above, black, sloped]  {\huge $3$};
			\draw[line width=0.8mm,cyan,->] (14.875,-0.3) -- ++(-2.85,-1.5) coordinate (p4prime) node[pos = 0.65,below, black, sloped] { \huge $4$};
			
			%
			
			\draw[thick, line width = 0.6mm, black, -]  (15,-4.25) --  (15,-5.5) node  [black, right] {\huge (ii) Cross $2 \leftrightarrow 3$};
			\draw[thick, line width = 0.6mm,black, ->] (15,-5.5)-- (15,-6.75);

			\coordinate (origin3) at (15,0);
			\coordinate (pprimeprime) at (33,2);
			\coordinate (pprimeprimeminus) at (27,-2);
			%
			%
			%
			
			\draw[line width=0.8mm,red,->] (15.125,-10) -- ++(3.25,0) coordinate (p1primeprime) node[pos = 0.65,below, black, sloped]  {\huge $1$};
			\draw[line width=0.8mm,magenta,->] (14.875,-10) -- ++(-3.25,-0) coordinate (p3primeprime) node[pos = 0.65,above, black, sloped] { \huge $3$};
			
			\draw[line width=0.8mm,cyan,->] (14.875,-10.125) -- ++(-2.85,-1.5) coordinate (p4primeprime) node[pos = 0.65,below, black, sloped]  {\huge $4$};
			\draw[line width=0.8mm,blue,->] (15.125,-9.875) -- ++(2.85,1.5) coordinate (p2primeprime) node[pos = 0.65,above, black, sloped] { \huge $2$};
			\draw[] (15,-13) node[] {\huge t channel COM frame};
			

		\end{tikzpicture}
	}
	\caption{Schematic picture of (23) crossing for COM frame amplitudes. }
	\label{fig:crossing_schematic}
\end{figure}

Once the left-hand sides of \eqref{eq_1} and \eqref{eq_2} are evaluated in the $12\rightarrow34$ COM frame  which respects \eqref{eq:COM_property} we do not get closed expressions  since the right-hand side is not in the $1\bar 3 \rightarrow \bar 2 4$ COM frame but rather in the (23) crossed COM frame\footnote{Recall that we defined the amplitudes as functions of Mandelstam invariants in the COM frame.}. We therefore need an additional Lorentz transformation.
It turns out to be simpler to bring the left-hand side to the (23) crossed COM frame. This way upon (23) crossing we end up in the COM frame for the  $1\bar 3\rightarrow \bar 24$. This is illustrated in figure \ref{fig:crossing_schematic}.

We now consider the Lorentz transformation $\Lambda$ such that
\begin{equation}
\label{eq:transformation_requirement}
\begin{aligned}
&p_1^{\text{com}}=(E_1,0,0,+\myP)\\
&p_2^{\text{com}}=(E_2,0,0,-\myP)\\
&p_3^{\text{com}}=(E_3,+\myP'\sin \theta ,0,+\myP'\cos \theta)\\
&p_4^{\text{com}}=(E_4,-\myP'\sin \theta ,0,-\myP'\cos \theta)
\end{aligned}
\quad\Longrightarrow\quad
\begin{aligned}
&\hat p_1^{\text{com}}=(\hat E_1,0,0,+\hat\myP),\\
&\hat p_2^{\text{com}}=(\hat E_2,-\hat\myP'\sin \hat\theta ,0,-\hat\myP'\cos\hat \theta),\\
&\hat p_3^{\text{com}}=(\hat E_3,0,0,\hat\myP),\\
&\hat p_4^{\text{com}}=(\hat E_4,-\hat\myP'\sin \hat\theta ,0,-\hat\myP'\cos\hat \theta),
\end{aligned}
\end{equation}
where $\hat E_i$, $\hat\myP$, $,\hat\myP'$ and $\hat{\theta}$ are the components of the 4-momenta in the new frame.  The frame described by the right-hand side \eqref{eq:transformation_requirement} is precisely the (23) crossed frame \eqref{eq:crossed_frame} since
\begin{equation}
\label{eq:transformation_requirement_2}
\begin{aligned}
\hat p_1^{\text{com}}-\hat p_3^{\text{com}} &= (\hat E_1-\hat E_3,0,0,0),\\
\hat p_4^{\text{com}}-\hat p_2^{\text{com}} &= (\hat E_4-\hat E_2,0,0,0).
\end{aligned}
\end{equation}
Due to the results of appendix \ref{sec:four_point_amplitudes_COM} we have
\begin{align}
E_1 = \frac{s+m_1^2-m_2^2}{2\sqrt{s}},\;\;
E_2 = \frac{s-m_1^2+m_2^2}{2\sqrt{s}}&,\;\
E_3 = \frac{s+m_3^2-m_4^2}{2\sqrt{s}},\;\;
E_4 = \frac{s-m_3^2+m_4^2}{2\sqrt{s}},\\
\myP  = \frac{\mathcal{L}_{12}(s)}{2\sqrt s}&,\;\;\;\;
\myP^{\prime} = \frac{\mathcal{L}_{34}(s)}{2\sqrt s}.
\end{align}
These are originally defined in the physical range of parameters of the $12\rightarrow34$ process, where in the case of identical particles we have
\begin{equation}
\label{eq:s-channel}
12\rightarrow 34:\qquad s\geq 4m^2,\quad t\in[4m^2-s\;,0].
\end{equation}
Remember that $s$ is regularized as $s+i\epsilon$, where $\epsilon>0$. We can unambiguously continue all the formulas valid in the domain \eqref{eq:s-channel} to the physical domain of the $1\bar 3\rightarrow \bar 2 4$ process  which for identical particles read as
\begin{equation}
\label{eq:t-channel}
1\bar 3\rightarrow \bar 2 4:\qquad t\geq 4m^2,\quad s\in[4m^2-t\;,0].
\end{equation}
The Lorentz transformation which allows for \eqref{eq:transformation_requirement} and \eqref{eq:transformation_requirement_2} has the following form
\begin{equation}
\label{eq:transformation_Lambda}
\Lambda = R_2(\psi_2)B_3(\chi)R_2(\psi_1),
\end{equation}
where $R_2$ and $B_3$ are finite rotations around the y-axis and boost in the z-direction. Their form in the vector representation is given in appendix \ref{app:representations}. By requiring \eqref{eq:transformation_requirement_2} we can determine the unknown parameters $\psi_1$, $\psi_2$ and $\chi$.
First, using the rotation $R_2(\psi_1)$ we make the x-components of $p^{\text{com}}_1$ and $p^{\text{com}}_3$ equal. This is achieved for example with
\begin{equation}
\label{eq:first_rotation}
\cos\psi_1 = \frac{\myP-\myP'\cos\theta}{\sqrt{\myP^2+\myP^{\prime 2}-2 \myP \myP' \cos\theta}},\qquad
\sin\psi_1 = \frac{\myP'\sin\theta}{\sqrt{\myP^2+\myP^{\prime 2}-2 \myP \myP' \cos\theta}}.
\end{equation}
Second, using the boost $B_3(\chi)$ along the z-axis we make  the z-components of the rotated vectors $p^{\text{com}}_1$ and $p^{\text{com}}_3$ equal. This is achieved with
\begin{align}
\label{eq:transformation_outcome}
\cosh\chi &= 
\frac{E_1-E_3}{\sqrt{(E_1-E_3)^2-\myP^2-\myP^{\prime 2}+2\myP\myP'\cos\theta}} =
\frac{m_1^2-m_2^2-m_3^2+m_4^2}{2\sqrt{s}\sqrt{t}},\\
\sinh\chi &=-
\frac{\sqrt{\myP^2+\myP^{\prime 2}-2 \myP \myP'\cos\theta}}
{\sqrt{(E_1-E_3)^2-\myP^2-\myP^{\prime 2}+2\myP\myP'\cos\theta}}=- \frac{\sqrt{s^{-1}(m_1^2-m_2^2-m_3^2+m_4^2)^2-4t}}{2\sqrt{t}}.
\nn
\end{align}
Third, the rotation $R_2(\psi_2)$ brings us to the desired frame. However, we will not need the explicit expression for the angle $\psi_2$. As a result we get the following expressions
\begin{align}
\label{eq:energies_hat}
\hat E_1 = \frac{m_1^2-m_3^2+t}{2\sqrt{t}},\;\;
\hat E_2 = \frac{m_4^2-m_2^2-t}{2\sqrt{t}}&,\;\;
\hat E_3 = \frac{m_1^2-m_3^2-t}{2\sqrt{t}},\;\;
\hat E_4 = \frac{m_4^2-m_2^2+t}{2\sqrt{t}},\\
\hat\myP^2 = \frac{(\mathcal{L}_{13}(t))^2}{4t}&,\;\;\;\;
\hat\myP^{\prime\,2} = \frac{(\mathcal{L}_{24}(t))^2}{4t}.
\end{align}
There were several choices of signs in \eqref{eq:transformation_outcome}. We picked one such that the energy $\hat E_1$ in \eqref{eq:energies_hat} is non-negative in the physical domain \eqref{eq:t-channel} of the process $1\bar 3\rightarrow \bar 2 4$.

Assuming that we have determined the correct values of parameters $\psi_1$, $\psi_2$ and $\chi$, we can then straightforwardly compute the Wigner angles using the results of appendix \ref{app:wigner_angles}, which read
\begin{equation}
\label{eq:w_angles}
\cos\alpha_i = \frac{E_i\, \hat E_i-m_i^2\cosh\chi}
{\myP_i\, \hat \myP_i},\quad
\sin\alpha_i = \frac{m_i\sinh\chi}
{\hat\myP_i}\,\sin(\theta_i+\psi_1).
\end{equation}
Here the Wigner angles with the subscript $i=1,2,3,4$ correspond to Lorentz transformations of $p_i^\text{com}$ in \eqref{eq:transformation_requirement}. The spherical angles of the four-particles in \eqref{eq:transformation_requirement} before the Lorentz transformation read as
\begin{equation}
\myP_1 = \myP_2 = \myP,\quad
\theta_1=0,\quad\theta_2=\pi,\quad
\myP_3 = \myP_4 = \myP',\quad
\theta_3=\theta,\quad
\theta_4=\theta+\pi.
\end{equation}
Using the transformation property \eqref{eq:T_matrix_amplitude_transformation_general} of scattering amplitudes we can then write
\begin{multline}
\label{eq:transformation_23}
T_{12\rightarrow 34}{\;}_{\lambda_1, \lambda_2}^{\lambda_3,\lambda_4}
(p_1^\text{com},
p_2^\text{com},
p_3^\text{com},
p_4^\text{com})
=\\
\sum_{\lambda_i'}
d^{(j_1)}_{\lambda'_1\lambda_1}(\alpha_1)
d^{(j_2)}_{\lambda'_2\lambda_2}(\alpha_2)
d^{(j_3)}_{\lambda'_3\lambda_3}(\alpha_3)
d^{(j_4)}_{\lambda'_4\lambda_4}(\alpha_4)
T_{12\rightarrow 34}{\;}_{\lambda'_1, \lambda'_2}^{\lambda'_3,\lambda'_4}
(\hat p_1^\text{com},
\hat p_3^\text{com},
\hat p_2^\text{com},
\hat p_4^\text{com}).
\end{multline}

\section*{Analytic continuation 1}
Having performed the Lorentz transformations \eqref{eq:transformation_requirement} we ended up with \eqref{eq:energies_hat}. These expressions contain an ambiguity on how to take a square root. This is related to the choice of the analytic continuation discussed in section \ref{app:crossing_equations}.
For the choice \eqref{eq:ac_1} in the domain \eqref{eq:t-channel} we have
\begin{equation}
\hat\myP  = +\frac{\mathcal{L}_{13}(t)}{2\sqrt{t}},\qquad
\hat\myP' = +\frac{\mathcal{L}_{24}(t)}{2\sqrt{t}}.
\end{equation}
The spherical angles of the four-momenta $\hat p_i^\text{com}$ after the Lorentz transformation read as
\begin{equation}
\hat\myP_1 = \hat\myP_3 = \hat\myP,\quad
\hat\theta_1=\hat\theta_3=0,\quad
\hat\myP_2 = \hat\myP_4 = \hat\myP',\quad
\hat\theta_2=\hat\theta_4=\hat\theta+\pi.
\end{equation}
Using these, the boost \eqref{eq:transformation_outcome} and \eqref{eq:w_angles} we  obtain in full generality the cosines\footnote{The values computed here match precisely the ones of formula (13) in \cite{Hara:1971kj}. Notice that our process $12\rightarrow 34$ corresponds to their $ac\rightarrow bd$. Thus, in order to see the equivalence one needs to identify the labels as $1=a$, $2=c$, $3=b$ and $4=d$.}
\begin{equation}
\label{eq:cos_alpha}
\begin{aligned}
\cos\alpha_1 &= \frac{+(s+m_1^2-m_2^2)(t+m_1^2-m_3^2)+2m_1^2\,\Delta_t}
{\mathcal{L}_{12}(s)\mathcal{L}_{13}(t)},\\
\cos\alpha_2 &= \frac{-(s-m_1^2+m_2^2)(t+m_2^2-m_4^2)+2m_2^2\,\Delta_t}
{\mathcal{L}_{12}(s)\mathcal{L}_{24}(t)},\\
\cos\alpha_3 &= \frac{-(s+m_3^2-m_4^2)(t-m_1^2+m_3^2)+2m_3^2\,\Delta_t}
{\mathcal{L}_{34}(s)\mathcal{L}_{13}(t)},\\
\cos\alpha_4 &= \frac{+(s+m_4^2-m_3^2)(t+m_4^2-m_2^2)+2m_4^2\,\Delta_t}
{\mathcal{L}_{34}(s)\mathcal{L}_{24}(t)},
\end{aligned}
\end{equation}
together with sines
\begin{equation}
\label{eq:sin_alpha}
\begin{aligned}
\sin\alpha_1 &=  -\frac{2m_1\,\sqrt{\Phi}}
{\mathcal{L}_{12}(s)\mathcal{L}_{13}(t)},\quad
\sin\alpha_2 =  +\frac{2m_2\,\sqrt{\Phi}}
{\mathcal{L}_{12}(s)\mathcal{L}_{24}(t)},\\
\sin\alpha_3 &= -\frac{2m_3\,\sqrt{\Phi}}
{\mathcal{L}_{34}(s)\mathcal{L}_{13}(t)},\quad
\sin\alpha_4 =  +\frac{2m_4\,\sqrt{\Phi}}
{\mathcal{L}_{34}(s)\mathcal{L}_{24}(t)},
\end{aligned}
\end{equation}
where the objects $\Phi$ and $\Delta_t$ were defined \eqref{eq:definition_Phi} and \eqref{eq:definition_delta}.

Using the crossing relation \eqref{eq_1}, where we set $\phi_2=\pi$ and $\phi_3=0$, and plugging it into \eqref{eq:transformation_23} we get
\begin{multline}
T_{12\rightarrow 34}{\;}_{\lambda_1, \lambda_2}^{\lambda_3,\lambda_4}
(p_1^\text{com},
p_2^\text{com},
p_3^\text{com},
p_4^\text{com})
=
\epsilon_{23}^{(1)}\sum_{\lambda_i'}
e^{i\pi(\lambda_2'+\lambda_3')}\\
d^{(j_1)}_{\lambda'_1\lambda_1}(\alpha_1)
d^{(j_2)}_{\lambda'_2\lambda_2}(\alpha_2)
d^{(j_3)}_{\lambda'_3\lambda_3}(\alpha_3)
d^{(j_4)}_{\lambda'_4\lambda_4}(\alpha_4)
T_{1\bar{3}\rightarrow \bar{2}4}{\;}_{\lambda_1', \lambda_3'}^{\lambda_2',\lambda_4'}(\hat p_1^\text{com},-\hat p_3^\text{com}, -\hat p_2^\text{com}, \hat p_4^\text{com}).
\label{eq:crossing_23}
\end{multline}
There is one last important subtlety we need to take into account. Consider the four-momentum  of particle 3 in the left- and right-hand side of \eqref{eq:transformation_requirement}. Before the Lorentz transformation the spherical angles of the $p_3^\text{com}$ are $(\theta_3,\phi_3)=(\theta,0)$ by definition of the center of mass frame. After the Lorentz transformation the spherical angles of $-\hat p_3^\text{com}$ are $(\hat\theta_3,\hat\phi_3)$.\footnote{We discuss here $-\hat p_3^\text{com}$ because it is the quantity which enters the right-hand side of \eqref{eq:crossing_23}.}
They can take one of the two option
\begin{equation}
\label{eq:options_planes}
(\hat{\theta},0)\quad{\text{or}}\quad(\hat{\theta},\pi)=(-\hat{\theta},0).
\end{equation}
It is hard to see which option is correct from the above arguments. To check this we pick random values of $s$ and $t$ from the domain \eqref{eq:t-channel}. We observe that $\hat\theta\in[-\pi,0]$ which favors the second option in \eqref{eq:options_planes} is correct.
As a result the amplitude in the right-hand side of \eqref{eq:crossing_23} depends on the $-\hat{\theta}$. In order to rewrite it in terms of $+\hat{\theta}$ we use \eqref{eq:partial_wave_decomposition_T} and the following properties of the Wigner d-matrix 
\begin{equation}
d^{(j)}_{\lambda'\lambda}(-\omega)=(-1)^{\lambda'-\lambda}d^{(j)}_{\lambda'\lambda}(+\omega)
\end{equation}
which lead to
\begin{align}
\label{eq:relation_minus_theta}
T_{\lambda_1', \lambda_3'}^{\lambda_2',\lambda_4'}(S,T(S,-\hat\theta),U(S,-\hat\theta))= 
e^{i\pi(\lambda_1'-\lambda_3'-\lambda_2'+\lambda_4')}
T_{\lambda_1', \lambda_3'}^{\lambda_2',\lambda_4'}(S,T(S,+\hat\theta),U(S,+\hat\theta)).
\end{align}
Plugging \eqref{eq:relation_minus_theta} in the right-hand side of \eqref{eq:crossing_23} and using the definition \eqref{eq:scalar_Tamplitude_definition} we arrive at the final crossing equation
\begin{multline}
\label{eq:com_crossing_23_ac1}
T_{12\rightarrow 34}{\;}_{\lambda_1, \lambda_2}^{\lambda_3,\lambda_4}
(s,t,u)
=
\epsilon_{23}^{(1)}\sum_{\lambda_i'}
e^{i\pi(\lambda_1'+\lambda_4')}\\
d^{(j_1)}_{\lambda'_1\lambda_1}(\alpha_1)
d^{(j_2)}_{\lambda'_2\lambda_2}(\alpha_2)
d^{(j_3)}_{\lambda'_3\lambda_3}(\alpha_3)
d^{(j_4)}_{\lambda'_4\lambda_4}(\alpha_4)
T_{1\bar{3}\rightarrow \bar{2}4}{\;}_{\lambda_1', \lambda_3'}^{\lambda_2',\lambda_4'}(t,s,u).
\end{multline} 
Focusing on the case of identical particles one can check that applying \eqref{eq:com_crossing_23_ac1} twice gives back the initial process $12\rightarrow34$. If we chose the other option in \eqref{eq:options_planes} the resulting crossing equations would not have  passed this   consistency check.

\section*{Analytic continuation 2}
For the choice of the analytic continuation \eqref{eq:ac_2} in the domain \eqref{eq:t-channel} we have instead
\begin{equation}
\hat\myP  = \pm\frac{\mathcal{L}_{13}(t)}{2\sqrt{t}},\qquad
\hat\myP' = \pm\frac{\mathcal{L}_{24}(t)}{2\sqrt{t}},
\end{equation}
where the minus is taken for the crossed particles 2 and 3, and the plus is taken for the unchanged particles 1 and 4. In other words after the Lorentz transformation \eqref{eq:transformation_requirement} we get
\begin{equation}
\hat\myP_1 = +\frac{\mathcal{L}_{13}(t)}{2\sqrt{t}},\quad
\hat\myP_2 = -\frac{\mathcal{L}_{24}(t)}{2\sqrt{t}},\quad
\hat\myP_3 = -\frac{\mathcal{L}_{13}(t)}{2\sqrt{t}},\quad
\hat\myP_4 = +\frac{\mathcal{L}_{24}(t)}{2\sqrt{t}}.
\end{equation}
Then the cosine and sine of Wigner angles will be given by \eqref{eq:cos_alpha} and \eqref{eq:sin_alpha} with an additional overall sign in both sines and cosines for particle 2 and 3. Denoting the angles for the second analytic continuation by $\alpha_i^{(2)}$ in other words we see that\footnote{Since Wigner d-matrices are $4\pi$ periodic we might make an overall sign mistake in the final crossing equation by choosing \eqref{eq:angles_ac2}.}
\begin{equation}
\label{eq:angles_ac2}
\alpha_1^{(2)}=\alpha_1,\quad
\alpha_2^{(2)}=\alpha_2+\pi,\quad
\alpha_3^{(2)}=\alpha_3+\pi,\quad
\alpha_4^{(2)}=\alpha_4.
\end{equation}
Using the crossing relation \eqref{eq_2} and plugging it into \eqref{eq:transformation_23} we simply get
\begin{multline}
T_{12\rightarrow 34}{\;}_{\lambda_1, \lambda_2}^{\lambda_3,\lambda_4}
(p_1^\text{com},
p_2^\text{com},
p_3^\text{com},
p_4^\text{com})
=
\epsilon_{23}^{(2)}\sum_{\lambda_i'}
d^{(j_1)}_{\lambda'_1\lambda_1}(\alpha_1^{(2)})
d^{(j_2)}_{\lambda'_2\lambda_2}(\alpha_2^{(2)})\\
d^{(j_3)}_{\lambda'_3\lambda_3}(\alpha_3^{(2)})
d^{(j_4)}_{\lambda'_4\lambda_4}(\alpha_4^{(2)})
T_{1\bar{3}\rightarrow \bar{2}4}{\;}_{+\lambda_1', -\lambda_3'}^{-\lambda_2',+\lambda_4'}(\hat p_1^\text{com},-\hat p_3^\text{com}, -\hat p_2^\text{com}, \hat p_4^\text{com}).
\label{eq:crossing_23_ac_2}
\end{multline} 
Plugging \eqref{eq:relation_minus_theta} and \eqref{eq:angles_ac2} in the right-hand side of \eqref{eq:crossing_23_ac_2} and using the definition \eqref{eq:scalar_Tamplitude_definition} we arrive at the final equation
\begin{multline}
\label{eq:com_crossing_23_ac2}
T_{12\rightarrow 34}{\;}_{\lambda_1, \lambda_2}^{\lambda_3,\lambda_4}
(s,t,u)
=
\epsilon_{23}^{(2)}\sum_{\lambda_i'}
e^{i\pi(\lambda_1'-\lambda_3'-\lambda_2'+\lambda_4')}
d^{(j_1)}_{\lambda'_1\lambda_1}(\alpha_1)
d^{(j_2)}_{\lambda'_2\lambda_2}(\alpha_2+\pi)\\
d^{(j_3)}_{\lambda'_3\lambda_3}(\alpha_3+\pi)
d^{(j_4)}_{\lambda'_4\lambda_4}(\alpha_4)
T_{1\bar{3}\rightarrow \bar{2}4}{\;}_{+\lambda_1', -\lambda_3'}^{-\lambda_2',+\lambda_4'}(t,s,u).
\end{multline} 
Renaming the dummy indices $\lambda_2'\rightarrow -\lambda_2'$ and $\lambda_3'\rightarrow -\lambda_3'$ and using the following property 
\begin{equation}
e^{i\pi j} d^{(j)}_{\lambda'\lambda}(\omega)=
e^{i\pi \lambda'} d^{(j)}_{-\lambda'\lambda}(\omega+\pi)
\end{equation}
it is straightforward to see that \eqref{eq:com_crossing_23_ac2} is equivalent to \eqref{eq:com_crossing_23_ac1} up to an overall phase.

\section*{Concluding remarks}
Let us conclude by stressing that Wigner small d-matrices obey the following property
\begin{equation}
d^{(\ell)}_{\lambda'\lambda}(\omega+2\pi)=
e^{2\pi i\, \ell} d^{(\ell)}_{\lambda'\lambda}(\omega).
\end{equation}
This means that for bosonic particles Wigner small d-matrices are $2\pi$ periodic instead for fermionic particles they are $4\pi$ periodic.
As a result the  knowledge of \eqref{eq:cos_alpha} and \eqref{eq:sin_alpha} does not always fix the Wigner angles $\alpha_i$ uniquely since the sine and cosine are $2\pi$ periodic functions. This issue appears only for fermionic particles and causes an ambiguity in the overall phase in the crossing equations. Extra input is needed to fix this ambiguity.

\subsubsection{$s-u$ crossing equation}
We now consider the $s-u$ crossing equation due to (2-4) permutation in the center of mass frame. It is given by
\begin{multline}
\label{eq:final_su}
T_{12\rightarrow 34}{\;}_{\lambda_1, \lambda_2}^{\lambda_3,\lambda_4}
(s,t,u)
=
\epsilon_{24}^{(1)}\sum_{\lambda_i'}
e^{i\pi(\lambda_1'+\lambda_3')}\\
d^{(j_1)}_{\lambda'_1\lambda_1}(\beta_1)
d^{(j_2)}_{\lambda'_2\lambda_2}(\beta_2)
d^{(j_3)}_{\lambda'_3\lambda_3}(\beta_3)
d^{(j_4)}_{\lambda'_4\lambda_4}(\beta_4)
T_{1\bar{4}\rightarrow 3\bar{2}}{\;}_{\lambda_1', \lambda_4'}^{\lambda_3',\lambda_2'}(u,t,s).
\end{multline} 
The cosines of Wigner angles read as\footnote{The values computed here match precisely the ones of formula (26) in \cite{Hara:1971kj}. Notice that our process $12\rightarrow 34$ corresponds to their $ac\rightarrow bd$. Thus, in order to see the equivalence one needs to identify the labels as $1=a$, $2=c$, $3=b$ and $4=d$.}
\begin{equation}
\label{eq:cos_beta}
\begin{aligned}
\cos\beta_1 &= \frac{+(s+m_1^2-m_2^2)(u+m_1^2-m_4^2)+2m_1^2\,\Delta_u}
{\mathcal{L}_{12}(s)\mathcal{L}_{14}(u)},\\
\cos\beta_2 &= \frac{-(s+m_2^2-m_1^2)(u+m_2^2-m_3^2)+2m_2^2\,\Delta_u}
{\mathcal{L}_{12}(s)\mathcal{L}_{23}(u)},\\
\cos\beta_3 &= \frac{+(s+m_3^2-m_4^2)(u+m_3^2-m_2^2)+2m_3^2\,\Delta_u}
{\mathcal{L}_{34}(s)\mathcal{L}_{23}(u)},\\
\cos\beta_4 &= \frac{-(s+m_4^2-m_3^2)(u+m_4^2-m_1^2)+2m_4^2\,\Delta_u}
{\mathcal{L}_{34}(s)\mathcal{L}_{14}(u)}.
\end{aligned}
\end{equation}
The sines of Wigner angles read as
\begin{equation}
\label{eq:sin_beta}
\begin{aligned}
\sin\beta_1 &=  +\frac{2m_1\,\sqrt{\Phi}}
{\mathcal{L}_{12}(s)\mathcal{L}_{14}(u)},\quad
\sin\beta_2 =  -\frac{2m_2\,\sqrt{\Phi}}
{\mathcal{L}_{12}(s)\mathcal{L}_{23}(u)},\\
\sin\beta_3 &= -\frac{2m_3\,\sqrt{\Phi}}
{\mathcal{L}_{34}(s)\mathcal{L}_{23}(u)},\quad
\sin\beta_4 = +\frac{2m_4\,\sqrt{\Phi}}
{\mathcal{L}_{34}(s)\mathcal{L}_{14}(u)},
\end{aligned}
\end{equation}
where the objects $\Phi$ and $\Delta_u$ are defined in \eqref{eq:definition_Phi} and \eqref{eq:definition_delta}.

\section{Perturbative amplitudes}
\label{app:perturbative_amplitudes}

To complement the discussion of the main text we derive several perturbative results in this section. The computations done in this section closely follow Part II of \cite{Srednicki:2007qs}. Let us start by considering the following free Lagrangian density
\begin{equation}
\mathcal{L}^\Psi_{free} \equiv \frac{i}{2}\overline \Psi \gamma^\mu \partial_\mu \Psi - \frac{1}{2}m \overline \Psi \Psi,
\end{equation}
where $\Psi$ is the four component Majorana field. It obeys the Majorana condition
\begin{equation}
\label{eq:majorana_condition}
\overline \Psi = \Psi^{T} \mathcal C,
\end{equation}
where $\mathcal{C}$ is the charge conjugation matrix defined as
\begin{equation}
\mathcal C \equiv \begin{pmatrix}
0 & -1 & 0 & 0 \\
+1 & 0 & 0 & 0 \\
0 & 0& 0 & +1 \\
0 & 0 & -1 & 0 
\end{pmatrix}.
\end{equation}
When acting on the vacuum Majorana field creates a neutral spin $\frac{1}{2}$ particle. The intrinsic parity, defined in \eqref{eq:parity_1PS_main}, for such a particle can only be
\begin{equation}
\label{eq:intrinsic}
\eta=\pm i.
\end{equation} 
This is the only option compatible with the Majorana condition \eqref{eq:majorana_condition}.

In what follows we will compute scattering amplitudes of spin $\frac{1}{2}$ particles in Fermi, Yukawa and pseudo-Yukawa theories to the leading order. We will conclude with a brief discussion on 
counting interaction terms in the effective Lagrangian using scattering amplitudes.

\subsection{Fermi theory}
\label{app:fermi_theory}
Consider the Fermi theory defined by the following Lagrangian density
\begin{equation}
\label{eq:fermi_lagrangian}
\mathcal{L} = \mathcal{L}^\Psi_{free} +
\frac{\lambda}{8 m^2}\,(\overline \Psi \Psi) (\overline \Psi \Psi),
\end{equation}
where $\lambda$ is the coupling known as the Fermi constant. After performing a standard computation one can obtain the following expressions for the scattering amplitudes of Majorana particles to the leading order
\begin{multline}
\label{eq:amplitude_fermi}
T_{\lambda_1, \lambda_2}^{\lambda_3, \lambda_4} (p_1, p_2, p_3, p_4) =
 \frac{\lambda}{m^2} \times\Big( 
[\bar u_{\lambda_3} (p_3) u_{\lambda_1}(p_1)] [\bar u_{\lambda_4} (p_4) u_{\lambda_2}(p_2)]\\
-[\bar u_{\lambda_4} (p_4) u_{\lambda_1}(p_1)] [\bar u_{\lambda_3} (p_3)u_{\lambda_2}(p_2)]
+ [\bar v_{\lambda_2} (p_2) u_{\lambda_1}(p_1)] [\bar u_{\lambda_3} (p_3) v_{\lambda_4}(p_4)] \Big),
\end{multline} 
where $u$ and $v$ are the spinor solutions of the Dirac equation. Their explicit form is given in \eqref{eq:Dirac_solutions}. By using the five tensor structures \eqref{tensor_structures_4_majorana} and Fierz identities one can bring the result \eqref{eq:amplitude_fermi}, as expected from \eqref{eq:tensor_structures}, to the following form
\begin{equation}
\label{eq:decomposition_structures}
T_{\lambda_1, \lambda_2}^{\lambda_3,\lambda_4}(p_1,p_2,p_3,p_4) = 
\sum_{I=1}^{5} H_I(s,t,u)\times \mathbb{T}_I{}_{\lambda_1, \lambda_2}^{\lambda_3,\lambda_4}(p_1,p_2,p_3,p_4),
\end{equation}
where the functions $H_I$ (with $I=1,\ldots,5$) denoted collectively by $\vec H$  read as
\begin{equation}
\label{eq:amplitude_fermi_final}
\vec H(s, t , u) = \frac{\lambda}{m^2} \times
\begin{pmatrix}
1 \\
0\\
0\\
1\\
-1
\end{pmatrix}.
\end{equation}

\subsection{Yukawa theory}
We consider now the Yukawa theory defined by the following Lagrangian density 
\begin{equation}
\label{eq:Yukawa_lagrangian}
\mathcal{L}=\mathcal{L}^\Psi_{free} -\frac{1}{2}
\partial_\mu\varphi\partial^\mu\varphi - \frac{1}{2}M^2 \varphi^2
	+\frac{1}{2}g \varphi \overline \Psi  \Psi.
\end{equation}
Due to \eqref{eq:intrinsic} the interaction is parity invariant only if the scalar field has the intrinsic parity
\begin{equation}
\eta_\phi=+1.
\end{equation}
Again performing a standard computation one gets the following scattering amplitude of neutral spin $\frac{1}{2}$ particles to the leading order
\begin{multline}
\label{eq:amplitude_yukawa}
T_{\lambda_1, \lambda_2}^{\lambda_3, \lambda_4} (p_1, p_2, p_3, p_4) =  g^2\times
\Bigg(
\frac{[\bar u_{\lambda_3} (p_3) u_{\lambda_1}(p_1)] [\bar u_{\lambda_4} (p_4) u_{\lambda_2}(p_2)]}{-t + M^2}\\
-\frac{[\bar u_{\lambda_4} (p_4) u_{\lambda_1}(p_1)] [\bar u_{\lambda_3} (p_3)u_{\lambda_2}(p_2)]}{-u + M^2}
+\frac{[\bar v_{\lambda_2} (p_2) u_{\lambda_1}(p_1)] [\bar u_{\lambda_3} (p_3) v_{\lambda_4}(p_4)]}{-s+M^2}
\Bigg).
\end{multline} 
By using the five tensor structures \eqref{tensor_structures_4_majorana} one can bring the above expression to the form \eqref{eq:decomposition_structures}, where
\begin{equation}
\label{eq:Yukawa_pert}
\vec H(s, t , u) = \frac{g^2}{2} \times
\begin{pmatrix}
-\frac{4}{s-M^2} + \frac{1}{t-M^2} + \frac{1}{u-M^2} \\
\frac{1}{t-M^2} - \frac{1}{u-M^2} \\
-\frac{1}{t-M^2} + \frac{1}{u-M^2}\\
-\frac{1}{t-M^2} - \frac{1}{u-M^2}\\
\frac{1}{t-M^2} + \frac{1}{u-M^2}
\end{pmatrix}.
\end{equation}

\subsection{Pseudo-Yukawa theory}
The pseudo-Yukawa theory is defined by the Lagrangian density 
\begin{equation}
\label{eq:pseudoYukawa_lagrangian}
\mathcal{L}=\mathcal{L}^\Psi_{free} - \frac{1}{2}
\partial_\mu \tilde \varphi\partial^\mu \tilde \varphi - \frac{1}{2}M^2 \tilde \varphi^2
+\frac{1}{2}\tilde g \tilde \varphi \overline \Psi \gamma^5 \Psi.
\end{equation}
Due to \eqref{eq:intrinsic} the interaction is parity invariant only if the scalar field has the intrinsic parity
\begin{equation}
\eta_{\tilde{\phi}}=-1,
\end{equation}
hence we refer to $\tilde{\phi}$ as  pseudo-scalar.
As before we compute the scattering amplitude to the leading order and obtain
\begin{multline}
\label{eq:amplitude_pseudo}
T_{\lambda_1, \lambda_2}^{\lambda_3, \lambda_4} (p_1, p_2, p_3, p_4) =  \tilde g^2\times
\Bigg(
\frac{[\bar u_{\lambda_3} (p_3) \gamma^5 u_{\lambda_1}(p_1)] [\bar u_{\lambda_4} (p_4)\gamma^5 u_{\lambda_2}(p_2)]}{-t + M^2}\\
-\frac{[\bar u_{\lambda_4} (p_4)\gamma^5 u_{\lambda_1}(p_1)] [\bar u_{\lambda_3} (p_3)\gamma^5u_{\lambda_2}(p_2)]}{-u + M^2} +\frac{[\bar v_{\lambda_2} (p_2) \gamma^5 u_{\lambda_1}(p_1)] [\bar u_{\lambda_3} (p_3) \gamma^5 v_{\lambda_4}(p_4)]}{-s+M^2} 
\Bigg).
\end{multline} 
Again by using the five tensor structures \eqref{tensor_structures_4_majorana} one can bring the above expression to the form \eqref{eq:decomposition_structures}, where
\begin{equation}
\label{eq:pseudoYukawa_pert}
\vec H(s, t , u) = \frac{\tilde g^2}{2} \times
\begin{pmatrix}
\frac{1}{t-M^2} + \frac{1}{u-M^2} \\
-\frac{1}{t-M^2}+ \frac{1}{u-M^2} \\
-\frac{1}{t-M^2} +\frac{1}{u-M^2}\\
\frac{1}{t-M^2} + \frac{1}{u-M^2}\\
-\frac{4}{s-M^2} + \frac{1}{t-M^2} + \frac{1}{u-M^2}
\end{pmatrix}.
\end{equation}

\subsection{Counting couplings at a given order in EFT}
Let us consider the effective Lagrangian density of a single Majorana field $\Psi$ describing the two to two scattering process schematically denoted by $\Psi\Psi\rightarrow \Psi\Psi$. It reads
\begin{equation}
\mathcal{L}=\mathcal{L}^\Psi_{free} +
\mathcal{L}_4+\mathcal{L}_5+\mathcal{L}_6+\ldots,
\end{equation}
where $\mathcal{L}_n$ with $n=4,5,6$ are the dimension 4, 5 and 6 terms. The question we will address now is how to count the number of linearly independent terms in such an effective Lagrangian density at each order $n$.

We start with $n=4$. It is well known that there is only one linearly independent term $(\overline \Psi \Psi) (\overline \Psi \Psi)$ as was used in \eqref{eq:fermi_lagrangian}. Naively, one can write however many more terms by appropriately combining
\begin{equation}
\overline \Psi \Psi, \quad \overline \Psi \gamma^5 \Psi, \quad \overline \Psi \gamma^\mu \Psi, \quad \overline \Psi \gamma^\mu \gamma^5 \Psi, \quad \overline\Psi \sigma^{\mu\nu} \Psi. 
\end{equation}
Let us now rewrite the Majorana field in terms of a two component left-handed Weyl spinor $\chi$, one has 
\begin{equation}
\Psi = \begin{pmatrix}
\chi_{\alpha} \\
\chi^\dagger{}^{\dot\alpha}
\end{pmatrix}.
\end{equation}
It is then straightforward to show that\footnote{Note that all the Weyl indices here have been contracted appropriately.}
\begin{eqnarray}
\overline \Psi \gamma^\mu \Psi &=& \chi^\dagger \sigma^\mu \chi - \chi^\dagger \sigma^\mu \chi = 0, \\
\frac{i}{2} \overline \Psi \sigma^{\mu\nu} \Psi &=& \chi \sigma^{\mu \nu} \chi - \chi^\dagger \overline \sigma^{\mu \nu}\chi = 0.
\end{eqnarray}
As a result at the $n=4$ level we can write only four terms
\begin{equation}
\mathcal{L}_4 \ni \left\{
(\overline \Psi \Psi) (\overline \Psi \Psi),\;
(\overline \Psi \gamma^5 \Psi) (\overline \Psi \gamma^5 \Psi),\;
(\overline \Psi \gamma^\mu \gamma^5 \Psi)(\overline \Psi \gamma_\mu \gamma^5 \Psi),\; 
(\overline \Psi \Psi)(\overline \Psi \gamma^5 \Psi)
\right\}.
\end{equation} 
Rewriting these in terms of the Weyl spinor $\chi$ we see that they are either proportional to each other or vanish
\begin{eqnarray}
(\overline \Psi \Psi) (\overline \Psi \Psi) &=& \left(\chi \chi + \chi^\dagger \chi^\dagger\right)^2 = 2(\chi \chi )(\chi^\dagger  \chi^\dagger),\nonumber \\
(\overline \Psi \gamma^5 \Psi) (\overline \Psi \gamma^5 \Psi)&=& \left(-\chi \chi + \chi^\dagger \chi^\dagger\right)^2 = -2(\chi \chi )(\chi^\dagger  \chi^\dagger), \nonumber\\
(\overline \Psi \gamma^\mu \gamma^5 \Psi)(\overline \Psi \gamma_\mu \gamma^5 \Psi)&=& \left(-  2\chi^\dagger \sigma^\mu \chi\right)\left(-  2\chi^\dagger \sigma_\mu \chi\right) = 4(\chi \chi )(\chi^\dagger  \chi^\dagger),\nonumber \\
(\overline \Psi \Psi)(\overline \Psi \gamma^5 \Psi)&=&\left(\chi \chi + \chi^\dagger \chi^\dagger\right)\left(-\chi \chi + \chi^\dagger \chi^\dagger\right)= 0.
\end{eqnarray}

If one is interested in the counting of independent terms only, instead of performing the above algebra one could notice that $\mathcal{L}_4$ terms in the leading order generate the improved scattering amplitudes $\vec H(s,t,u)$ at the crossing symmetric point \eqref{eq:crossing_symmetric}. Crossing equations severely restrict the form of the improved amplitudes at the crossing symmetric point, see \eqref{eq:csp}. The latter contains only one independent parameter. One concludes that there should be only a single parameter in the effective Lagrangian density $\mathcal{L}_4$.

As another example let us consider the $n=6$ part. One can write
\begin{equation}
\label{eq:terms_6}
\mathcal{L}_6 \ni \left\{
(\overline \Psi \partial^2\Psi) (\overline \Psi\Psi),\;
(\overline \Psi \partial^\mu\Psi) (\overline \Psi\partial_\mu \Psi),\;
(\overline \Psi \partial^\mu\partial^\nu\Psi) (\overline \Psi\gamma_\mu\gamma_\nu \Psi),\ldots
\right\}.
\end{equation}
The improved amplitudes generated by such terms at the leading order will have the following most general form
\begin{equation}
\label{eq:general_form}
\vec H(s,t,u) = (s-4m^2/3)\times \vec A + (t-4m^2/3)\times \vec B 
+ (u-4m^2/3)\times \vec C,
\end{equation}
where $\vec A$, $\vec B$ and $\vec C$ are some real constants. They are constrained by the crossing equations \eqref{eq:crossing_st_fermions_amp_improved} and \eqref{eq:crossing_su_fermions_amp_improved} which require \eqref{eq:general_form} to take the following form
\begin{equation}
\vec H(s,t,u) = \frac{1}{m^4}
\begin{pmatrix}
a\times(s-4 m^2/3) \\
b \times(s+2t-4m^2) \\
\frac{1}{3}(a + 2b)\times(s+2t-4m^2)\\
b\times(s-4m^2/3)\\
(a + 4 b)\times(s-4m^2/3)
\end{pmatrix},
\end{equation}
where $a$ and $b$ are the undetermined parameters. One concludes that there are only two linearly independent terms in \eqref{eq:terms_6}.

\section{Bound state close to the two-particle threshold}
\label{app:close_to_threshold}

In section \ref{sec:non-perturbative_couplings} we studied the non-perturbative structure of the scattering amplitude of neutral spin $\frac{1}{2}$ particles with mass $m$ in the presence of a scalar particle with mass $M$. Such a particle can be interpreted as a bound state of two fermions. The structure of the improved scattering amplitude is given by \eqref{eq:object_P_scalar_part} and \eqref{eq:object_P_scalar} for the parity even scalar particle and by \eqref{eq:object_P_pseudoscalar_part} and \eqref{eq:object_P_pseudoscalar} for parity odd scalar particle. The structure of the center of mass amplitudes is obtained by plugging these into \eqref{eq:ksfa_definition}.

In this appendix we study the behavior of the center of mass amplitudes in the presence of bound states in the limit when $M\rightarrow 2m$. We follow the analysis presented in appendix E of \cite{Paulos:2017fhb}. The leading behavior of the COM amplitudes is obtained by taking the following limit
\begin{equation}
\label{eq:limit}
M=(2-\epsilon)\,m,\quad
s=(2m+E \epsilon)^2,
\end{equation}
where $E\ge 0$ is kept  fixed as we take the limit $\epsilon\rightarrow 0$. Applying it to \eqref{eq:ksfa_definition} combined with \eqref{eq:object_P_scalar_part} and \eqref{eq:object_P_pseudoscalar_part} in the leading order in $\epsilon$ we get
\begin{align}
\label{eq:threshold_bound_state}
\vec\Phi_\text{scalar}(E) = 
\frac{g^2}{4}\frac{E}{E + m}\times
\begin{pmatrix}
-1 \\
+1 \\
0\\
0\\
0
\end{pmatrix},\quad
\vec\Phi_\text{pseudoscalar}(E) = \frac{\tilde g^2}{4}\frac{m}{E + m}\frac{1}{\epsilon}\times\begin{pmatrix}
1 \\
1\\
0\\
0\\
0
\end{pmatrix}.
\end{align}
We see that the amplitude with the scalar particle is completely finite in this limit, however the amplitude with the pseudoscalar particle diverges. The partial amplitudes were defined in \eqref{eq:partial_amplitudes}. Plugging there the expressions \eqref{eq:threshold_bound_state}, replacing $s$ by $M^2$ and only then taking the limit \eqref{eq:limit} we get
\begin{equation}
\label{eq:threshold_bound_state_partial}
\begin{aligned}
\vec\Phi^{\ell=0}_\text{scalar}(E) &= 
\frac{ig^2}{64\pi}\frac{1}{1 + m/E}\,\sqrt{\epsilon}\times
\begin{pmatrix}
-1 \\
+1 \\
0\\
0\\
0
\end{pmatrix},\\
\vec\Phi^{\ell=0}_\text{pseudoscalar}(E) &= \frac{i\tilde g^2}{64\pi}\frac{1}{1 + E/m}\frac{1}{\sqrt\epsilon}\times\begin{pmatrix}
1 \\
1\\
0\\
0\\
0
\end{pmatrix}.
\end{aligned}
\end{equation}
Notice that poles coming from scalar particles can appear only in $\ell=0$ partial amplitudes as is explicitly stated here. 

According to \eqref{eq:objects_parity} we can take combination of partial amplitude components to define  parity even $\Phi_+^\ell(s)$ and parity odd  $\Phi_-^\ell(s)$ partial amplitudes. In terms of the objects \eqref{eq:objects_parity} unitarity takes a very simple form
\begin{equation}
\label{eq:unitarity_bound_states}
|1+i\Phi_+^\ell(s)|\leq 1,\quad
|1+i\Phi_-^\ell(s)|\leq 1.
\end{equation}
Using \eqref{eq:threshold_bound_state_partial} in our context we can the write
\begin{align}
\label{eq:scalars}
\text{scalar exchange:}&\qquad
\Phi_+^{\ell=0}(E)= -\frac{g^2}{32\pi}\frac{1}{1 + m/E}\,\sqrt{\epsilon},\quad
\Phi_-^{\ell=0}(E)= 0,\\
\label{eq:pseudoscalars}
\text{pseudoscalar exchange:}&\qquad
\Phi_-^{\ell=0}(E)= \frac{\tilde g^2}{32\pi}\frac{1}{1 + E/m}\frac{1}{\sqrt\epsilon},\quad\;\;\,
\Phi_+^{\ell=0}(E)= 0.
\end{align}

In the limit $\epsilon\rightarrow 0$ the partial amplitudes \eqref{eq:scalars} vanish and we cannot say anything interesting about the coupling $g^2$. Instead \eqref{eq:pseudoscalars} diverges.
In order to have a partial amplitudes which is able to satisfy unitarity \eqref{eq:unitarity_bound_states} we need the scaling
\begin{equation}
\label{eq:requirement}
\tilde g^2 = a \sqrt{\epsilon},
\end{equation}
with $a$ finite as $\epsilon \to 0$. It is also convenient to make a change of variables from $E$ to $z$ variable which are related via
\begin{equation}
E=-m\,\frac{(z-1)^2}{(z+1)^2}.
\end{equation}
This maps the cut in the $E$ plane to the boundary of the unit disk in $z$. 
Plugging the above into \eqref{eq:pseudoscalars} we get
\begin{equation}
\Phi_-^{\ell=0}(z)= \frac{ia}{128\pi} \frac{(1+z)^2}{z}.
\end{equation}
Using it and taking the leading behavior in small $z$, the unitarity conditions \eqref{eq:unitarity_bound_states} leads to
\begin{equation}
\label{eq:final_condition}
\left|\frac{a}{128\pi} \frac{1}{z}\right|\leq 1,
\end{equation}
which should be satisfied on the boundary of the disc described by $z=e^{i \phi}$, where $\phi\in[0,2\pi]$. It is then straightforward to see that the maximally allowed value of $a$ which obeys the unitarity condition \eqref{eq:final_condition} is
\begin{equation}
a=128\pi.
\end{equation}
Plugging it into \eqref{eq:requirement} and expressing $\epsilon$ in terms of $m$ and $M$ from \eqref{eq:limit} we get the analytic upper bound
\begin{equation}
\label{eq:analytic_bound}
\tilde g^2 \leq 128 \pi \sqrt{\frac{2m - M}{m}}.
\end{equation}

\section{General spin tensor structures}
\label{sec:general_spin_tensor_structures}
We have introduced the notion of tensor structures in \eqref{eq:tensor_structures}. Even though one can completely avoid talking about them, it is sometimes beneficial to know a basis of tensor structures explicitly. In the case of Majorana fermions the detailed discussion of tensor structures was given in section \ref{sec:crossing_alternative}. In this appendix we will briefly explain how to construct tensor structures for amplitudes with generic spin. There are several possible ways of doing this. One way is to treat particles with generic spin as multi-spinors simply described by tensor products of $u$ and $v$ objects defined in section \ref{sec:crossing_alternative}. This approach was employed in \cite{deRham:2017zjm}. Here we describe another approach used in \cite{Arkani-Hamed:2017jhn}. (In the massless case it reduces to the well known spinor-helicity formalism, see for example \cite{Cheung:2017pzi}.) We chose the latter because it closely resembles various approaches used in the CFT literature \cite{Costa:2011mg,Costa:2011dw,Iliesiu:2015qra,SimmonsDuffin:2012uy,Elkhidir:2014woa,Cuomo:2017wme}.

\subsection*{Index free formalism}
In section \ref{sec:states}, more precisely in equation \eqref{poincare_irreps}, we have chosen the basis of states to be $|c, \vec p;  \spin, \hel\>$, where $\lambda=-\spin,\ldots,+\spin$ are the helicity labels. It is convenient to move to another basis where instead of helicities $\lambda$ we use a symmetrized set of indices
\begin{equation}
(a_1\ldots a_{2\spin}),
\end{equation}
where $a_1,\,a_2,\ldots$ are the indices in the fundamental representation of the $SU(2)$ Little group.\footnote{Here we simply use the fact that any generic irreducible representation $j$ of the $SU(2)$ can be represented as $\ell = \left(\frac{1}{2}\otimes \ldots \otimes \frac{1}{2}\right)_{\text{sym}}$.} In other words we can have two equivalent bases
\begin{equation}
\label{eq:two_bases}
|c, \vec p;  \spin, \hel\>
\quad\leftrightarrow\quad
|c, \vec p;  \spin\>^{(a_1\ldots a_{2\spin})}.
\end{equation}
It is then extremely convenient to introduce the index free notation by contracting the states with a complex vector (spinor polarization)
\begin{equation}
\label{eq:components_of_s}
s^{a}= \begin{pmatrix}
\xi\\
\eta
\end{pmatrix},
\end{equation}
where $\xi$ and $\eta$ are simply  the components of the spinor polarization. With the help of \eqref{eq:components_of_s} one can define the Little group index-free states
\begin{align}
\label{eq:index-free_states}
|c, \vec p;  \spin \>(s) &\equiv
|c, \vec p; \spin \>^{(a_1\ldots a_{2\spin})}\times
s_{a_1}\ldots s_{a_{2\spin}}.
\end{align}
The relation between the two bases \eqref{eq:two_bases} can be determined by requiring
\begin{align}
|c, \vec p;  \spin \>(s) =
\sum_{\lambda=-\spin}^{\spin}|c,\vec p;\spin,\lambda\rangle
\times \xi^\lambda \eta^{\spin - \lambda}.
\end{align}

We have defined in \eqref{eq:scattering_amplitude_interacting} the interacting part of the scattering amplitude of four particles. In index free notation it reads
\begin{multline}
(2\pi)^4\delta^{(4)}(p_1^\mu + p_2^\mu - p_3^\mu - p_4^\mu) \times
T_{12\rightarrow 34}(p_i,s_i)\equiv\\
\Big((s_3)\<m_3,\vec p_3;\ell_3|\otimes(s_4)\<m_4,\vec p_4;\ell_4|\Big)
T
\Big(|m_1,\vec p_1;\ell_1\>(s_1)\otimes|m_2,\vec p_2;\ell_2\>(s_2)\Big).
\end{multline}
Analogously to \eqref{eq:tensor_structures} we can perform the decomposition of the index free interacting scattering amplitudes
\begin{equation}
\label{eq:tensor_structure_decomposition}
T_{12\rightarrow 34}(p_i,s_i) = \sum_{I=1}^{(2j_1+1)\ldots(2j_4+1)} T_I(s,t,u)\;
\mathbb{T}_I(p_i,s_i),
\end{equation}
where $\mathbb{T}_I(p_i,s_i)$ are the index-free tensor structures.

\subsubsection*{Auxiliary objects}
\label{sec:auxiliary}
The index-free tensor structures appearing in \eqref{eq:tensor_structure_decomposition} are kinematic objects constructed from the 4-momenta $p_i^\mu$ and the spinor polarizations $s_i^a$. The former have Lorentz indices and the latter have  Little group indices. In order to contract them we need to somehow introduce an auxiliary object which has both   Lorentz and   Little group indices and can be an intermediary in the contraction of the two. In what follows we will define such an auxiliary object.

We use the two-component spinor notation of Wess and Bagger \cite{Wess:1992cp}. Given a 4-momentum $p^\mu$ one can define the standard $SL(2,C)$ matrices
using \eqref{4Pauli},
\begin{equation}
\label{momentum_SL2}
\mat p_{\alpha\dot\alpha}\equiv p\cdot \sigma_{\alpha\dot\alpha},
\qquad
\bar{\mat p}^{\dot\beta\beta}\equiv p\cdot \bar\sigma^{\dot\beta\beta}.
\end{equation}
The indices are raised and lowered by the $\epsilon$-symbols $\epsilon_{\alpha\beta}$ and $\epsilon_{\dot\alpha\dot\beta}$, where $\epsilon^{12}=-\epsilon_{12}=+1$.
The representation \eqref{momentum_SL2} leads to
\begin{align}
\label{eq:determinant}
p^2 &= -\det\mat p = -\det\bar{\mat p},\\
p_i\cdot p_j &= -\frac{1}{2}\,\text{tr}\,[\mat p_i\bar{\mat p}_j].
\end{align}
One can introduce the following two objects which have one Lorentz index $\beta$ or $\dot{\beta}$ and one Little group index $b=1,2$ (in the fundamental representation)\footnote{These ``spinor-helicity" variables are denoted by $\lambda$ and $\tilde \lambda$ in \cite{Arkani-Hamed:2017jhn}.}
\begin{equation}
\label{eq:lambda}
\aux_\beta{}^b,\quad
\bar\aux_b{}^{\dot\beta}
\end{equation}
related by hermitian conjugation
\begin{equation}
\Big(\aux_\beta{}^b\Big)^\dagger = \bar\aux{}_{b\dot\beta},\qquad
\Big(\bar\aux_b{}^{\dot\beta}\Big)^\dagger = \aux{}^{\beta b}.
\end{equation}
The Little group indices are raised and lowered by the $\epsilon$-symbol $\epsilon^{ab}=-\epsilon_{ab}$, where $\epsilon^{12}=+1$.
By taking these objects and contracting their Little group indices one can represent the $SL(2,C)$ matrices as
\begin{equation}
\label{eq:expressing_momentum_as_lambdas}
\mat p_{\alpha\dot\beta} =
\aux_{\alpha}{}^b\bar\aux_{b\,\dot\beta},\qquad
\bar{\mat p}^{\dot\alpha\beta} = 
\aux^{\beta b}\bar\aux_b{}^{\dot\alpha}.
\end{equation}
By definition the Little group transformations leave \eqref{eq:determinant} invariant. Using the representation \eqref{eq:expressing_momentum_as_lambdas} and this invariance one can compute the actual expressions of $\aux$ and $\bar\aux$. They read
\begin{equation}
\label{eq:explicit_expressions_lambda_lambda_bar}
\text{matrix}\; \aux_{\beta}{}^b =
\text{matrix}\; \bar\aux_{b\dot\beta} =
\frac{1}{\sqrt{2}\,\sqrt{m+p^0}}\;
\begin{pmatrix}
m+p^0+p^3 & p^1-i\,p^2 \\
p^1+i\,p^2  & m+p^0-p^3
\end{pmatrix}.
\end{equation}

\subsubsection*{Tensor invariants}
\label{sec:tensor_invariants}
We can build tensor structures in \eqref{eq:tensor_structure_decomposition} as products of elementary tensor invariants. These tensor invariants are in turn built out of
\begin{equation}
\label{eq:list_elements}
\mat p_i{}_{\alpha\dot\beta},\quad
\aux_i{}_\beta{}^b,\quad
\bar{\aux}_i{}_b{}^{\dot\beta},\quad
s_i^b
\end{equation}
by fully contracting their indices with all possible invariant objects such as the Kronecker and the Levi-Civita symbols. Notice also that one can pair $\aux_i$ and $s_j$ by contracting their Little group indices only if $i=j$ because each particle has its own little group. For transparency we indicate this contraction by ``$\cdot$''. Below we make a summary of all the possible invariants.

\subparagraph{Type I} consists of invariant objects with an even number of 4-momenta
\begin{equation}
\label{eq:type1}
\begin{aligned}
\<\pmb i \pmb j\> &\equiv
s_{i} \cdot \aux^{\alpha}_{i}
\;\delta_{\alpha}^{\beta}\; \aux_{j\,\beta}\cdot s_{j},\\
\<\pmb i m n\pmb j\> &\equiv s_{i}\cdot \aux^{\alpha}_{i}\;
{}_\alpha(\mat p_m \bar{\mat p}_n)^{\beta}\;
\aux_{j\,\beta}\cdot s_{j},\\
\<\pmb i m n p r\pmb j\> &\equiv s_{i}\cdot \aux^{\alpha}_{i}\;
{}_\alpha(\mat p_m \bar{\mat p}_n \mat p_p \bar{\mat p}_r)^{\beta}\;
\aux_{j\,\beta}\cdot s_{j},\\
\ldots & \ldots
\end{aligned}
\end{equation}

\subparagraph{Type I*} consists of structures related by complex conjugation to the ones of type I
\begin{equation}
\label{eq:type1*}
\begin{aligned}
[\pmb i \pmb j] &\equiv
s_{i}\cdot \bar\aux_{i\,\dot\alpha}
\;\delta_{\dot\beta}^{\dot\alpha}\; \bar\aux_{j}^{\dot\beta}\cdot s_{j},\\
[\pmb i m n \pmb j] &\equiv
s_{i}\cdot \bar\aux_{i\,\dot\alpha}
\;{}^{\dot\alpha}(\bar{\mat p}_m \mat p_n)_{\dot\beta}\; \bar\aux_{j}^{\dot\beta}\cdot s_{j},\\
[\pmb i m n p r \pmb j] &\equiv
s_{i}\cdot \bar\aux_{i\,\dot\alpha}
\;{}^{\dot\alpha}(\bar{\mat p}_m \mat p_n \bar{\mat p}_p \mat p_r)_{\dot\beta}\; \bar\aux_{j}^{\dot\beta}\cdot s_{j},\\
\ldots & \ldots
\end{aligned}
\end{equation}

\subparagraph{Type II} consists of invariant objects with an odd number of 4-momenta
\begin{equation}
\label{eq:type2}
\begin{aligned}
\<\pmb i m \pmb j] &\equiv
s_{i}\cdot \aux^{\alpha}_{i}
\;\mat p_{m\;\alpha\dot\beta}\; \bar\aux_{j}^{\dot\beta}\cdot s_{j},\\
\<\pmb i m n p \pmb j] &\equiv
s_{i}\cdot \aux^{\alpha}_{i}
\;{}_{\alpha}(\mat p_m \bar{\mat p}_n \mat p_p)_{\dot\beta}\; \bar\aux_{j}^{\dot\beta}\cdot s_{j},\\
\ldots & \ldots
\end{aligned}
\end{equation}

\subparagraph{Type II*} consists of structures related by complex conjugation to the ones of type II
\begin{equation}
\label{eq:type2*}
\begin{aligned}
[\pmb i m \pmb j\> &\equiv
s_{i}\cdot \bar\aux_{i\,\dot\alpha}
\;\bar{\mat p}_m^{\dot\alpha\beta}\; \aux_{j\,\beta}\cdot s_{j},\\
[\pmb i mnp \pmb j\> &\equiv
s_{i}\cdot \bar\aux_{i\,\dot\alpha}
\;{}^{\dot\alpha}(\bar{\mat p}_m \mat p_n \bar{\mat p}_p)^{\beta}\; \aux_{j\,\beta}\cdot s_{j},\\
\ldots & \ldots
\end{aligned}
\end{equation}

\subparagraph{Type III and III*} consists of invariant objects involving the $\epsilon_{\mu\nu\rho\sigma}$ or $\epsilon^{\mu\nu\rho\sigma}$ symbols
\begin{equation}
\label{eq:type3}
\begin{aligned}
\<\pmb i \overline{m n p} \pmb j ]
&\equiv  s_{i}\cdot\aux^{\alpha}_{i}\;
(p_{m\, \mu}\,p_{n\,\nu}\,p_{p\,\rho}\epsilon^{\mu \nu \rho \kappa}\sigma_\kappa)_{\alpha \dot\beta}\;
\bar \aux_j^{\dot \beta}\cdot s_{j},\\
[\pmb i \overline{mnp} \pmb j \>
&\equiv  s_{i}\cdot \bar\aux_{i\,\dot\alpha}\;
(p_{m\, \mu}\,p_{n\,\nu}\,p_{p\,\rho}\epsilon^{\mu \nu \rho \kappa}
\bar\sigma_\kappa)^{\dot\alpha\beta}\;
\aux_{j\,\beta}\cdot s_{j}.
\end{aligned}
\end{equation}

\subsection*{Basis of tensor structures}
There are a large number of relations among tensor invariants. For instance due to the following properties of $\sigma$-matrices
\begin{equation}
\begin{aligned}
\sigma^\mu \bar\sigma^\nu \sigma^\rho &= - \eta^{\mu\nu}\sigma^\rho
+\eta^{\mu\rho}\sigma^\nu - \eta^{\nu\rho}\sigma^\mu -
i\epsilon^{\mu \nu \rho \kappa}\sigma_\kappa,\\
\bar\sigma^\mu \sigma^\nu \bar\sigma^\rho &= - \eta^{\mu\nu}\bar\sigma^\rho
+\eta^{\mu\rho}\bar\sigma^\nu - \eta^{\nu\rho}\bar\sigma^\mu +
i\epsilon^{\mu \nu \rho \kappa}\bar\sigma_\kappa,
\end{aligned}
\end{equation}
see for example \cite{Dreiner:2008tw} for details, any invariant with many products of 4-momenta $(\mat p_m\,\bar{\mat p}_n\,\mat{p}_p\ldots)$ can be reduced to the ones involving at most two 4-momenta. As a result the most generic tensor structure can be represented by
\begin{equation}
\label{eq:covariant_structure}
\mathbb{T}_I=
\<\pmb i \pmb j\>^{A_{ij}}\;
[\pmb i \pmb j]^{B_{ij}}\;
\<\pmb i m \pmb j]^{C_{ij}}\;
\<\pmb i m n \pmb j\>^{D_{ij}}\;
[\pmb i m n \pmb j]^{E_{ij}}\;
\<\pmb i \overline{m n p} \pmb j ]^{F_{ij}},
\end{equation}
where $A,\,B,\,C\,D,\,E$ and $F$ are exponents fixed by the requirement\footnote{Notice that the exponent $F$ is either $0$ or $1$ since any pair of $\epsilon$-symbols can be written in terms of the metric.}${}^,$\footnote{Notice also that $\ell_1+\ell_2+\ell_3+\ell_4$ must be even otherwise one will never be able to fully contract all the Little group indices and form tensor invariants. This is a standard selection rule which comes out naturally from this formalism.}
\begin{eqnarray}
\mathbb{T}_I\propto s_1^{2\spin_1}s_2^{2\spin_2}s_3^{2\spin_3}s_4^{2\spin_4}.
\end{eqnarray}
The latter is simply the statement that the amplitude must be a polynomial in each $s_i$ with the degree fixed by the spin of the $i$th particle. This directly follows from the definition of the index-free states \eqref{eq:index-free_states}.

Constructing all the possible structures according to \eqref{eq:covariant_structure} still gives a set of linearly dependent objects. Eliminating all the dependent structures and forming the basis is the most challenging part of the formalism. It can be done for particles with low spin, but it does not seem to be a viable procedure for higher spin particles. Below we simply give a taste of what kind of relations one could expect.

First, one can show that
\begin{equation}
\<\pmb i m \pmb j] = + [\pmb j m \pmb i\>,\qquad
\<\pmb i \overline{m n p} \pmb j ] = +
[\pmb j \overline{mnp} \pmb i \>.
\end{equation}
This is the reason why type II* and type III* invariants have not being included in \eqref{eq:covariant_structure}. Second, due to
\begin{equation}
\sigma^\mu \bar\sigma^\nu + \sigma^\nu \bar \sigma^\mu = 2 \eta^{\mu\nu}
\end{equation}
one can show that
\begin{equation}
\begin{aligned}
\<\pmb i \pmb j\> &= - \<\pmb j \pmb i\>,\quad
\<\pmb i m n\pmb j\> = - \<\pmb j n m\pmb i\>,\\
[\pmb i \pmb j] &= - [\pmb j \pmb i],\quad\;\,
[\pmb i m n\pmb j] = - [\pmb j n m\pmb i]
\end{aligned}
\end{equation}
together with
\begin{equation}
\begin{aligned}
\<\pmb i m n\pmb j\> &+ \<\pmb i n m\pmb j\> = 2\,(k_m\cdot k_n)\, \<\pmb i \pmb j\>,\\
[\pmb i m n\pmb j] &+ [\pmb i n m\pmb j] \;= 2\,(k_m\cdot k_n)\, [\pmb i \pmb j].
\end{aligned}
\end{equation}
These relations enforce for instance that type I and I* structures must vanish unless $\pmb i \neq \pmb j$ and that without loss of generality one can choose $m<n$.
Third, one can write a number of Schouten identities. Some of them are 
\begin{equation}
\begin{aligned}
\< \pmb i \pmb j\>\<\pmb k \pmb l\>  + \< \pmb i \pmb k\>\<\pmb j \pmb l\> + \<\pmb i \pmb l\>\<\pmb j \pmb k\> &= 0,\\
\<\pmb i \pmb j \> \< \pmb k l \pmb m{]} + \<\pmb i \pmb k \> \<\pmb j l \pmb m{]} + \<\pmb j \pmb l \> \<\pmb i l \pmb m{]} &= 0, \\
\<\pmb i j \pmb k{]} \<\pmb l m \pmb n {]} + \<\pmb i \pmb l \> {[}\pmb k j m \pmb n {]}  +\<\pmb l j \pmb k{]}  \< \pmb i m \pmb n {]}  &= 0.
\end{aligned}
\end{equation}
Finally, one should take into account  the conservation of 4-momenta
\begin{equation}
p_1^\mu+p_2^\mu=p_3^\mu+p_4^\mu
\end{equation}
and its consequences.

\subsection*{Partial amplitudes}
For completeness let us mention that using tensor structures one can also compute partial amplitudes.

In \eqref{eq:decomposition} we have shown how to decompose the scattering amplitudes into partial amplitudes by injecting a complete set of states. The main objects in this decomposition to be determined is the following matrix element
\begin{equation}
\label{eq:CG_repeated}
\<\kappa_1,\kappa_2|c,\vec p,\spin,\lambda\>,
\end{equation} 
where $|\kappa_1\>$ and $|\kappa_2\>$ are the 1PS and $|c,\vec p,\spin,\lambda\>$ is a generic irrep with spin $\ell$ and helicity $\lambda$. The objects \eqref{eq:CG_repeated} are the Clebsch-Gordan coefficients of the decomposition. They were computed in appendix \ref{app:clebsch_gordan_coefficients} in full generality using group-theoretic arguments. In the COM frame they are basically the Wigner d-matrices.

We can repeat this procedure in the index-free formalism by injecting a complete set of states in the following form
\begin{equation}
\label{eq:2PS_decomposition_general_index_free}
\mathbb{I}=\int \frac{d^4p}{(2\pi)^4} \, \theta(p^0)\sum_\gamma \sum_{\ell} 
|c,\vec p;\spin;\gamma\rangle(s)\times \overset{\leftrightarrow}{D}_s\times
(s)\langle c,\vec p;\spin;\gamma|,
\end{equation}
where $\gamma$ are all the additional indices characterizing the state and $\overset{\leftrightarrow}{D}_s$ is the ``gluing'' operator defined as
\begin{equation}
\label{eq:glueing}
\overset{\leftrightarrow}{D}_s\equiv \frac{1}{(2\ell)!^2}
(\overset{\leftarrow}{\partial}_s^{a_1}\ldots
\overset{\leftarrow}{\partial}_s^{a_{2\ell}})
(\overset{\rightarrow}{\partial}_{s,\,a_1}\ldots
\overset{\rightarrow}{\partial}_{s,\,a_{2\ell}}).
\end{equation}
It simply contracts all the Little group indices of two states. The Clebsch-Gordon coefficient \eqref{eq:CG_repeated} then becomes
\begin{equation}
\label{eq:CG_repeated_if}
C_\ell(s_1,s_2,s)\equiv\big((s_1)\<m_1,\vec p_1;\ell_1|\otimes(s_2)\<m_2,\vec p_2;\ell_2|\big)
|c,\vec p,\spin\>(s_3).
\end{equation} 
One can explicitly construct tensor structures for \eqref{eq:CG_repeated_if}. The partial amplitudes is given then by gluing left- and right-hand sides of the amplitudes after the injection of the identity \eqref{eq:2PS_decomposition_general_index_free}, namely
\begin{equation}
\label{eq:gluing}
C_\ell(s_1,s_2,s)\times \overset{\leftrightarrow}{D}_s\times C_\ell(s_3,s_4,s).
\end{equation}
Here we keep the expressions slightly schematic by dropping the dependence of the 4-momenta and  focusing only on the spin dependence.
The resulting expression \eqref{eq:gluing} should encode the Wigner d-matrix. For instance we have explicitly checked that for the scalar particles (when there is no dependence on $s_1$, $s_2$, $s_3$ and $s_4$) the expression \eqref{eq:gluing} is proportional to the Legendre polynomial $P_\ell$. For a similar discussion see \cite{Jiang:2020rwz}.

\bibliographystyle{JHEP}
\bibliography{refs}

\end{document}